%% file: dmondal_PhD_Thesis.tex
\documentclass[12pt, 
english, 
onehalfspacing, 
liststotoc, 
headsepline, 
]{MastersDoctoralThesis} 

\usepackage[utf8]{inputenc} 
\usepackage[T1]{fontenc} 
\usepackage{mathpazo} 

\usepackage[autostyle=true]{csquotes} 
\usepackage{amssymb}
\usepackage{amsmath}
\usepackage{graphicx}
\usepackage{subfigure}
\usepackage[backend=bibtex,style=nature,sorting=nyt,natbib=true]{biblatex}
\usepackage{hyperref}
\usepackage{float}
\usepackage{enumitem}

\thesistitle{\huge \textsc{Chaotic Propellers of Barred Galaxies and Central Explosions}} 
\supervisor{\Large \textsc{Prof. (Dr.) Tanuka Chattopadhyay}} 
\degree{Doctor of Philosophy (Science)} 
\author{\Large \href{https://sites.google.com/view/debasishmondalastrophysics}{\textsc{Debasish Mondal}}} 

\subject{\textsc{Applied Mathematics}} 
\university{\href{https://www.caluniv.ac.in}{University of Calcutta}} 
\department{\href{https://www.caluniv.ac.in/academic/Apl-Math.html}{Department of Applied Mathematics}} 
\faculty{Faculty of Science} 

\AtBeginDocument{
\hypersetup{pdftitle=\ttitle} 
\hypersetup{pdfauthor=\authorname} 
\hypersetup{pdfkeywords=\keywordnames} 
}

\begin{document}
\frontmatter 
\pagestyle{plain}


\begin{titlepage}
\begin{center}

\HRule \\[0.2cm] 
{\Large \bfseries \ttitle\par}\vspace{0.2cm} 
\HRule \\[0.2cm] 
\vspace{0.3cm}
  
\Large \textit{Thesis Submitted to University of Calcutta\\ for the Degree of \degreename\\ in Applied Mathematics} 
\vspace{0.3cm}
 
\emph{\Large \textsc{By}}\\
\Large \textsc{\authorname} 
\vspace{0.2cm}

\emph{\Large \textsc{Supervisor:}}\\
\Large \href{https://www.researchgate.net/profile/Tanuka-Chattopadhyay}{\supname} 
\vspace{0.1cm}

\begin{figure}[H]	
\centering
\includegraphics[width=0.4\columnwidth,height=0.4\columnwidth]{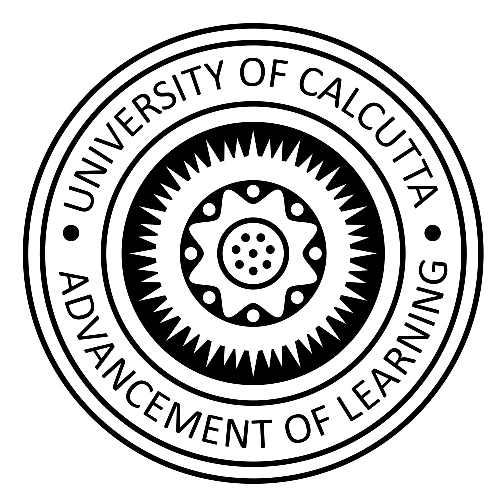}
\label{fig:logo}
\end{figure} 

\Large \textsc{\deptname\\ \univname}\\
\vspace{1cm}

{\Large 2023}\\[4cm] 
\vfill
\end{center}
\end{titlepage}


\dedicatory{\LARGE Dedicated to my beloved family and my Ph.D. Supervisor late Prof. (Dr.) Tanuka Chattopadhyay \ldots} 


\begin{acknowledgements}
\addchaptertocentry{\acknowledgementname} 
First of all, I would like to convey sincere gratitude and respect to my Ph.D. supervisor, Prof. Tanuka Chattopadhyay (Department of Applied Mathematics, University of Calcutta), for her unprecedented guidance throughout my Ph.D. journey, without which my research work would not have been completed. I am fortunate to have a supervisor like her who constantly encouraged me with exciting research ideas and gave me the confidence and liberty to deal with them. Her careful analyses and valuable suggestions on various aspects of research problems improved the quality of my research work to a great extent. Since my M.Sc. days, she has always stood behind me like a gigantic epitome of knowledge. Unfortunately, she died on 16th October 2023, much before my Ph.D. thesis defence, but her valuable advice is still guiding me in my post-Ph.D. research life.  

I wholeheartedly thank my M.Sc. teacher and our present HOD, Prof. Samiran Ghosh (Department of Applied Mathematics, University of Calcutta), for his sincere guidance and constant motivation since my M.Sc. days. I also express my thanks to our ex-Vice Chancellor, Prof. Asis Kumar Chattopadhyay (Department of Statistics, University of Calcutta), our ex-HODs Prof. Suma Debsarma and Prof. Debasis Sarkar (both from Department of Applied Mathematics, University of Calcutta), for their cordial cooperation for both academic and administrative purposes at various stages of my Ph.D. tenure. I am also thankful to all the other academic and non-academic staff of the Department of Applied Mathematics, University of Calcutta, for their sincere cooperation in every possible way.

I want to convey my thanks to Prof. Gulab Chand Dewangan (IUCAA, Pune) for selecting me for the IUCAA-NCRA graduate school (January 2018 to July 2018), which helped me to complete my Ph.D. course work. In this regard, I would also like to thank Prof. Kanak Saha (IUCAA, Pune) for guiding me on my Ph.D. course work project during this visit to IUCAA, Pune.

My earnest thanks are extended to Dr. Sukanta Das and Dr. Pradip Karmakar, two senior researchers of our department, for their warm support and encouragement in all aspects of my research. I cheerfully acknowledge my fellow labmates, especially Mousumi Di, Mrinal Da, Kalyanashis Da, Sumana Di, Avipsita Di, Sourav Da, Aktar Da, Aman Da, Pallabi Di, Ashok Da, Jyotirmoy Da, Debkumar, Bivas, Suman, Suparna, Akash, Susmita, Selina and Jayita for making my Ph.D. journey a memorable one for a lifetime. Moreover, I would also like to acknowledge my friends Abhishek, Prakash, Shreejit, Ashish, Uday (all from IUCAA Pune), Anshuman (Tezpur University), Silpa (NCRA Pune), Priyanka (SRTM University Nanded), Richa (St. Joseph's College Bengaluru), late Mayur (SSGM College Kopargaon), Anupam (IIEST Shibpur) and my M.Sc. classmates Arnab, Sayani, Sudeshna, Avijit and Nibedita for playing essential parts in my journey from M.Sc. to Ph.D. 

I sincerely thank the University Grant Commission, Government of India and the Department of Higher Education, Government of West Bengal, for providing me with Junior and Senior Research Fellowships at various stages of my Ph.D.

Finally, I would like to acknowledge the support from my family members (my father, mother and brother), which motivated me to choose a research career. They are my biggest sources of inspiration. Without their unwavering love and support, I would not have been able to progress to where I am now in my career.

\begin{flushright}
Debasish Mondal\\
Department of Applied Mathematics\\
University of Calcutta, India\\
\end{flushright}

\vspace{13cm}
\hrulefill
\end{acknowledgements}
\cleardoublepage



\section*{}
\addchaptertocentry{Abstract}
\begin{center}
\huge \textit{Abstract}
\end{center}
The formation and evolution of galaxies are complex dynamical processes. Many theories, like monolithic collapse, hierarchical fragmentation, etc., have been developed to describe galaxy formation mechanisms across galactic morphologies. However, each has some limitations, so more is needed to describe the formation mechanism fully. Although these theories provide a sufficient understanding of elliptical galaxies' formation and evolution mechanisms, more is needed for disc galaxies like spirals and lenticulars. In this regard, one important aspect is the formation of spiral arms inside the disc region. The central theme of this thesis work is to explore the possibilities of spiral arm formations from instabilities formed inside the central region. These instabilities originate from the central baryonic feedback and have many prospects regarding the evolution of disc galaxies. They can trigger the gravitational collapse inside the dense molecular clouds that lead to the formation of stars under suitable astrophysical circumstances. In the present work, the role of parameters like the molecular cloud's magnetic field, rotation, etc., has been investigated behind this explosion-triggered star formation process with the help of Jeans instability analysis. From this study, our essential observation is that the formation of star clusters is favoured by a strong magnetic field ($\sim 10 \; \mu$G), and the effect is enhanced at a more considerable distance from the centre. Again, this instability also contributes to the formation of stellar bars. This dense rotating component may drive out the chaotic stellar orbits from the disc through its two ends like a propeller. This has been modelled in the thesis work from the viewpoint of chaotic scattering in open Hamiltonian systems. From this analysis, we conclude that this bar-driven chaotic motion (or simply escaping motion) may lead to forming spiral arms or inner disc rings, depending on the bar strength. Our study also found that, compared to NFW dark haloes, the oblate dark haloes offer a more cohesive evolutionary framework for generating bar-driven escape structures in giant spiral and dwarf galaxies. Moreover, the formation of spiral arms via bar-driven escaping motion is only encouraged in galaxies with NFW dark haloes if they have highly energetic centres, like active galaxies.

\vspace{1.5cm}
\hrulefill
\cleardoublepage

\section*{}
\addchaptertocentry{List of Publications}
\begin{center}
\Large \textbf{List of Publications}
\end{center}
\begin{center}
\large \textit{Ph.D. Thesis Related Publications}
\end{center}
\begin{enumerate} 
\item \textit{Star formation under explosion mechanism in a magnetized medium}, \textbf{Mondal, D.}, Chattopadhyay, T., 2019, \textit{Bulgarian Astronomical Journal}, 31, 16–29, \href{https://astro.bas.bg/AIJ/issues/n31/DMondal.pdf}{https://astro.bas.bg/AIJ/issues/n31/DMondal.pdf}

\item \textit{Role of galactic bars in the formation of spiral arms: a study through orbital and escape dynamics—I}, \textbf{Mondal, D.}, Chattopadhyay, T., 2021, \textit{Celestial Mechanics and Dynamical Astronomy}, 133(9), 43:1–29, \href{https://doi.org/10.1007/s10569-021-10037-5}{https://doi.org/10.1007\\/s10569-021-10037-5}

\item \textit{Effect of dark matter haloes on the orbital and escape dynamics of barred galaxies}, \textbf{Mondal, D.}, Chattopadhyay, T., 2023, \textit{The European Physical Journal Plus}, 138(12), 1144:1–16, \href{https://doi.org/10.1140/epjp/s13360-023-04715-6}{https://doi.org/10.1140/epjp/s13360-023-04715-6}

\item \textit{Orbital and escape dynamics in double-barred galaxies}, \textbf{Mondal, D.}, Chattopadhyay, T., 2023, \textit{Monthly Notices of the Royal Astronomical Society}, To be Submitted
\end{enumerate}

\begin{center}
\large \textit{Other Publications}
\end{center}
\begin{enumerate}
\item \textit{Fate of escaping orbits in barred galaxies}, \textbf{Mondal, D.}, Chattopadhyay, T., 2020, \textit{Proceedings of the International Astronomical Union}, 16(S362), 122–127, \href{https://doi.org/10.1017/S1743921322001338}{https://doi.org/10.1017/S1743921322001338}
\end{enumerate}

\vspace{7.5cm}
\hrulefill


\tableofcontents 

\listoffigures 

\listoftables 

\mainmatter 

\pagestyle{thesis} 


\include{Chapter_1}
\include{Chapter_2}
\include{Chapter_3}
\include{Chapter_4}
\include{Chapter_5}
\include{Chapter_6}
\include{Chapter_7}





\printbibliography[heading=bibintoc]
\end{document}

%% file: Chapter_1.tex
\chapter{Introduction}
\label{chap:1}

\section{History of Astronomy}
\label{sec:1.1}
From early civilizations to present-day modern societies, humans have always been inquisitive about the sightings of astrophysical objects in the sky. These objects have fascinated them for centuries and ignited the pursuit of knowledge about their paths, properties, and working mechanisms. Incidents of many ancient astrophysical objects have been recorded in ancient cave paintings, literature, and other historical artefacts. A few such notable examples include the sighting of supernova HB9 in the Burzahama region of Kashmir on dates between 4100 BC and 2100 BC \cite{Joglekar2011}; the observation of supernova SN 1054 (Crab Nebula) in 1054 AD \cite{Stephenson2003}. In pre-medieval times, all such astrophysical objects were considered stars and were part of our Milky Way galaxy. As time progressed, people started to acquire a better understanding of their properties and were able to classify them according to their motion and appearance. Philosopher Immanuel Kant first proposed the concept of the island universe in 1755, regarding observing nebulae as separate worlds or outer galactic objects similar to our Milky Way. Due to the emergence of modern telescopes, state-of-the-art sophisticated astronomical instruments, large-scale computational facilities, etc., the last century of the second millennium saw remarkable growth in this field. Thus, many outer galactic or extragalactic astrophysical objects come into the picture, which enriches the field of astronomy and astrophysics in many scientific aspects.

\section{Overview of Galaxies}
\label{sec:1.2}
The majority of objects that are seen in the sky are galaxies, like our Milky Way - the home of our solar system. Galaxies are vast gravitationally bound systems composed of stars, molecular gas clouds, dust particles, and unseen dark matter particles pervaded by the magnetic field, cosmic rays, etc. They are the universe's building blocks and evolve under gravity's influence. Stars, molecular gas clouds, and dust are its components visible in electromagnetic bands. However, the dark matter particles are not visible as they do not appear to interact with the electromagnetic radiation. In galaxies, stars form from molecular gas clouds, and when they die, they enrich the interstellar environment with chemical elements such as hydrogen, helium, and other heavy metals. 

Edwin Powell Hubble first classified the galaxies in 1926 \cite{Hubble1926} based on their morphological structures. Hubble's classification scheme only considers the prominent galactic structural features like bulges, bars, and spiral arms. According to this scheme, galaxies are mainly classified into the following categories: elliptical, lenticular, spiral and irregular. The famous Hubble's tuning fork diagram represents this entire classification framework. There are two most important things about this classification scheme. First, this is only valid for the galaxies in the local universe. Far-distant galaxies do not fall under this classification scheme. Second, this is not a galaxy evolutionary diagram from left to right; instead, this only shows the morphological distinction.
\begin{figure}[H]	
\centering
\includegraphics[width=0.6\columnwidth,height=0.4\columnwidth]{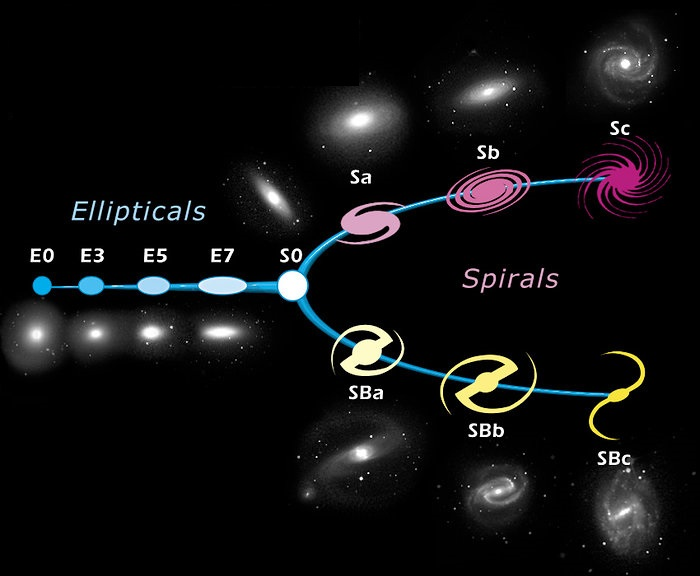}
\caption{The Hubble's tuning fork diagram. \\\textit{Image Credit: European Space Agency (ESA)}}
\label{fig:1.1}
\end{figure} 

\noindent Later, Gérard de Vaucouleurs in 1959 \cite{De1959} provided a more accurate extension of Hubble's classification scheme by classifying spiral galaxies into three types rather than two of earlier Hubble's scheme based on morphological features like bar, rings and spiral arms.
\begin{figure}[H]	
\centering
\includegraphics[width=0.6\columnwidth,height=0.4\columnwidth]{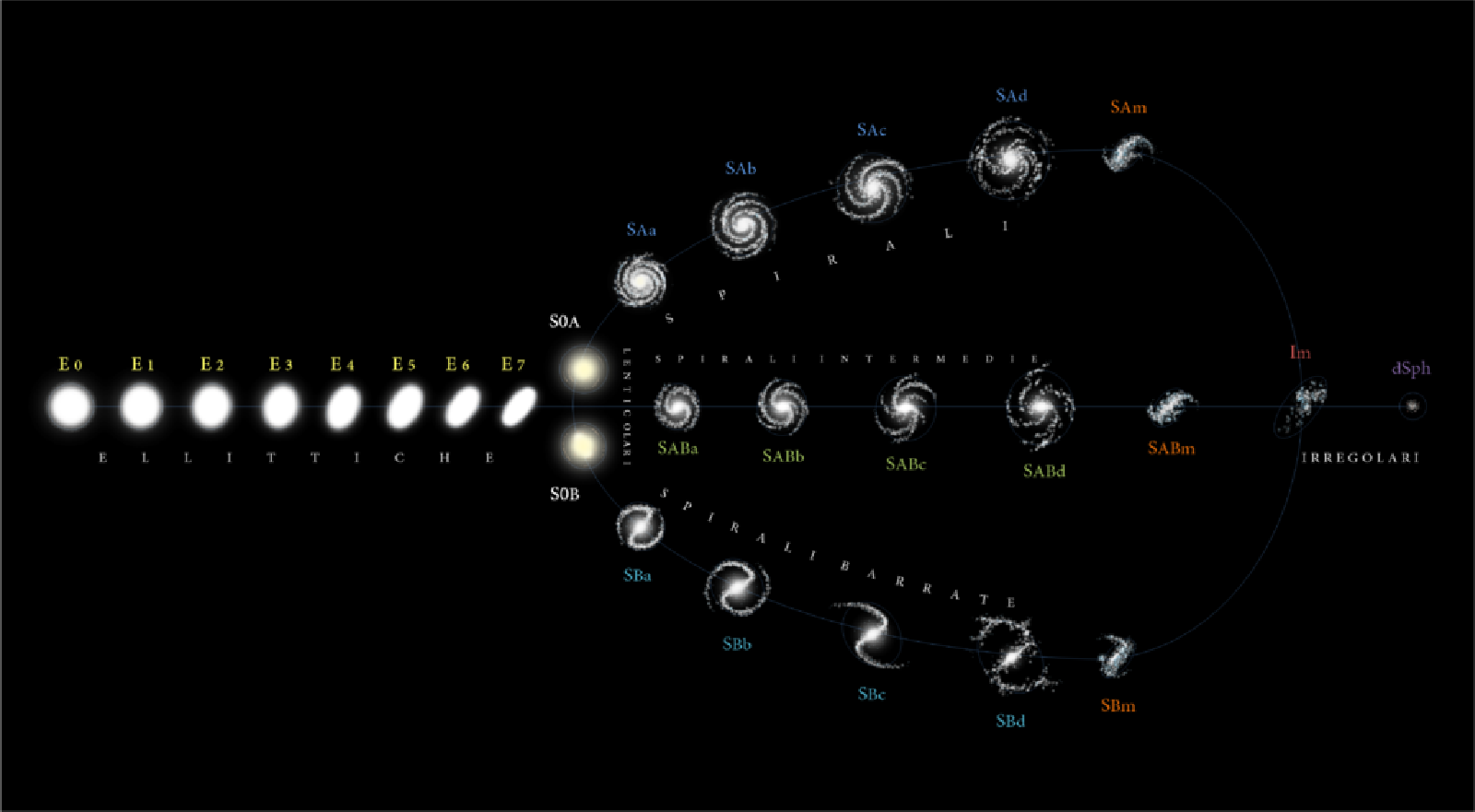}
\caption{The Hubble – de Vaucouleurs diagram. \\\textit{Image Credit: Antonio Ciccolella / M. De Leo}}
\label{fig:1.2}
\end{figure}

\begin{itemize}[leftmargin=*]
\item \textbf{Elliptical galaxies:} The elliptical galaxies have ellipsoidal structures and lack other morphological features like discs, spiral arms, rings, etc. In Hubble's classification scheme, elliptical galaxies are identified by $E0$ to $E7$ depending on their degree of flattening.
\begin{equation*}
\text{degree of flattening} = 10 \times (1 - \frac{b}{a}),
\end{equation*}
where $a$ and $b$ are the lengths of the semi-major and semi-minor axes of the galaxy's isophotes, respectively. $E0$ ellipticals are roughly spheroidal and $E7$ ellipticals are the most flattened ones. About 10-15\% of the galaxies in the local universe are elliptical.

Elliptical galaxies are 'early-type' galaxies. They experience very little ongoing star formation activity due to the presence of a sparse interstellar medium. That is why old red stars dominate them and contain mostly globular clusters. Their dynamical characteristics resemble the bulges of disc galaxies. The orbital dynamics of stars inside the ellipticals are dominated by random motion rather than rotation \cite{De1991}. Ellipticals are mainly found in galaxy clusters and galaxy groups. The mass of the ellipticals typically ranges from $10^5 M_\odot$ to $10^7 M_\odot$. Massive ellipticals host supermassive black holes (SMBHs) at their centres (e.g., M87). There is also a particular sub-class of elliptical galaxies known as 'cD galaxies', which are extremely large and only found near the centre of massive galaxy clusters. IC 1101 is an example of a CD galaxy.

\item \textbf{Lenticular galaxies:} In Hubble's classification scheme, lenticular galaxies are identified by SB0 or S0, depending upon the presence or absence of the bar. Lenticular galaxies sometimes have a bar along with a prominent disc and bulge. These galaxies have no spiral arms, which distinguishes them from spiral galaxies. Their bulge-to-disc ratio is much bigger than usual spirals \cite{Laurikainen2005}. The spectral properties and scaling relations of lenticulars and ellipticals are nearly identical. Lenticulars, like ellipticals, are 'early-type' galaxies with low ongoing star formation activity. Also, like ellipticals, lenticulars are dominated by an older population of stars and contain mostly globular clusters, but they have a significant amount of dust in their disc. Giant lenticulars host SMBHs at their centres. NGC 2787 is an example of a barred lenticular, and PGC 2248 (Cartwheel galaxy) is an example of an unbarred lenticular.

\item \textbf{Spiral galaxies:} Spiral galaxies are the most abundant type of galaxy in the Universe. Roughly 60\% of all galaxies are spirals. They are distinguished from others by the spiral arms visible outside the disc \cite{Sellwood2022}. Like lenticulars, spirals have a prominent disc and bulge, mostly along with a bar. The bulge-to-disc ratio of spirals is less than that of lenticulars. Spiral galaxies are 'late-type' galaxies and experience significant ongoing star formation activity. Their spiral arms are the primary source of the ongoing star formation process. Spirals are mostly field galaxies and are rarely found in galaxy clusters. The orbital dynamics of stars inside the spirals are dominated by rotation rather than random motion. The mass of the spirals typically ranges from $10^9 M_\odot$ to $10^{12} M_\odot$. Giant spirals host SMBHs at their centres. Hubble's classification scheme divides spirals into two broad categories: barred and unbarred. 

\begin{enumerate}[label=(\roman*)]
\item \textbf{Barred spiral:} Most spiral galaxies have an elongated stellar structure at the centre known as the bar. In Hubble's classification scheme, barred spirals are identified by either SBa, SBb, or SBc. Barred spirals are classified according to the tightness of spiral arms and bulge prominence \cite{Savchenko2013, Masters2019}. SBa galaxies are early-type spirals with the most prominent bulge and tightly wrapped spiral arms. SBc galaxies are late-type spirals with a less prominent bulge and loosely wrapped spiral arms. NGC 1300, NGC 1345, Milky Way, etc. \cite{Helou1991} are some examples of barred spirals.

\item \textbf{Unbarred spiral:} In Hubble's classification scheme, unbarred spirals are identified by either Sa, Sb, or Sc. The classification of unbarred spirals is similar to that of barred spirals, i.e., according to the pitch angle of spiral arms and bulge prominence. Some examples of unbarred spirals are M33 (Triangulum), NGC 300, NGC 4414, etc. \cite{Helou1991}.
\end{enumerate}
	
\indent According to the Hubble–de Vaucouleurs diagram, there is also an additional class of spiral galaxies: intermediate spirals. They are mainly weakly barred systems and are identified by either SABa, SABb, or SABc. The classification scheme of intermediate spirals is precisely the same as that of barred or unbarred spirals. M65, M106, NGC 4725, etc., are some examples of intermediate spirals. In the same manner, SAB0 represents the intermediate lenticulars. NGC 1387 is an example of an intermediate lenticular.

\item \textbf{Irregular galaxies:} In Hubble's classification, galaxies that do not fit into any classes are known as irregular galaxies. These galaxies mostly have no organised structure, i.e., do not have any particular shape. The mass of irregulars typically ranges from $10^8 M_\odot$ to $10^{10} M_\odot$. Roughly 25\% of galaxies in the local universe are irregular. A few irregular galaxies show some signatures of structures, like Large Magellanic Cloud (LMC), which has an off-centred stellar bar \cite{Zhao2000}, NGC 4625 has only one spiral arm, etc.
\end{itemize}

Our galaxy, the Milky Way, is a member of the galaxy group known as the \textit{Local Group}. Milky Way, Andromeda (M31) and Triangulum (M33) are this group's three most prominent members. The Milky Way is a giant barred spiral with four major and two minor spiral arms. Our solar system lies on the Orion arm at a radius of 8.2 kpc from the Galactic Centre. The stars and gas in the Milky Way rotate differently than a solid body's rotation, known as `differential galactic rotation'. Near the Galactic Centre, galaxies rotate faster than at the outskirts. Furthermore, the spiral arm, disc, bar, and bulge all have individual pattern speeds that differ from this galaxy's differential rotation \cite{Helmi2020}. 

\section{Galaxy Formation Theories}
\label{sec:1.3}
Galaxy formation is a complex dynamic process that has yet to be fully understood. Major theories regarding the galaxy formation process are as follows:

\begin{enumerate}[label=(\roman*)]
\item \textbf{Monolithic collapse model:} The monolithic collapse model for galaxy formation was first proposed by three astrophysicists, Eggen, Lynden-Bell, and Sandage \cite{Eggen1962}. According to their names, this model is also popularly known as the ELS model. As per this model, galaxies were created by the gravitational collapse of a large primordial gas cloud. This suggests that the gas cloud was initially homogeneous and began to collapse due to gravity. It fragmented into smaller, denser regions as it collapsed, eventually forming protogalactic clumps. After that, these clumps continued collapsing and merging to create a single, massive galaxy.  

One of the challenges this model faces is difficulty explaining the diversity of galaxy morphologies observed in the present day universe. This model disagrees with the globular cluster distributions in elliptical galaxies. As per this model, elliptical galaxies should originate from a massive gas cloud at high redshift, and thus, the colour distribution in its globular clusters should be unimodal, which was also supported earlier by many authors \cite{Peebles1969, Calberg1984, Arimoto1987}. However, later, bimodal distributions in metallicity have been found among globular clusters of the majority of the giant elliptical galaxies \cite{Kundu2001, Kundu2007}. Regardless, the monolithic collapse model was one of the simple earlier models of galaxy formation, and later, this has been replaced mainly by hierarchical models that better fit the observational data \cite{Kampakoglou2008}..

\item \textbf{Major merger model:} The major merger model is one of the leading theories for galaxy formation. This suggests that, galaxies are formed through the merger of smaller gas-rich galaxies. When two galaxies merge, their gas clouds collide and are compressed, which triggers the star formation process. The energy released causes a burst of star formation, and the merging galaxy becomes more massive and has a higher stellar concentration than either of the original galaxies \cite{Ashman1992, Zepf2000}. This also suggests that many of the massive galaxies that are seen today, including ellipticals, were formed through a series of mergers between smaller galaxies. In this scenario, then younger metal-rich globular clusters of elliptical galaxies should originate from the shock-compressed layer of dense molecular clouds inside the discs, and older metal-poor globular clusters should come from the haloes of the merging galaxies. Thus, this has well agreement with the bimodal metallicity distribution in elliptical galaxies \cite{Bekki2002}. 

Despite the success, this model has some limitations. For example, the kinematical properties of the merger-driven globular clusters should depend on the orbital configuration of the merging galaxies. Because in merging galaxies, the metal-rich globular clusters are usually located in the inner region and the metal-poor ones in the outer region. So, the metal-rich globular clusters are speculated to be produced in the merger of gaseous discs, and a strong correlation is expected between their mean metallicity and specific orbital frequency. However, in practice, no such correlation exists. Also, in many elliptical galaxies, higher mean metallicity does not reflect the higher proportion of metal-rich globular clusters \cite{Cote1998}. Besides that, it does not account for the formation of spiral galaxies, which have distinct features and are thought to be more abundant than elliptical galaxies in the local universe. Despite these drawbacks, this model provides a valuable foundation for understanding the galaxy formation process.

\item \textbf{Multiphase dissipational collapse model:} The idea of multiphase dissipational collapse in the context of the galaxy formation framework was first proposed by \citeauthor{Forbes1997}, \citeyear{Forbes1997} \cite{Forbes1997}. This attempts to explain how galaxies form from the initial distribution of matter in the universe. The model proposes that galaxies were formed through the process of gravitational collapse of gas clouds, which are triggered by a variety of astrophysical processes, such as shocks from supernova explosions, stellar radiation, and turbulence.

This model suggests that the globular clusters have been formed in distinct star formation episodes via dissipational collapse, and tidal stripping of globular clusters from satellite galaxies is present. As a result, blue metal-poor globular clusters will form in the initial stage, and red metal-rich globular clusters will form later. This supports the bimodality in metallicity. Blue accreted globular clusters have no rotation, but red globular clusters show rotation depending on the degree of dissipation \cite{Cote1998}. In this respect, the modelling of galaxy formation theory still needs to be improved, and studies of its various constituents like stellar populations (primarily globular clusters), their star formation history, and bulges in greater detail can throw light only on their formation theory. Overall, this model can reproduce many of the observed properties of galaxies, such as their size, mass, and star formation rates. However, the model requires the inclusion of supplementary physical processes, such as feedback from supernova explosions and the effects of magnetic fields, to fully explain the observed properties of galaxies.
 
\item \textbf{Dissipation-less or dry merger model:} The dissipation-less merger model is another galaxy formation framework that seeks to explain how galaxies, particularly massive elliptical galaxies, formed through the merging of smaller galaxies without gas dissipation. According to this model, galaxies are thought to form through a series of mergers between smaller galaxies, mainly composed of dark matter and stars. The dark matter in the merging galaxies interacts gravitationally, causing them to merge and eventually into a single, more giant galaxy. Because there is no dissipation of gas, the newly formed galaxy is essentially made up of stars and dark matter from the merging galaxies. In this model, the key to forming massive elliptical galaxies is the frequent merging of smaller galaxies. As galaxies merge and grow in size, they become more massive and can attract even more galaxies, leading to a self-reinforcing cycle of mergers and growth \cite{Boylan2005, Lackner2010}.

One challenge for the dissipation-less merger model is explaining the observed properties of massive elliptical galaxies, such as their high-velocity dispersions and old stellar populations. Recent simulations and observations have suggested that some gas dissipation may be necessary to explain the formation of the central regions of these galaxies \cite{Ji2023}. However, the overall mass assembly through merging may still be the dominant process. Regardless of this challenge, this is another essential galaxy formation framework, particularly in massive elliptical galaxies. However, ongoing observations and simulations are necessary to refine and test the model and explore gas dissipation's role in galaxy formation.

\item \textbf{Accretion and in-situ hierarchical merger model:} The accretion and in-situ hierarchical merger model is one of the most widely accepted theories of galaxy formation. This suggests that galaxies are formed through the accumulation of smaller structures, such as gas clouds and smaller galaxies, and the mergers of galaxies of similar size. The accretion component of the model refers to the idea that galaxies grow by the accretion of gas from their surrounding environment. The in-situ component of the model refers to the formation of stars within the galaxy itself. The hierarchical merger component of the model suggests that galaxies grow through the merger of smaller galaxies of similar size. When galaxies come into proximity, their gravitational forces can cause them to merge, creating a more giant, more massive galaxy. The occurrence of this process multiple times results in the formation of giant galaxies \cite{Lackner2012, Gu2016}. This model provides a comprehensive galaxy formation framework over cosmic time and is well supported by observations from several astronomical surveys and numerical simulations. It continues to be an active area of research in galactic astrophysics.
\end{enumerate}  
 
In summary, no single theory is perfect for explaining the galaxy formation framework across all galactic morphologies. From all these theories, the astrophysical community has a sufficient understanding of the galaxy formation mechanism, at least in early-type galaxies like ellipticals. However, that is different for late-type galaxies like spirals. More comprehensive frameworks are required to explain them.

\section{Order and Chaos in Galaxies}
\label{sec:1.4}
The formation and evolution scenario in the form of the merger or hierarchical models needs to be better-sufficient in rotation-dominated galaxies like spirals and lenticulars, which are disc-supported systems. Structures like the stellar bar, spiral arms, dark matter haloes, etc., significantly influence the formation and evolution of these galaxies. Unlike ellipticals, which are primarily random motion-dominated systems, these galaxies can exhibit both ordered and chaotic motions in various ways. The ordered motion is characterised by the rotation of stars and gas in a well-defined direction around the galactic centre. This rotation is due to the gravitational attraction of the matter in the galaxy, and this motion of stars and gas is generally stable and predictable. On the other hand, the chaotic motion is characterised by the random motions of individual stars and gas clouds within the galaxy. The gravitational interactions between stars and gas clouds and other factors, such as supernova explosions and the presence of dark matter, cause this type of motion. The ordered motion provides a sense of overall structure to the galaxy. In contrast, chaotic motion contributes to forming new structures and the distribution of matter within the galaxy over time. Chaotic motions can disrupt the ordered motion and cause the formation of structures such as bars, rings and spiral arms. Thus, both the ordered and chaotic motions are essential regarding the evolution of disc galaxies \cite{Contopoulos2002p, Vandervoort2011, Vandervoort2018}.

\subsection{Importance of stellar bar}
\label{sec:1.4.1} 
One of the significant implications of chaotic motion in the ordered and chaotic dynamics of disc galaxies is the formation of a stellar bar. This bar structure is an elongated thin structure made of stars that extends from the central regions of disc galaxies. They are thought to result from gravitational instabilities produced in the rotational disc. These instabilities result in the formation of stars in the central region, which move in non-circular orbits. Bar can create regions of high gas density due to stronger gravitational forces that pull more stars into the central region, which can trigger the formation of new stars. The stellar bar plays an essential role in the evolutionary dynamics of galaxies. In the long-term evolution, this robust feature can affect how stars and gas are distributed within a galaxy, which can influence the rate at which new stars are formed and the overall shape of the galaxy. So, understanding bar properties and dynamics can help us to understand the evolution of disc galaxies better as a whole \cite{Bournaud2002, Vera2016}.

The occurrence of stellar bars among disc galaxies is quite common. However, not all disc galaxies clench the bar structure. Nearly $\frac{2}{3}$rd of the disc galaxies (spirals and lenticulars) in the local universe have a stellar bar feature in their central region \cite{Eskridge2000}. Besides that, some irregular galaxies like LMC possess an off-centred stellar bar \cite{Monteagudo2018}. Also, many galactic discs have a secondary bar feature that has been detected in the recent past due to the emergence of state-of-the-art astronomical imaging facilities \cite{Friedli1993, Shaw1993}.
\begin{figure}[H]	
\centering
\includegraphics[width=0.6\columnwidth]{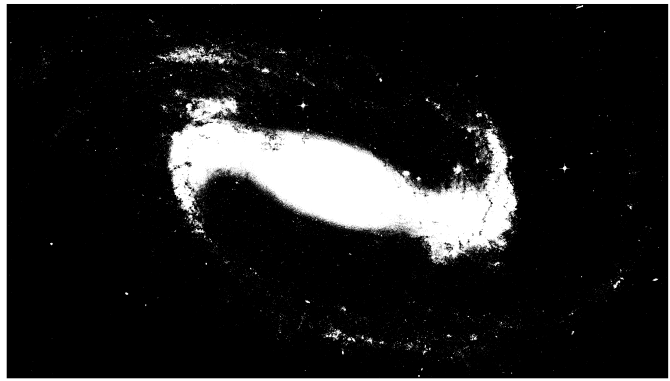}
\caption{NGC 1300 -- a grand design barred spiral galaxy.\\ \textit{Image Credit: Hubble Space Telescope (HST)}}
\label{fig:1.3}
\end{figure}

\subsection{Dynamical astronomy as a tool}
\label{sec:1.4.2} 
Dynamic astronomy can be an essential modelling tool for studying the stellar orbital dynamics in galaxies. Dynamical astronomy is a branch of astronomy that studies the motion of celestial objects, such as stars, planets, moons, asteroids, and comets, under the influence of gravitational and other forces. It involves using mathematical and computational techniques to model and analyse the complex orbital dynamics of these objects and the interactions between them. These are important for various reasons, including predicting the future positions and movements of celestial objects, determining the shape and orientation of their orbits, and understanding the formation and evolution of planetary systems. Some key concepts and techniques used in dynamical astronomy include celestial mechanics, numerical integration, perturbation theory, and the use of computer simulations \cite{Contopoulos2002, Zhang2023}.

\indent The effect of stellar bars in the orbital dynamics of disc galaxies can be studied from the viewpoint of ordered and chaotic motions in the phase space of Hamiltonian systems \cite{Aguirre2001, Aguirre2003}. In this regard, modelling can be done using the concepts used in dynamical astronomy \cite{Jung2016, Zotos2018}. In the upcoming chapters, detailed discussions have been provided about several bar dynamics-related problems that were solved from the perspective of dynamical astronomy.

\section{Thesis Motivation and Outline}
\label{sec:1.5} 
The research works presented in this thesis are motivated by the various aspects of chaotic motions in the central region of barred galaxies, which are violence-prone (or chaotic motion dominated) due to the baryonic feedback from supernovas, black holes, etc. and leave a substantial impact on the stellar dynamics therein. Baryonic feedback refers to the processes by which the baryons (ordinary matter) affect the galaxy's formation and evolution process \cite{Rovskar2014}. It is a crucial factor in understanding the observed properties of galaxies, such as their morphology, stellar content, and gas content. One of the primary forms of baryonic feedback in galaxies is through supernova explosions. It occurs when massive stars reach the end of their lives and collapse, releasing tremendous energy and ejecting material (in the form of shock waves) into the surrounding interstellar medium. The energy and ejected material from supernovae can heat and compress the surrounding gas, triggering the formation of new stars and regulating the growth of existing ones \cite{Efstathiou2000}. Another form of baryonic feedback is through the accretion of SMBHs at the centres of galaxies. Large amounts of gas are accreted in the process, which can become highly energised and emit radiation observed in quasars or active galactic nuclei. This radiation can ionize and heat the surrounding gas, affecting its ability to cool and form new stars \cite{Silk2010}. Overall, baryonic feedback is essential for understanding the complex interplay between the different components of galaxies and their evolution over time. By incorporating the effect of baryonic feedback into galaxy formation models, astrophysicists can gain insights into the physical processes that govern the properties of galaxies and their evolution in the universe. Now, one of the significant implications of baryonic feedback is the propagation of matter from the centre in an outward manner, which is known as density waves. Shock waves generated from such feedback are the reason behind this. These density waves have many aspects regarding the galaxy's evolution. One such aspect is the process of star formation. Density waves induce instabilities inside the molecular gas clouds in the neighbouring medium. That results in the gravitational collapse of gas clouds and leads to the path of star formation. Another aspect is the process of structure formation. In long-time evolution, the density waves redistribute the trajectory of stellar orbits, leading to a stellar bar formation. So, the insatiabilities or chaotic motions produced in the central disc region have two prospects -- (i) star formation (termed as a small-scale effect) and (ii) structure formation (termed as a large-scale effect).

\subsection{Star formation}
\label{sec:1.5.1}
Baryonic feedback, such as supernova explosions in the central disc region, can trigger the star formation process. When a massive star explodes as a supernova, it releases much energy and material into its surrounding environment. This energy radiates as shock waves and can compress neighbouring gas and dust, creating regions of higher density and pressure that can collapse under their gravity and form new stars. These shock waves can also stir up the gas and dust in a galaxy, increasing its overall star formation rate \cite{Vasiliev2023}. However, it is worth noting that while explosions can contribute to star formation, they are not the only factor determining galaxies' star formation rate. Other factors, such as the density and composition of the interstellar medium, the presence of magnetic fields, turbulence, and the overall gravitational potential of the galaxy, can all play a significant role in determining how many stars a galaxy forms and how quickly it forms them \cite{Byleveld1996, Krause2023}. This resulting system may evolve hierarchically into higher-order stellar systems via coagulation and gas accretion from their ambient medium \cite{Toonen2016}. 

\subsection{Structure formation}
\label{sec:1.5.2}
Instabilities originating from the central explosions evolve outwards from the galactic core as density waves. In due course, these density waves redistribute stellar trajectories near the centre, making their motion chaotic. It leads to the formation of a self-stabilising, non-axisymmetric solid, dense structure known as a stellar bar, which can act as a propeller that helps regulate the chaotic movement of stars from the centre to the outer disc. This chaotic motion may escape from the central region through the bar ends. In this way, the escape of stars results in the formation and evolution of other structures in the galaxy, such as spiral arms, inner disc rings, and star clusters \cite{Regan2004, Jung2016, Sellwood2016}. Now, the dark matter haloes, too, substantially influence this bar-driven stellar escaping phenomena \cite{Lieb2022}. Thus, the chaotic motion (or simply chaos) that originates in the central region is closely associated with the structure formation. In other words, galactic components like spiral arms, inner disc rings, etc., are outcomes of the propulsion of chaos from the galactic centre. 

\indent Next, Chapter \ref{chap:2} gives an overview of the modelling tools used in the entire study. The star formation process due to central explosions has been discussed in greater detail in Chapter \ref{chap:3}. The process of bar-driven structure formation under several astrophysical circumstances and also the role of the dark matter haloes in this process have been discussed in greater detail in Chapters \ref{chap:3} - \ref{chap:6}. Overall conclusions of the thesis work and a future research plan have been provided in Chapter \ref{chap:7}.

%% file: Chapter_2.tex
\chapter{Some Preliminaries}
\label{chap:2}

\section{Jeans Instability}
\label{sec:2.1}
The motion of stars and gas in the central region of disc galaxies is primarily dominated by chaos or instabilities, which mainly originated due to the explosions occurring therein. These instabilities play an essential role in the galaxy's evolution. Inside the rotating disc, these instabilities move radially outward like density waves. In the process, they trigger the formation of stars in the molecular gas clouds of the neighbouring medium. In this wave propagation scenario, the Jeans insatiability is crucial to figuring out how these density waves propagate in the neighbouring medium and under what physical circumstances they trigger the star formation process. 

\indent The Jeans instability is an astrophysical phenomenon that describes the gravitational collapse of gas clouds in space. It is named after the British physicist James Hopwood Jeans, who first derived the mathematical conditions for its occurrence in 1902 \cite{Jeans1902}. The Jeans instability arises when the thermal pressure within a gas cloud is not strong enough to counteract the force of gravity. It can happen when the cloud is sufficiently large and dense and the temperature is sufficiently low. Under these conditions, small density fluctuations within the cloud can grow and amplify, eventually leading to the collapse of the entire cloud into a denser structure, such as a star or a planet \cite{Whitworth1998, Heitsch2005, Longarini2023}. The Jeans instability is essential in many areas of astrophysics, including the formation of stars, galaxies, and other astronomical objects. It is also relevant in studying dark matter and the universe's large-scale structure.

\indent The Jeans parameters like Jeans length and Jeans mass are essential in describing the conditions necessary for the instability to occur. The Jeans mass is derived from the dispersion relation for the Jeans instability analysis of a self-gravitating gas. In the star formation context, due to density waves, one needs to analyse Jeans instability in dense isothermal star-forming gas clouds and derive the dispersion relation for small amplitude density perturbations. Let us begin this study with the help of continuity, Euler, and Poisson's equations, respectively:
\begin{equation}
\label{eq:2.1}
\frac{\partial \rho}{\partial t} + \vec{\nabla}.(\rho \vec{V}) = 0,
\end{equation}

\begin{equation}
\label{eq:2.2}
\frac{\partial \vec{V}}{\partial t} + (\vec{V}.\nabla) \vec{V} = -\frac{1}{\rho} \vec{\nabla} p - \vec{\nabla} \Phi,
\end{equation}

\begin{equation}
\label{eq:2.3}
\nabla^2 \Phi = 4 \pi G \rho,
\end{equation}
 
\noindent where $\rho$ is the density, $\vec{V}$ is the velocity, $p$ is the pressure, and $\Phi$ is the gravitational potential. Now, for an isothermal gas cloud, the pressure is related to the density by 
\begin{equation}
\label{eq:2.4}
p = c_s^2 \rho,
\end{equation}

\noindent where $c_s$ is the isothermal sound speed. Let linearize the Eqs. (\ref{eq:2.1}), (\ref{eq:2.2}), (\ref{eq:2.3}), and (\ref{eq:2.4}) under the assumption of small perturbations, i.e., $\rho = \rho_0 + \delta \rho$, $\vec{V} = \vec{V_0} + \delta \vec{V}$, $p = p_0 + \delta p$, and $\Phi = \Phi_0 + \delta \Phi$, where $\rho_0$, $\vec{V_0}$, and $p_0$ are the background density, velocity, pressure, and gravitational potential, respectively. The perturbed versions of Eqs. (\ref{eq:2.1}), (\ref{eq:2.2}), (\ref{eq:2.3}), and (\ref{eq:2.4}) are as follows, respectively:
\begin{equation}
\label{eq:2.5}
\frac{\partial (\delta \rho)}{\partial t} + \rho_0 \vec{\nabla}.(\delta \vec{V}) = 0,
\end{equation}

\begin{equation}
\label{eq:2.6}
\frac{\partial (\delta \vec{V})}{\partial t} = -\frac{1}{\rho_0} \vec{\nabla} \delta p - \vec{\nabla} \delta \Phi,
\end{equation}

\begin{equation}
\label{eq:2.7}
\nabla^2 \delta \Phi = 4 \pi G \delta \rho,
\end{equation}

\begin{equation}
\label{eq:2.8}
\delta p = c_s^2 \delta \rho.
\end{equation}

\noindent A combination of the above four equations yields,
\begin{equation}
\label{eq:2.9}
\frac{\partial^2 (\delta \rho)}{\partial t^2} = c_s^2 \nabla^2 \delta \rho + 4 \pi G \rho_0 \delta \rho.
\end{equation}

\noindent This is the wave equation for the density perturbations with a wave speed of $c_s$. The corresponding dispersion relation can be evaluated by assuming solutions of the form: $\delta \rho \propto e^{i(\vec{k} \cdot \vec{r} - \omega t)}$, where $k \; (= |\vec{k}|)$ is the wave number, $\vec{r}$ is the position vector, and $\omega$ is the angular frequency. Substituting this into the wave equation Eq. (\ref{eq:2.9}) gives the dispersion relation as,
\begin{equation}
\label{eq:2.10}
\omega^2 = c_s^2 k^2 - 4 \pi G \rho_0.
\end{equation}

\noindent This is the dispersion relation for small amplitude density perturbations in an isothermal gas. It shows that the frequency of the waves is proportional to the wave number, with a proportionality constant of $c_s$. This means shorter wavelength waves have higher frequencies and propagate faster than longer wavelength waves. For the perturbations to grow in amplitude and lead to gravitational collapse, the second term on the right-hand side of the Eq. (\ref{eq:2.10}) must dominate over the first term, i.e., the wave number ($k$) is less than a critical value. This happens when $\omega <0$, and the expression of that critical wave number is given by, 
\begin{equation}
\label{eq:2.11}
k^2 < \frac{4 \pi G \rho_0}{c_s^2} = k_J^2 \;\; \text{(say),}
\end{equation}

\noindent where $k_J$ is the Jeans wave number and $\lambda_J = \frac{2 \pi}{k_J}$ is the corresponding Jeans wavelength. Substituting the above critical value of $k$, i.e., $k_J$ in the expression for $\lambda_J$ gives, 
\begin{equation}
\label{eq:2.12}
\lambda_J = \frac{\sqrt{\pi} c_s}{\sqrt{G \rho_0}}.
\end{equation} 

\noindent Thus, any density fluctuations with wavelength $\lambda > \lambda_J$ will grow exponentially in time and will be unstable. A molecular gas cloud with dimensions greater than $\lambda_J$ will be gravitationally unstable and contract continuously under gravitational pull. The corresponding critical mass, i.e., Jeans mass ($M_J$) for a spherical molecular gas cloud with diameter $\lambda_J$ is given by, 
\begin{equation}
\label{eq:2.13}
M_J = \frac{4}{3} \pi {(\frac{\lambda_J}{2})}^3 \rho_0 = \frac{\pi}{6} \lambda_J^3 \rho_0 = \frac{c_s^3}{6} (\frac{\pi^5}{G^3 \rho_0})^\frac{1}{2} \; (\text{using Eq. (\ref{eq:2.12}})).
\end{equation}

\indent This concept of the Jeans instability is used in the upcoming chapter \ref{chap:3} in the context of the explosion-triggered star formation mechanism in the central region of disc galaxies.

\section{Hamiltonian Systems}
\label{sec:2.2}
Instabilities produced inside the central region of disc galaxies also play an essential role in forming and evolving structures like bars, spiral arms, rings, etc. These structure formations can be modelled from the viewpoint of the chaotic scattering (or escape of chaotic stellar orbits) phenomena in open Hamiltonian systems. Let us discuss the characteristics of Hamiltonian systems, especially open Hamiltonian systems, and also discuss some popular chaos detection methods to identify chaotic motions in the system's phase space.

\indent Hamiltonian systems are a specific class of dynamical systems where dynamical properties are described by a scalar function called Hamiltonian, usually denoted by $H$. This $H$ is the function of generalised coordinates, generalised momentums and time. Let us explore a Hamiltonian system with a degree of freedom $n$. Then, the dimension of the phase space is $2n$. Furthermore, consider that $(q_1(t), q_2(t), ... , q_n(t))$ and $(p_1(t), p_2(t), ... , p_n(t))$ are the generalised coordinates and generalised momentums, respectively. In this setup, the system's equations of motion, or Hamilton's equations of motion, are as follows:
\begin{equation}
\label{eq:2.14}
\begin{array}{lr}
\begin{split}
\dot{q}_i = & \; \frac{\partial}{\partial p_i}H (q_i, p_i, t),\\
\dot{p}_i = & - \frac{\partial}{\partial q_i}H (q_i, p_i, t),
\end{split}
\end{array}\Bigg\} \text{for} \;\; i = 1,2, ... , n
\end{equation}

\noindent and the expression of $H$ is,
\begin{equation}
\label{eq:2.15}
H = \frac{1}{2}(p_1^2 + p_2^2 + ... + p_n^2) + \Phi(q_1, q_2, ... , q_n, p_1, p_2, ... , p_n, t),
\end{equation}
where $\Phi$ is the potential energy of the system; for conservative Hamiltonian systems, $H$ is an integral or constant of motion and equivalent to the system's total energy ($E$); and for dissipative Hamiltonian systems: $H \leq E$. Moreover, suppose $H$ explicitly does not contain the $t$ coordinate, i.e., $H \equiv H (q_1, q_2, ... , q_n, p_1, p_2, ... , p_n)$. In that case, the system is called an autonomous Hamiltonian system, or else it will be called a non-autonomous Hamiltonian system. Some examples of physical systems where Hamiltonian formalism can be applied are simple harmonic motions, planetary motions, galactic motions, etc.

\subsection{Open Hamiltonian systems and chaotic scattering}
\label{sec:2.2.1}
Open Hamiltonian systems refer to physical systems that are not isolated from their environment and are governed by the principles of Hamiltonian mechanics. In contrast to closed or isolated Hamiltonian systems, open Hamiltonian systems can exchange energy with their environment \cite{Sanjuan2003}. The dynamics of open Hamiltonian systems are more complicated and challenging to study than their closed counterparts due to their complex phase space geometry. The dynamical behaviour of open Hamiltonian systems can exhibit a variety of exciting and complex phenomena, such as chaos, turbulence, and non-equilibrium steady states \cite{Nieto2020}.

\indent Regarding the phase space geometry of closed or isolated Hamiltonian systems, orbits are always confined within a region constrained by the total energy. However, for open Hamiltonian systems, the energy shell becomes non-compact for energies above certain escape thresholds. As a result, a part of the orbits explores an infinite part of the position space. This phenomenon is termed as chaotic scattering \cite{Lai1991, Aguirre2003}. In such a scenario, the potential boundaries in the phase space are generally fractals \cite{Aguirre2001}. A fractal is a mathematical idea describing a pattern or object with self-similarity at different scales. Examples of open Hamiltonian systems include the motion of stellar orbits under the gravitational influence of different galactic components. In this case, stellar orbits can escape from the potential interior beyond a certain energy threshold through the exit channels, i.e., chaotic scattering occurs. This phenomenon of chaotic scattering in open Hamiltonian systems has a wide range of applications in various aspects of astrophysics \cite{Zhang2023}. It is essential to understand the dynamics of comets, asteroids, and stellar orbits, which can be scattered chaotically by the gravitational fields of other celestial bodies \cite{Zotos2014}.

\subsection{Chaos detection techniques}
\label{sec:2.2.2} 
Now, the goal is to characterise the notion of chaos in open Hamiltonian systems. In order to distinguish between ordered and chaotic motions in the phase space of a dynamical system, many dynamical indicators are available in the literature, like the maximal Lyapunov exponent (MLE), the fast Lyapunov indicator (FLI), the smaller alignment index (SALI), etc. Among these, MLE and SALI will be used in the upcoming chapters to analyse the phase space properties of barred galaxy models. The analytical descriptions of MLE and SALI are as follows:
 
\begin{itemize}[leftmargin=*]
\item \textbf{Maximal Lyapunov exponent:} The maximal Lyapunov exponent (MLE) is one of the most popular dynamical indicators \cite{Sandri1996} that helps to differentiate ordered and chaotic motions in phase space for systems of any number of degrees of freedom. The central idea of MLE is to track the evolution of the separation rate (either growth or decay) of two infinitesimally close trajectories or orbits (as shown in Fig. \ref{fig:2.1}). To define MLE, let us consider a dynamical system with degree of freedom $n$ ($n$ is a natural number). Now consider two neighbouring orbits starting with their initial separation $\vec{\delta x}(t_0)$ and after time $t$, let $\vec{\delta x}(t)$ be their separation. Then MLE is defined as,
\begin{equation}
\label{eq:2.16}
\text{MLE} = \lim_{t \to \infty} \lim_{{\parallel \delta \vec{x}(t_0) \parallel}_2 \to 0} \frac{1}{t} \ln \frac{{\parallel \delta \vec{x}(t) \parallel}_2}{{\parallel \delta x(t_0) \parallel}_2},
\end{equation}
where ${\parallel \cdot \parallel}_2$ stands for the Euclidean norm of a vector. If $\text{MLE} = 0$, then the orbit is periodic and if $\text{MLE} > 0$, then the orbit is chaotic. MLE is a handy and sufficient tool if one wants only the binary distinction between periodic and chaotic motion for any orbit in the phase space. 
\begin{figure}	
\centering
\includegraphics[width=0.6\columnwidth]{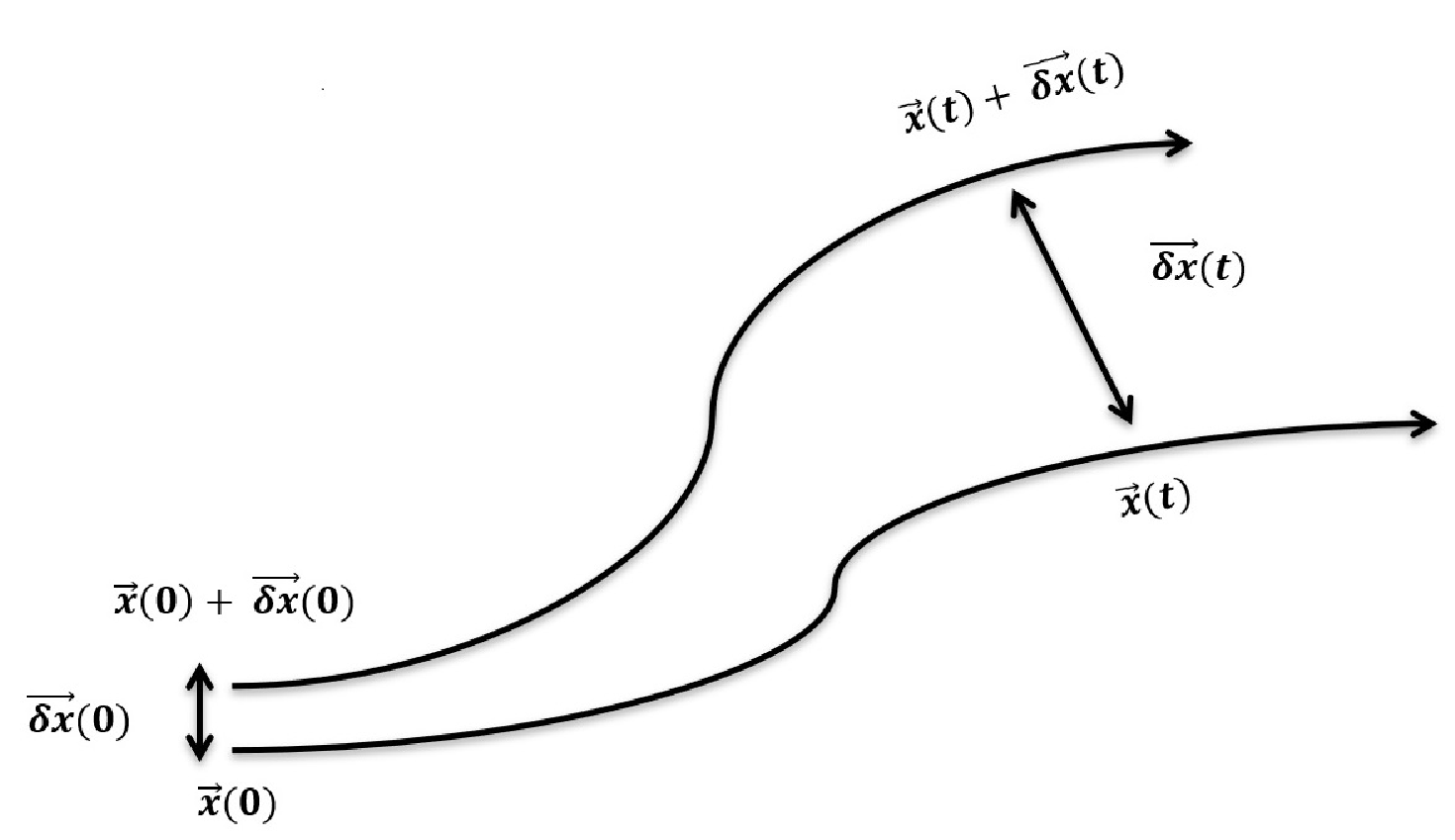}
\caption{Evolution of the separation vector of two neighbouring orbits.}
\label{fig:2.1}
\end{figure}

\item \textbf{Smaller alignment index:} Smaller alignment index (SALI) is an efficient dynamical indicator \cite{Skokos2001} that also helps to distinguish ordered and chaotic motions in the phase space of Hamiltonian flows or symplectic maps. SALI has many advantages over other dynamical indicators. The working algorithm of SALI is quite simple, and it can even figure out every possible tiny chaotic region (weak chaos) of the phase space. It also works much faster than other dynamical indicators. To define SALI, let us consider a conservative Hamiltonian dynamical system with degree of freedom $n$ ($n$ is a natural number). Then, the dimension of its phase space becomes $2n$. Furthermore, consider an orbit with an initial condition vector $\vec{v}(0) \equiv (v_1(0), v_2(0), ... , v_{2n}(0))$. Along with these two initial deviation vectors $\vec{v}^{'}(0) \equiv (dv^{'}_1(0), dv^{'}_2(0), ... , dv^{'}_{2n}(0))$ and $\vec{v}^{''}(0) \equiv (dv^{''}_1(0), dv^{''}_2(0), ... , dv^{''}_{2n}(0))$ are defined. The central idea behind SALI is to track down the evolution of these two deviation vectors. The evolution of the initial condition vector can be tracked down by the system Hamilton's equation of motion:
\begin{equation}
\label{eq:2.17}
\vec{\dot{v}} = J \cdot \Bigg[\frac{\partial H}{\partial v_1} \; \frac{\partial H}{\partial v_2} \; ... \; \frac{\partial H}{\partial v_{2n}}\Bigg]^T,
\end{equation}
where $J = \Bigg[\begin{matrix} 0_n & I_n\\ - I_n & 0_n \end{matrix}\Bigg]$, $I_n$ and $0_n$ are the identity and null matrices or order $n$ respectively. The evolution of the deviation vectors can be tracked down by the variational form of Eq. (\ref{eq:2.17}),

\begin{equation}
\label{eq:2.18}
\vec{\dot{v}} = J \cdot P \cdot [v_1 \; v_2 \; ... \; v_{2n}]^T
\end{equation}
where $P = \bigg(\frac{\partial^2 H}{\partial v_i \partial v_j}\bigg)_{2n \times 2n}; \; i, j = 1, 2, ..., n$. Now, the computation of SALI for the above initial condition and deviation vectors is associated with the following two alignment indices:

\begin{equation}
\label{eq:2.19}
\begin{split}
\text{parallel alignment index} & \rightarrow \text{ALI}_{-}(t) = {\parallel \frac{v^{'}(t)}{\parallel v^{'}(t) \parallel}_2 - \frac{v^{''}(t)}{\parallel v^{''}(t) \parallel}_2 \parallel_2},\\
\text{antiparallel alignment index} & \rightarrow \text{ALI}_{+}(t) = {\parallel \frac{v^{'}(t)}{\parallel v^{'}(t) \parallel}_2 + \frac{v^{''}(t)}{\parallel v^{''}(t) \parallel}_2 \parallel_2}.
\end{split}
\end{equation}

The SALI is defined as: 
\begin{equation}
\label{eq:2.20}
\text{SALI}(t) = min\{\text{ALI}_{-}(t), \text{ALI}_{+}(t)\}
\end{equation}
 
\noindent From this definition, it is evident that if, during the evolution two deviation vectors coincide with each other then $\text{ALI}_{-}(t) = 0$ and $\text{ALI}_{+}(t) \neq 0$, similarly when they are in a straight line but along the opposite direction then $\text{ALI}_{+}(t) = 0$ and $\text{ALI}_{-}(t) \neq 0$. Also, the values of $\text{SALI}(t)$ always lie in the interval $[0, \sqrt{2}]$. Let us discuss how the SALI helps to distinguish between ordered and chaotic motions for Hamiltonian flows with different values of the degree of freedom:
\begin{enumerate}[label=(\roman*)] 
\item $n = 1$: In this case, orbital motions in the phase space are restricted to a 1D torus (as shown in Fig. \ref{fig:2.2}), and both deviation vectors become tangent to the invariant curve (1D torus), i.e., they either coincide or opposite to each other. It means one of the alignment indexes $\to 0$ (smaller alignment index) and another $\to \sqrt{2}$ (larger alignment index). So, eventually, SALI tends to zero. For both the ordered (periodic or quasi-periodic) and chaotic orbits $\text{SALI}(t) \to 0$ but in different time rates, that provides the distinction between these two types of motions.
\begin{figure}
\centering
\includegraphics[width=0.6\columnwidth]{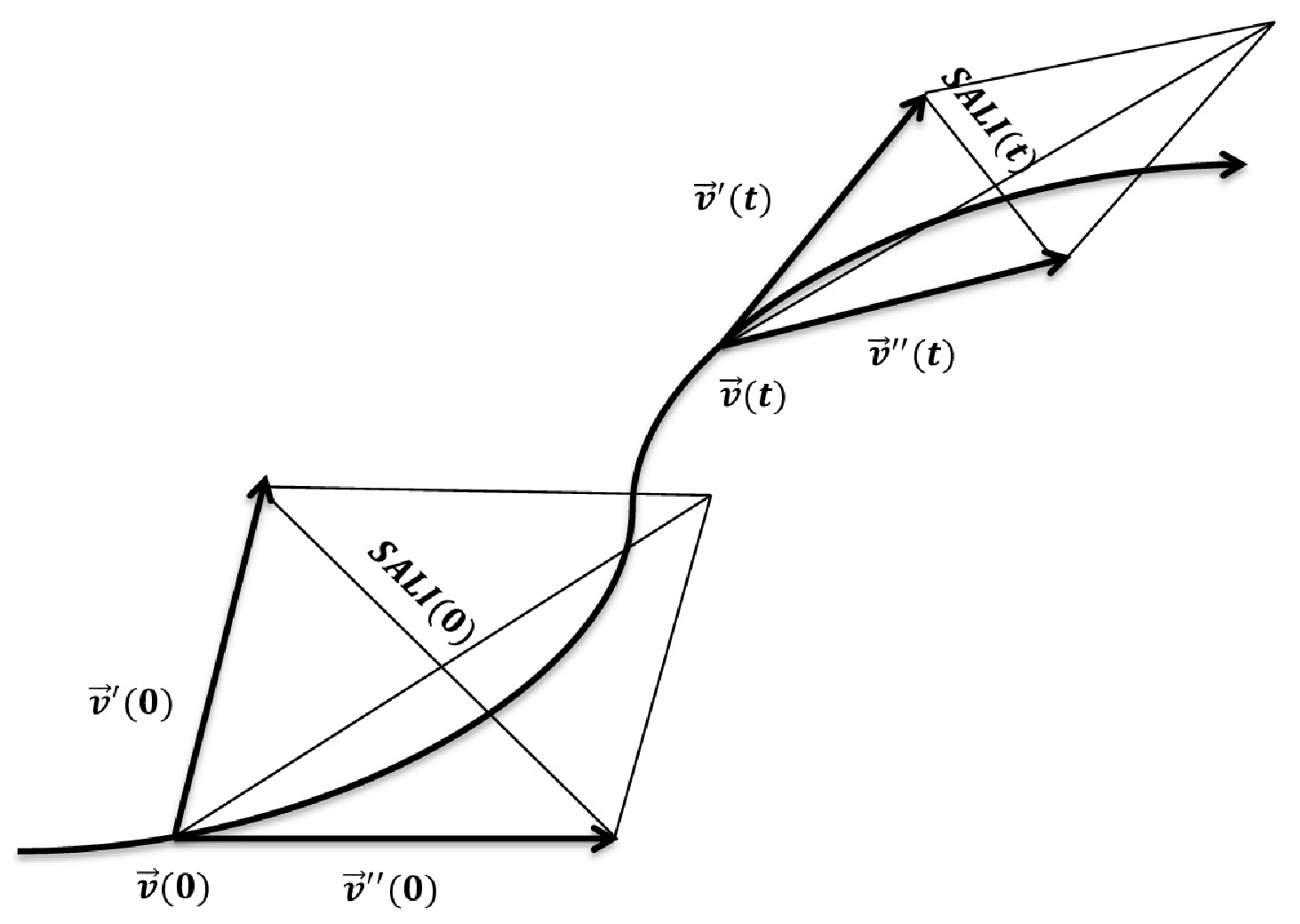}
\caption{Evolution of the deviation vectors on a 1D torus.}
\label{fig:2.2}
\end{figure}
  
\item $n = 2$: In this case, the motion of alignment indices is quite different from $n = 1$. Here, orbital motions in the phase space are restricted to a 2D torus (as shown in Fig. \ref{fig:2.3}). For the evolution of periodic or quasi-periodic orbits, both alignment indices lie on the tangent plane of the 2D torus but do not coincide or are opposite. Thus, both the alignment indices become non-zero for ordered motions. On the other hand, for the evolution of chaotic orbits, both deviation vectors become tangent to the invariant curve (2D torus). Thus, any alignment indices become zero, ultimately $\text{SALI}(t) \to 0$. This is how SALI helps to distinguish between ordered and chaotic motion in the case of a 2D Hamiltonian flow.
\begin{figure} 
\centering
\includegraphics[width=0.6\columnwidth]{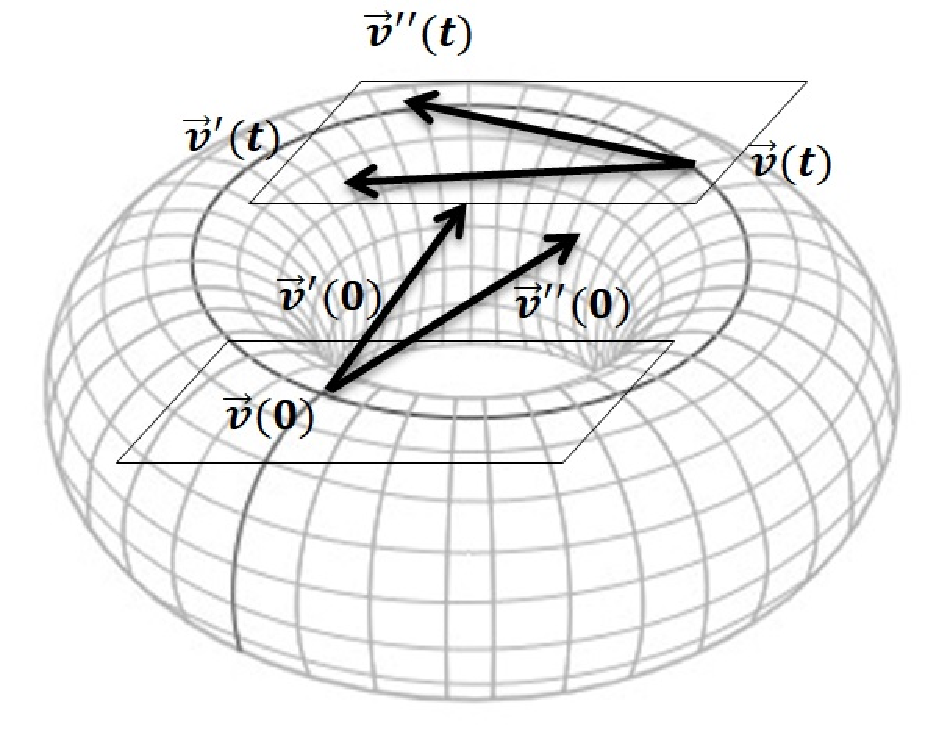}
\caption{Evolution of the deviation vectors on a 2D torus.}
\label{fig:2.3}
\end{figure}

\item $n > 2$: In this case, the distinction between ordered and chaotic motions with the help of SALI is just a higher-dimensional extension of the concepts discussed above for $n = 2$. In practice, the SALI is a handy and efficient tool if one wants more characterization for any orbit in phase space other than only the notions of periodic and chaotic motion.
\end{enumerate}
\end{itemize} 

\indent The above concepts about Hamiltonian systems are used in the upcoming Chapters \ref{chap:3} - \ref{chap:6} in the context of the evolution of bar-driven escaping orbits in disc galaxies.

%% file: Chapter_3.tex
\chapter{Explosion Triggered Star Formation}
\label{chap:3}

\section{Introduction}
\label{sec:3.1}
The intriguing topic of star formation still plagues current astrophysical research. Most astronomers now believe that the gravitational collapse of dense molecular clouds triggers star formation. However, more information needs to be gathered about the long path leading from diffuse interstellar matter to emerging main sequence stars \cite{McKee2007, Kennicutt2012}. Various theories arising out of the observations as well as from simulations have been developed to demonstrate the mechanism of the growth and evolution of the molecular clouds that lead to the formation of stars \cite{Lada2003, Khoperskov2013}. It is widely accepted that activities occurring in the nuclei of galaxies, including our Milky Way, greatly influence the growth and evolution of the molecular clouds \cite{Burbidge1970, Morris1996}.

\indent The evidence of violent activities near the centre is revealed by the images of galaxies like M82, M87, NGC 5128, etc. \cite{Lynds1963, Solinger1969}. Besides photometric studies, spectroscopic observations also reveal similar violent phenomena in the nuclei of some Seyfert galaxies. From their nuclei, vast clouds of gas are ejected with a speed of 2000 km s$^{-1}$. NGC 1068 is an excellent example of observing such a high-velocity ejection. Many other authors have found similar types of observations \cite{Van1970, Van1971, Sanders1974, Defouw1976}. Shock fronts or streams of charged particles are produced due to nuclear explosions and propagate in the ambient medium like density waves. Due to the propagation of shock fronts, the neighbouring medium is compressed into a thin, dense shell \cite{Aluzas2012, Aluzas2014}. Further, this dense shell undergoes gravitational instability and fragments into molecular clouds. There are several numerical studies about gravitational fragmentation of the dense shell, which several authors have done \cite{Dale2009, Dale2011, Iwasaki2011}. Stars are born out of the gravitational collapse of these molecular clouds.

\indent The actual reason for violent activities near the galactic centre has yet to be established, but many authors have suggested different mechanisms. One such mechanism is the black hole hypothesis \cite{Vanbeveren1978}. Some authors (e.g., \cite{Sargent1978}) have proposed that the existence of a black hole of mass $5 \times 10^ 9 M_{\odot}$ in the nucleus of M87 is responsible for the high energy activities near its centre. The shock waves generated via nuclear explosions have a more significant impact on the formation and evolution of galaxies. It is now well accepted that the formation of the galactic disc, including the molecular rings and spiral arms, is closely associated with generating shock waves due to nuclear explosions \cite{Kato1977, Saito1977}. Molecular rings found in the Milky Way's central region formed as a result of explosion shocks have been modelled by many authors \cite{Saito1980, Saha1985, Basu1989}. Moreover, the star formation process in the shock-compressed molecular clouds has been modelled by several authors \cite{Elmegreen1977, Elmegreen1978, Whitworth1994, Balfour2015}. In this aspect, it is essential to investigate the role of parameters like molecular cloud's magnetic field strength, rotation, etc. These studies are essential to better understanding the internal mechanisms of the gravitational collapse of molecular clouds. A clear grasp of how these parameters contribute to the growth and evolution of molecular clouds has yet to be understood.

\indent One important parameter in the study of star formation in shock-compressed molecular clouds is the molecular cloud's magnetic field. The role of this fundamental quantity is important for the growth and evolution of the shock-compressed molecular clouds. The strength of the magnetic field varies across the galaxy morphologies. The average strength of the magnetic field for a sample of 74 spiral galaxies studied by \citeauthor{Niklas1995}, \citeyear{Niklas1995} \cite{Niklas1995} is about $9 \; \mu$G - 11 $\mu$G, whereas gas-rich spirals with high star-formation rates like M51, M83, NGC 6946, etc. have magnetic field strengths of $20 \; \mu$G - 30 $\mu$G in their spiral arms. The strongest magnetic field strength is about $50 \; \mu$G - 100 $\mu$G and is found in starburst galaxies like M82, NGC 253 \cite{Heesen2011, Adebahr2013} and in barred galaxies like NGC 1097, NGC 1365 \cite{Beck2005}. Radio-faint galaxies like M31, M33, etc. have weaker magnetic field strengths of about $6 \; \mu$G. Spiral galaxies like our Milky Way have a comparatively weaker total magnetic field of about $10 \; \mu$G at $r$ = 3 kpc \cite{Beck2004}.

\indent The upcoming sections of this chapter discuss the magnetic field's role in the star formation process inside shock-compressed molecular clouds under different physical assumptions.

\section{Shock Wave Propagation Model}
\label{sec:3.2}
\begin{figure}[H]
\centering
\subfigure[Before explosion]{\label{fig:3.1a}\includegraphics[width=0.48\textwidth]{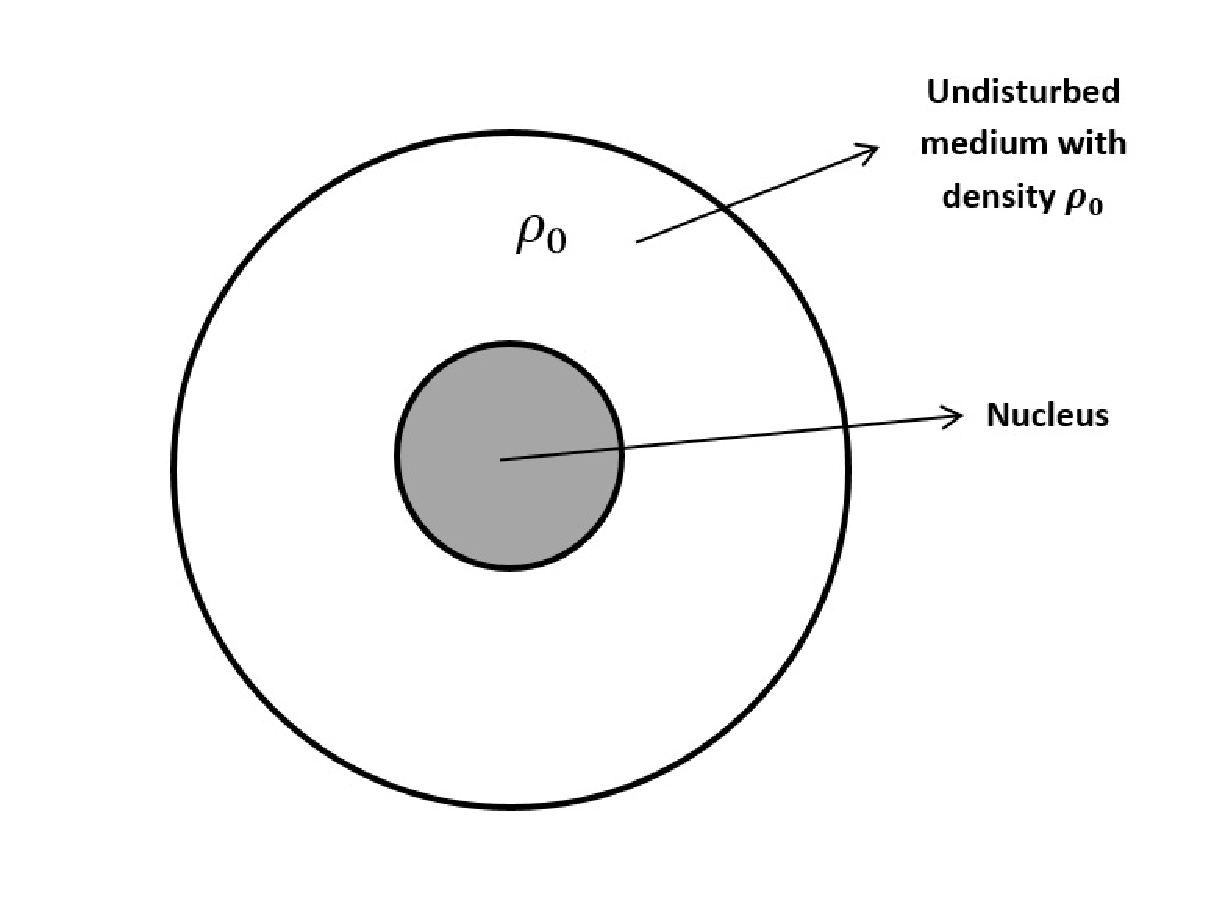}}
\subfigure[After explosion]{\label{fig:3.1b}\includegraphics[width=0.48\textwidth]{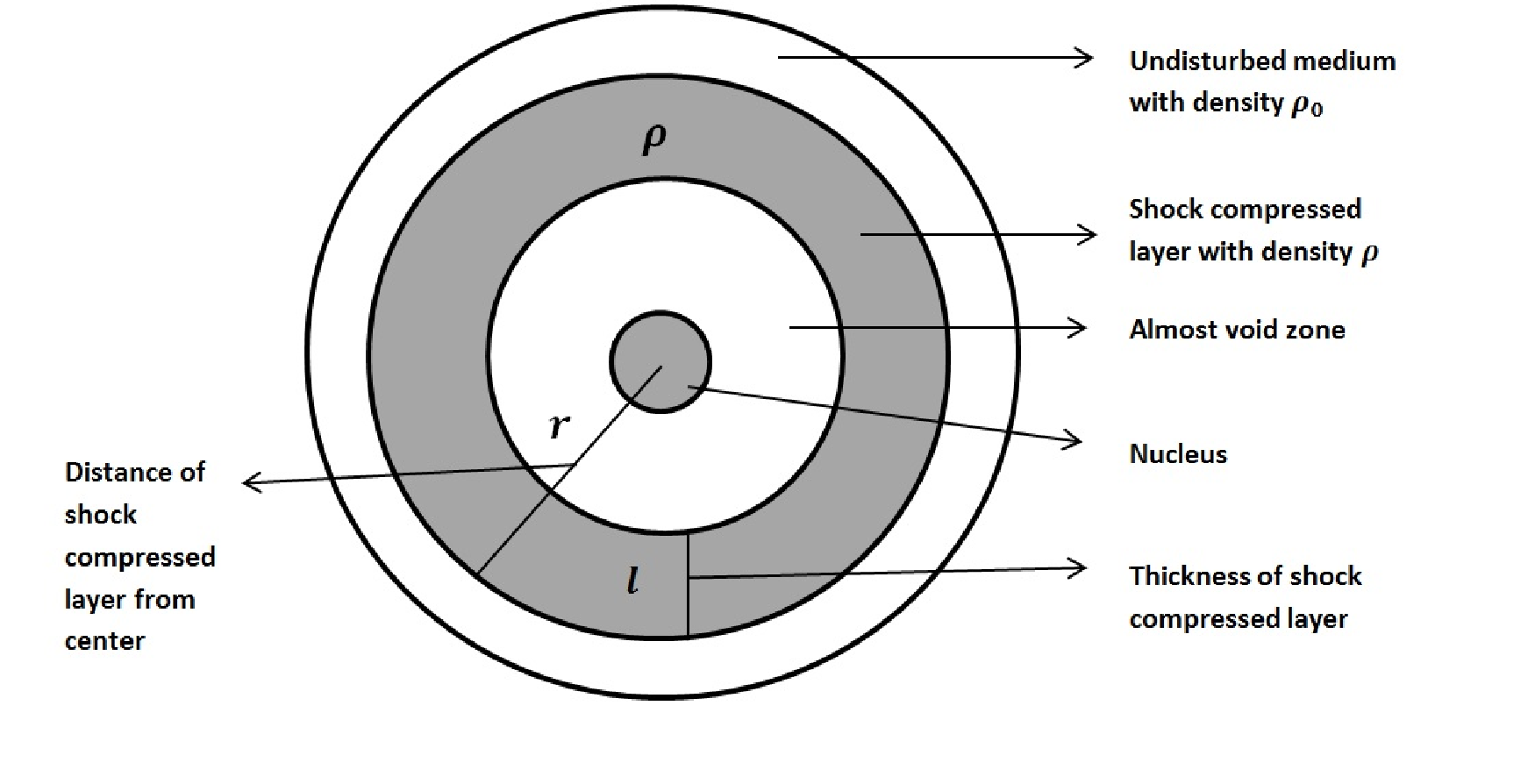}}
\caption{Schematic diagram of shock wave propagation in the central region of galaxies.}
\label{fig:3.1}
\end{figure}

Let us consider a spherical distribution of molecular clouds inside the bulge of a disc galaxy with the assumption that only a negligible fraction of gas is blown away. Shock waves generated by nuclear explosions move gradually outwards and compress the neighbouring medium into a thin, dense spherical shell (see Fig. \ref{fig:3.1}).
\noindent Before a nuclear explosion occurs, let $\rho_0$ be the gas density, and after that, let $\rho$ be the gas density, $l$ be the spherical shell's thickness, and $r$ be the radial distance of this shell's outer edge from the galactic centre at time instant $t$. In this astrophysical context, the principle of conservation of mass can be written as,
\begin{equation}
\label{eq:3.1}
\int_{0}^{r} 4 \pi r^2 \rho_{0} dr = \frac{4}{3}\pi [{r^3 -(r-l)^3}] \rho,
\end{equation}

\begin{figure}[H]
\centering
\includegraphics[height=0.6\columnwidth,width=1\columnwidth]{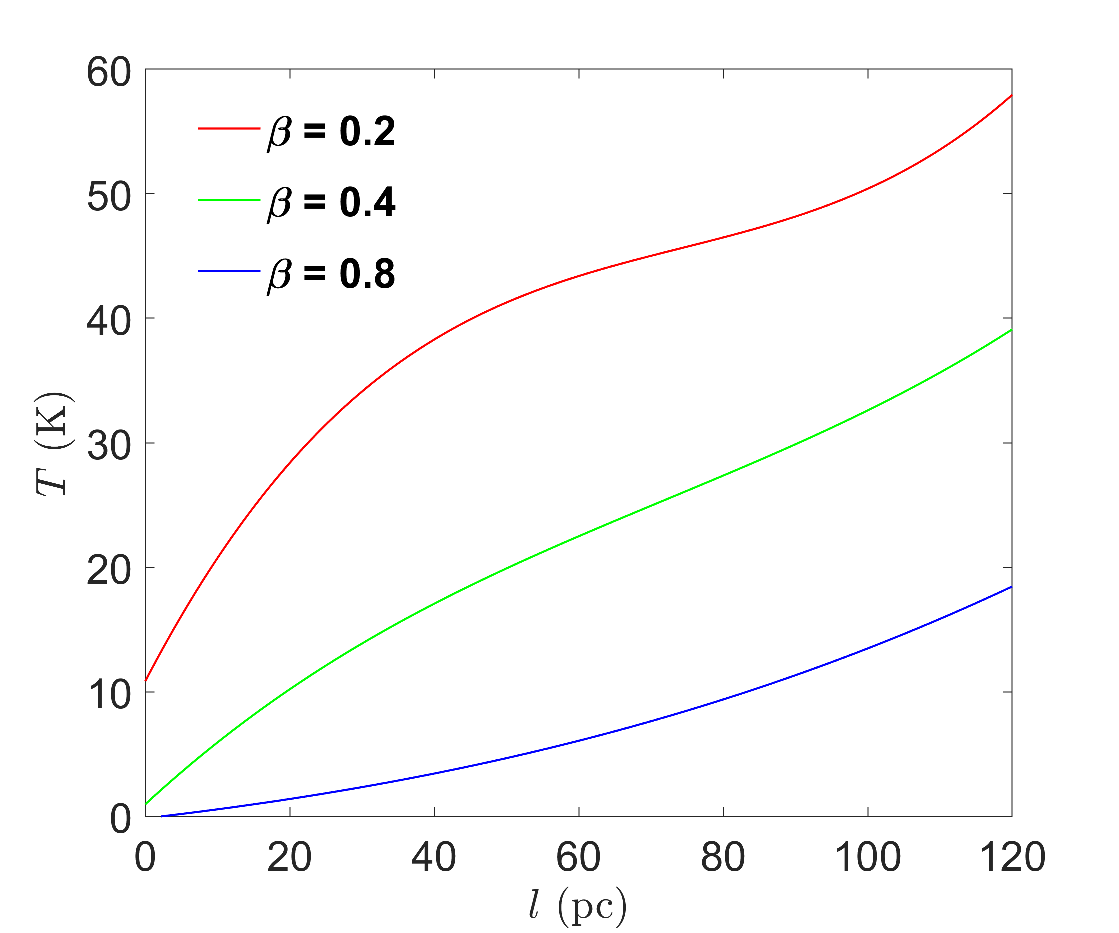}
\caption{Temperature ($T$) versus shell thickness ($l$) for different values of $\beta$.}
\label{fig:3.2}
\end{figure}

\noindent The distribution of initial density ($\rho_0$) for the unperturbed disc is considered by a power-law formula,
\begin{equation}
\label{eq:3.2}
\rho_0 = sr^{-\alpha},
\end{equation} 
 
\noindent where $\alpha$ is the power-law index and $s$ is the normalization constant. For strong non-radiating shock the value of $\alpha$ is $1.8$ and $s$ can be determined from the boundary condition: $\rho_0 = 460 \; \text{cm}^{-3}$ at $r = 1$ kpc \cite{Saito1980}. On the assumption of adiabatic expansion of molecular clouds, Eq. (\ref{eq:3.1}) can be simplified with the help of Eq. (\ref{eq:3.2}) as,
\begin{equation}
\label{eq:3.3}
\begin{split}
\int_{0}^{r} 4 \pi r^{2-\alpha} \ s \ dr & \simeq 4 \pi r^2 l \rho \;\; (\because r \gg l)\\
\Rightarrow \rho & = \frac{s r^{1-\alpha}}{(3-\alpha)l}\\
\Rightarrow l & = \frac{r}{3-\alpha}\frac{\rho_0}{\rho} = \frac{r}{3-\alpha}\frac{(\gamma - 1)}{(\gamma +1)} \; \left(\because \frac{\rho_0}{\rho} = \frac{(\gamma - 1)}{(\gamma +1)} \; \text{\cite{Zel2002}} \right),
\end{split}
\end{equation}

\noindent where $\gamma$ is the adiabatic constant, and for an ideal gas, its value is $\frac{4}{3}$. Here, molecular clouds are considered ideal gas. So, Eq. (\ref{eq:3.3}) is further simplified into $l \simeq 0.119 r$. Again, the temperature ($T$) distribution in the shock-compressed layer is also considered by a power-law formula, 
\begin{equation}
\label{eq:3.4}
T = \bar{B} \rho^{-\beta}, 
\end{equation}

\noindent where $\beta \; (0 < \beta < 1)$ is the power-law index and $\bar{B}$ is the normalisation constant, which is determined from the boundary condition $T = 80$ K at $\rho = 460 \; \text{cm}^{-3}$ \cite{Saito1980} for a specific value of $\beta$. The variation of temperature ($T$) with the thickness of the spherical shell ($l$) is shown in Fig. \ref{fig:3.2}. This reveals that $T$ steadily increases with $l$ and falls rapidly as $\beta$ increases.

\noindent Moreover, consider $p$, $\Phi$, $\vec{V}$, $\vec{\Omega}$ and $\vec{B}$ as the pressure, gravitational potential, linear velocity, angular velocity and magnetic field terms for the molecular clouds. Under these considerations, the governing equations of this shock wave propagation scenario are as follows:

\noindent equation of state $\rightarrow$
\begin{equation}
\label{eq:3.5}
\begin{split}
p & = R \rho T\\
  & = R \bar{B} \rho^{1 - \beta} \; (\text{using Eq. (\ref{eq:3.4}})),
\end{split} 
\end{equation}

\noindent equation of continuity $\rightarrow$ 
\begin{equation}
\label{eq:3.6}
\frac{\partial \rho}{\partial t} + \vec{\nabla}.\{\rho \vec{V} + \rho (\vec{\Omega} \times \vec{r})\} = 0,
\end{equation}

\noindent equation of motion $\rightarrow$
\begin{equation}
\label{eq:3.7}
\frac{\partial \vec{V}}{\partial t} + (\vec{V}.\vec{\nabla}) \vec{V} + 2 (\vec{\Omega} \times \vec{V})+\vec{\Omega} \times(\vec{\Omega} \times \vec{r}) = - \frac{\vec{\nabla} p}{\rho} + \frac{1}{\rho}(\vec{\nabla}\times \vec{B})\times \vec{B} - \vec{\nabla}\Phi + \vec{f}(r),
\end{equation}

\noindent Poisson equation $\rightarrow$
\begin{equation}
\label{eq:3.8}
\frac{1}{r^2 \sin \theta} \left[\sin \theta \frac{\partial}{\partial r}(r^2 \frac{\partial \Phi}{\partial r}) + \frac{\partial}{\partial \theta}(\sin \theta \frac{\partial \Phi}{\partial \theta}) + \frac{1}{\sin \theta} \frac{\partial^2 \Phi}{\partial \phi^2}\right] = 4 \pi G \rho,
\end{equation}

\noindent where $R$ is the universal gas constant and $G$ is the gravitational constant. Now, as the dense molecular clouds undergo compression due to shock waves, the quantities $p$, $\Phi$, $\vec{V}$, $\vec{\Omega}$ and $\vec{B}$ have been perturbed around their equilibrium values. So, these quantities can be written as,
\begin{equation*}
\rho = \rho_0 + \rho_1, \; p = p_0 + p_1, \; \Phi = \Phi_0 + \Phi_1, \; \vec{V} = \vec{V}_0 + \vec{V}_1, \; \vec{B} = \vec{B}_0 + \vec{B}_1,
\end{equation*} 

\noindent where suffixes $`0'$ and $`1'$ signifies the unperturbed and perturbed terms receptively. Under this perturbation scenario, the linearised versions of Eqs. (\ref{eq:3.6}), (\ref{eq:3.7}), and (\ref{eq:3.8}) are as follows:

\noindent equation of continuity $\rightarrow$ 
\begin{equation}
\label{eq:3.9}
\frac{\partial \rho_1}{\partial t} + \rho_0 (\vec{\nabla}.\vec{V}_1) = 0,
\end{equation}

\noindent equation of motion $\rightarrow$ 
\begin{equation}
\label{eq:3.10}
\begin{array}{lr}
\begin{split}
\frac{\partial u_1}{\partial t} & = - \frac{\partial \Phi}{\partial r} - \frac{d p}{d \rho_1}\frac{1}{\rho_0}\frac{\partial \rho_1}{\partial r} -\frac{B_0}{\rho_0}\frac{\partial B_1}{\partial r}+ f(r) + \Omega^2 r + 2\Omega v_1\\
\frac{\partial v_1}{\partial t} & = - \frac{1}{r}\frac{\partial \Phi}{\partial \theta} - 2\Omega u_1\\
\frac{\partial w_1}{\partial t} & = - \frac{1}{r \sin \theta}\frac{\partial \Phi}{\partial \phi}
\end{split}
\end{array}\Biggl\},
\end{equation}

\noindent Poisson equation $\rightarrow$ 
\begin{equation}
\label{eq:3.11}
\frac{1}{r^2 \sin \theta} \left[\sin \theta \frac{\partial}{\partial r}(r^2 \frac{\partial \Phi_1}{\partial r}) + \frac{\partial}{\partial \theta}(\sin \theta \frac{\partial \Phi_1}{\partial \theta}) + \frac{1}{\sin \theta} \frac{\partial^2 \Phi_1}{\partial \phi^2}\right] = 4 \pi G (\rho_0 + \rho_1),
\end{equation}

\noindent where $\vec{V}_0 = \vec{0}$ ($\because$ the spherical shell initially is at rest), $\vec{V}_1 \equiv (u_1, v_1, w_1)$, $\vec{\Omega} \equiv (0,0,\Omega)$, $\vec{B} \equiv (0,0,B(r) = B_0(r) + B_1(r))$ \cite{Ferriere2015}, $\vec{f}(r) \equiv (0,0,f(r) = \frac{P}{l \rho} = \frac{8}{25}\frac{(3 - \alpha)}{(1 + \gamma)}\frac{E}{sr^{4 - \alpha}})$ is the shock force per unit mass, $P = \frac{8}{25} \frac{1}{(1 + \gamma)}\frac{E}{r^3}$ is the shock pressure and $E$ is the explosion energy \cite{Basu1989, Zel2002}. Furthermore, the form of perturbed terms (i.e., terms with suffix $'1'$) is $X = A e^{i(\frac{2\pi r}{\lambda} - \omega t)}$, where $A$, $\lambda$ and $w$ are the amplitude, wavelength and frequency of the associated wave propagation, respectively. Now, a combination of Eqs. (\ref{eq:3.10}) and (\ref{eq:3.11}) gives,
\begin{equation*}
\begin{split}
\vec{\nabla}.(\frac{\partial \vec{V_1}}{\partial t}) = & - 4\pi G (\rho_0 + \rho_1) - \vec{\nabla}.(\frac{dp}{d \rho_1}\frac{1}{\rho_0}\frac{\partial \rho_1 }{\partial r},0,0) - \vec{\nabla}.(\frac{B_0}{\rho_0}\frac{\partial B_1}{\partial r},0,0)\\ &
+ \vec{\nabla}.(f(r),0,0) + 2\Omega\vec{\nabla}.(v_1,0,0) - 2\Omega\vec{\nabla}.(0,u_1,0) + \Omega^2.
\end{split}
\end{equation*}
Now substituting Eq. (\ref{eq:3.9}) in the above equation yields,
\begin{equation*}
\begin{split}
-\frac{1}{\rho_0}\frac{\partial^2 \rho_1}{\partial t^2} = & - 4\pi G (\rho_0 +\rho_1)-\frac{\partial}{\partial r}(\frac{R\bar{B}(1-\beta)(\rho_0+\rho_1)^{-\beta}}{\rho_0} \frac{\partial \rho_1}{\partial r}) -\frac{B_0}{\rho_0}\frac{\partial^2 B_1}{\partial r^2}\\ & + f^{\prime}(r)+\Omega^2.
\end{split}
\end{equation*}
Moreover, using Eqs. (\ref{eq:3.2}) and (\ref{eq:3.5}) with the perturbation form of the magnetic field, i.e., $B_1 = B_0 e^{i(\frac{2\pi r}{\lambda} - \omega t)}$ in the above equation results in the dispersion relation as,
\begin{equation}
\label{eq:3.12}
\begin{split}
w^2 & = \frac{4 \pi^2}{\lambda^2}\left[\frac{B_0}{\rho_1}B_1 + R\bar{B}(1-\beta)\rho_0^{-\beta} - R\bar{B}\beta(1-\beta)\rho_0^{-\beta}\frac{\rho_1}{\rho_0}\right]\\ & + \frac{\rho_0}{\rho_1}\left[f^{\prime}(r)-4\pi G\rho_0+ \Omega^2\right],
\end{split}
\end{equation}

\noindent where the following approximations: $\frac{\rho_1}{\rho_0} < 1$, $\frac{{B}_1}{B_0} < 1$ and $0 < 1 - \beta\frac{\rho_1}{\rho_0} < 1$ are used. Inside the dense molecular clouds, instability arises due to shock compression, and they undergo gravitational collapse. In this aspect, the theory of Jeans instability is very relevant, as it describes the path of star formation from these collapsing molecular clouds. According to this theory, a certain mass limit exists -- known as Jeans mass ($M_\text{J}$), below which Jeans instability causes star formation due to the gravitational collapse. Beyond this mass limit, gravity dominates, and molecular clouds become unstable to collapse. The Jeans mass ($M_\text{J}$) depends on the radius of the molecular clouds -- known as Jeans length. Now $w^2 > 0$ is essential for propagating shock waves inside these clouds. Furthermore, these clouds are gravitationally unstable against the growth of linear perturbation only for $\lambda^2 \le \lambda_\text{J}^2$, where $\lambda_\text{J}$ is the Jeans length. Thus the analytical form of $\lambda_\text{J}$ is derived from the dispersion relation (Eq. (\ref{eq:3.12})) is as follows: 
\begin{equation}
\label{eq:3.13}
\lambda_\text{J}^2 = \frac{\pi R\bar{B}(1-\beta)\rho_0^{-\beta-1} + \pi \frac{B_0^2}{\rho_0^2}}{G(1-\frac{f^{\prime}(r)}{4\pi G\rho_0}-\frac{\Omega^2}{4\pi G\rho_0})}.
\end{equation}

\noindent Moreover, with aid of the above expression of $\lambda_\text{J}$, the analytical form of $M_\text{J}$ is,
\begin{equation}
\label{eq:3.14}
\begin{split}
M_\text{J} & = \frac{\pi}{6}\rho_0 \lambda_\text{J}^3\\
& = \frac{\pi}{6}\rho_0 \left\{\frac{\pi R\bar{B}(1-\beta)\rho_0^{-\beta-1} + \pi\frac{B_0^2}{\rho_0^2}}{G(1-\frac{f^{\prime}(r)}{4\pi G\rho_0} - \frac{\Omega^2}{4\pi G\rho_0})}\right\}^\frac{3}{2}.
\end{split}
\end{equation}

\section{Role of Magnetic Field}
\label{sec:3.3}
The values of Jeans mass ($M_\text{J}$) for gravitationally collapsing molecular clouds have been derived for the following cases to investigate the influence of the magnetic field on the formation of stars due to gravitational collapse.

\subsection{Non-rotating clouds with constant magnetic field}
\label{sec:3.3.1}
Consider the gravitational collapse of non-rotating molecular clouds with constant magnetic field, i.e., $\Omega = 0$ and $B_0(r) = B_0$. For these considerations the expression of $M_\text{J}$ is derived from the Eq. (\ref{eq:3.14}) as,
\begin{equation}
\label{eq:3.15}
M_\text{J} = \frac{\pi}{6}\rho_0 \left\{\frac{\pi R\bar{B}(1-\beta)\rho_0^{-\beta-1} + \pi\frac{B_0^2}{\rho_0^2}}{G(1-\frac{f^{\prime}(r)}{4\pi G\rho_0})}\right\}^{\frac{3}{2}}.
\end{equation}

\begin{figure}
\centering
\subfigure[$E = 10^{54}$ erg]{\label{fig:3.3a}\includegraphics[width=0.49\columnwidth]{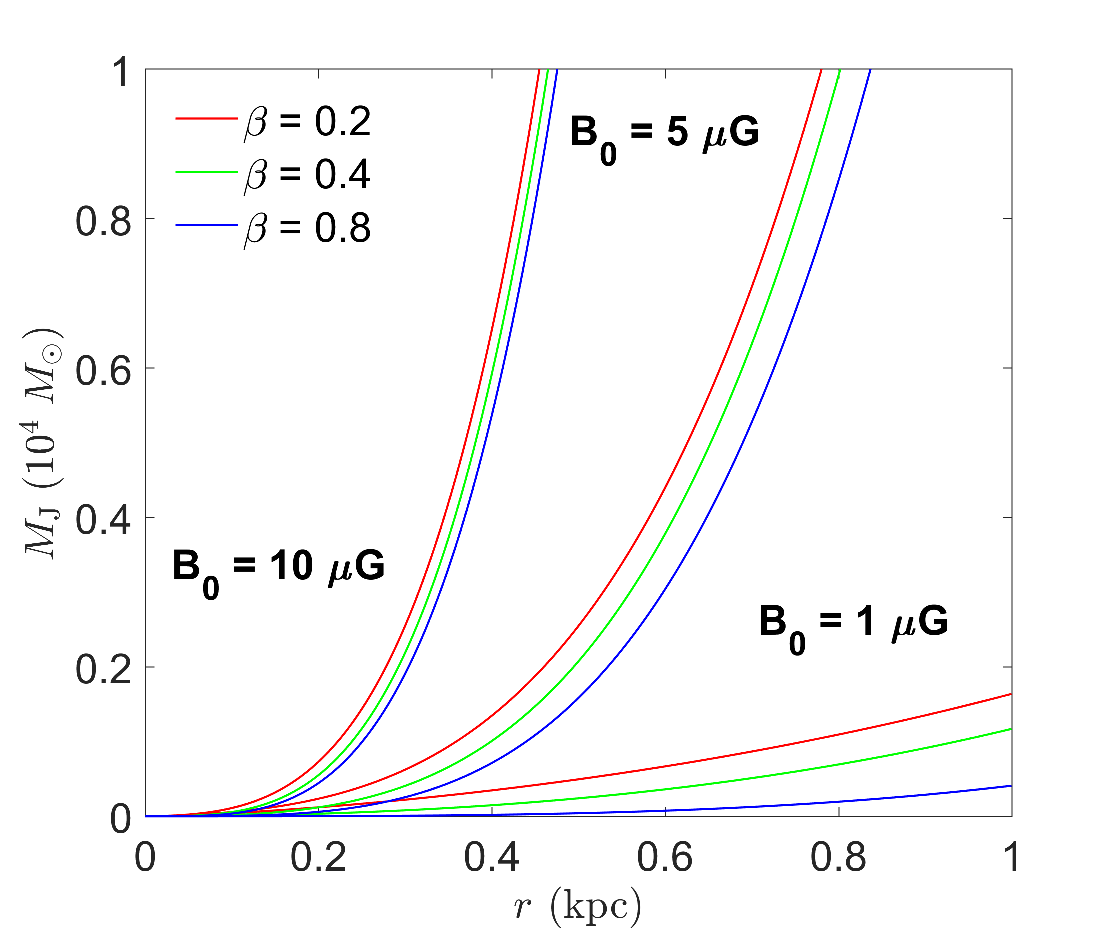}}
\subfigure[$E = 10^{56}$ erg]{\label{fig:3.3b}\includegraphics[width=0.49\columnwidth]{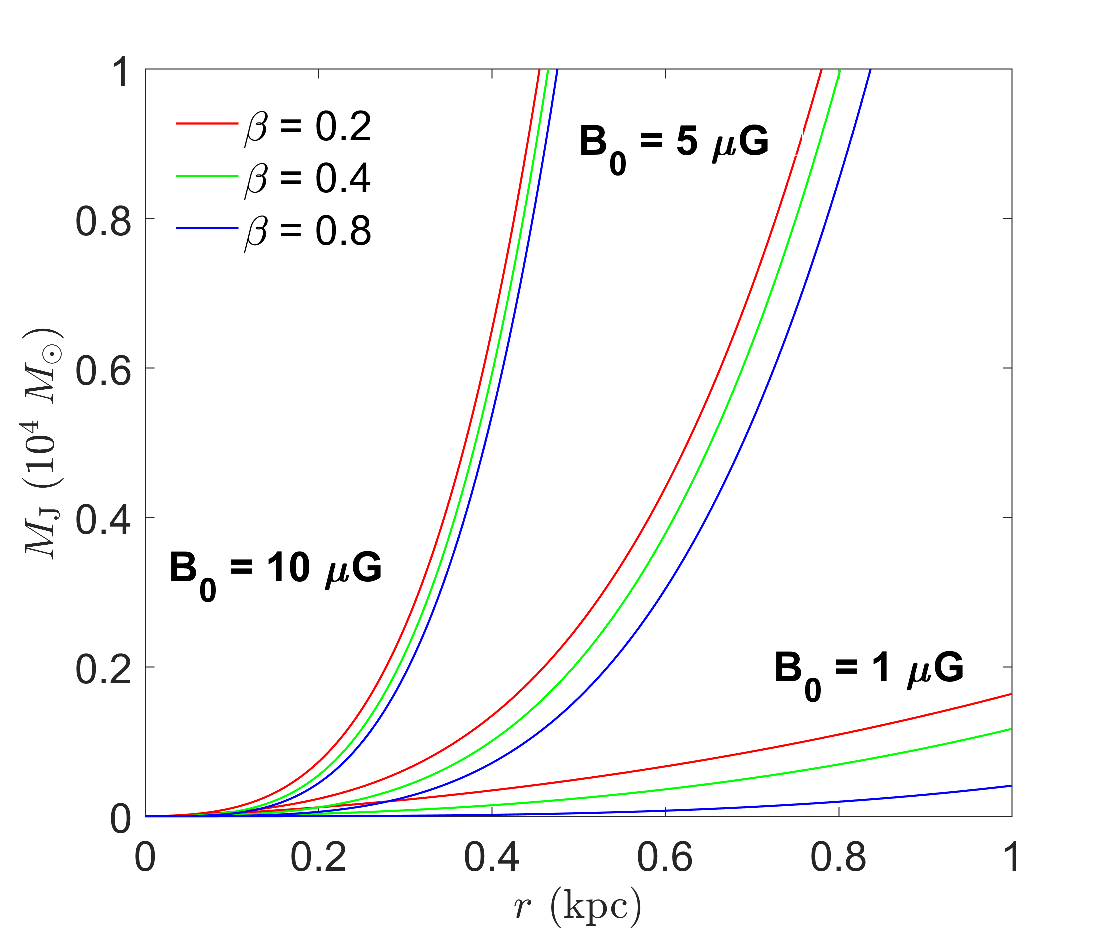}}
\caption{Jeans mass ($M_\text{J}$) values for non-rotating molecular clouds with a constant magnetic field.}
\label{fig:3.3}
\end{figure}

\noindent The values of $M_\text{J}$ for different parameter values are obtained from the above formula (Eq. (\ref{eq:3.15})) and provided in Fig. \ref{fig:3.3} and Table \ref{tab:3.1}.

\begin{table}
\scriptsize
\centering 
\parbox{\columnwidth}{
\begin{tabular}{|c|c c c c c||c|c c c c c|}
\hline
$r$ (pc) & $B_0$ ($\mu$G) & $E$ (erg) & $\beta$ & $T$ (K) & $M_\text{J} \; (M_\odot)$ & $r$ (pc) & $B_0$ ($\mu$G) & $E$ (erg) & $\beta$ & $T$ (K) & $M_\text{J} \; (M_\odot)$\\
\hline
\hline
		 &  & & & & &       &     & &  &    &  \\
		 &  1             & $10^{54}$ & 0.2     & 13.26   & 4.24   &       &  1             & $10^{54}$ & 0.2     &  27.38  &  78.96\\
         &                &           & 0.4     &  2.20   & 0.33    &     &                &           & 0.4     &   9.37  &  19.13\\
         &                &           & 0.8     &  0.06   & 0.0013  &       &                &           & 0.8     &   1.10  &   0.82\\
         & 10             & $10^{54}$ & 0.2     & 13.26   & 4.80     &    & 10             & $10^{54}$ & 0.2     &  27.38  & 322.01\\
         &                &           & 0.4     &  2.20   & 0.60     &    &                &           & 0.4     &   9.37  & 213.55\\
         &                &           & 0.8     &  0.06   & 0.12     &    &                &           & 0.8     &   1.10  & 159.79\\
20       &  1             & $10^{56}$ & 0.2     & 13.26   & 3.82 & 150      &  1             & $10^{56}$ & 0.2     &  27.38  &  78.46\\
         &                &           & 0.4     &  2.20   & 0.30 &          &                &           & 0.4     &   9.37  &  19.01\\
         &                &           & 0.8     &  0.06   & 0.0012 &          &                &           & 0.8     &   1.10  &   0.81\\
         & 10             & $10^{56}$ & 0.2     & 13.26   & 4.33 & & 10             & $10^{56}$ & 0.2     &  27.38  & 319.97\\
         &                &           & 0.4     &  2.20   & 0.54 & &                &           & 0.4     &   9.37  & 212.20\\
         &                &           & 0.8     &  0.06   & 0.10 & &                &           & 0.8     &   1.10  & 158.78\\
		 &  & & & & &       &     & &  &    &  \\       
\hline
		 &  & & & & &       &     & &  &    &  \\
         &  1             & $10^{54}$ & 0.2     & 18.44   & 15.92 & &  1             & $10^{54}$ & 0.2     & 42.24   & 496.99\\
		 &                &           & 0.4     &  4.25   &  2.08 &  &                &           & 0.4     & 22.30   & 241.93\\
         &                &           & 0.8     &  0.22   &  0.02 & &                &           & 0.8     &  6.22   &  41.56\\
         & 10             & $10^{54}$ & 0.2     & 18.44   & 24.24 & & 10             & $10^{54}$ & 0.2     & 42.24   &   1.38 $\times 10^{4}$\\
         &                &           & 0.4     &  4.25   &  7.05 & &                &           & 0.4     & 22.30   &   1.30 $\times 10^{4}$\\
         &                &           & 0.8     &  0.22   &  3.10 & &                &           & 0.8     &  6.22   &   1.21 $\times 10^{4}$\\
50       &  1             & $10^{56}$ & 0.2     & 18.44   & 15.46 & 500      &  1             & $10^{56}$ & 0.2     & 42.24   & 496.41\\
         &                &           & 0.4     &  4.25   &  2.02 &  &                &           & 0.4     & 22.30   & 241.65\\
         &                &           & 0.8     &  0.22   &  0.02 & &                &           & 0.8     &  6.22   &  41.52\\
         & 10             & $10^{56}$ & 0.2     & 18.44   & 23.54 & & 10             & $10^{56}$ & 0.2     & 42.24   &   1.38 $\times 10^{4}$\\
         &                &           & 0.4     &  4.25   &  6.84 & &                &           & 0.4     & 22.30   &   1.30 $\times 10^{4}$\\
         &                &           & 0.8     &  0.22   &  3.02 & &                &           & 0.8     &  6.22   &   1.20 $\times 10^{4}$\\
		 &  & & & & &       &     & &  &    &  \\
\hline
		 &  & & & & &       &     & &  &    &  \\
         &  1             & $10^{54}$ & 0.2     & 23.66   & 43.58 &          &  1             & $10^{54}$ & 0.2     &  54.21  &   1.64 $\times 10^{3}$\\
         &                &           & 0.4     &  7.00   &  8.39 & &                &           & 0.4     &  36.73  &   1.18 $\times 10^{3}$\\
         &                &           & 0.8     &  0.61   &  0.22 & &                &           & 0.8     &  16.87  & 412.32\\
         & 10             & $10^{54}$ & 0.2     & 23.66   & 111.62 & & 10             & $10^{54}$ & 0.2     &  54.21  &   1.52 $\times 10^{5}$\\
         &                &           & 0.4     &  7.00   &  57.67 & &                &           & 0.4     &  36.73  &   1.50 $\times 10^{5}$\\
         &                &           & 0.8     &  0.61   &  37.29 & &                &           & 0.8     &  16.87  &   1.46 $\times 10^{5}$\\
100      &  1             & $10^{56}$ & 0.2     & 23.66   &  43.10 & 1000     &  1             & $10^{56}$ & 0.2     &  54.21  &   1.64 $\times 10^{3}$\\
         &                &           & 0.4     &  7.00   &   8.29 & &                &           & 0.4     &  36.73  &   1.17 $\times 10^{3}$\\
         &                &           & 0.8     &  0.61   &   0.22 & &                &           & 0.8     &  16.87  & 412.14\\
         & 10             & $10^{56}$ & 0.2     & 23.66   & 110.37 & & 10             & $10^{56}$ & 0.2     &  54.21  &   1.52 $\times 10^{5}$\\
         &                &           & 0.4     &  7.00   &  57.02 & &                &           & 0.4     &  36.73  &   1.50 $\times 10^{5}$\\
         &                &           & 0.8     &  0.61   &  36.88 & &                &           & 0.8     &  16.87  &   1.46 $\times 10^{5}$\\
         &  & & & & &       &     & &  &    &  \\ 
\hline  
\end{tabular}
}
\caption{Jeans mass ($M_\text{J}$) values for non-rotating molecular clouds with a constant magnetic field.}
\label{tab:3.1}
\end{table}

\subsection{Rotating clouds with constant magnetic field}
\label{sec:3.3.2}
Consider the gravitational collapse of rotating molecular clouds with constant magnetic field, i.e., $\Omega = 4 \; \text{km s}^{-1} \text{kpc}^{-1}$ \cite{Turner1984} and $B_0(r) = B_0$. For these considerations the expression of $M_\text{J}$ is derived from the Eq. (\ref{eq:3.14}) as, 
\begin{equation}
\label{eq:3.16}
\begin{split}
M_\text{J} = \frac{\pi}{6}\rho_0 \left\{\frac{\pi R\bar{B}(1-\beta)\rho_0^{-\beta-1} + \pi\frac{B_0^2}{\rho_0^2}}{G(1-\frac{f^{\prime}(r)}{4\pi G\rho_0} - \frac{\Omega^2}{4\pi G\rho_0})}\right\}^\frac{3}{2}.
\end{split}
\end{equation} 

\begin{figure}
\centering
\subfigure[$E = 10^{54}$ erg]{\label{fig:3.4a}\includegraphics[width=0.49\columnwidth]{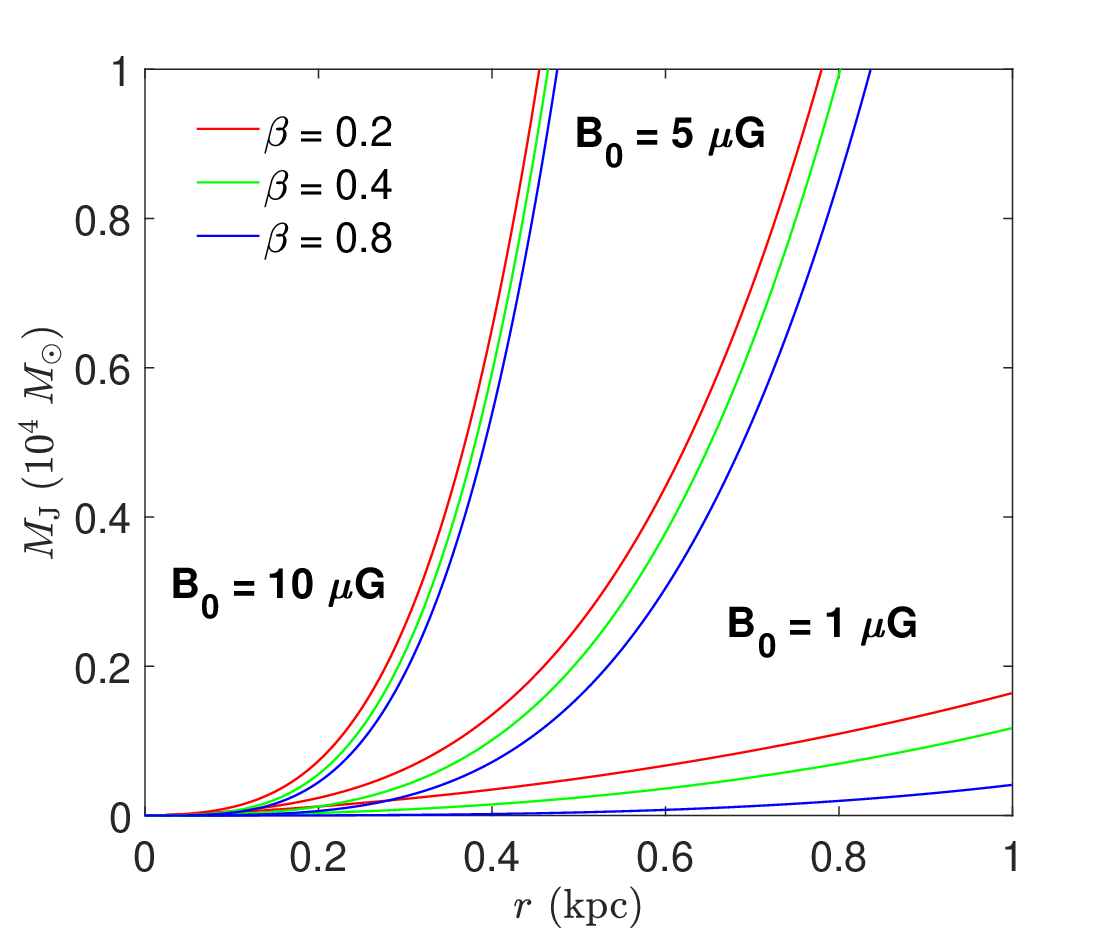}}
\subfigure[$E = 10^{56}$ erg]{\label{fig:3.4b}\includegraphics[width=0.49\columnwidth]{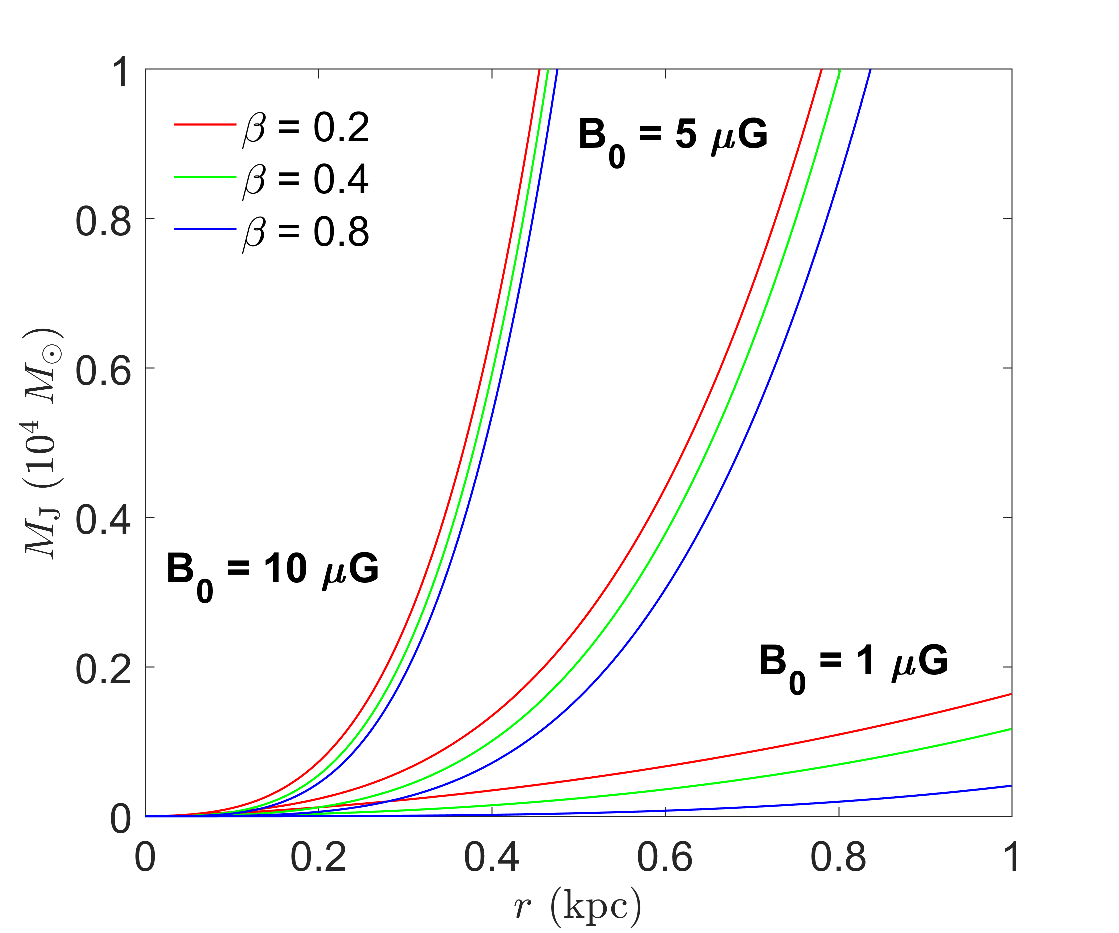}}
\caption{Jeans mass ($M_\text{J}$) values for rotating molecular clouds with a constant magnetic field.}
\label{fig:3.4}
\end{figure}

\noindent The values of $M_\text{J}$ for different parameter values are obtained from the above formula (Eq. (\ref{eq:3.16})) and provided in Fig. (\ref{fig:3.4}) and Table \ref{tab:3.2}.

\begin{table}
\scriptsize
\centering 
\parbox{\columnwidth}{
\begin{tabular}{|c|c c c c c||c|c c c c c|}
\hline
$r$ (pc) & $B_0$ ($\mu$G) & $E$ (erg) & $\beta$ & $T$ (K) & $M_\text{J} \; (M_\odot)$ & $r$ (pc) & $B_0$ ($\mu$G) & $E$ (erg) & $\beta$ & $T$ (K) & $M_\text{J} \; (M_\odot)$\\
\hline
\hline
         &               &            &         &         & &          &               &            &         &         &\\
         &  1            & $10^{54}$  & 0.2     & 13.26   & 4.24 &          &  1             & $10^{54}$ & 0.2     & 27.38   &  78.96\\
         &               &            & 0.4     &  2.20   & 0.33 &          &                &           & 0.4     &  9.37   &  19.13\\
         &               &            & 0.8     &  0.06   & 0.0013 &          &                &           & 0.8     &  1.10   &   0.82\\
         & 10            & $10^{54}$  & 0.2     & 13.26   & 4.80 & & 10             & $10^{54}$ & 0.2     & 27.38   & 322.01\\
         &               &            & 0.4     &  2.20   & 0.60 &          &                &           & 0.4     &  9.37   & 213.55\\
         &               &            & 0.8     &  0.06   & 0.12 &          &                &           & 0.8     &  1.10   & 159.79\\	
20       &  1            & $10^{56}$  & 0.2     & 13.26   & 3.82 & 150      &  1             & $10^{56}$ & 0.2     & 27.38   &  78.46\\
         &               &            & 0.4     &  2.20   & 0.30 & &                &           & 0.4     &  9.37   &  19.01\\
         &               &            & 0.8     &  0.06   & 0.0012 & &                &           & 0.8     &  1.10   &   0.81\\
         & 10            & $10^{56}$  & 0.2     & 13.26   & 4.33 & & 10             & $10^{56}$ & 0.2     & 27.38   & 319.97\\
         &               &            & 0.4     &  2.20   & 0.54 &          &                &           & 0.4     &  9.37   & 212.20\\
         &               &            & 0.8     &  0.06   & 0.10 & &                &           & 0.8     &  1.10   & 158.78\\
         &               &            &         &         & & &               &            &         &         &\\        
\hline
         &               &            &         &         & & &               &            &         &         &\\
         &  1            & $10^{54}$  & 0.2     & 18.44   & 15.93 & &  1             & $10^{54}$ & 0.2     &  42.24  & 497.00\\
         &               &            & 0.4     &  4.25   &  2.08 & &                &           & 0.4     &  22.30  & 241.94\\
         &               &            & 0.8     &  0.22   &  0.02 & &                &           & 0.8     &   6.22  &  41.57\\
         & 10            & $10^{54}$  & 0.2     & 18.44   & 24.24 & & 10             & $10^{54}$ & 0.2     &  42.24  &   1.38 $\times 10^{4}$\\
         &               &            & 0.4     &  4.25   &  7.05 & &                &           & 0.4     &  22.30  &   1.30 $\times 10^{4}$\\
         &               &            & 0.8     &  0.22   &  3.10 & &                &           & 0.8     &   6.22  &   1.21 $\times 10^{4}$\\
50       &  1            & $10^{56}$  & 0.2     & 18.44   & 15.46 & 500      &  1             & $10^{56}$ & 0.2     &  42.24  & 496.41\\
         &               &            & 0.4     &  4.25   &  2.02 & &                &           & 0.4     &  22.30  & 241.65\\
         &               &            & 0.8     &  0.22   &  0.02 & &                &           & 0.8     &   6.22  &  41.52\\
         & 10            & $10^{56}$  & 0.2     & 18.44   & 23.54 & & 10             & $10^{56}$ & 0.2     &  42.24  &   1.38 $\times 10^{4}$\\
         &               &            & 0.4     &  4.25   &  6.84 & &                &           & 0.4     &  22.30  &   1.30 $\times 10^{4}$\\
         &               &            & 0.8     &  0.22   &  3.02 & &                &           & 0.8     &   6.22  &   1.20 $\times 10^{4}$\\
         &               &            &         &         &  &                &           &         &         &\\        
\hline
         &               &            &         &         & & &                &           &         &         &\\
         &  1            & $10^{54}$  & 0.2     & 23.66   &  43.58 & &  1             & $10^{54}$ & 0.2     & 54.21   &   1.64 $\times 10^{3}$\\
         &               &            & 0.4     &  7.00   &   8.39 & &                &           & 0.4     & 36.73   &   1.18 $\times 10^{3}$\\
         &               &            & 0.8     &  0.61   &   0.22 & &                &           & 0.8     & 16.87   & 412.34\\
         & 10            & $10^{54}$  & 0.2     & 23.66   & 111.62 & & 10             & $10^{54}$ & 0.2     & 54.21   &   1.52 $\times 10^{5}$\\
         &               &            & 0.4     &  7.00   &  57.67 & &                &           & 0.4     & 36.73   &   1.50 $\times 10^{5}$\\
         &               &            & 0.8     &  0.61   &  37.29 & &                &           & 0.8     & 16.87   &   1.46 $\times 10^{5}$\\
100      &  1            & $10^{56}$  & 0.2     & 23.66   &  43.10 & 1000     &  1             & $10^{56}$ & 0.2     & 54.21   &   1.64 $\times 10^{3}$\\
         &               &            & 0.4     &  7.00   &   8.29 & &                &           & 0.4     & 36.73   &   1.17 $\times 10^{3}$\\
         &               &            & 0.8     &  0.61   &   0.22 & &                &           & 0.8     & 16.87   & 412.15\\
         & 10            & $10^{56}$  & 0.2     & 23.66   & 110.37 & & 10             & $10^{56}$ & 0.2     & 54.21   &   1.52 $\times 10^{5}$\\
         &               &            & 0.4     &  7.00   &  57.02 & &                &           & 0.4     & 36.73   &   1.50 $\times 10^{5}$\\
         &               &            & 0.8     &   0.61  &  36.88 & &                &           & 0.8     & 16.87   &   1.46 $\times 10^{5}$\\ 
         &               &            &         &         & & &                &           &         &         &\\  
\hline    	
\end{tabular}
}
\caption{Jeans mass ($M_\text{J}$) values for rotating molecular clouds with a constant magnetic field.}
\label{tab:3.2}
\end{table}

\subsection{Rotating clouds with density-varying magnetic field}
\label{sec:3.3.3}
Consider the gravitational collapse of rotating molecular clouds with density-dependent magnetic field, i.e., $\Omega = 4 \; \text{km s}^{-1} \text{kpc}^{-1}$ \cite{Turner1984} and $B_0(r) = B_0 {\left(\frac{\rho (r)}{\rho_0}\right)}^\kappa$ \cite{Boss1999}. This dependence of magnetic field strength on density has a fairly robust outcome for the asymptotic value of $\kappa = \frac{1}{2}$ \cite{Tomisaka1990}. For these considerations the expression of $M_\text{J}$ is derived from the Eq. (\ref{eq:3.14}) as,
\begin{equation}
\label{eq:3.17}
M_\text{J} = \frac{\pi}{6}\rho_0 \left\{\frac{\pi R\bar{B}(1-\beta)\rho_0^{-\beta-1}+\pi\frac{B_0^2}{2\rho_0^2}}{G(1-\frac{f^{\prime}(r)}{4\pi G\rho_0}-\frac{\Omega^2}{4\pi G\rho_0})}\right\}^{\frac{3}{2}}.
\end{equation}

\begin{figure}
\centering
\subfigure[$E = 10^{54}$ erg]{\label{fig:3.5a}\includegraphics[width=0.49\columnwidth]{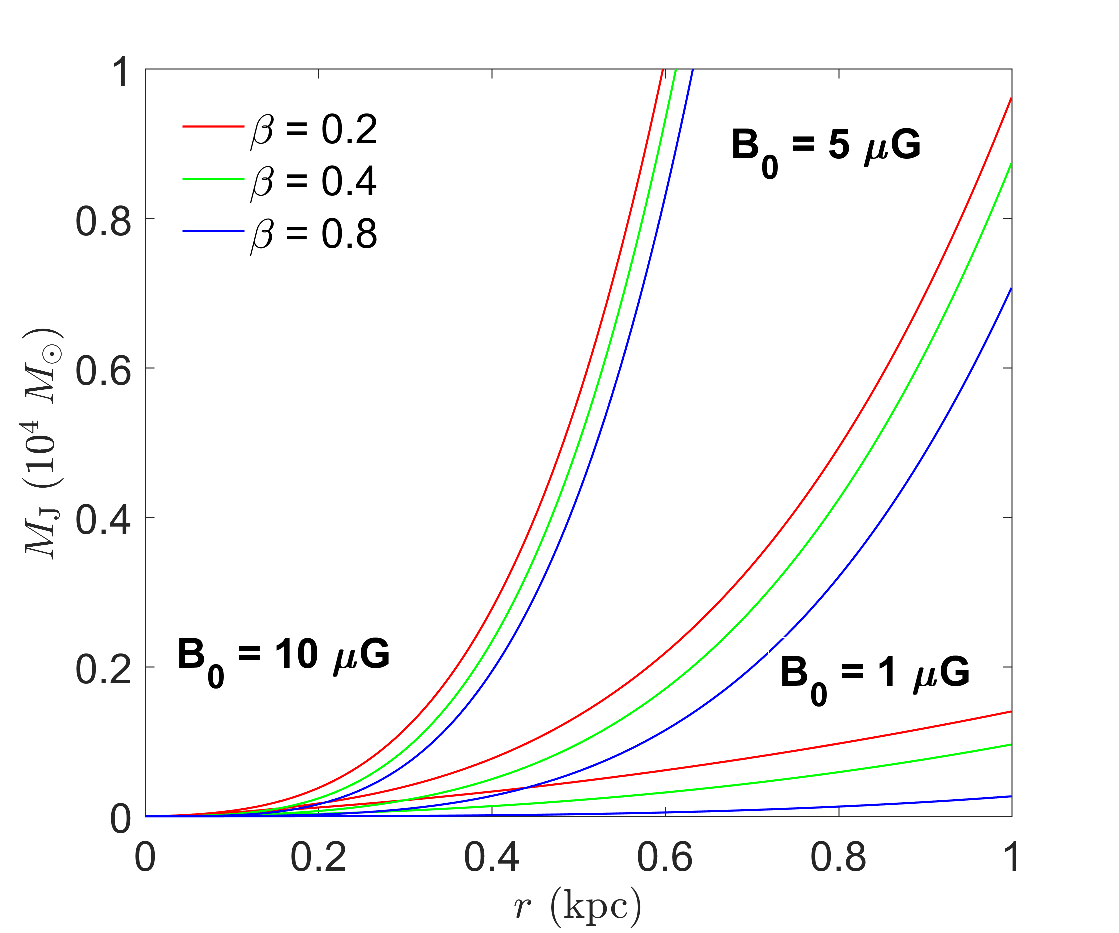}}
\subfigure[$E = 10^{56}$ erg]{\label{fig:3.5b}\includegraphics[width=0.49\columnwidth]{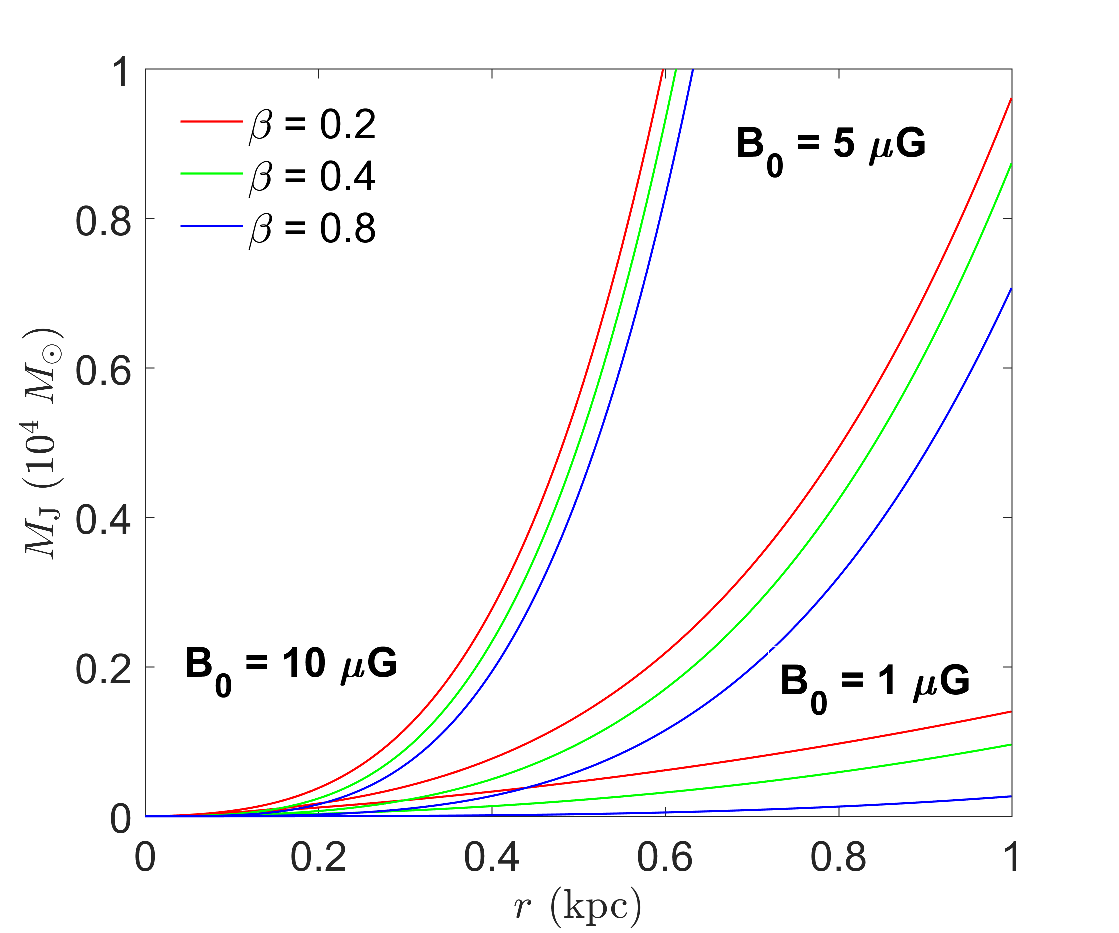}}
\caption{Jeans mass ($M_\text{J}$) values for rotating molecular clouds with a density varying magnetic field.}
\label{fig:3.5}
\end{figure}

\noindent The values of $M_\text{J}$ for different parameter values are obtained from the above formula (Eq. (\ref{eq:3.17})) and provided in Fig. \ref{fig:3.5} and Table \ref{tab:3.3}.

\begin{table}
\scriptsize
\centering 
\parbox{\columnwidth}{
\begin{tabular}{|c|c c c c c||c|c c c c c|}
\hline
$r$ (pc) & $B_0$ ($\mu$G) & $E$ (erg) & $\beta$ & $T$ (K) & $M_\text{J} \; (M_\odot)$ & $r$ (pc) & $B_0$ ($\mu$G) & $E$ (erg) & $\beta$ & $T$ (K) & $M_\text{J} \; (M_\odot)$\\
\hline
\hline
         &                &           &         &         & &          &                &           &         &         &\\
         &  1             & $10^{54}$ & 0.2     & 13.26   & 4.24 & &  1             & $10^{54}$ & 0.2     &  27.38  &  78.04\\
         &                &           & 0.4     &  2.20   & 0.33 & &                &           & 0.4     &   9.37  &  18.55\\
         &                &           & 0.8     &  0.06   & 0.0011 & &                &           & 0.8     &   1.10  &   0.62\\
         & 10             & $10^{54}$ & 0.2     & 13.26   & 4.51 & & 10             & $10^{54}$ & 0.2     &  27.38  & 185.69\\
         &                &           & 0.4     &  2.20   & 0.46 & &                &           & 0.4     &   9.37  &  98.27\\
         &                &           & 0.8     &  0.06   & 0.04 & &                &           & 0.8     &   1.10  &  58.19\\
20       &  1             & $10^{56}$ & 0.2     & 13.26   & 3.82 & 150      &  1             & $10^{56}$ & 0.2     &  27.38  &  77.54\\
         &                &           & 0.4     &  2.20   & 0.30 & &                &           & 0.4     &   9.37  &  18.44\\
         &                &           & 0.8     &  0.06   & 0.0010 & &                &           & 0.8     &   1.10  &   0.62\\
         & 10             & $10^{56}$ & 0.2     & 13.26   & 4.07 & & 10             & $10^{56}$ & 0.2     &  27.38  & 184.52\\
         &                &           & 0.4     &  2.20   & 0.41 & &                &           & 0.4     &   9.37  &  97.65\\
         &                &           & 0.8     &  0.06   & 0.04 & &                &           & 0.8     &   1.10  &  57.82\\
\hline
         &                &           &         &         & & &                &           &         &         &\\
         &  1             & $10^{54}$ & 0.2     & 18.44   & 15.89 & & 1              & $10^{54}$ & 0.2     & 42.24   & 466.48\\
         &                &           & 0.4     &  4.25   &  2.06 & &                &           & 0.4     & 22.30   & 218.09\\
         &                &           & 0.8     &  0.22   &  0.02 & &                &           & 0.8     &  6.22   &  28.84\\
         & 10             & $10^{54}$ & 0.2     & 18.44   & 19.90 & & 10             & $10^{54}$ & 0.2     & 42.24   &   5.65 $\times 10^{3}$\\
         &                &           & 0.4     &  4.25   &  4.30 & &                &           & 0.4     & 22.30   &   5.02 $\times 10^{3}$\\
         &                &           & 0.8     &  0.22   &  1.15 & &                &           & 0.8     &  6.22   &   4.35 $\times 10^{3}$\\
50       &  1             & $10^{56}$ & 0.2     & 18.44   & 15.43 & 500      &  1             & $10^{56}$ & 0.2     & 42.24   & 465.93 \\
         &                &           & 0.4     &  4.25   &  2.00 & &                &           & 0.4     & 22.30   & 217.83\\
         &                &           & 0.8     &  0.22   &  0.02 & &                &           & 0.8     &  6.22   &  28.80\\
         & 10             & $10^{56}$ & 0.2     & 18.44   & 19.32 & & 10             & $10^{56}$ & 0.2     & 42.24   &   5.64 $\times 10^{3}$\\
         &                &           & 0.4     &  4.25   &  4.17 & &                &           & 0.4     & 22.30   &   5.01 $\times 10^{3}$\\
         &                &           & 0.8     &  0.22   &  1.11 & &                &           & 0.8     &  6.22   &   4.34 $\times 10^{3}$\\
\hline
         &                &           &         &         & & &                &           &         &         &\\
         & 1              & $10^{54}$ & 0.2     & 23.66   & 43.29 & &  1             & $10^{54}$ & 0.2     & 54.21   &   1.41 $\times 10^{3}$\\
         &                &           & 0.4     &  7.00   &  8.22 & &                &           & 0.4     & 36.73   & 965.39\\
         &                &           & 0.8     &  0.61   &  0.17 & &                &           & 0.8     & 16.87   & 269.27\\
         & 10             & $10^{54}$ & 0.2     & 23.66   & 74.65 & & 10             & $10^{54}$ & 0.2     & 54.21   &   5.70 $\times 10^{4}$\\
         &                &           & 0.4     &  7.00   & 29.16 & &                &           & 0.4     & 36.73   &   5.54 $\times 10^{4}$\\
         &                &           & 0.8     &  0.61   & 13.64 & &                &           & 0.8     & 16.87   &   5.23 $\times 10^{4}$\\
100      &  1             & $10^{56}$ & 0.2     & 23.66   & 42.81 & 1000     &  1             & $10^{56}$ & 0.2     & 54.21   &   1.41 $\times 10^{3}$\\
         &                &           & 0.4     &  7.00   &  8.13 & &                &           & 0.4     & 36.73   & 964.96\\
         &                &           & 0.8     &  0.61   &  0.17 & &                &           & 0.8     & 16.87   & 269.15\\
         & 10             & $10^{56}$ & 0.2     & 23.66   & 73.82 & & 10             & $10^{56}$ & 0.2     & 54.21   &   5.70 $\times 10^{4}$\\
         &                &           & 0.4     &  7.00   & 28.83 & &                &           & 0.4     & 36.73   &   5.54 $\times 10^{4}$\\
         &                &           & 0.8     &  0.61   & 13.49 &          &                &           & 0.8     & 16.87   &   5.22 $\times 10^{4}$\\
\hline
\end{tabular}
}
\caption{Jeans mass ($M_\text{J}$) values for rotating molecular clouds with a density varying magnetic field.}
\label{tab:3.3}
\end{table}

\section{Star Formation}
\label{sec:3.4}
Let us summarise the overall scenario. Nuclear explosions near the galactic centre generate shock waves propagating in the ambient medium and compressing the gas into a thin, dense shell. Such a cooled and compressed high-density medium undergoes gravitational instability and produces fragments called molecular clouds. Stars are formed out of these molecular clouds due to gravitational collapse. The theory of Jeans instability well describes the collapse of molecular clouds and subsequent star formation. As per this theory, star formation is only possible for a threshold mass known as the Jeans mass ($M_\text{J}$), above which molecular clouds become unstable to gravitational collapse. The values of $M_\text{J}$ have been calculated To investigate the role of magnetic field on the formation of stars from shock-induced collapsing molecular clouds for the following cases -- (i) non-rotating molecular clouds with a constant magnetic field (see Table \ref{tab:3.1}), (ii) rotating molecular clouds with a constant magnetic field (see Table \ref{tab:3.2}) and (iii) rotating molecular clouds with a density-varying magnetic field (see Table \ref{tab:3.3}). These analyses lead to the following important observations:

\begin{enumerate}[label=(\roman*)]
\item The formation of field stars is preferred near the galactic centre ($r \sim$ 20 pc), whereas massive stars or starbursts are preferred at more considerable distances. So, the formation of a star cluster is improbable near the galactic centre. This star formation tendency is consistent with the radio and infrared observations near Sgr A$^{*}$ -- the supermassive black hole in the Galactic Centre of our Milky Way \cite{Yusef2015a, Yusef2015b}.

\item The temperature-density relationship gives another critical insight into the star formation process. As the temperature-density relationship steepens (i.e., $\beta$ rises), the temperature falls rapidly with increasing density, and the Jeans mass ($M_\textbf{J}$) decreases. Moreover, very small Jeans masses ($M_\textbf{J} \sim 10^{-2} M_\odot$) can result in the formation of Jupiter-like objects but not stars. Thus, star formation is preferred for a moderate rate of temperature change with density.

\item \citeauthor{Crutcher2010}, \citeyear{Crutcher2010} \cite{Crutcher2010} revised the scaling relation of magnetic fields in interstellar clouds from Zeeman observations - with exponent $\kappa \sim 0$ at densities less than $10^{3}$ cm$^{-3}$. The weak magnetic field (viz., $B_0 \sim 1 \mu$G) favours the formation of field stars, whereas the strong magnetic field (viz., $B_0 \sim 10 \mu$G) is suitable for bursts of stars, and the effect is enhanced at a more considerable distance (viz., $r \sim 1$ kpc).

\item First-generation star formation can result from high explosion energy, but a fragmentation hierarchy is not always required \cite{Hoyle1953}. The moderate explosion energy (viz., $E \sim 10 ^{54}$ erg) is better suited for star formation than the most vigorous explosion energy (viz., $E$ $\sim 10 ^{56}$ erg). More substantial shock pressure causes higher compression, which partially inhibits radiant energy transfer, preventing a smooth and continuous collapse.

\item Rotation always tends to stabilise a system \cite{Fall1980}. So, instability growth is only possible for slow rotation, i.e., supersonic radial shock should predominate over rotation to create a consistent environment for star formation. That's why $\Omega$ is only an order of a few km s$^{-1}$ kpc$^{-1}$ (viz., $\sim 4$ km s$^{-1}$ kpc$^{-1}$ \cite{Turner1984}). For the same reason, Jeans mass increases in the presence of rotation. This rotation of molecular clouds ($\Omega$) differs from the differential galactic rotation, which can be seen at a slightly greater distance $> 2$ kpc \cite{Chemin2015}. The kinematics and star formation mechanisms are very complex close to the Galactic Centre and still need to be clearly understood \cite{Kauffmann2016}.
\end{enumerate}

\section*{What's next?} 
This chapter focuses on how the magnetic field impacts the explosion-triggered star formation process in the rotating molecular clouds inside the central region of disc galaxies. Now, instabilities or chaos originating from the central explosion impact not only the star formation process but are also related to the formation and evolution of several galactic components like bars, spiral arms, etc. In the upcoming chapter, the relationship between the escape of these instabilities from the central barred region and the bar strengths is investigated in detail, impacting the structural evolution.

%% file: Chapter_4.tex
\chapter{Bar-driven Structure Formation}
\label{chap:4}

\section{Introduction}
\label{sec:4.1}
In Hubble's classification scheme of galaxies, central stellar bar structure is mainly observed in spirals (e.g., M58, NGC 1300, NGC 1365, etc.) and lenticulars (e.g., NGC 1460, NGC 1533, NGC 2787, etc.) galaxies. Lenticular and spirals are disc-supported systems, but their disc may not always support the bar structure. About $\frac{2}{3}$rd of the local disc galaxies possess stellar bars. Among them, only $\frac{1}{3}$rd have a prominent or strong bar feature, and the remaining $\frac{1}{3}$rd have an indeterminable type or weak bar feature \cite{Eskridge2000, Yoon2019}. The bars' presence in disc galaxies strongly depends on redshift, stellar mass, colour, and bulge prominence \cite{Sheth2008, Nair2010}. Other than disc galaxies, some irregular galaxies like the Large Magellanic Cloud (LMC) have an off-centred stellar bar \cite{Bekki2009, Monteagudo2018} and a bar-like feature has also been observed in the Small Magellanic Cloud (SMC), but whether that is a genuine bar or a temporary starburst region has not been confirmed yet \cite{Strantzalis2019}.  

\indent The stellar bar is one of the robust galactic structures. These dense stellar bodies revolve around the galactic core like a solid body rotating. The pattern speed of the bar is different than that of the disc. There are many theories behind the bar's origin \cite{Bournaud2002, Petersen2019, Polyachenko2020}. It is widely accepted that the galactic bars result from density waves radiating outward from the galactic core. Baryonic shocks are produced due to explosive phenomena near the galactic centre propagating in the ambient medium and creating dense shells of molecular clouds like a density wave. The role of shock-driven instabilities in a rotationally supported disc is not limited to inducing star formation inside the molecular clouds but also redistributing stellar trajectories as they evolve over time. These reshaped orbits produce a self-stabilising solid, dense stellar structure in the form of a bar \cite{Raha1991, Sellwood2016, Lokas2019}. Generally, disc galaxies mostly have a single bar structure embedded inside the bulge. However, with the advent of modern high-resolution imaging facilities, many double-barred galaxies (e.g., NGC 1291, NGC 1326, NGC 1543, etc.) have been discovered in recent years. In double-barred discs, the secondary bar (also known as the inner bar) is wrapped inside the larger primary bar \cite{Erwin2004, Debattista2006} in a nested manner. It is speculated that our Milky Way is also a double-barred system \cite{Nishiyama2006}. 

\indent From the viewpoint of the evolutionary dynamics of disc galaxies, the stellar bar is significant as it influences the entire stellar and gas dynamics inside the disc. Many realistic potential models have been used in the literature to model this robust component. Such potentials include three-dimensional spherical, homeoidal, triaxial, ellipsoidal potentials, etc. \cite{Ferrers1877, De1972, Dehnen2000, Caranicolas2002, Jung2015}. Among them, Ferrers' triaxial potential \cite{Ferrers1877} is the most commonly used bar potential. However, its functional form is very complex and also computationally very expensive to deal with. Potentials like the three-dimensional homeoidal potential \cite{De1972}, two-dimensional ad hoc potential \cite{Dehnen2000}, etc. have simpler functional forms than Ferrers', but they are still very rigorous to handle numerically. There are some simple, realistic bar models, too, like the two-dimensional anharmonic mass-model potential of \cite{Caranicolas2002}, the three-dimensional potential of \cite{Jung2015}, etc., which are less computationally expensive.

\indent Due to galactic bars' influence, some stellar orbits remain trapped inside the potential interior, while others escape from that inner boundary after sufficient time. This problem can suitably be modelled from the viewpoint of escape in open Hamiltonian dynamical systems \cite{Pfenniger1984, Contopoulos2004, Ernst2014, Jung2016}. An open Hamiltonian system is a system where, for energies above an escape threshold, the energy shell becomes non-compact and, as a result, a part of the stellar orbits explores (here from potential holes to saddles) an infinite part of the position space. Also, this system is time reversal invariant \cite{Jung2016}. For a conservative system, the Hamiltonian (or the total energy) is a constant of motion. In such a case, all the stellar orbits are confined inside the five-dimensional energy hypersurface of the six-dimensional phase space for a system with degree of freedom 3. According to their initial condition, the nature of stellar orbits is either regular or chaotic. Orbits with initial energies below the escape threshold remain trapped inside the potential interior and exhibit bounded motion (regular or chaotic). Again, orbits with initial energy above the escape threshold can exhibit bounded or escaping motion. Escape from the potential interior is possible only along the open zero velocity curves (also known as the escape channels). For bounded motions, there are chaotic orbits that do not escape within the predefined time interval and eventually escape to infinity. These orbits are known as trapped chaotic orbits. Such orbits in phase space make the orbital dynamics more geometrically complicated. For escaping motions, orbits are generally chaotic. Regular or chaotic classification of orbits is possible with the aid of many dynamical indicators, like Lyapunov exponents, the smaller alignment index (SALI), etc. Among them, the Lyapunov exponent is one effective dynamical indicator, and its working mechanism is quite simple \cite{Sandri1996}. Calculating Lyapunov exponent values of orbits is sufficient for binary classification purposes (regular or chaotic). It calculates the separation rate of two neighbouring trajectories during the specified time interval, and the corresponding maximal Lyapunov exponent (MLE) value is used to specify the dynamical nature of the orbits (regular or chaotic). Geometrically, MLE is the highest separation between two neighbouring trajectories starting from the same initial condition in a designated time interval. If the value of MLE is positive, it indicates that orbits are chaotic, while MLE $= 0$ indicates that orbits are periodic \cite{Strogatz2018}. To analyse the escape properties of the orbits in the vast sea of initial conditions of the phase space, one needs to visualise Poincaré surface section maps \cite{Birkhoff1927} in different two-dimensional phase planes. Under suitable physical conditions, escaping orbits further fuel the formation of spiral arms, i.e., the mechanism of bar-driven structure formation is an essential aspect of studying the orbital and escape dynamics of barred galaxies.

\indent The effects of the dynamical chaos of the stellar orbits behind the formation of the spiral arms have been extensively studied in the recent past \cite{Lindblad1947, Lynden1972, Patsis2012, Mestre2020}. The majority of the earlier studies were focused on the computation of the chaotic invariant manifolds of hyperbolic orbits around the escape saddles, which govern the general dynamics in the central barred region \cite{Romero2007, Sanchez2016, Efthymiopoulos2019}. Besides that, the role of these chaotic invariant manifolds in the subsequent structure formations has also been discussed. In this process, escaping stars leave tidal trails near the ends of the bar, forming patterns like spiral arms via non-axisymmetric perturbations. This is the overall framework about how the fate of escaping stars further relates with the subsequent structure formations \cite{Quillen2011, Onghia2013, Ernst2014, Jung2016}.

\indent In the upcoming sections of this chapter, the bar-driven structure formation process has been discussed from the viewpoint of stellar motions that may escape from the central region. Primarily, discussions revolve around the underlying relationship between the fate of escaping stars and the bar strength.

\section{Barred Galaxy Model}
\label{sec:4.2}
A four-component, three-dimensional gravitational model has been discussed here to study the orbital motion of stars inside the central barred region of disc galaxies. The model components are as follows: (i) a spherical bulge, (ii) a bar embedded inside the bulge, (iii) a flat disc, and (iv) a dark matter halo (or simply a dark halo). The entire model has been studied twice for the following bar strengths: (i) strong and (ii) weak. Also, the entire formalisation is done in Cartesian coordinates ($x, y, z$). Now, let $\Phi_\text{t}(x, y, z)$ and $\rho_\text{t}(x, y, z)$ be the gravitational potential and associated volume density terms for a test particle (star) of unit mass moving under the influence of the model's gravitational potential. The Poisson equation relates this potential-density pair as,
\begin{equation}
\label{eq:4.1}
\nabla^2 \Phi_\text{t}(x, y, z) = 4 \pi G \rho_\text{t}(x, y, z),
\end{equation}
where $G$ is the gravitational constant, again, let $\Phi_\text{B}$, $\Phi_\text{b}$, $\Phi_\text{d}$, $\Phi_\text{h}$ be the potentials for the galactic components like the bulge, bar, disc, and dark halo, respectively. So $\Phi_\text{t}(x, y, z)$ can be written as,

\begin{equation*}
\Phi_\text{t}(x,y,z) = \Phi_\text{B}(x,y,z) + \Phi_\text{b}(x,y,z) + \Phi_\text{d}(x,y,z) + \Phi_\text{h}(x,y,z).
\end{equation*}

\noindent Now, let $\vec{\Omega}_\text{b} \equiv (0, 0, \Omega_\text{b})$ be the pattern speed of the bar (in a clockwise sense along the $z$ - axis). Then in the rotating reference frame of the bar, the effective potential ($\Phi_\text{eff}$) and Hamiltonian ($H$) of the system are as follows:
\begin{equation}
\label{eq:4.2}
\Phi_\text{eff}(x,y,z) = \Phi_\text{t}(x,y,z) - \frac{1}{2} \Omega_\text{b}^2 (x^2 + y^2),
\end{equation}

\begin{equation}
\label{eq:4.3}
H = \frac{1}{2}(p_x^2 + p_y^2 + p_z^2) + \Phi_\text{t}(x,y,z) - \Omega_\text{b} L_z,
\end{equation}
where $\vec{r} \equiv (x,y,z)$, $\vec{p} \equiv (p_x,p_y,p_z)$ and $\vec{L} \; (= \vec{r} \times \vec{p}) \equiv (0,0,L_z = x p_y - y p_x)$ are the position, linear momentum and angular momentum of the test particle at time $t$ respectively. This is a conservative system, and for that, $H = E$ (the system's total energy). Now, the governing equations, i.e., Hamilton's equations of motion, are as follows:
\begin{equation}
\label{eq:4.4}
\begin{split}
\dot{x}  = p_x + \Omega_\text{b} y, \; 
\dot{y} = p_y - \Omega_\text{b} x, \;
\dot{z}  = p_z,\\
\dot{p}_x = - \frac{\partial \Phi_\text{t}}{\partial x} + \Omega_\text{b} p_y, \;
\dot{p}_y  = - \frac{\partial \Phi_\text{t}}{\partial y} - \Omega_\text{b} p_x, \;
\dot{p}_z = - \frac{\partial \Phi_\text{t}}{\partial z},
\end{split}
\end{equation} 
where $`\cdot$' represents the time derivative ($\frac{\mathrm{d}}{\mathrm{dt}}$). The Lagrangian (or equilibrium) points of the system are the solutions of,
\begin{equation}
\label{eq:4.5}
\frac{\partial \Phi_\text{eff}}{\partial x} = 0, \;  
\frac{\partial \Phi_\text{eff}}{\partial y} = 0, \;
\frac{\partial \Phi_\text{eff}}{\partial z} = 0.
\end{equation}

In the upcoming parts of this chapter, this barred galaxy model has been analysed separately for the following two bar profiles: (i) strong bar (model $1$) and (ii) weak bar (model $2$). This helps us investigate the underlying relationship between the bar-driven escaping mechanism and bar strength.

\subsection{Gravitational potentials}
\label{sec:4.2.1}
The potential forms of the bulge, bar, disc and dark halo are as follows:
\begin{itemize}[leftmargin=*]
\item Bulge: The old red stars dominate the central bulge component, which is spherically symmetric. This can be modelled using Plummer potential \cite{Plummer1911}, $$\Phi_\text{B}(x,y,z) = - \frac{G M_\text{B}}{\sqrt{x^2 + y^2 + z^2 + c_\text{B}^2}},$$ where $M_\text{B}$ is the bulge mass and $c_\text{B}$ is the scale length.

\item Bar: A bar-like feature has been found in the central region of many disc galaxies. This component is non-axisymmetric and can be modelled using one of the following potentials depending on its strength.  
\begin{itemize}[leftmargin=*]
\item Strong bar: Strong bars (i.e., bar density is of cuspy type; see Fig. \ref{fig:4.2a}) can be modelled using the three-dimensional extension of an anharmonic mass-model bar potential \cite{Caranicolas2002}, $$\Phi_\text{b}(x,y,z) = - \frac{G M_\text{b}}{\sqrt{x^2 + \alpha^2 y^2 + z^2 + c_\text{b}^2}},$$ where $M_\text{b}$ is the bar mass, $\alpha$ is the flattening parameter and $c_\text{b}$ is the scale length.

\item Weak bar: Weak bars (i.e., bar density is of flat type; see Fig. \ref{fig:4.2a}) can be modelled using the Zotos bar potential \cite{Jung2015}, $$\Phi_\text{b}(x,y,z) = \frac{G M_\text{b}}{2 a} \ln \left(\frac{x - a + \sqrt{{(x - a)}^2 + y^2 + z^2 + {c_\text{b}^2}}}{x + a + \sqrt{{(x + a)}^2 + y^2 + z^2 + {c_\text{b}^2}}}\right),$$ where $a$ is the semi-major axis length.
\end{itemize} 

\item Disc: The Miyamoto and Nagai potential can model the flattened axisymmetric disc component \cite{Miyamoto1975}, $$\Phi_\text{d}(x,y,z) = - \frac{G M_\text{d}}{\sqrt{x^2 + y^2 + (k + \sqrt{h^2 + z^2})^2}},$$ where $M_\text{d}$ is the disc mass and $k$, $h$ are the horizontal and vertical scale lengths respectively.

\item Dark halo: For modelling of the dark halo component a logarithmic potential \cite{Zotos2012}, $$\Phi_\text{h}(x,y,z) = \frac{v_0^2}{2} \; \ln(x^2 + \beta^2 y^2 + z^2 + c_\text{h}^2)$$ can be used, where $v_0$ is the circular velocity, $\beta$ is the flattening parameter and $c_\text{h}$ is the scale length.
\end{itemize}

\subsection{Parameter values}
\label{sec:4.2.2}
Without loss of any generality, let us set $G = 1$ and adopt the following scaling relations \cite{Jung2016}: unit of length - $1$ kpc, unit of mass - $2.325 \times 10^7 M_\odot$, unit of time - $0.9778 \times 10^8$ yr, unit of velocity - $10$ km $\text{s}^{-1}$, unit of angular momentum per unit mass - $10$ km $\text{s}^{-1}$ $\text{kpc}^{-1}$ and unit of energy per unit mass - $100$ $\text{km}^2$ $\text{s}^{-2}$ are used. Values of all physical parameters as per these scaling relations \cite{Zotos2012, Jung2016} are given in Table \ref{tab:4.1}. 
\begin{table}
\centering	
\begin{tabular}{|c|c||c|c|}
\hline
Parameter         & Value & Parameter    & Value\\
\hline
\hline
$M_\text{B}$      & 400   & $M_\text{d}$ & 7000\\
$c_\text{B}$	  & 0.25  & $k$          & 3\\
$M_\text{b}$      & 3500  & $h$          & 0.175\\
$\alpha$          & 2     & $v_0$        & 15\\
$a$               & 10    & $\beta$      & 1.3\\ 
$c_\text{b}$      & 1     & $c_\text{h}$ & 20\\
$\Omega_\text{b}$ & 1.25  &              &\\ 
\hline
\end{tabular}
\caption{Physical parameter values.}
\label{tab:4.1}
\end{table}

\begin{figure}
\centering
\subfigure[Model $1$ (strong bar)]{\label{fig:4.1a}\includegraphics[width=0.49\columnwidth]{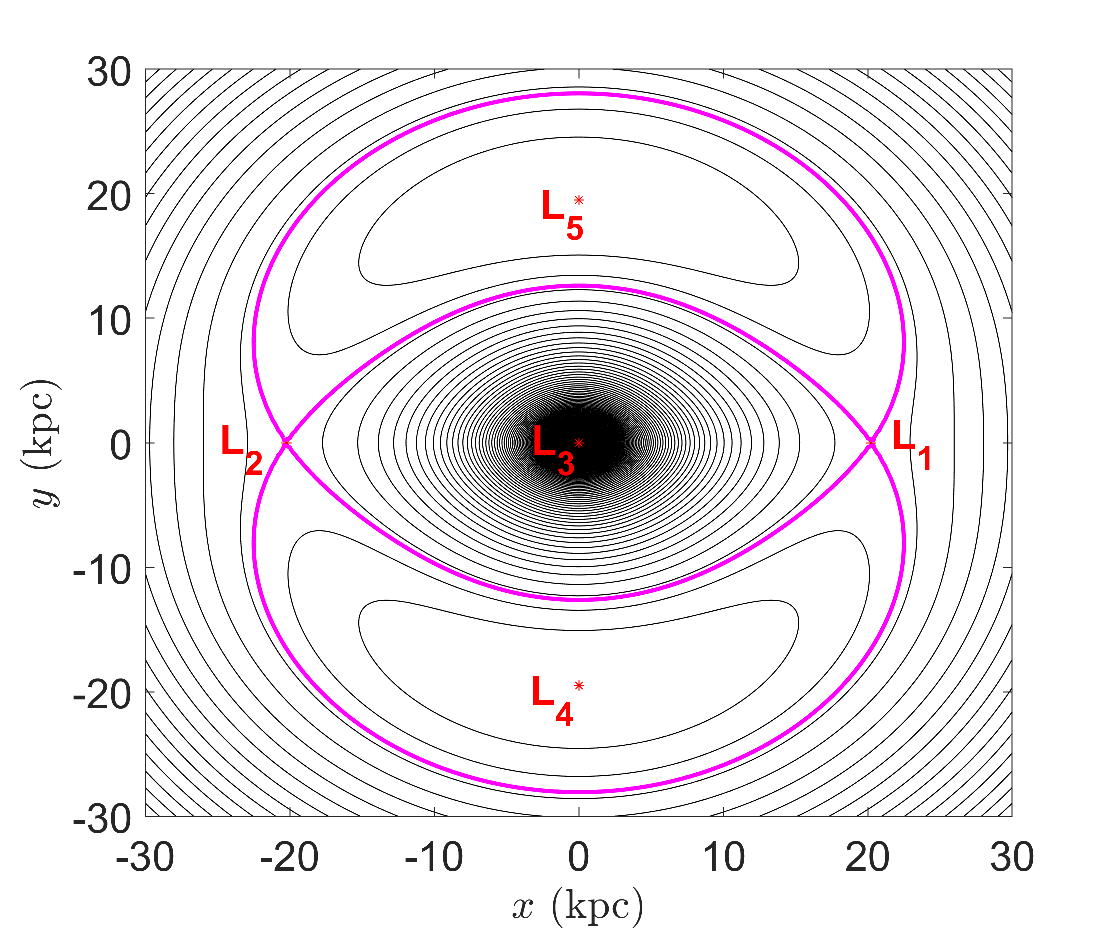}}
\subfigure[Model $2$ (weak bar)]{\label{fig:4.1b}\includegraphics[width=0.49\columnwidth]{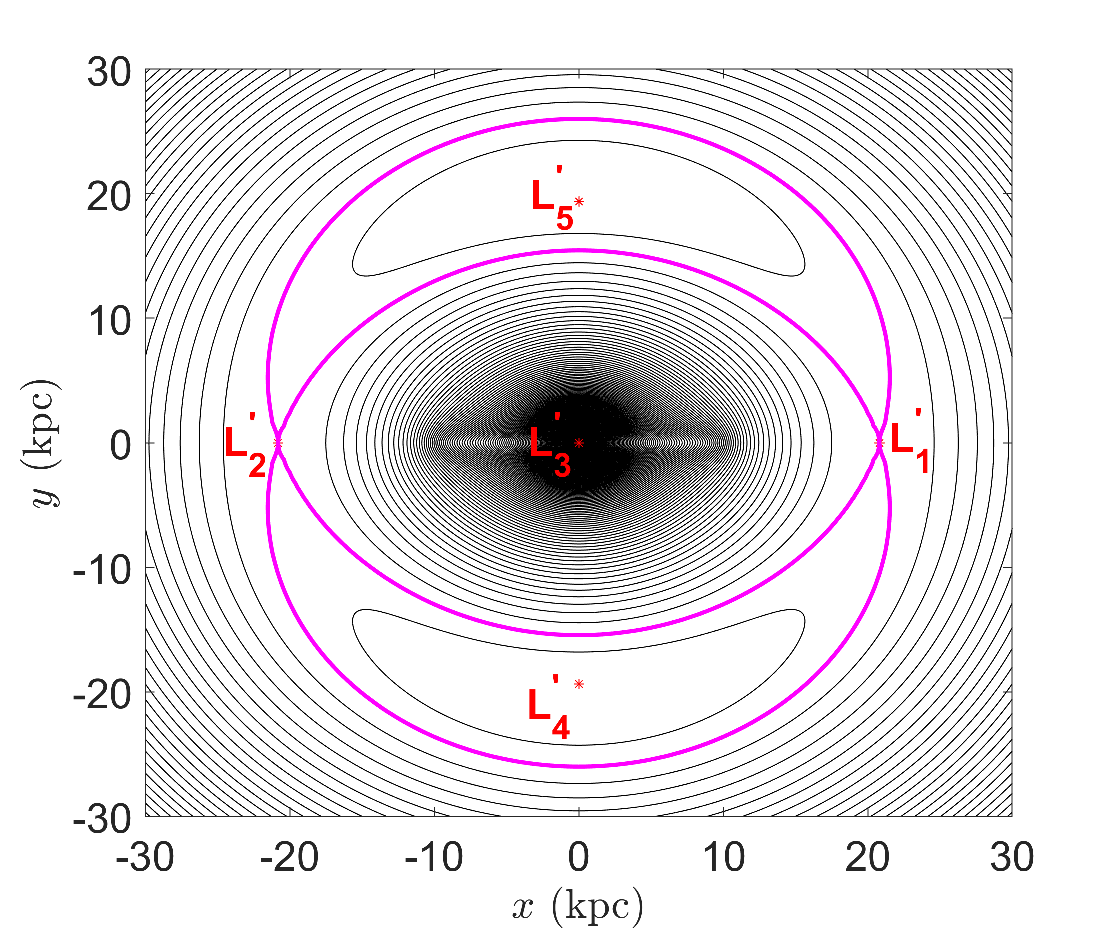}}
\caption{The isoline contours of $\Phi_\text{eff}(x,y,z)$ in the $x - y$ plane for $z = 0$, where the locations of the five Lagrangian points are marked in red, while the contours in magenta correspond to the energy values of the index-1 saddle point $L_1$ (and $L_1^{'}$).}
\label{fig:4.1}
\end{figure}

Here, the system has five Lagrangian points (solutions of the system of Eqs. (\ref{eq:4.5})), namely $L_1$, $L_2$, $L_3$, $L_4$, $L_5$ for model $1$, and $L_1^{'}$, $L_2^{'}$, $L_3^{'}$, $L_4^{'}$, $L_5^{'}$ for model $2$. The Lagrangian point locations with their types for both models are given in Fig. \ref{fig:4.1} and Table \ref{tab:4.2}.

\begin{table}	
\centering
\begin{tabular}{|c|c|c|c|}
\hline  
Lagrangian Point            & Lagrangian Point                & Type\\
(Model $1$)                 & (Model $2$)                     &\\                 
\hline
\hline
$L_1 \equiv (20.2311,0,0)$  & $L_1^{'} \equiv (20.8298,0,0)$  & Index-1 Saddle\\
$L_2 \equiv (-20.2311,0,0)$ & $L_2^{'} \equiv (-20.8298,0,0)$ & Index-1 Saddle\\
$L_3 \equiv (0,0,0)$        & $L_1^{'} \equiv (0,0,0)$        & centre\\
$L_4 \equiv (0,-19.4982,0)$ & $L_4^{'} \equiv (0,-19.3672,0)$ & Index-2 Saddle\\
$L_5 \equiv (0,19.4982,0)$  & $L_5^{'} \equiv (0,19.3672,0)$  & Index-2 Saddle\\
\hline
\end{tabular}
\caption{Lagrangian point locations and their types.}
\label{tab:4.2}
\end{table} 

\noindent Due to the structural symmetry, it is apparent that the energy values, i.e., the values of the Jacobi integral of motion at $L_1$ (or $L_1^{'}$) and $L_2$ (or $L_2^{'}$), i.e., $E_{L_1}$ (or $E_{L_1^{'}}$) and $E_{L_2}$ (or $E_{L_2^{'}}$) respectively, are identical. Similarly, those values at $L_4$ (or $L_4^{'}$) and $L_5$ (or $L_5^{'}$), i.e., $E_{L_4}$ (or $E_{L_4^{'}}$) and $E_{L_5}$ (or $E_{L_5^{'}}$) respectively, are identical (see Table \ref{tab:4.3}).

\begin{table}
\centering
\begin{tabular}{|c|c|c|}
\hline
Energy Value                    & Energy Value\\
(Model $1$)                     & (Model $2$)\\
\hline
\hline
$E_{L_1} / E_{L_2} = -100.8231$ & $E_{L_1^{'}} / E_{L_2^{'}} = -116.4614$\\
$E_{L_3} = -6630.6846$          & $E_{L_3^{'}} = -4180.0627$\\
$E_{L_4} / E_{L_5} = 20.2140$   & $E_{L_4^{'}} / E_{L_5^{'}} = -60.7659$\\
\hline
\end{tabular}
\caption{Energy values at the Lagrangian points.}
\label{tab:4.3}
\end{table} 

\noindent The nature of stellar orbits in different energy ranges is given as follows:
\begin{enumerate}[label=(\roman*)]
\item $E_{L_3} \leq E < E_{L_1}$ (or $E_{L_3^{'}} \leq E < E_{L_1^{'}}$): In this energy range, stellar orbits exhibit bounded motion inside the barred region.
\item $E = E_{L_1}$ (or $E = E_{L_1^{'}}$): This is the threshold energy for orbital escapes through the bar ends.
\item $E > E_{L_1}$ (or $E > E_{L_1^{'}}$): In this energy range, escape is possible for stellar orbits from the barred region through the two symmetrical exit channels near ${L_1}$ and ${L_2}$. Also, orbits are bounded or escaped depending on their initial starting point.
\end{enumerate}

\subsection{Strong versus weak bar}
\label{sec:4.2.3}
\begin{figure}
\centering
\subfigure[Bar density ($\rho_\text{b}$) versus $R = \sqrt{x^2 + y^2}$ \newline for $z = 0$]{\label{fig:4.2a}\includegraphics[width=0.49\columnwidth]{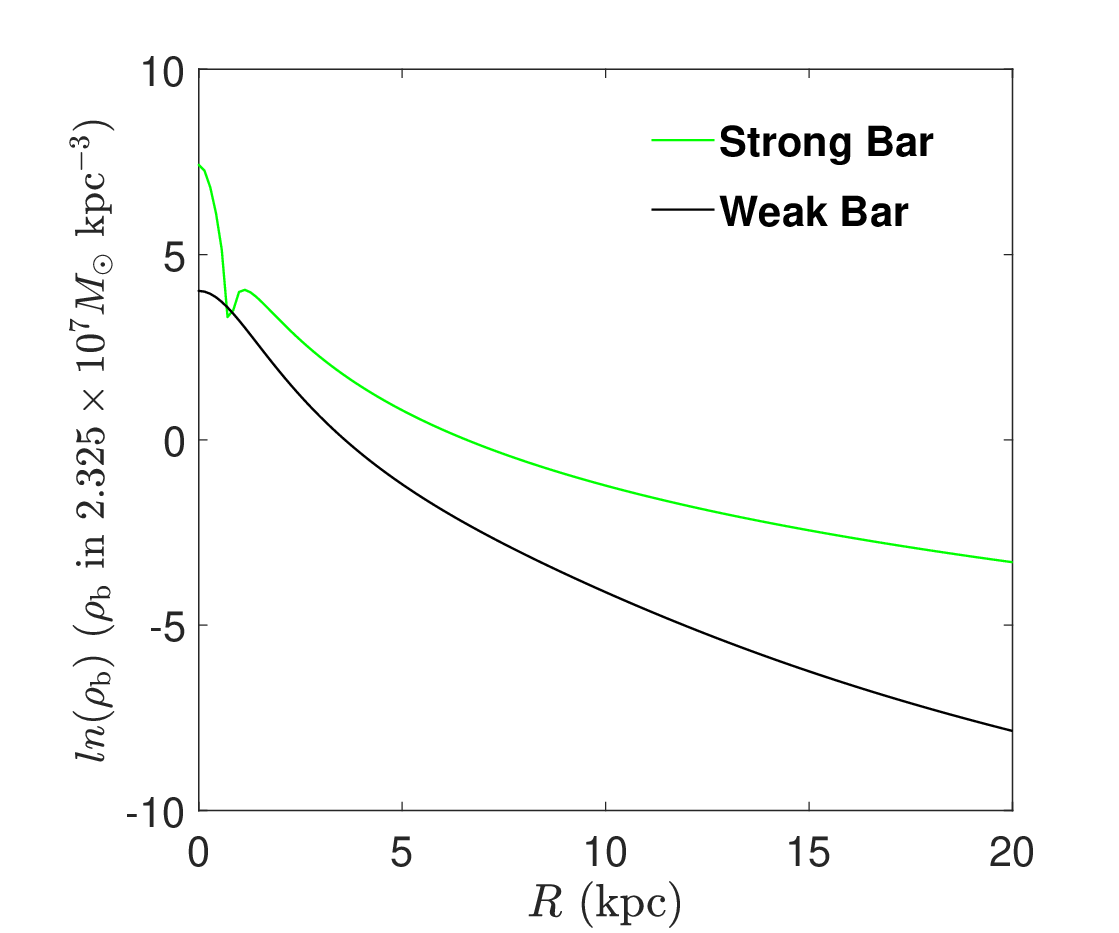}}
\subfigure[Rotation curve]{\label{fig:4.2b}\includegraphics[width=0.49\columnwidth]{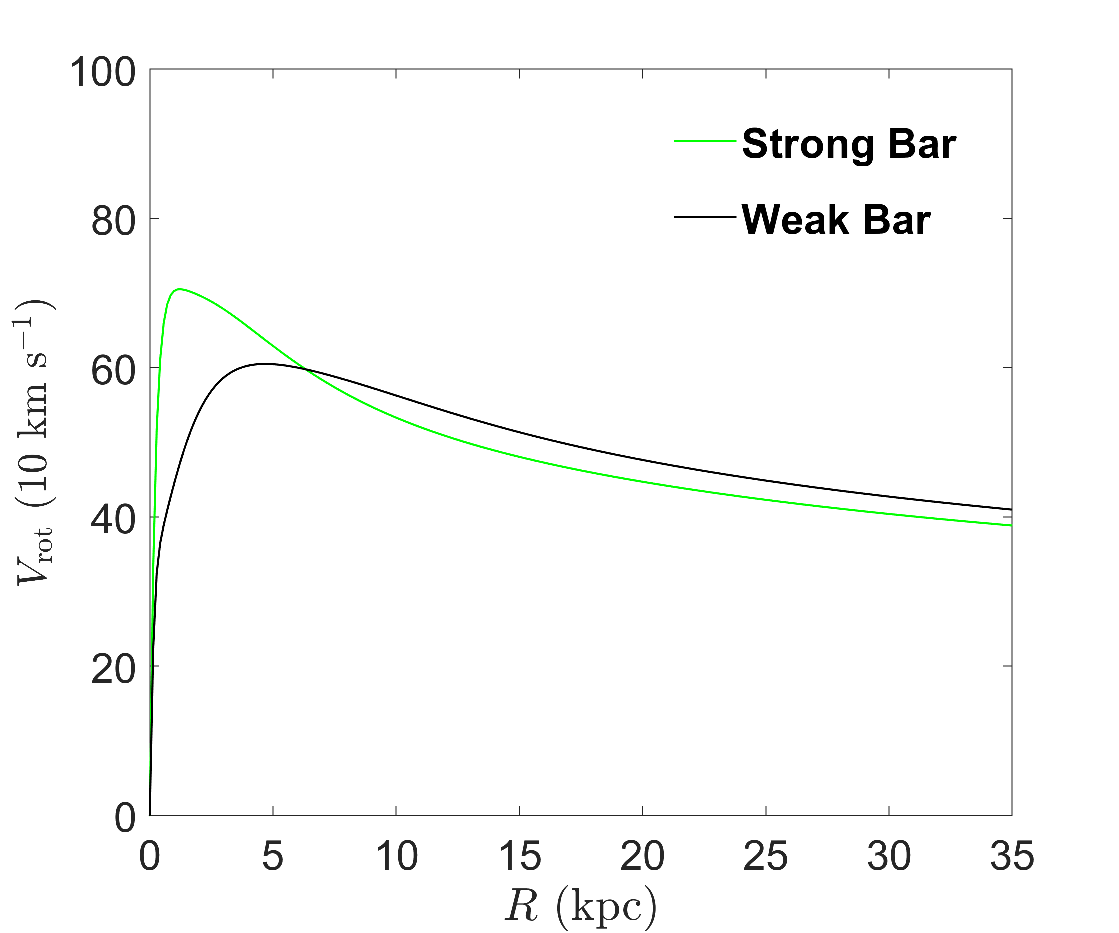}}
\subfigure[Radial force ($F_R$) versus $R = \sqrt{x^2 + y^2}$ \newline for $z = 0$]{\label{fig:4.2c}\includegraphics[width=0.49\columnwidth]{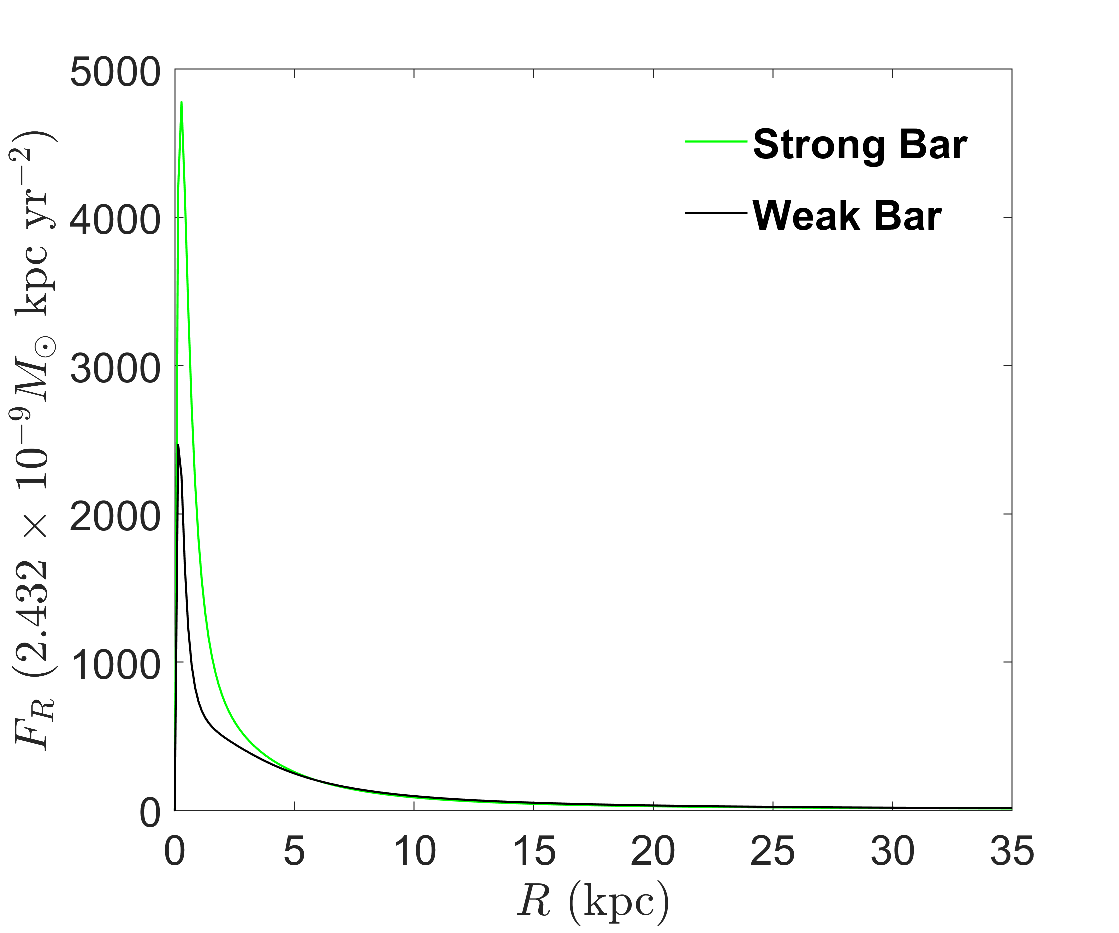}}
\subfigure[Tangential force ($F_\theta$) versus $R = \sqrt{x^2 + y^2}$ for $z = 0$]{\label{fig:4.2d}\includegraphics[width=0.49\columnwidth]{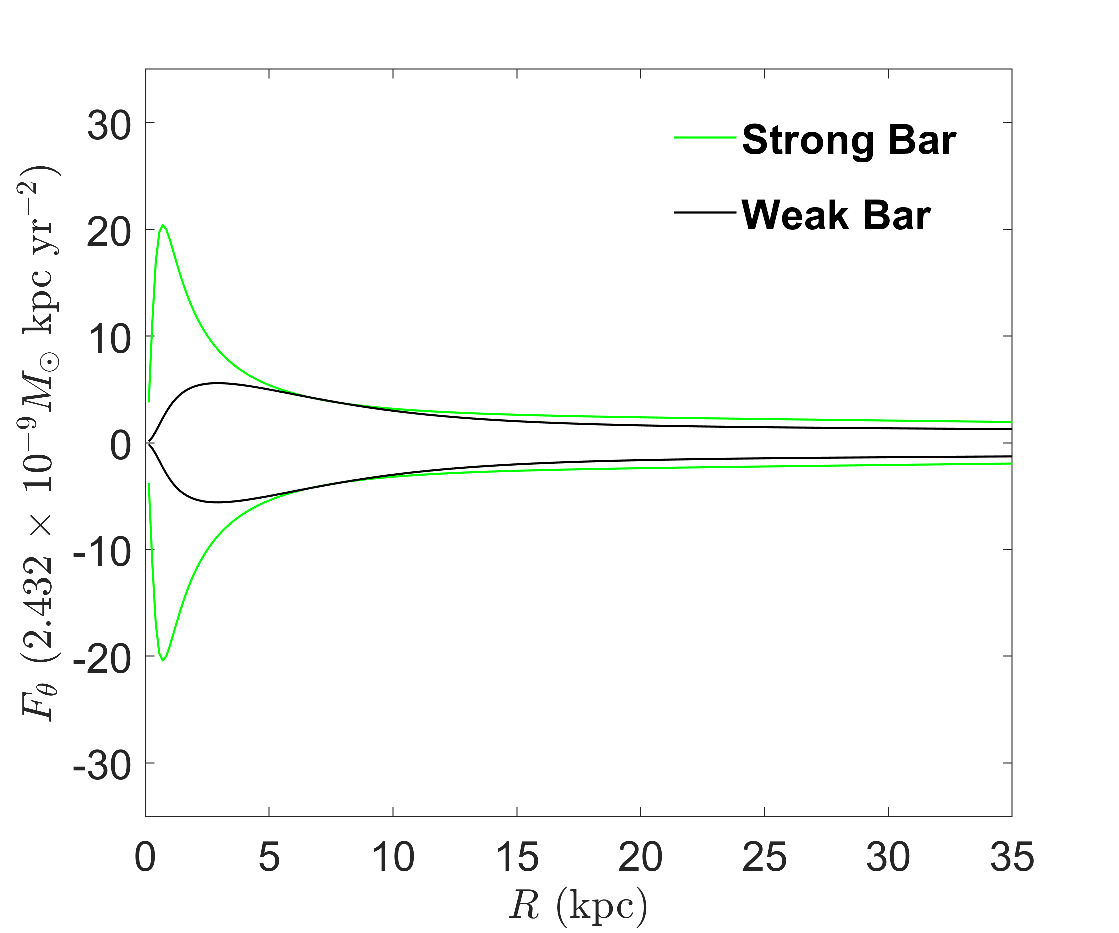}}
\caption{Characteristics: Model $1$ (strong bar) versus model $2$ (weak bar).}
\label{fig:4.2}
\end{figure}

Here are some comparisons between the bar types given in the two models.
\begin{itemize}[leftmargin=*]
\item Bar density: The distribution of bar density ($\rho_\text{b}$) for $z = 0$ is shown in Fig. \ref{fig:4.2a}. This shows that the density of the strong bar profile is way cuspier than that of the weak bar profile. Such strong bars are mainly observed in early-type disc galaxies.

\item Rotation curve: The rotation curve ($V_{\text{rot}} = \sqrt{R \frac{\partial \Phi_\text{t}}{\partial R}}$ versus $R =\sqrt{x^2 + y^2}$ for $z = 0$) is given in Fig. \ref{fig:4.2b}. This shows that the rotational velocities of the strong bar profile are higher inside the bulge than the weak bar profile, while the reverse trend is observed outside the bulge. Higher kinetic energy transport mechanisms within the bulge of massive bars are the reason for such excess rotational velocities.

\item Radial force: The variation of the radial force component ($F_R = \frac{\partial \Phi_\text{t}}{\partial R}$) with $R \; (= \sqrt{x^2 + y^2})$ for $z = 0$ is given in Fig. \ref{fig:4.2c}. This indicates that the radial force components are significant only inside the bulge and that these distributions become steeper as $R \to 0$. This steepness died out completely outside the barred region. This steepness is higher for the strong bar profile due to the presence of a massive bar than for the weak bar profile. Again, for the weak bar profile, radial force near the galactic centre has less strength than the strong bar profile, and that is why it has less tendency to escape through the bar ends.

\item Tangential force: The variation of the tangential force component ($F_\theta = \frac{1}{R} \frac{\partial \Phi_\text{t}}{\partial \theta}$) with $R \; (= \sqrt{x^2 + y^2})$ for $z = 0$ is given in Fig. \ref{fig:4.2d}, where $\theta = tan^{-1} \frac{y}{x}$. This indicates that the tangential force component of the strong bar profile is significantly near the central bulge region and dies out for larger values of $R$ than the weak bar profile, while outside, the trends are almost similar for both bar profiles. For the weak bar profile, tangential force strength is higher than the strong bar profile in the region between the bulge and the bar, and that is why, for the weak bar profile, orbits tend to form structures within the disc.
\end{itemize}

\section{Phase Space Analysis}
\label{sec:4.3}
Let us set $z = 0 = p_z$ to study the stellar orbital and escape dynamics along the plane of the bar. In the energy domain: $E \geq E_{L_1}$ (or $E_{L_1^{'}}$), stellar orbits may escape through the symmetrical exit channels that exist near the bar ends, i.e., $L_1$, $L_2$ (or $L_1^{'}$, $L_2^{'}$) (see Fig. \ref{fig:4.1}). Now, the effective potential term (see Eq. (\ref{eq:4.2})) is symmetric about the $y$ axis and $E_{L_1} = E_{L_2}$, where $E_{L_1}$ and $E_{L_2}$ denote the energy values of $L_1$ and $L_2$, respectively. Hence, studying the dynamics near either $L_1$ or $L_2$ is sufficient to analyse the system. Now, the escape of stars from the central barred region is only possible in the energy range $E > E_{L_1}$, and for $E \le E_{L_1}$, orbits are bounded inside the central barred region. To investigate the escaping motion, the system of differential equations, i.e., Eqs. (\ref{eq:4.4}) is integrated over a time scale of $10^2$ time units, equivalent to 10 Gyr, the typical age of the bars \cite{James2018}. The $\tt{ode45}$ package of $\tt{MATLAB}$ is used for this orbit integration. Further, in order to simplify the calculations, the following dimensionless energy parameter ($C$) \cite{Jung2016} is adopted,
\begin{equation}
\label{eq:4.6}
C = \frac{E_{L_1} (\text{or} \; E_{L_1^{'}}) - E}{E_{L_1} (\text{or} \; E_{L_1^{'}})} = \frac{E_{L_2} (\text{or} \; E_{L_2^{'}}) - E}{E_{L_2} (\text{or} \; E_{L_2^{'}})}.
\end{equation} 
Hence, orbital escapes are possible for $C > 0$. The chaotic nature of stellar orbits is determined using the chaos detector MLE \cite{Strogatz2018}, which is defined in Eq. (\ref{eq:2.16}), where the initial separation between two infinitesimal trajectories is taken as $\delta x(t_0) = 10^{-8}$.

\subsection{Orbital maps}
\label{sec:4.3.1}
To study the nature of orbits near the Lagrangian point $L_1$, an initial condition $x_0 = 5$, $y_0 = 0$, $p_{x_0} = 15$, where $p_{y_0}$ is evaluated from Eq. (\ref{eq:4.3}) and the corresponding trajectories in the $x - y$ plane are drawn for several values of $C>0$ (see Figs. \ref{fig:4.3} and \ref{fig:4.4}). Any other initial condition in the suitable neighbourhoods of the aforesaid initial condition has followed a similar trend.

\begin{itemize}[leftmargin=*]
\item Model $1$: In Fig. \ref{fig:4.3}, stellar orbits in the $x - y$ plane have been plotted for values $C = 0.01$ and $0.1$, respectively, with the initial condition: $(x_0,y_0,p_{x_0},p_{y_0}) \equiv (5,0,15,p_{y_0})$, where $p_{y_0}$ value is evaluated from Eq. (\ref{eq:4.3}). In both cases, escaping chaotic orbits are observed. Again, for the initial condition: $(x_0,y_0,p_{x_0},p_{y_0}) \equiv (-5,0,15,p_{y_0})$, non-escaping retrograde quasi-periodic rosette orbits are observed for both $C = 0.01$ and $0.1$. Now, for each figure, the associated MLE value is calculated from Eq. (\ref{eq:2.16}) and listed in Table \ref{tab:4.4}.
\begin{figure}
\centering
\subfigure[$C = 0.01$ and $(x_0,y_0,p_{x_0})$ $\equiv (5,0,15)$]{\label{fig:4.3a}\includegraphics[width=0.48\textwidth]{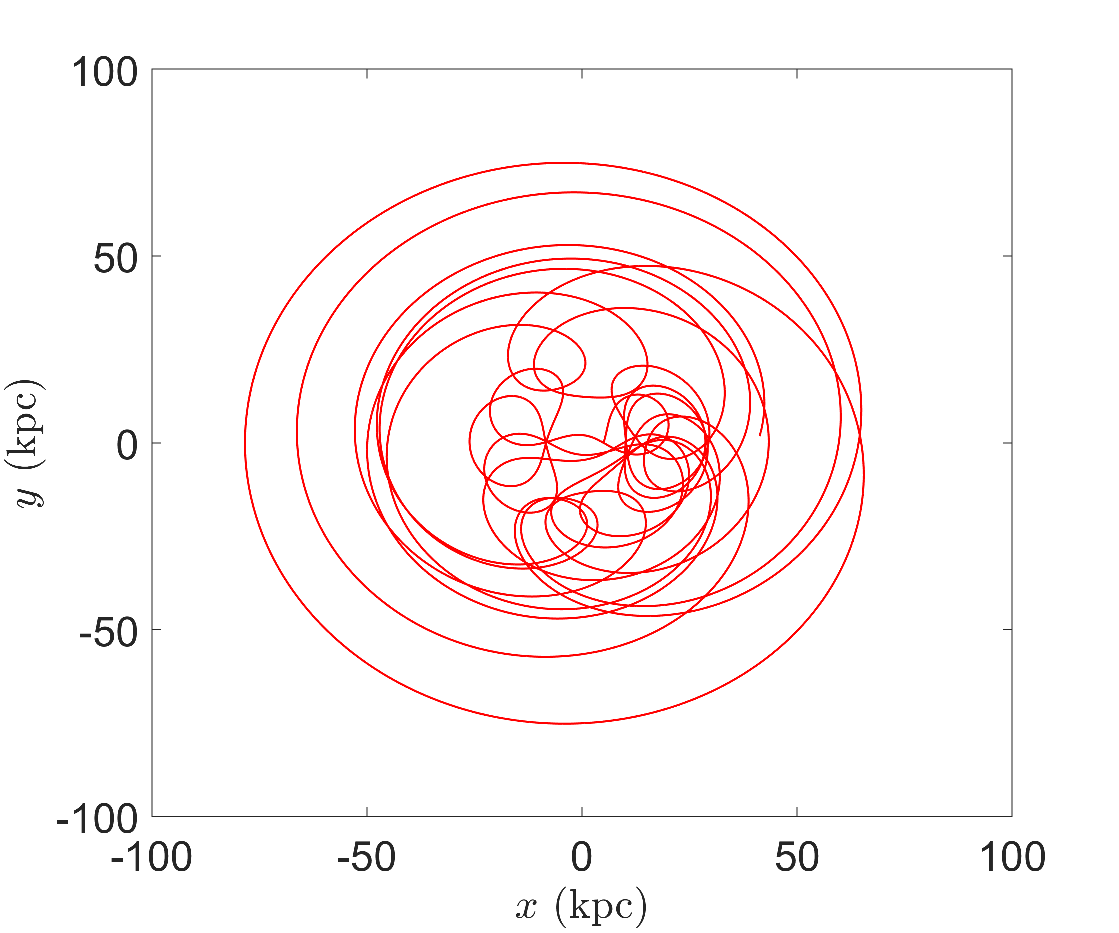}}
\subfigure[$C = 0.1$ and $(x_0,y_0,p_{x_0})$ $\equiv (5,0,15)$]{\label{fig:4.3b}\includegraphics[width=0.48\textwidth]{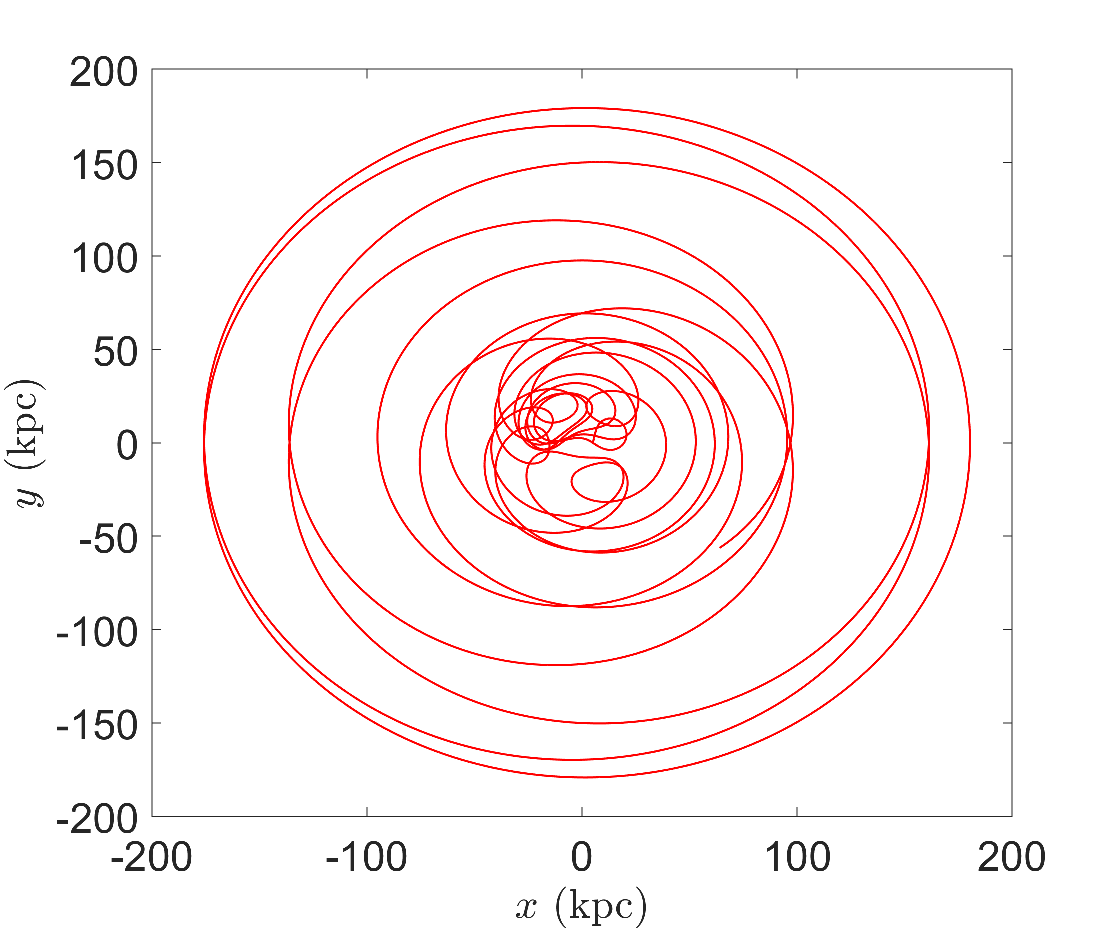}}
\subfigure[$C = 0.01$ and $(x_0,y_0,p_{x_0})$ $\equiv (-5,0,15)$]{\label{fig:4.3c}\includegraphics[width=0.48\textwidth]{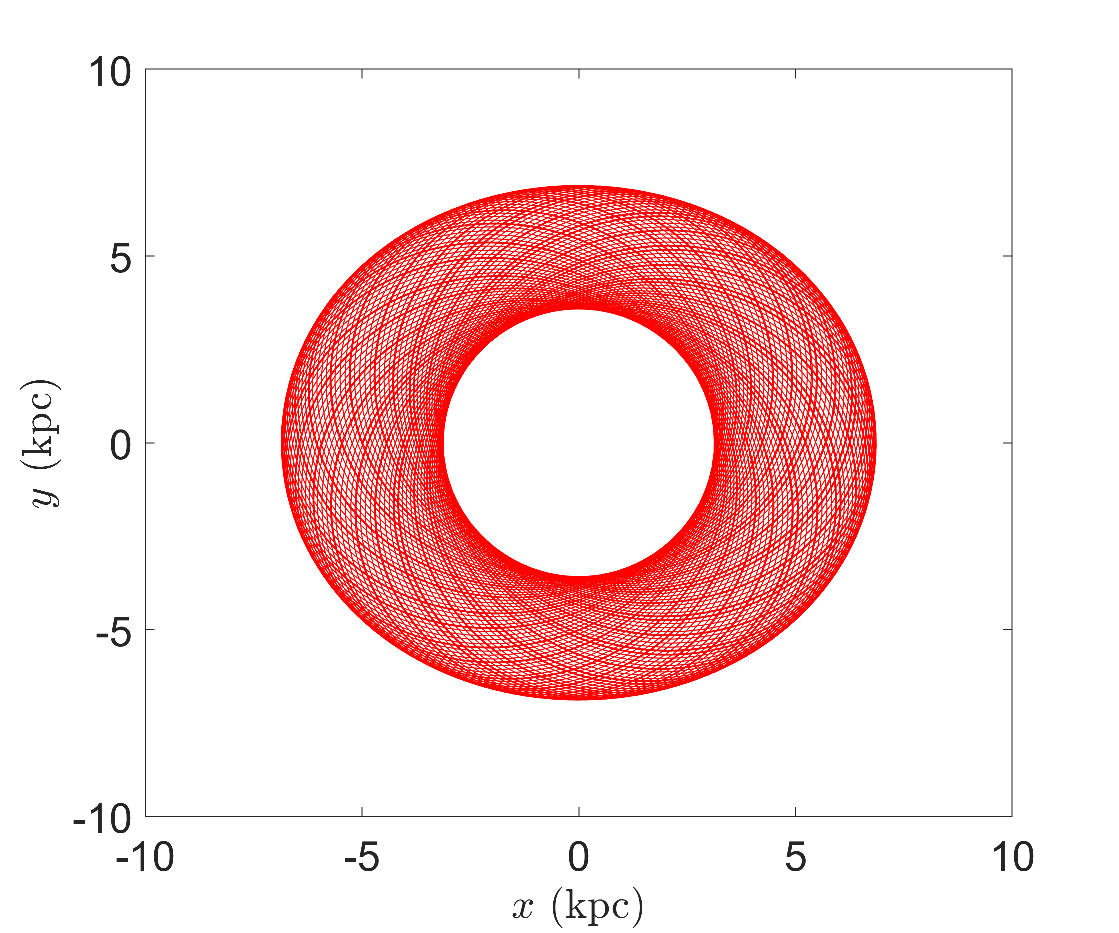}}
\subfigure[$C = 0.1$ and $(x_0,y_0,p_{x_0})$ $\equiv (-5,0,15)$]{\label{fig:4.3d}\includegraphics[width=0.48\textwidth]{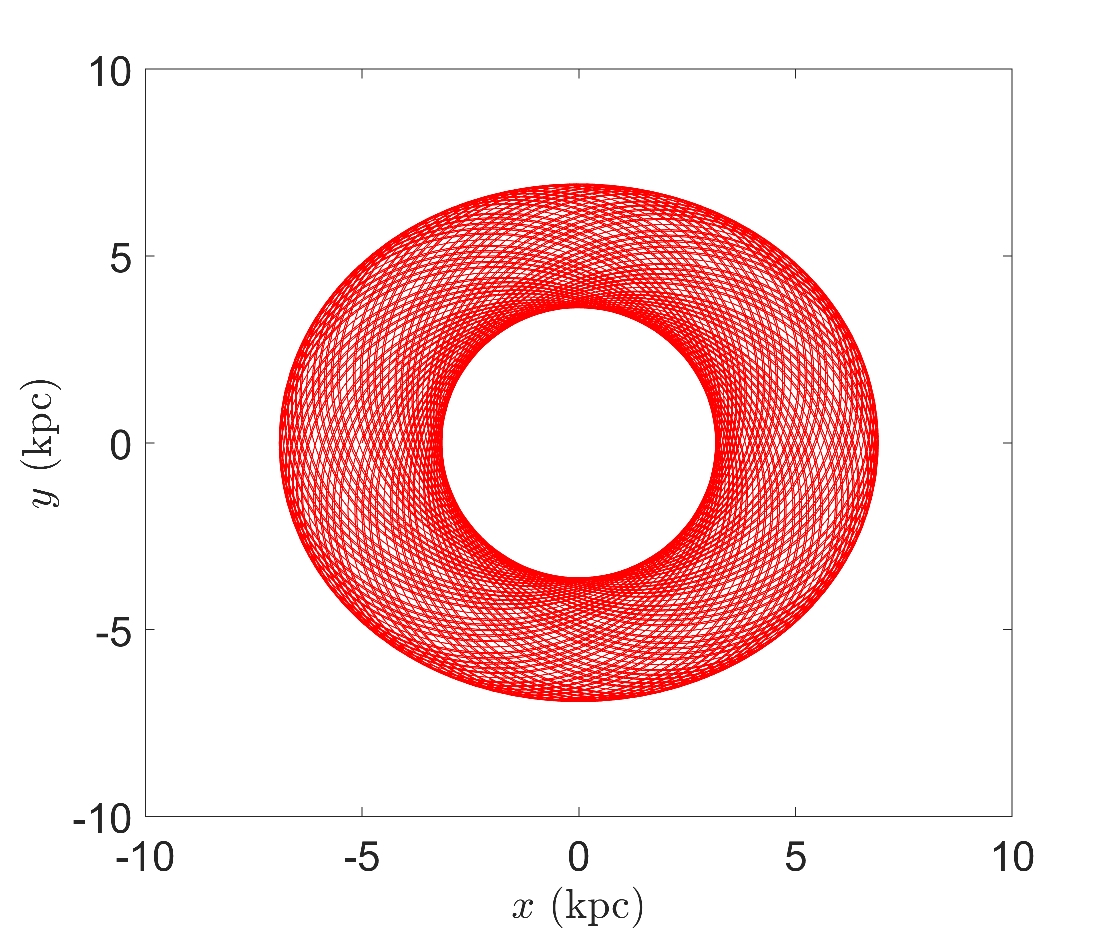}}
\caption{Model $1$: Orbits in the $x - y$ plane, where $p_{y_0}$ value is evaluated from Eq. (\ref{eq:4.3}); (a)-(b) escaping chaotic orbit, (c)-(d) non-escaping retrograde quasi-periodic rosette orbit.}
\label{fig:4.3}
\end{figure}

\begin{table}
\centering
\begin{tabular}{|c|c|c|}
\hline
Initial Condition                    & $C$    & MLE\\
\hline
\hline
$(x_0,y_0,p_{x_0}) \equiv (5,0,15)$  & $0.01$ & $0.1804$\\
                                     & $0.1$  & $0.1909$\\
\hline
$(x_0,y_0,p_{x_0}) \equiv (-5,0,15)$ & $0.01$ & $0.0854$\\
                                     & $0.1$  & $0.0895$\\
\hline
\end{tabular}
\caption{Model $1$: MLE value for different values of $C$, where $p_{y_0}$ value is evaluated from Eq. (\ref{eq:4.3}).}
\label{tab:4.4}
\end{table}

\item Model $2$: Similar to model $1$, here also in Fig. \ref{fig:4.4}, stellar orbits in the $x - y$ plane have been plotted for values $C = 0.01$ and $0.1$, respectively, with the initial condition: $(x_0,y_0,p_{x_0},p_{y_0}) \equiv (5,0,15,p_{y_0})$, where $p_{y_0}$ value is evaluated from Eq. (\ref{eq:4.3}). Non-escaping chaotic orbits are observed in both cases. Again, for the initial condition: $(x_0,y_0,p_{x_0},p_{y_0}) \equiv (-5,0,15,p_{y_0})$, non-escaping retrograde quasi-periodic rosette orbits are observed for both $C = 0.01$ and $0.1$. Now, for each figure, the associated MLE value is calculated from Eq. (\ref{eq:2.16}) and listed in Table \ref{tab:4.5}.
\begin{figure}
\centering
\subfigure[$C = 0.01$ and $(x_0,y_0,p_{x_0})$ $\equiv (5,0,15)$]{\label{fig:4.4a}\includegraphics[width=0.48\textwidth]{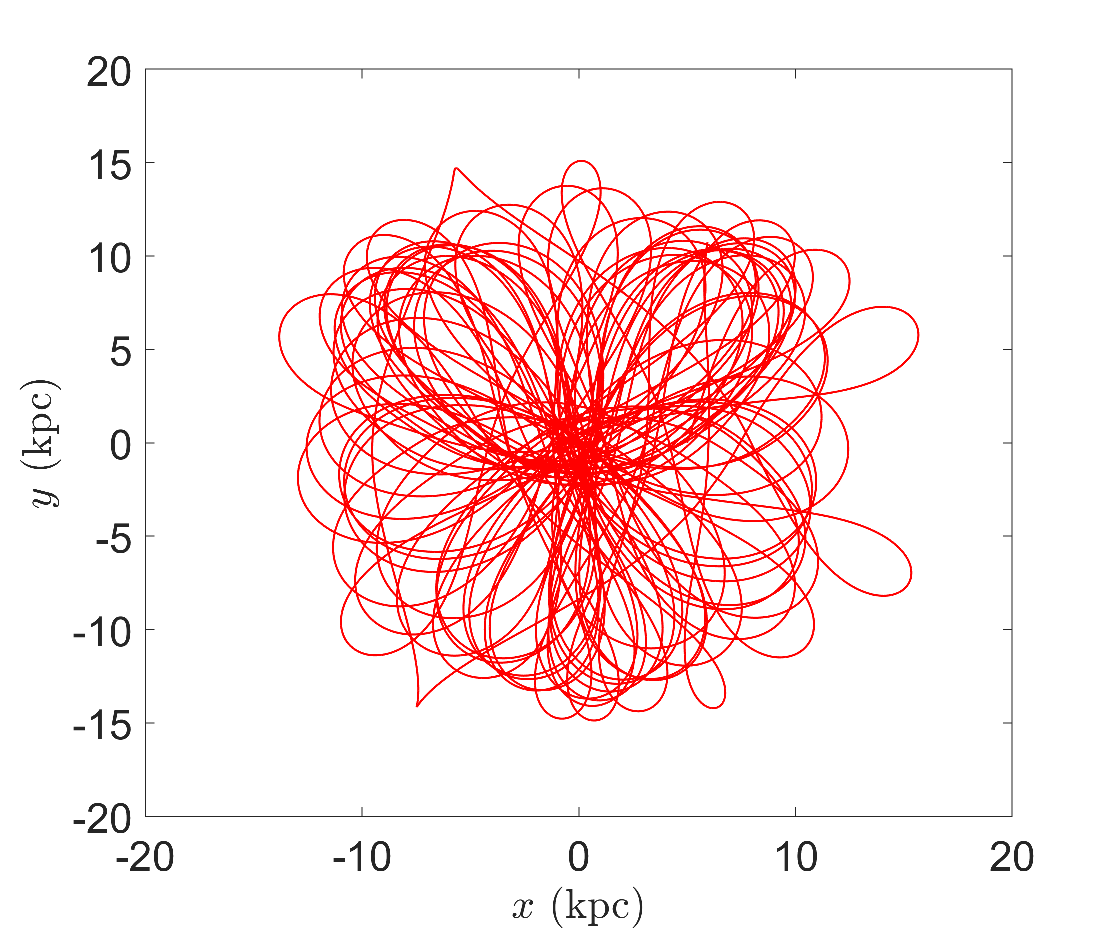}}
\subfigure[ $C = 0.1$ and $(x_0,y_0,p_{x_0})$ $\equiv (5,0,15)$]{\label{fig:4.4b}\includegraphics[width=0.48\textwidth]{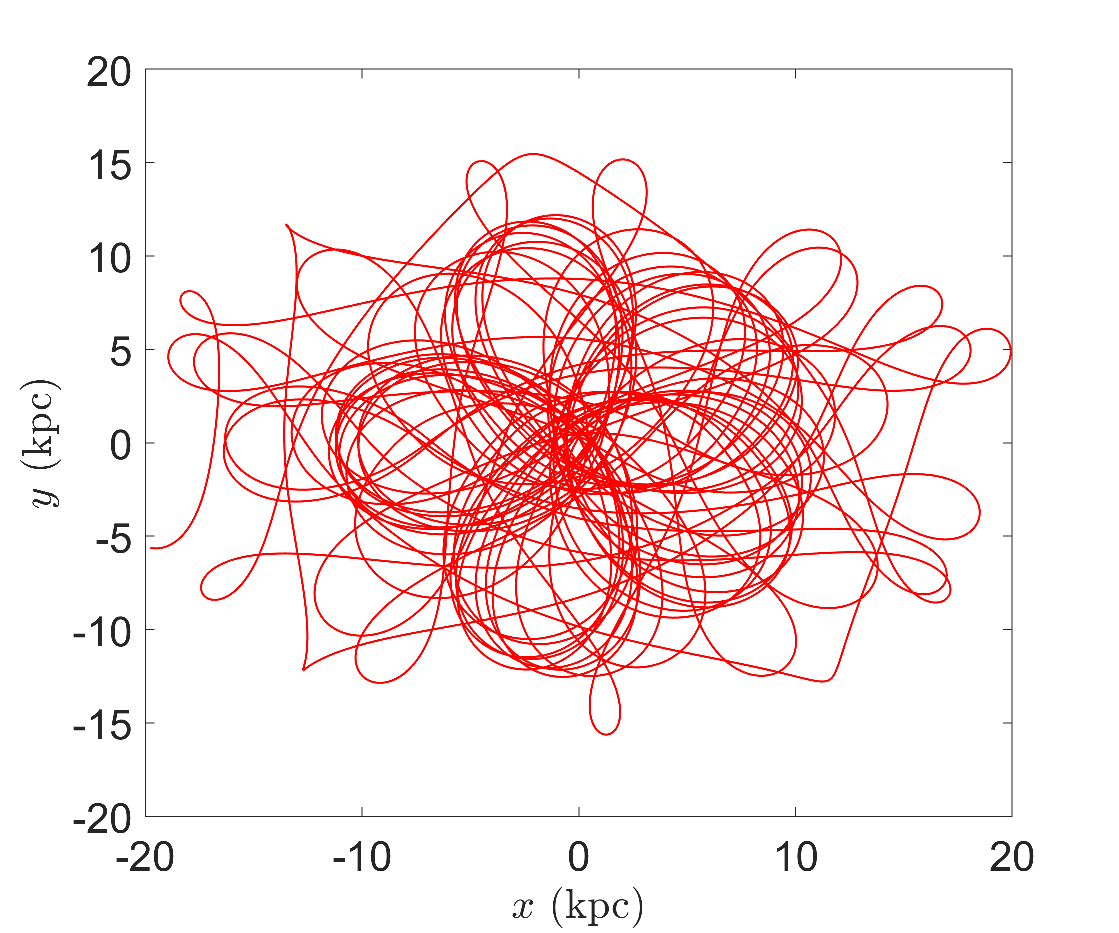}}
\subfigure[$C = 0.01$ and $(x_0,y_0,p_{x_0})$ $\equiv (-5,0,15)$]{\label{fig:4.4c}\includegraphics[width=0.48\textwidth]{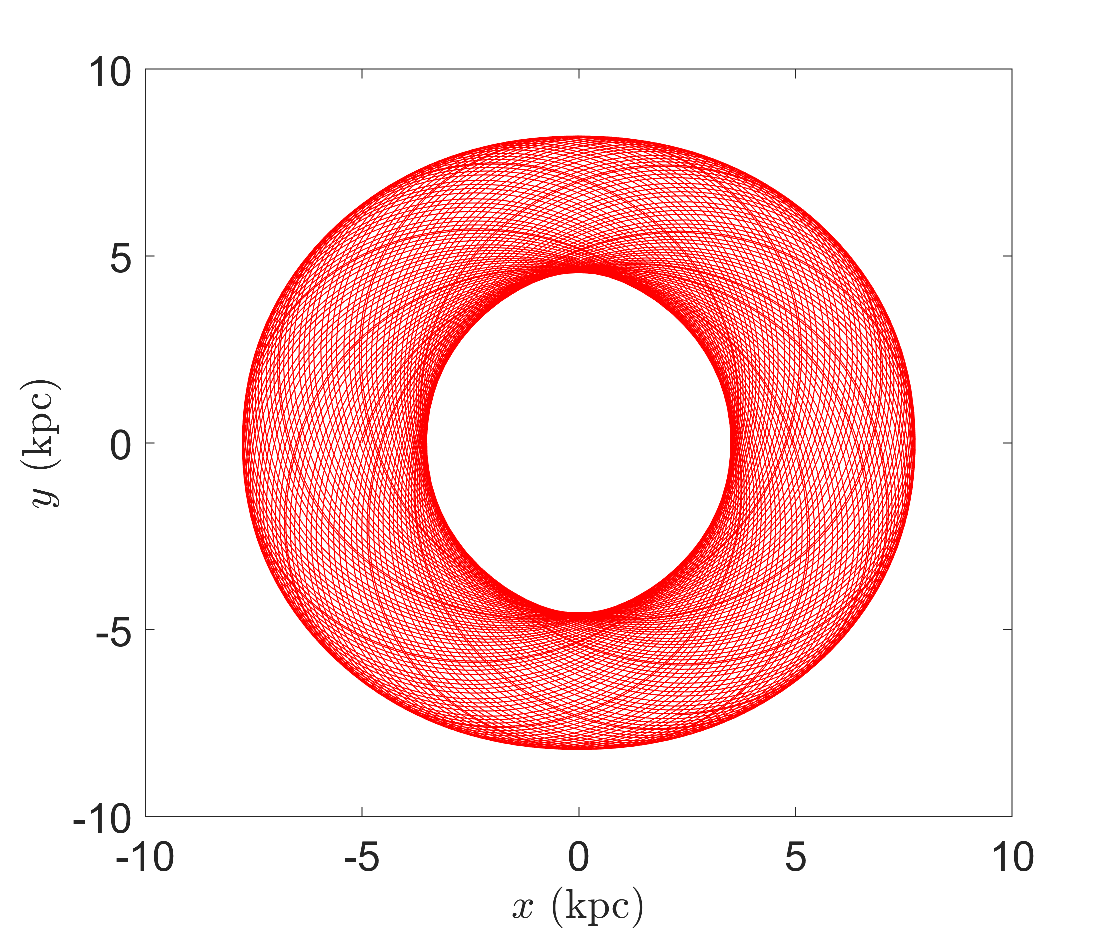}}
\subfigure[$C = 0.1$ and $(x_0,y_0,p_{x_0})$ $\equiv (-5,0,15)$]{\label{fig:4.4d}\includegraphics[width=0.48\textwidth]{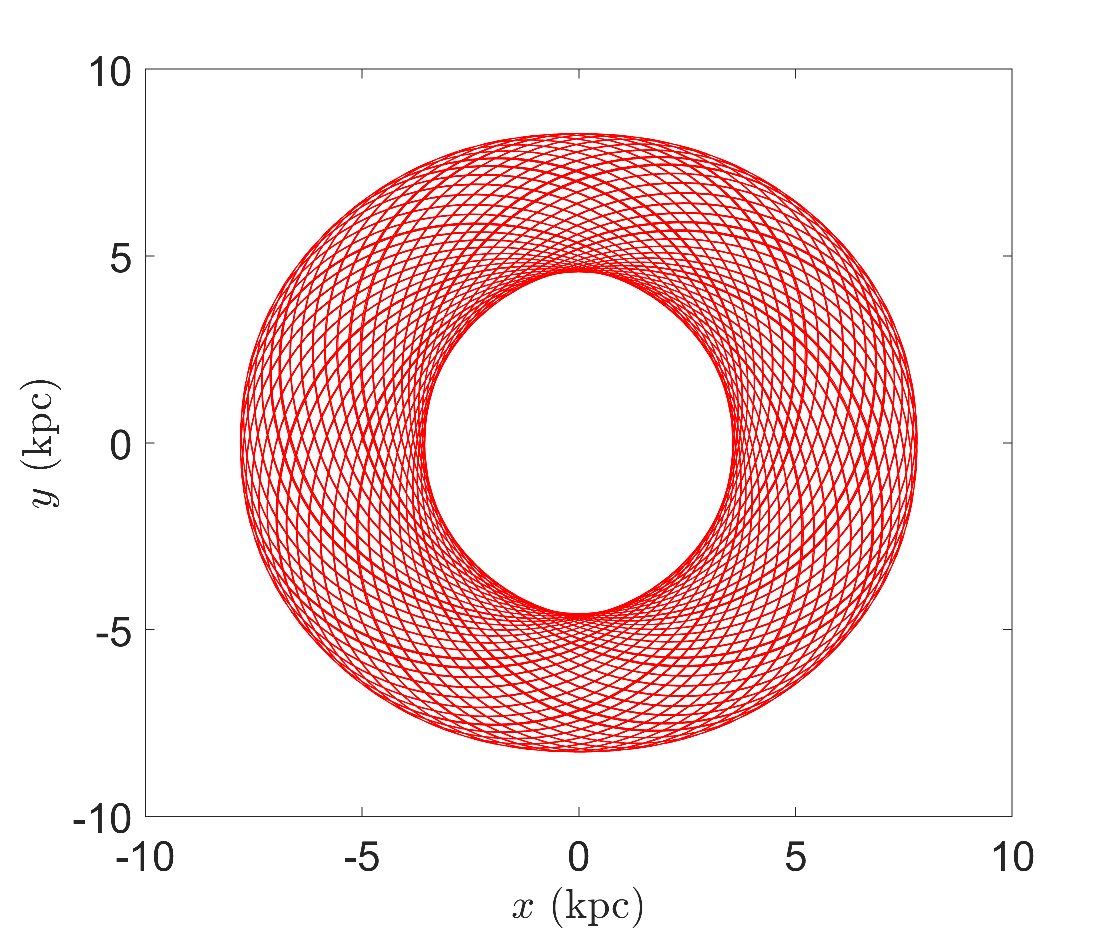}}
\caption{Model $2$: Orbits in the $x - y$ plane, where $p_{y_0}$ value is evaluated from Eq. (\ref{eq:4.3}); (a)-(b) non-escaping chaotic orbit, (c)-(d) non-escaping retrograde quasi-periodic rosette orbit.}
\label{fig:4.4}
\end{figure}

\begin{table}
\centering
\begin{tabular}{|c|c|c|}
\hline
Initial Condition                    & $C$    & MLE\\
\hline
\hline
$(x_0,y_0,p_{x_0}) \equiv (5,0,15)$  & $0.01$ & $0.1691$\\
                                     & $0.1$  & $0.1742$\\
\hline
$(x_0,y_0,p_{x_0}) \equiv (-5,0,15)$ & $0.01$ & $0.0761$\\
                                     & $0.1$  & $0.0769$\\
\hline
\end{tabular}
\caption{Model $2$: MLE value for different values of $C$, where $p_{y_0}$ value is evaluated from Eq. (\ref{eq:4.3}).}
\label{tab:4.5}
\end{table}
\end{itemize}

\subsection{Poincaré maps}
\label{sec:4.3.2}
\noindent Poincaré surface section maps in the $x - y$ and $x - p_x$ planes are plotted to visualise the bar-driven escapes in phase space. Now, for $z = 0 = p_z$, the phase space reduces from four-dimensional to six-dimensional. Here only initial conditions lying within the energetically allowed region: $\Phi_\text{eff}(x,y,z) < E_{L_1} (\text{or} \; E_{L_1^{'}})$ are taken into consideration for constructing Poincaré surface section maps. A $43\times43$ grid of initial conditions has been set up in the $x - y$ plane with step sizes: $\Delta x = 1$ kpc and $\Delta y = 1$ kpc to construct Poincaré maps in the $x - y$ plane (Figs. \ref{fig:4.5} and \ref{fig:4.7}). Only initial conditions within the Lagrange radius ($r_{L_1}$ or $r_{L_1^{'}}$) are considered from this grid. The initial value of $p_x$, i.e., $p_{x_0}$ is considered to be $p_{x_0} = 0$. Moreover, the initial value of $p_y$, i.e., $p_{y_0}$ is considered as $p_{y_0} > 0$ and further evaluated from Eq. (\ref{eq:4.3}). The chosen surface cross-sections for these Poincaré maps are $p_x = 0$ and $p_y \le 0$ \cite{Ernst2014}.

Similarly, for Poincaré maps in the $x - p_x$ plane (Figs. \ref{fig:4.6} and \ref{fig:4.8}), a $43\times31$ grid of initial conditions has been set up in the $x - p_x$ plane with step sizes: $\Delta x = 1$ kpc and $\Delta p_x = 10$ km $\text{s}^{-1}$. Here also, only initial conditions within the Lagrange radius ($r_{L_1}$ or $r_{L_1^{'}}$) are considered from this grid. The initial value of $y$, i.e., $y_0$ is considered to be $y_0 = 0$. Also, as in the case of the $x - y$ plane, the initial value of $p_y$, i.e., $p_{y_0}$ is considered as $p_{y_0} > 0$ and further evaluated from Eq. (\ref{eq:4.3}). The chosen surface cross-sections for these Poincaré surface section maps are $y = 0$ and $p_y \le 0$ \cite{Ernst2014}.

\begin{itemize}[leftmargin=*]
\item Model $1$: In Fig. \ref{fig:4.5a}, a primary stability island is seen near $(5,0)$ in the $x - y$ plane for an energy value of $C = 0.01$, formed due to the quasi-periodic motions. Again, this stability island diminishes when the energy value is increased to $C = 0.1$ (see Fig. \ref{fig:4.5b}). This is because of the enhancement of chaoticity in that region. There is also a hint about escaping motions as the number of cross-sectional points in this $x - y$ plane outside the corotation region (near $L_2$) increases with the increment of $C$. Similarly, a primary stability island has been observed in Figs. \ref{fig:4.6a} and \ref{fig:4.6b} which are located near $(5,0)$ in the $x - p_x$ plane. In these figures, orbital trends are similar to those observed in the $x - y$ plane.
\begin{figure}
\centering
\subfigure[$C = 0.01$]{\label{fig:4.5a}\includegraphics[width=0.48\columnwidth]{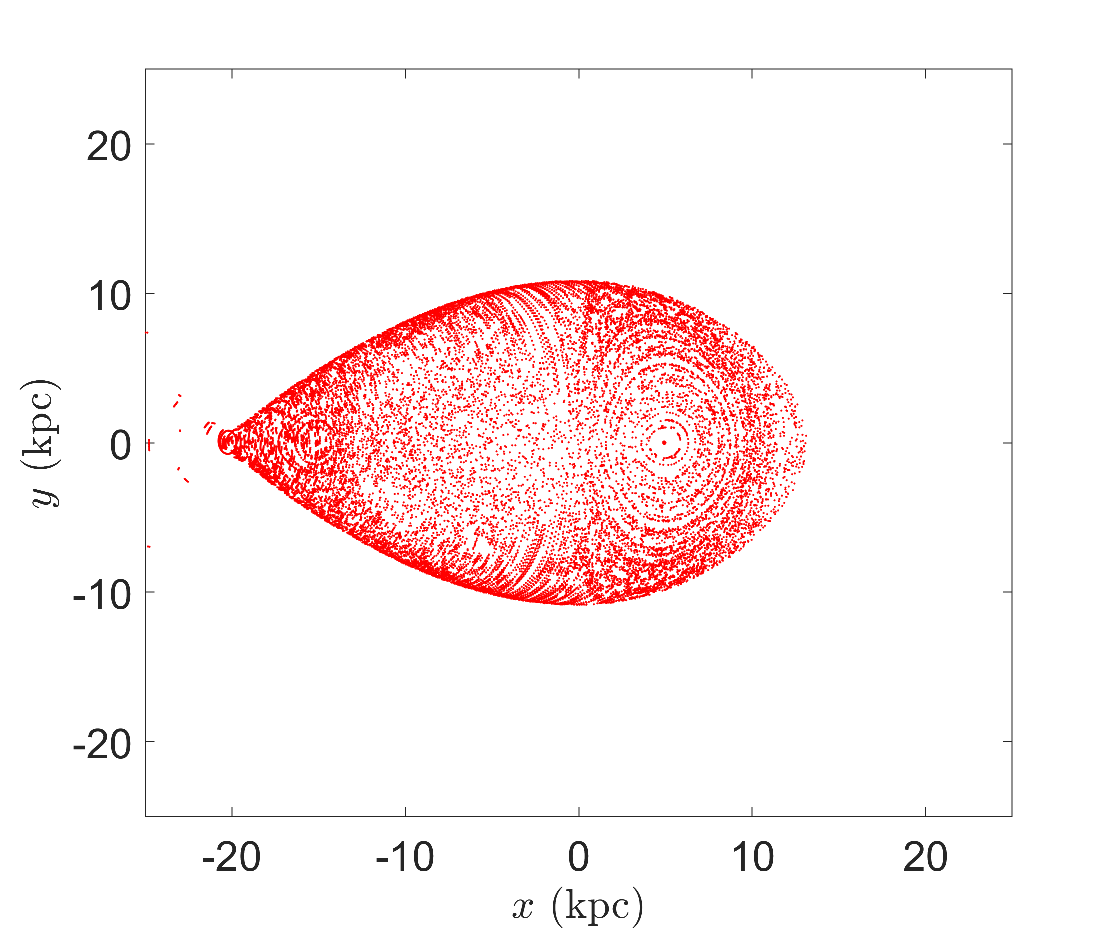}}
\subfigure[$C = 0.1$]{\label{fig:4.5b}\includegraphics[width=0.48\columnwidth]{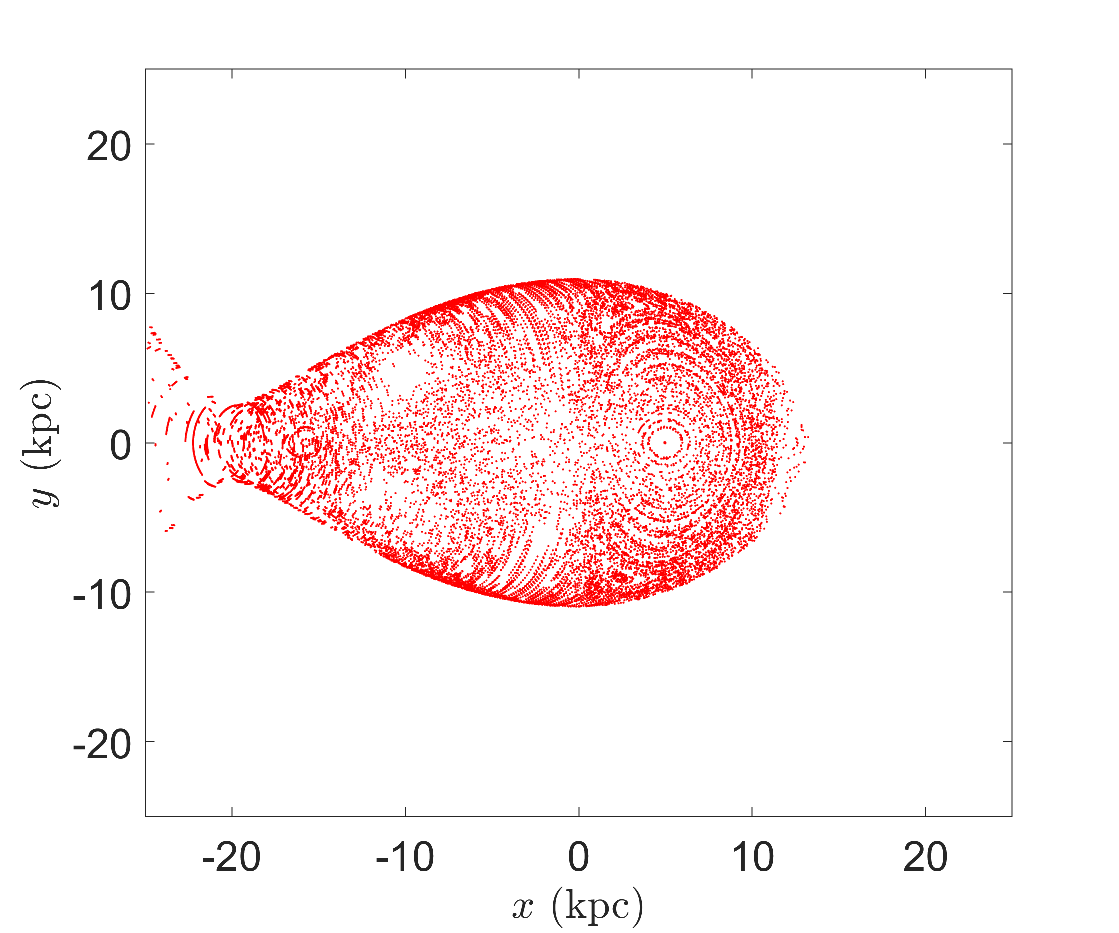}}
\caption{Model $1$: Poincaré maps in the $x - y$ plane with surface sections: $p_x = 0$ and $p_y \leq 0$.}
\label{fig:4.5}
\end{figure}

\begin{figure}
\centering
\subfigure[$C = 0.01$]{\label{fig:4.6a}\includegraphics[width=0.48\columnwidth]{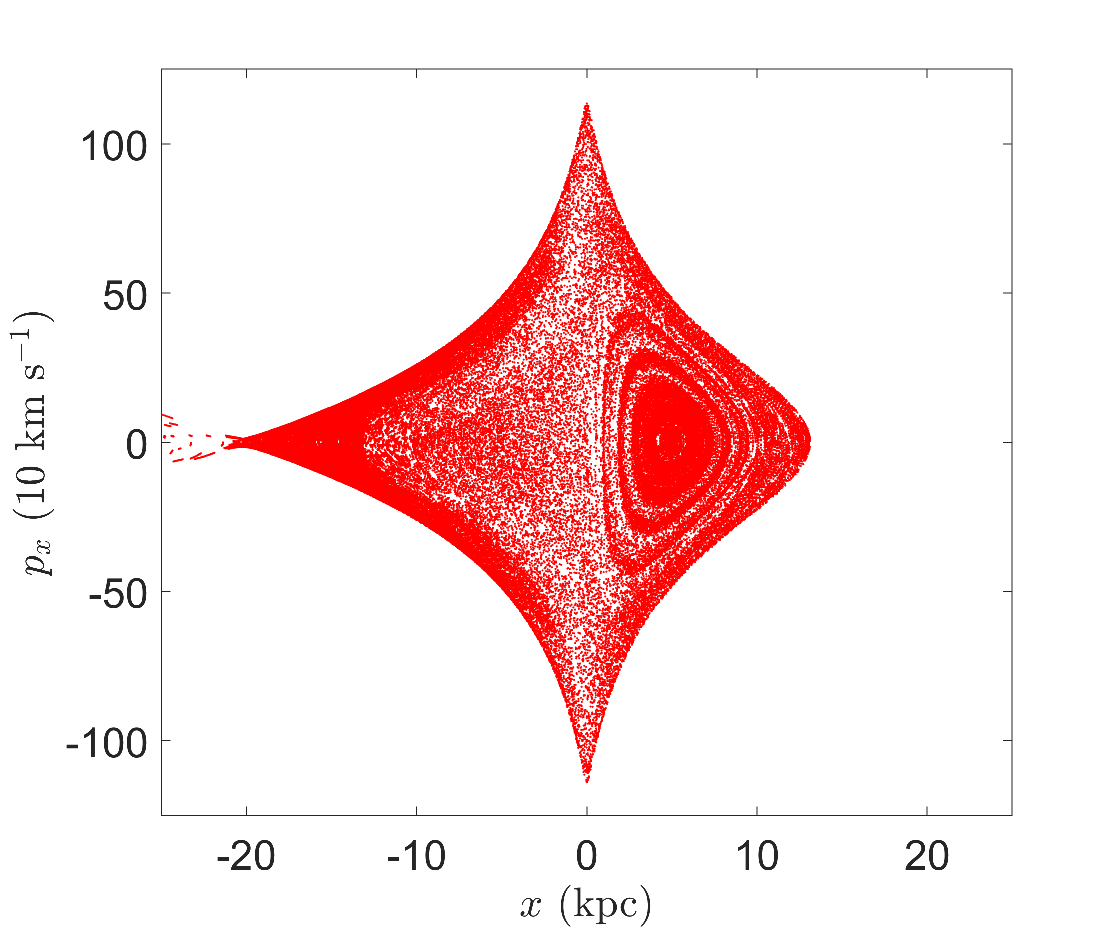}}
\subfigure[$C = 0.1$]{\label{fig:4.6b}\includegraphics[width=0.48\columnwidth]{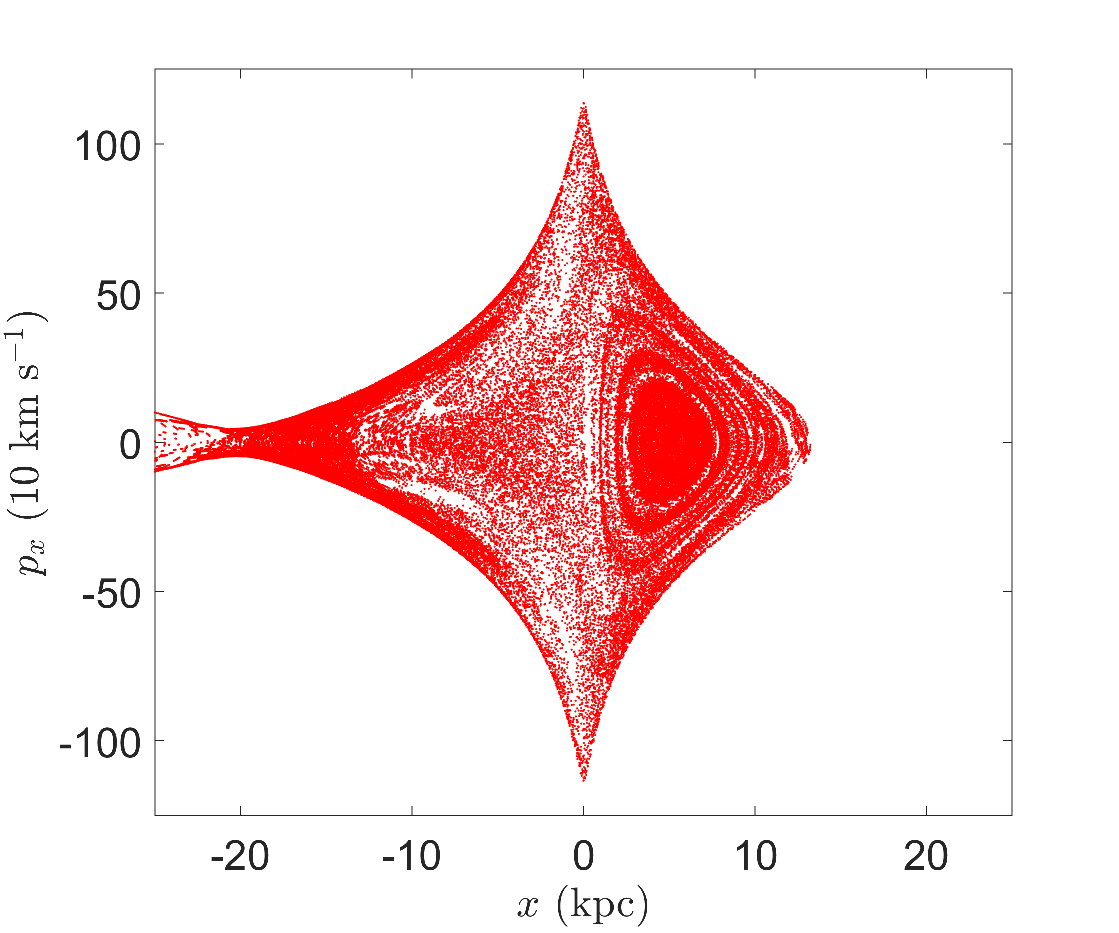}}
\caption{Model $1$: Poincaré maps in the $x - p_x$ plane with surface sections: $y = 0$ and $p_y \leq 0$.}
\label{fig:4.6}
\end{figure}

\item Model $2$: Same as model $1$, here in Figs. \ref{fig:4.7a} and \ref{fig:4.7b}, a primary stability island is seen near $(6,0)$ in the $x - y$ plane that is formed due to quasi-periodic motions. As the energy value increases from $C=0.01$ to $C=0.1$, this stability island diminishes due to an enhanced chaoticity in that region. Here is also a hint about escaping motions, as the number of cross-sectional points in this $x - y$ plane outside the corotation region (near $L_2^{'}$) increases with the increment of $C$. Similarly, in Figs. \ref{fig:4.8a} and \ref{fig:4.8b} a primary stability island has formed and is located near $(6,0)$ in the $x - p_x$ plane. Also, in this $x - p_x$ plane, some smaller stability islands are formed for $C = 0.01$; these islands diminish as the energy value increases to $C = 0.1$. This feature of smaller stability islands is absent in model $1$. This indicates that the bar of model $2$ is more dynamically stable than model $1$. Besides this, overall orbital trends are similar to model $1$.
\begin{figure}
\centering
\subfigure[$C = 0.01$]{\label{fig:4.7a}\includegraphics[width=0.48\columnwidth]{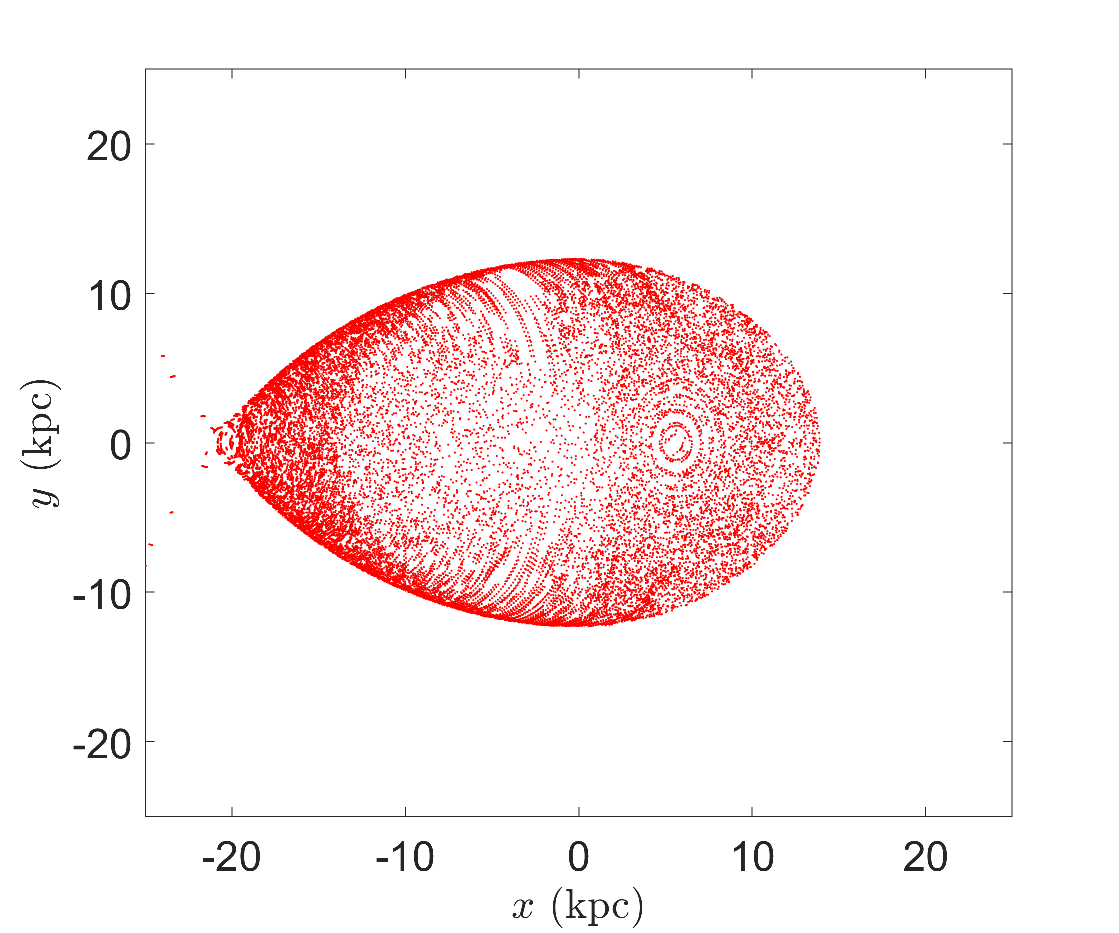}}
\subfigure[$C = 0.1$]{\label{fig:4.7b}\includegraphics[width=0.48\columnwidth]{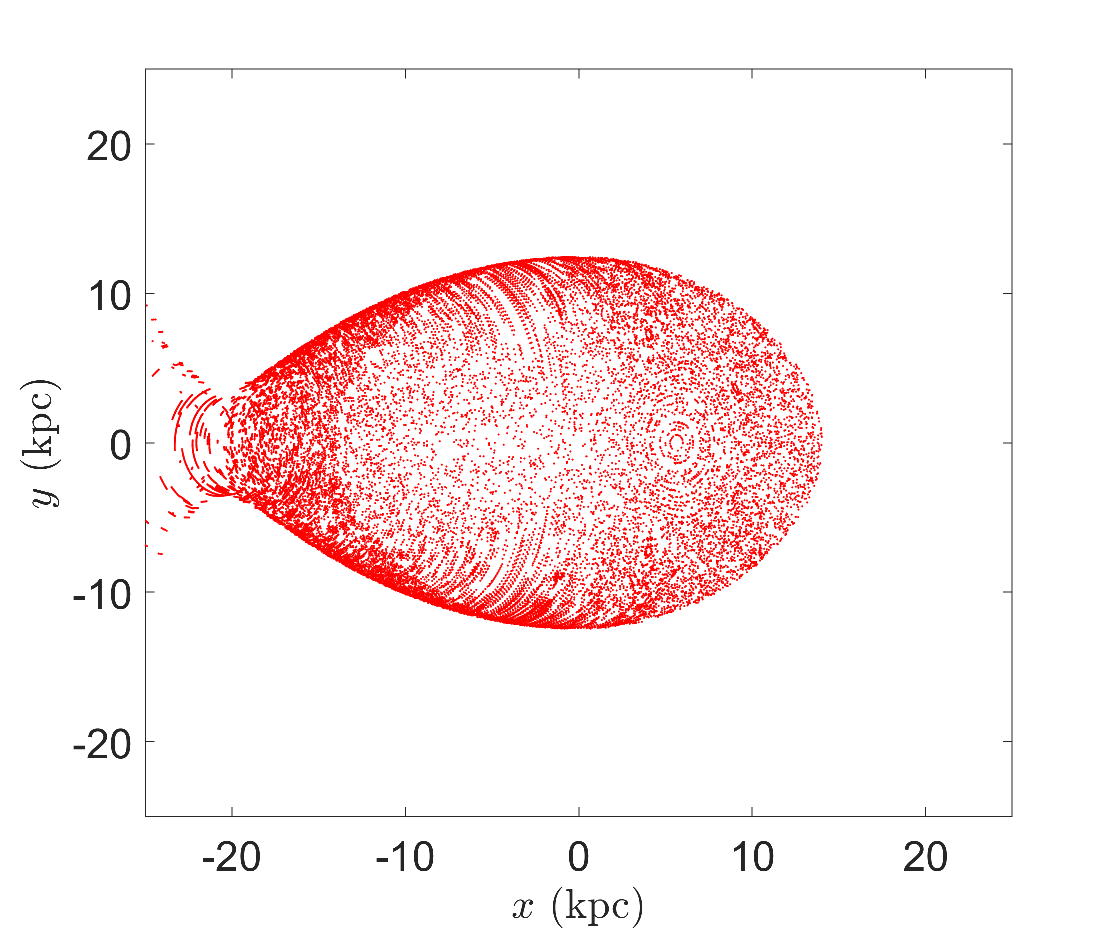}}
\caption{Model $2$: Poincaré maps in the $x - y$ plane with surface sections: $p_x = 0$ and $p_y \leq 0$.}
\label{fig:4.7}
\end{figure}

\begin{figure}
\centering
\subfigure[$C = 0.01$]{\label{fig:4.8a}\includegraphics[width=0.48\columnwidth]{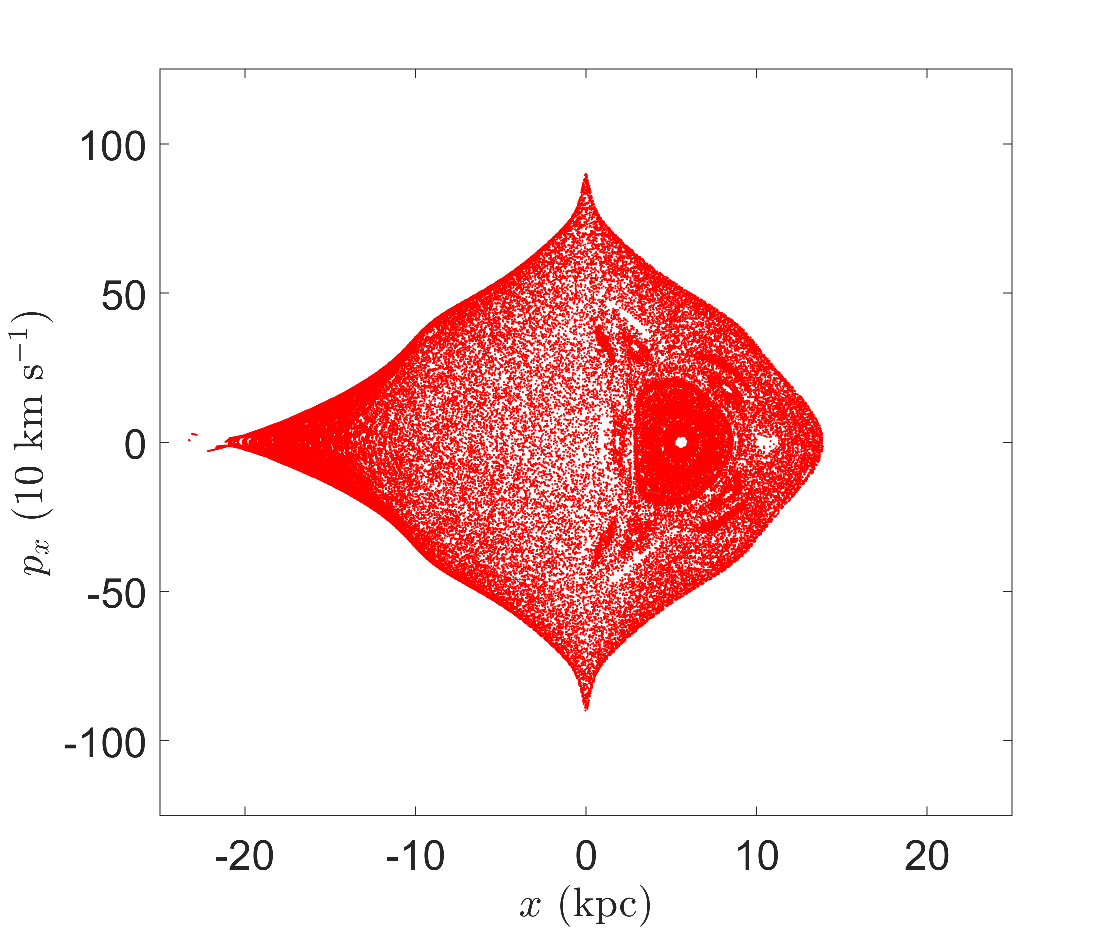}}
\subfigure[$C = 0.1$]{\label{fig:4.8b}\includegraphics[width=0.48\columnwidth]{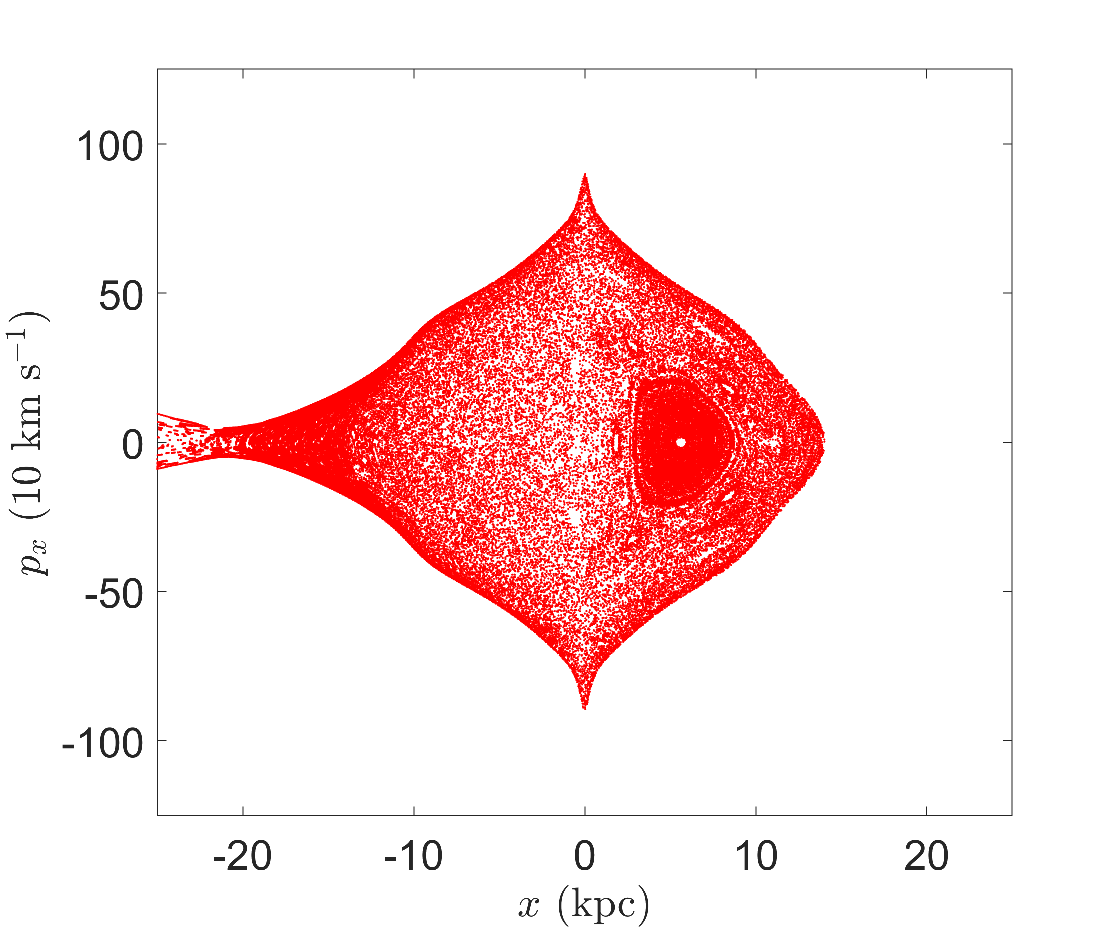}}
\caption{Model $2$: Poincaré maps in the $x - p_x$ plane with surface sections: $y = 0$ and $p_y \leq 0$.}
\label{fig:4.8}
\end{figure}
\end{itemize}

\subsection{Evolution of orbital chaos with the bar parameters}
\label{sec:4.3.3}
In this part of this chapter, how the chaotic dynamics (in terms of MLE) evolved over the vast integration time concerning bar parameters such as mass and length is discussed in detail. In order to do that, for each of the two bar potential models, MLE values have been calculated for different values of the bar mass and length. The primary goal is to investigate the influence of bar parameters on the orbital dynamics near the Lagrangian points $L_1$ and $L_1^{'}$. Due to this, the scope of this section is limited to orbits beginning with the initial condition: $(x_0,y_0,p_{x_0},p_{y_0}) \equiv (5,0,15,p_{y_0})$, where the $p_{y_0}$ value is calculated using Eq. (\ref{eq:4.3}). In the gravitational model, the total galaxy mass ($M$) is equal to $30,900$ mass units, and the total bar mass ($M_\text{b}$) is equal to $3500$ mass units, i.e., $M_\text{b}$ is nearly $11.33\%$ of $M$. In the study, the range of $M_\text{b}$ is considered to be between $3100$ and $4000$ mass units, i.e., nearly between $10\%$ and $13\%$ of $M$. Also, the range of the bar flattening parameter ($\alpha$) and bar semi-major axis ($a$) is considered between $1$ and $10$ length units. These parameter ranges are used to establish the relationship between them and the MLE value of the orbit starting from the aforesaid initial condition.

\begin{itemize}[leftmargin=*]
\item Model $1$: For model $1$, Fig. \ref{fig:4.9a} manifests the variation of the MLE value at the aforesaid initial condition with the bar flattening parameter ($\alpha$) and dimensionless energy parameter ($C$). Similarly, Fig. \ref{fig:4.9b} shows how that MLE value varies with the bar mass ($M_\text{b}$) and $C$. 
\begin{figure}
\centering
\subfigure[$\alpha$ versus MLE]{\label{fig:4.9a}\includegraphics[width=0.48\columnwidth]{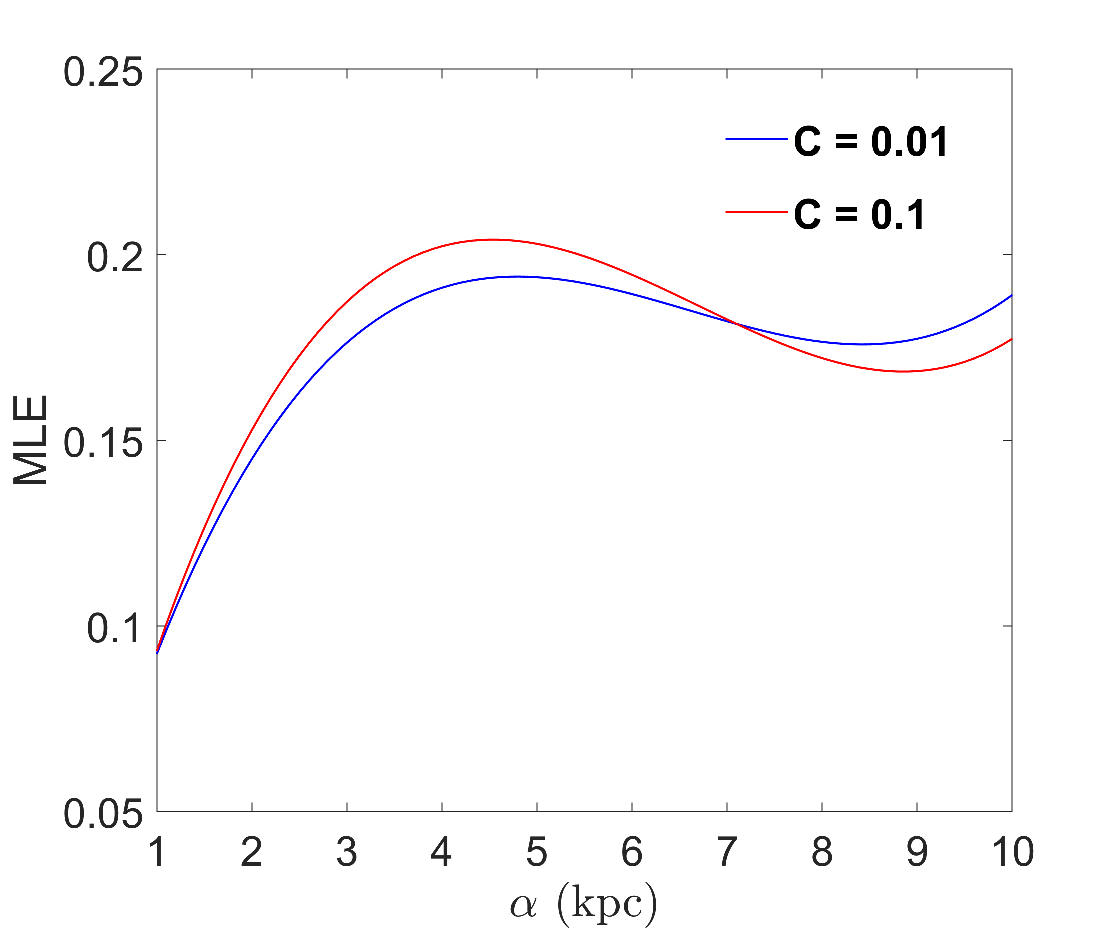}}
\subfigure[$M_\text{b}$ versus MLE]{\label{fig:4.9b}\includegraphics[width=0.48\columnwidth]{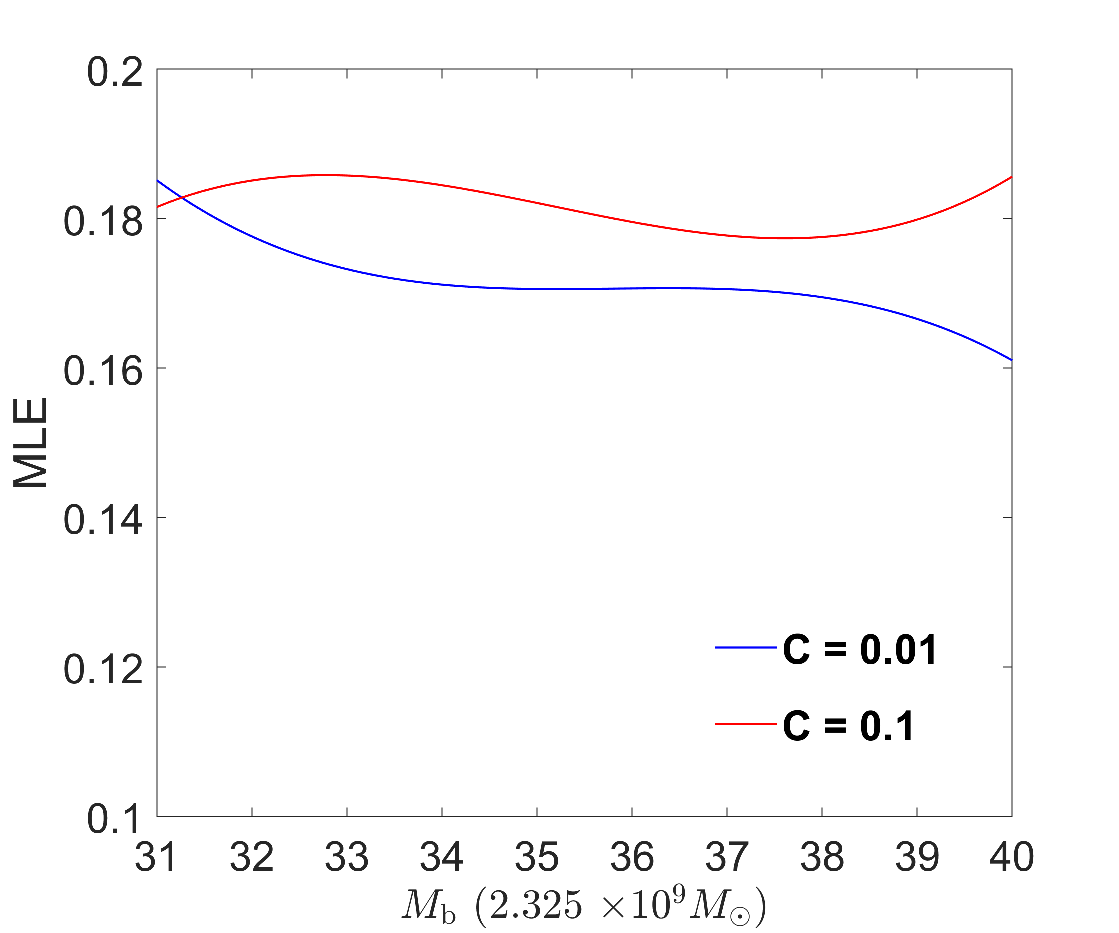}}
\caption{Variation of MLE with the strong bar parameters.}
\label{fig:4.9}
\end{figure}

\item Model $2$: For model $2$, Fig. \ref{fig:4.10a} manifests the variation of that MLE value with the bar semi-major axis ($a$) and dimensionless energy parameter ($C$). Fig. \ref{fig:4.10b} shows a similar type of analysis for the bar mass ($M_\text{b}$) and $C$.  
\begin{figure}
\centering
\subfigure[$a$ versus MLE]{\label{fig:4.10a}\includegraphics[width=0.48\columnwidth]{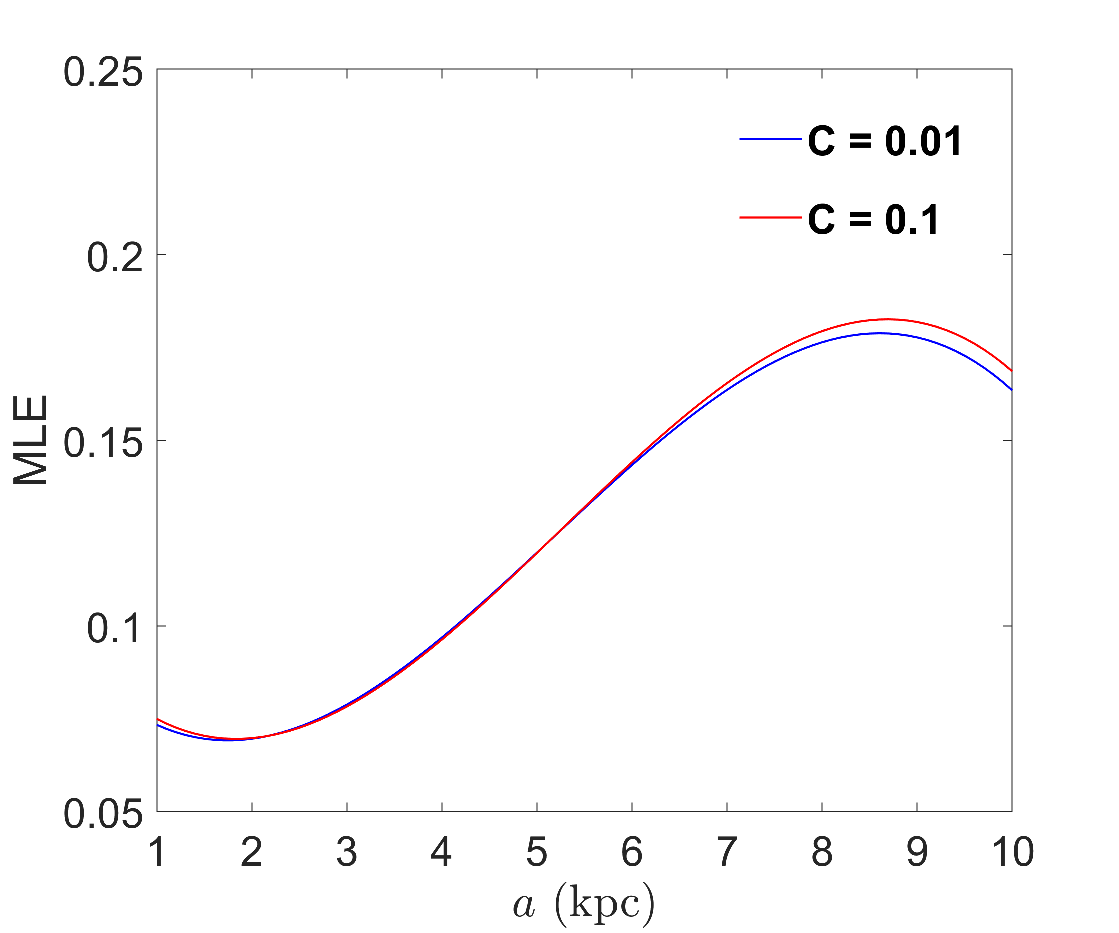}}
\subfigure[$M_\text{b}$ versus MLE]{\label{fig:4.10b}\includegraphics[width=0.48\columnwidth]{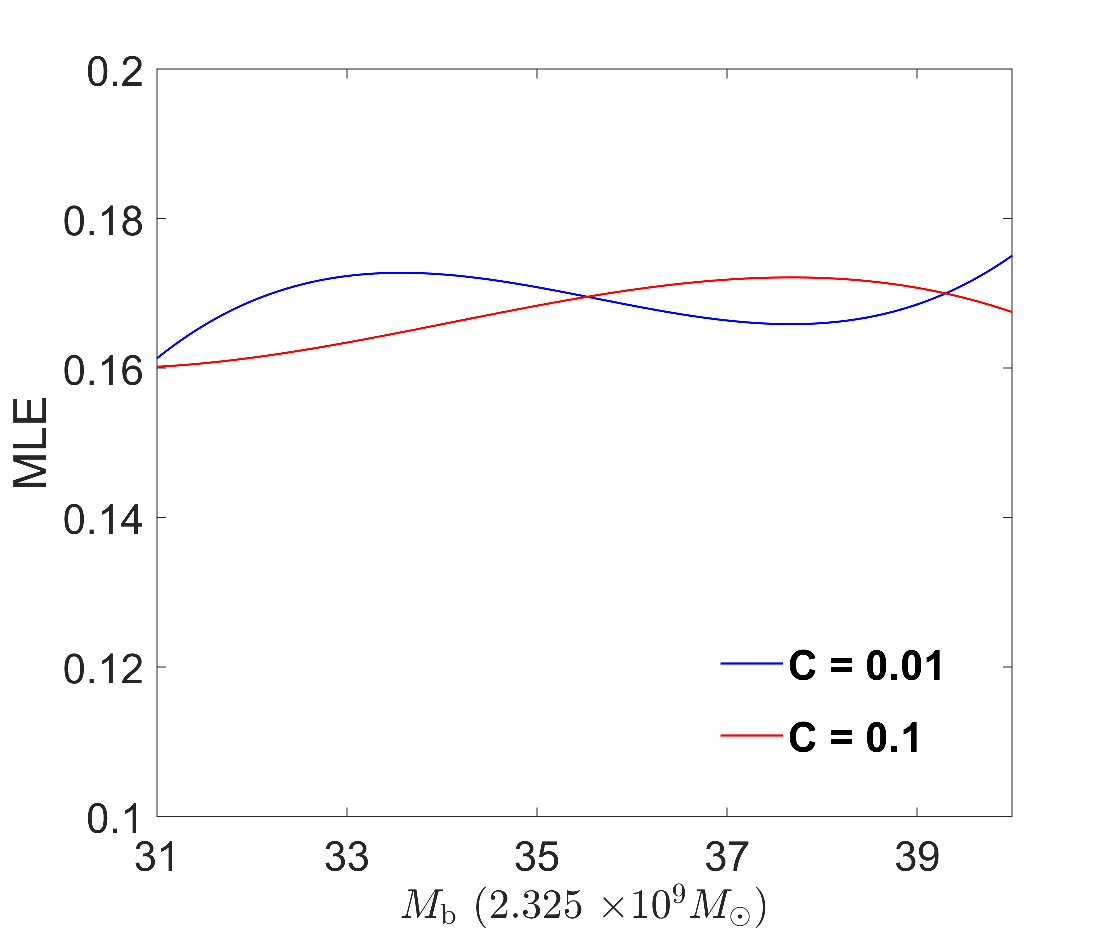}}
\caption{Variation of MLE with the weak bar parameters.}
\label{fig:4.10}
\end{figure}
\end{itemize}

\section{Fate of Bar-driven Escaping Motion}
\label{sec:4.4}
Until now, a comprehensive analytical analysis has been provided to describe the influence of bars on the nature of stellar orbits in barred galaxies, along with the development of spiral arms due to bar-driven stellar escape motions.

\indent In this chapter, the motion of stellar orbits has been studied in barred galaxies with four components: the bulge, bar, disc, and dark matter halo. Further, there are two types of bars, namely (i) strong bar (model $1$) and (ii) weak bar (model $2$), for a comparative study. It is clear from Fig. \ref{fig:4.2a} that the bar area of the latter is larger and more elongated in the $x$ - direction than the first one. Also, the density distribution for a bar in model $1$ is very steeply close to the centre, i.e., nature is one of the cuspy types compared to model $2$. Thus, the bar density profile in model $1$ justifies a strong bar. On the other hand, in model $2$, the density distribution is relatively flat in the central region. This kind of bar density profile justifies a weak bar.

\indent The nature of orbits depends upon the initial conditions and the bar potential. For model $1$, the orbit integrated for initial condition $(x_0,y_0,p_{x_0},p_{y_0}) \equiv (5,0,15,p_{y_0})$, is escaping and chaotic with a high MLE. Also, the orbit is non-escaping and quasi-periodic with a low MLE for $(x_0,y_0,p_{x_0},p_{y_0}) \equiv (-5,0,15,p_{y_0})$ (see Table \ref{tab:4.4}), where $p_{y_0}$ value is evaluated from Eq. (\ref{eq:4.3}). Again, the radial and tangential force components of this model are very high near the galactic centre and eventually die off at larger distances (see Figs. \ref{fig:4.2c} and \ref{fig:4.2d}). Hence, the bar potential of model $1$ promotes a bar-driven escape mechanism, and escape of orbits from the disc is possible through the bar ends for a suitable choice of initial condition. Moreover, escaping orbits may result in spiral arms. On the other hand, for model $2$, the orbit integrated for the initial condition $(x_0,y_0,p_{x_0},p_{y_0}) \equiv (5,0,15,p_{y_0})$ is non-escaping and chaotic with a high MLE. Also, the orbit is non-escaping and quasi-periodic with a low MLE for $(x_0,y_0,p_{x_0},p_{y_0}) \equiv (-5,0,15,p_{y_0})$ (see Table \ref{tab:4.5}). In both cases, MLE values are comparatively lower than model $1$. Again, the radial force component of this model is very high near the galactic centre. However, the tangential force component is very high in the annulus between the outer bulge and barred region (see Figs. \ref{fig:4.2c} and \ref{fig:4.2d}). Because the strength of the radial force component near the galactic centre in this model is less than that of model $1$, it does not promote any bar-driven escape mechanism through the bar ends. However, compared to model $1$, the tangential force strength is higher in an annulus between the outer bulge and barred region. As a result, orbits that cross from the bar ends do not leave the disc and tend to form ring-type structures. Such structures are known as inner disc rings. This is evident in S0 or ring galaxies where spiral arms are more or less absent \cite{Sil2018}. 

\indent Another aspect of the analysis is studying the nature of the chaotic orbits of stars in barred galaxies. Understanding the dynamics of barred galaxies relates to the chaotic motions of stars in the central region. The presence of chaos is a manifestation of unstable orbits. The chaos propagates further over a long time, influencing the evolution of galactic structures like bars, discs, and dark matter haloes. Also, integrating chaotic theory into the orbital and escaping stellar motions helps us figure out the actual shape of these components. In barred galaxies, these growing instabilities relate to the formation and strength of spiral arms.  

\indent Galactic bars are formed inside the disc by the outwardly propagating density waves. Spiral arms are also thought to be the result of these density waves. Observational studies in the blue and near-IR bands confirm that spiral arms are continuations of the bar, i.e., spiral arms are outcomes of a bar-driven mechanism. Theoretical studies also suggested that there is a correlation between bar pattern speed and that of spiral arms \cite{Buta2009}. These density waves and stellar orbits usually have different rotational speeds, but some corotation region exists between the disc and bar. Spiral arms may emerge from both ends of the bar in that corotation region, dominated by chaotic orbits. Theoretical studies confirm that these chaotic orbits in the corotation region are the building blocks of the spiral arms \cite{Contopoulos1989, Kaufmann1996, Efthymiopoulos2019}. A study by \cite{Patsis1997} have shown that these chaotic orbits are the reason behind the characteristic outer boxy isophotes of the nearly face-on bar of the barred spiral galaxy NGC 4314. Under suitable physical circumstances, spiral arms may survive the chaotic dynamics in the corotation region of the bar and disc and emerge from the bar ends. Regarding the occurrence of spiral arms in two types of spiral galaxies, barred and unbarred, nearly $70\%$ of the barred spirals in the field have tightly wound two-armed spiral patterns. In comparison, about $30\%$ of the unbarred spirals in the field have grand design spiral patterns \cite{Elmegreen1982}.  

\indent It follows from the preceding discussions that the spiral arms might be the continuation of stellar orbits that are escaping and emerging from the bar ends due to chaotic motion. Now, this escaping motion is only possible when the orbital energy exceeds the energy of Lagrangian point $L_1$ (or $L_1^{'}$), i.e., $E > E_{L_1}$ (or $E > E_{L_1^{'}}$). In the analysis, these chaotic motions are characterised by MLE values of stellar orbits calculated for various initial conditions and bar parameters. The following observations have been made:

\begin{itemize}[leftmargin=*]
\item Model $1$:
\begin{enumerate}[label=(\roman*)]
\item From the orbital structures (Fig. \ref{fig:4.3}) of model $1$, it is concluded that stellar orbits may escape from the disc through the bar ends (i.e., Lagrangian points $L_1$ and $L_2$). Also, the chaoticity of orbits inside the disc and bar's corotation region increases with the increment in energy. In the Poincaré surface section maps (Figs. \ref{fig:4.5} and \ref{fig:4.6}) of model $1$, trends are similar to those observed in the orbital maps.

\item For model $1$, MLE increases with the flattening parameter ($\alpha$) up to a threshold length and again slowly decreases (viz., Fig. \ref{fig:4.9a}). This threshold length decreases with an increase in the value of $C$. This implies that when the escape energy of $L_1$ is high ($C \sim 0.1$), the escape of stars is even encouraged at a smaller bar length, e.g., for $C = 0.01$, the threshold value of $\alpha = 8$ whereas for $C = 0.1$, $\alpha = 6$ (viz., Fig. \ref{fig:4.9a}). Also, there is a sudden increment in the MLE from $\alpha=1$ to $\alpha = 2$. When $\alpha = 1$, the bar of model $1$ has a bulge-like spherical structure rather than the usual elongated structure. For this reason, chaotic orbits in the central region are comparatively less due to the presence of a central black hole, which suppresses star formation more effectively inside that spherical region. The bar becomes flatter for $\alpha > 1$, and that suppression effect diminishes.

\item The following observations are made about Fig. \ref{fig:4.9b}: (i) The variation of $M_\text{b}$ with MLE follows a decaying oscillating pattern, i.e., the amplitude of oscillations gradually diminishes with $M_\text{b}$ for a lower escape energy value ($C = 0.01$) and (ii) the same variation follows a stable oscillating pattern for a higher escape energy value ($C = 0.1$). Now, the amount of chaos depends on the orbital energy value \cite{Zotos2017}. At a lower escape energy ($C = 0.01$), the gradual increment of $M_\text{b}$ diminishes the chaos because the stellar orbits spend a significant fraction of their time inside the barred region and do not fill up the phase space homogeneously, which suggests a weak chaotic motion. Again, at higher escape energy ($C = 0.1$), a stable oscillating pattern of MLE is followed along with an increment in $M_\text{b}$ because stellar orbits fill up the phase space homogeneously, which suggests a robust chaotic motion. Hence, for model $1$, the formation of prominent spiral arms for smaller values of escape energy is not favourable. However, it becomes favourable once the escape energy value increases (maybe due to central explosions, shocks, etc.).

\item Therefore, in the presence of a heavier (or stronger) bar, the formation of prominent spiral arms is more likely in those barred galaxies where violent activities occur in the central region. Giant spirals harbour super-massive black holes (SMBHs) in their nuclear regions, where violent activities are occurring \cite{Melia2001}, and the strength of the spiral arms is strongly correlated with the central black hole mass and the kinetic energy of random motions inside the bulge. Galaxies with SMBHs have higher kinetic energy transport inside their bulges and spiral arms with small pitch angles, which results in tightly wound grand design spiral patterns \cite{Seigar2008, Berrier2013, Al2014}. Many giant spirals containing SMBH, like NGC 3310, NGC 4303, NGC 4258, etc. \cite{Pastorini2007} have grand design spiral patterns. Hence, the bar potential used in model $1$ favours the formation of grand design spirals when SMBH is present at the core and spiral arms emerge from the bar ends.
\end{enumerate}

\item Model $2$:
\begin{enumerate}[label=(\roman*)]
\item From the orbital structures (Fig. \ref{fig:4.4}) of model $2$, it is concluded that stellar orbits are not encouraged to escape from the disc through the bar ends (i.e., Lagrangian points $L_1^{'}$ and $L_2^{'}$) but remain encapsulated within the disc. Also, the chaoticity of orbits inside the corotation region of the disc and bar increases with the increment in energy, but this increment in chaoticity is less comparable to model $1$. In the Poincaré surface section maps (Figs. \ref{fig:4.7} and \ref{fig:4.8}) of model $2$, trends are similar to those observed in the orbital maps.

\item The MLE increases with the semi-major axis of the bar ($a$) up to a threshold value ($a = 6$) and does not vary much further (viz., Fig. \ref{fig:4.10a}). Again, this variation is independent of the tested escape energy levels. As a result, it is concluded that the fate of the escaping orbits does not depend much on the escape energy values for weaker bars. Also, the escape of stars through the bar ends is only possible for an optimal bar length, and that length does not depend on the escape energy.

\item Weaker bars also aid the escape mechanism, but the increase in chaos with bar mass is very slow (viz., Fig. \ref{fig:4.10b}). It is also observed that the variation $M_\text{b}$ with MLE follows a stable oscillating pattern for both higher and lower escape energy values. Such a stable oscillating pattern is only observed for a higher escape energy value in model $1$. Also, the MLE values of model $2$ are lower than model $1$. Hence, for model $2$, chaotic motions escaping from the bar ends remain trapped inside the disc and become favourable for forming ring-type structures, irrespective of the escape energy values.

\item Thus, the formation of ring-type structures is more likely in the presence of weak bars. This might be why inner rings are observed in many disc galaxies \cite{Byrd2006, Proshina2019}. Observational evidence of ring structures has been found in NGC 1326 by \cite{Buta1995}. Similarly, there are studies \cite{Sakamoto1999, Sakamoto2000} that found such ring structures in NGC 5005. Hence, the bar potential used in model $2$ favours the formation of ring-type structures that will emerge from the bar ends. 
\end{enumerate}
\end{itemize}

Now, from all the above discussions for the two bar types, the major conclusions are summarised as follows:
\begin{enumerate}[label=(\roman*)]
\item Barred galaxies with more substantial bar potential may lead to the formation of grand design spirals only when violent activities occur inside their central region. SMBHs may be one of the reasons behind this. Again, galaxies with massive bars but without central SMBHs may lead to the formation of less prominent spiral arms.

\item On the contrary, barred galaxies with weaker bar potential may lead to the formation of ring-type structures inside the disc.
\end{enumerate}

\section*{What's next?} 
This chapter discusses how the bar strength decides the fate of the bar-driven escaping motions. Now, the dark haloes too substantially impact this escaping motion. This is the central theme of the upcoming chapter, where the role of the dark halo behind the formation of bar-driven structures has been studied.

%% file: Chapter_5.tex
\chapter{Role of Dark Haloes in Bar-driven Structure Formation}
\label{chap:5}

\section{Introduction}
\label{sec:5.1}
The halo of dark matter particles is observed around field galaxies, galaxy groups, and galaxy clusters. These dark matter haloes (or simply dark haloes) are essential drivers of the galaxy's evolution. The dark matter particles are mostly non-baryonic and influence the formation and evolution of other galactic components only via gravity. According to the Planck collaboration's findings, dark matter particles account for about $\frac{5}{6}\text{th}$ of the universe's mass \cite{Ade2016}. The inconsistencies between the anticipated mass and measured luminous mass of numerous galaxies bring them under the observation of the scientific community \cite{Kapteyn1922, Freeman1970, Rubin1980, Zwicky2009}. A vast dark halo is embedded over the visible components of giant spiral galaxies like the Milky Way, M31, NGC 1365, and others, and its density steadily decreases as one moves away from the centre. The radial span of such dark haloes is about 100 to 200 kpc, whereas the luminous portion of the galaxy only exists up to a radius of 30 to 35 kpc \cite{Klypin2002, Bhattacharjee2014, Huang2016}. In dwarf galaxies, the structure of dark haloes is generally core-dominated \cite{Blok2002, Simon2005, Swaters2011}. Thus, the dark halo structure varies across galaxy morphologies, and besides that, they have an immense influence on the kinematic and structural properties of galaxies \cite{Weinberg2002, Karmakar2015, Salucci2019, Thob2019}. So, it is crucial to investigate the role of dark haloes in forming and evolving various galactic structures, such as bars and spiral arms.

\indent Given the $\Lambda$-Cold Dark Matter ($\Lambda$CDM) model, dark matter particles are collision-less and evolve under the influence of gravity. In practice, cosmological \textit{N}-body simulations are widely used to model the structure of dark haloes \cite{Moore1998, Yoshida2000, Hahn2013, Fischer2021}. The results of \textit{N}-body simulations predict that dark haloes should have a structure akin to the Navarro-Frenk-White (NFW) profile \cite{Navarro1996}. The central density of this NFW profile is divergent (infinite) and has a cuspy distribution near the centre. This cuspy dark halo profile agrees well with giant galaxy clusters \cite{Umetsu2011} but not with many other galaxy types, particularly low-mass galaxies like dwarfs and LSBs. The rotation curves of these low-mass galaxies reveal flat dark halo distributions, i.e., core-dominated dark haloes. Later, the Einasto profile \cite{Einasto1965} emerges as an alternative to the NFW profile, but it can still not remove the central cusp \cite{Beraldo2013}. This discrepancy between the predicted dark halo profiles of \textit{N}-body simulations and the observed dark halo profiles of dwarf and LSB galaxies is a matter of debate in the scientific community. It is known as the `cuspy halo problem' or `core-cusp problem' \cite{Jing2000, Blok2010, Ogiya2011, Popolo2021}. Regarding the solution to this problem, Pontzen \& Governato, 2012 \cite{Pontzen2012} suggest the role of supernova-driven feedback that turns dark matter cusps into cores. Studies of the bar-driven structure formation mechanism similar to Chapter \ref{chap:4} may shed some light on this topic. Moreover, such analysis should be done separately for the following dark halo profiles: NFW \cite{Navarro1996} and oblate \cite{Zotos2012}. In this chapter, the NFW halo is used to model giant spiral galaxies, while the oblate halo is used to model dwarf galaxies.

\noindent Bar-driven structure formation is an essential aspect of the formation and evolution of disc galaxies. Under suitable physical conditions, the chaos that originates in the central galactic region leads to the formation of a bar and then escapes from the bar ends to form structures such as spiral arms, inner disc rings, etc. Such an escaping pattern resembles the chaotic scattering observed in open Hamiltonian systems and can be studied in greater detail if one investigates the trajectory of escaping stars. An open Hamiltonian system is a system where the energy shell is non-compact for energies above an escape threshold. As a result, a part of the stellar orbit explores an infinite part of the position space \cite{Aguirre2001}. There are numerous earlier works in the literature on bar-driven structure formation from the perspective of orbital and escape dynamics in phase space \cite{Navarro2001, Romero2006, Voglis2006, Romero2007, Zotos2011, Jung2016, Sanchez2016, Efthymiopoulos2019, Alrebdi2021, Alrebdi2022, Suraj2023}. These studies precisely identify that normally hyperbolic invariant manifolds in phase space are associated with such stellar escape mechanisms. Earlier, Mondal \& Chattopadhyay, 2020\cite{Mondal2020}; 2021 \cite{Mondal2021} investigated how the bar strength relates to the fate of bar-driven escaping stellar orbits. These works found that forming spiral arms is feasible for more robust bars and inner disc rings for weaker bars. Besides stellar bars, dark haloes also influence galaxy formation and evolution. However, the impact of dark haloes on bar-driven structure formation has received little attention so far \cite{Debattista2000}. Studies on the underlying relationship between the dark haloes and bar-driven structures are crucial for understanding the galaxy's evolutionary process. Also, such studies may shed some critical insight into the `cuspy halo problem'.

\indent In the upcoming sections of this chapter, the influence of dark haloes on the bar-driven structure formation process has been discussed from the viewpoint of stellar motions that may escape from the central region. The main focus of concern is the underlying relationships between the fate of escaping stars, the nature of dark haloes, and the amount of baryonic feedback produced therein. Also, the discussions are under the process of bar-driven structure formations in light of the `cuspy halo problem'.

\section{Barred Galaxy Model}
\label{sec:5.2}
Let us construct a three-dimensional gravitational model of barred galaxies to investigate the role of dark haloes on the orbital and escape dynamics of stars inside the central barred region. The model components are as follows: central black hole, bulge, bar, disc, and dark halo. All the modelling and numerical calculations are done in a Cartesian coordinate system and corrected up to four decimal places. Let $\Phi_\text{t}(x,y,z)$ be the total galactic potential and $\rho_\text{t}(x,y,z)$ be the associated density. The Poisson equation relates this potential-density pair as,
\begin{equation}
\label{eq:5.1}
\nabla^2 \Phi_\text{t}(x,y,z) = 4 \pi G \rho_\text{t}(x,y,z),
\end{equation}
 
\noindent where $G$ is the gravitational constant. Now $\Phi_\text{t}(x, y, z)$ can be written as,
\begin{equation*}
\Phi_\text{t}(x,y,z) = (\Phi_\text{bh} + \Phi_\text{B} + \Phi_\text{b} + \Phi_\text{d} + \Phi_\text{h})(x,y,z),
\end{equation*} 
where $\Phi_\text{bh}$, $\Phi_\text{B}$, $\Phi_\text{b}$, $\Phi_\text{d}$, $\Phi_\text{h}$ are the potentials of the central black hole, bulge, bar, disc and dark halo, respectively, and the galactic bar rotates at a constant pattern speed $\Omega_\text{b}$ in a clockwise sense around the z-axis. In this rotational reference frame of the bar, the effective potential is,
\begin{equation}
\label{eq:5.2}
\Phi_\text{eff}(x,y,z) = \Phi_\text{t}(x,y,z) - \frac{1}{2} \Omega_\text{b}^2 (x^2 + y^2).
\end{equation}

\noindent The Hamiltonian ($H$) or Jacobi integral of the given conservative system for a test particle (star) of unit mass is, 
\begin{equation}
\label{eq:5.3}
H = \frac{1}{2} (p_x^2 + p_y^2 + p_z^2) \; + \; \Phi_\text{t}(x,y,z) \; - \; \Omega_\text{b} L_z = E,
\end{equation}
where $\vec{r} \equiv (x,y,z)$, $\vec{p} \equiv (p_x,p_y,p_z)$ and $\vec{L} \; (=\vec{r} \times \vec{p}) \equiv (0,0,L_z = x p_y - y p_x)$ are the position, linear momentum vector, and angular momentum vector of test particle at time $t$, respectively. Because this is a conservative system, $H$ equals the system's total energy ($E$) and is also an integral of motion. For this, the system's governing equations, i.e., Hamilton's equations of motion, are as follows:

\begin{equation}
\label{eq:5.4}
\begin{split}
\dot{x} = p_x + \Omega_\text{b} y, \;
\dot{y} = p_y - \Omega_\text{b} x, \;
\dot{z} = p_z,\\
\dot{p}_x = - \frac{\partial \Phi_\text{t}}{\partial x} + \Omega_\text{b} p_y, \;
\dot{p}_y = - \frac{\partial \Phi_\text{t}}{\partial y} - \Omega_\text{b} p_x, \;
\dot{p}_z = - \frac{\partial \Phi_\text{t}}{\partial z},\\
\end{split}
\end{equation}
where `$\cdot$' is the time derivative `$\frac{\mathrm{d}}{\mathrm{dt}}$'. Now, the Lagrangian (or equilibrium) points of this system are solutions of the following equations,
\begin{equation}
\label{eq:5.5}
\frac{\partial \Phi_\text{eff}}{\partial x} = 0, \;\; 
\frac{\partial \Phi_\text{eff}}{\partial y} = 0, \;\; 
\frac{\partial \Phi_\text{eff}}{\partial z} = 0.
\end{equation}

In the upcoming parts of this chapter, this barred galaxy model has been analysed separately for the following two dark halo profiles: (i) NFW dark halo (model $1$) and (ii) oblate dark halo (model $2$). This helps us investigate the underlying relationship between the bar-driven escaping mechanism and dark halo strength.

\subsection{Gravitational potentials}
\label{sec:5.2.1} 
The potential forms of the central black hole, bulge, bar, disc and dark halo are as follows: 
\begin{itemize}[leftmargin=*]
\item Central black hole: A non-relativistic pseudo-Newtonian approach is adopted to model the central black hole to avoid relativistic calculations. For that, the Paczy{\'n}sky-Wiita potential \cite{Paczynsky1980, Abramowicz2009} is used, $$\Phi_\text{bh}(x,y,z) = - \frac{G M_\text{bh}}{\sqrt{x^2 + y^2 + z^2} - r_\text{s}},$$ where $M_\text{bh}$ is the black hole mass, $r_\text{s} = \frac{2 G M_\text{bh}}{c^2}$ is the Schwarzschild radius, and $c$ is the speed of light.

\item Bulge: To model a massive dense spherical bulge, the Plummer potential \cite{Plummer1911} is used, $$\Phi_\text{B}(x,y,z) = - \frac{G M_\text{B}}{\sqrt{x^2 + y^2 + z^2 + c_\text{B}^2}},$$ where $M_\text{B}$ is the bulge mass and $c_\text{B}$ is the scale length. 

\item Bar: For the central stellar bar, an anharmonic mass-model potential is considered, which resembles a strong bar \cite{Mondal2021}. This potential has the following form, $$\Phi_\text{b}(x,y,z) = - \frac{G M_\text{b}}{\sqrt{x^2 + \alpha^2 y^2 + z^2 + c_\text{b}^2}},$$ where $M_\text{b}$ is the bar mass, $\alpha$ is the flattening parameter, and $c_\text{b}$ is the scale length. 

\item Disc: The Miyamoto and Nagai potential \cite{Miyamoto1975} is used to model the flattened disc. This potential has the following form, $$\Phi_\text{d}(x,y,z) = - \frac{G M_\text{d}}{\sqrt{x^2 + y^2 + (k + \sqrt{h^2 + z^2})^2}},$$ where $M_\text{d}$ is the disc mass and $k$, $h$ are the corresponding horizontal and vertical scale lengths, respectively.

\item Dark halo (Model $1$): For the extended dark haloes as observed in massive disc galaxies, the Navarro-Frenk-White (NFW) potential \cite{Navarro1996} is considered, which resembles a cuspy dark halo profile (see Fig. \ref{fig:5.2a}), $$\Phi_\text{h}(x,y,z) = - \frac{G M_\text{vir}}{\ln (1 + c_\text{p}) - \frac{c_\text{p}}{1 + c_\text{p}}} \frac{\ln (1 + \frac{r}{c_\text{h}})}{r},$$ where $r^2 = x^2 + y^2 + z^2$, $M_\text{vir}$ is the virial mass of the dark halo, $c_\text{h}$ is the scale length, and $c_\text{p}$ is the concentration parameter.

\item Dark halo (Model $2$): For the core-dominated dark haloes as observed in low-mass disc galaxies, an oblate potential \cite{Binney2011, Zotos2012, Mondal2021} is considered, which resembles a flat dark halo profile (see Fig. \ref{fig:5.2b}), $$\Phi_\text{h}(x,y,z) = \frac{v_0^2}{2} \; \ln(x^2 + \beta^2 y^2 + z^2 + c_\text{h}^2),$$ where $v_0$ is the circular velocity of the dark halo, $\beta$ is the flattening parameter, and $c_\text{h}$ is the scale length.
\end{itemize}

\subsection{Parameter values}
\label{sec:5.2.2}
Without losing any generality, let us assume $G = 1$ and use the following unit system \cite{Jung2016}: unit of length - $1$ kpc, unit of mass - $2.325 \times 10^7 M_\odot$, unit of time - $0.9778 \times 10^8$ yr, unit of velocity - $10$ km $\text{s}^{-1}$, angular momentum unit per unit mass - $10$ km $\text{s}^{-1}$ $\text{kpc}^{-1}$, energy unit per unit mass - $100$ $\text{km}^2$ $\text{s}^{-2}$. Table \ref{tab:5.1} shows the values of all physical parameters \cite{Zotos2012, Jung2016}. For disc galaxies, the mass of the central black hole typically lies within the range: $10^8 M_\odot \le M_\text{bh} \le 10^{10} M_\odot$ \cite{Binney2011, Zotos2020a}. Here, the model $1$ is developed to represent massive disc galaxies, so for this,  $M_\text{bh} = 10^{10} M_\odot$ value is used, which is numerically equivalent to $430$ in the above-defined mass unit. Again, the model $2$ is developed to represent low-mass disc galaxies, so for this, $M_\text{bh} = 10^8 M_\odot$ value is used, which is numerically equivalent to $4.3$ in the above-defined mass unit.
\begin{table}
\centering	
\begin{tabular}{|c|c||c|c|}
\hline
Parameter          & Value           & Parameter      & Value\\
\hline
\hline
$M_\text{bh}$      & 430 (model $1$) & $M_\text{d}$   & 7000\\   
$M_\text{bh}$      & 4.3 (model $2$) & $k$            & 3\\       
$c$                & $3 \times 10^4$ & $h$            & 0.175\\
$M_\text{B}$       & 400             & $M_\text{vir}$ & $2 \times 10^4$\\
$c_\text{B}$       & 0.25            & $c_\text{p}$   & 15\\
$M_\text{b}$       & 3500            & $c_\text{h}$   & 20\\
$\alpha$           & 2               & $v_0$          & 15\\
$c_\text{b}$       & 1               & $\beta$        & 1.3\\
$\Omega_\text{b}$  & 1.25            &                &\\
\hline
\end{tabular}
\caption{Physical parameter values.}
\label{tab:5.1}
\end{table}

\begin{figure}
\centering
\subfigure[Model $1$ (NFW dark halo)]{\label{fig:5.1a}\includegraphics[width=0.49\columnwidth]{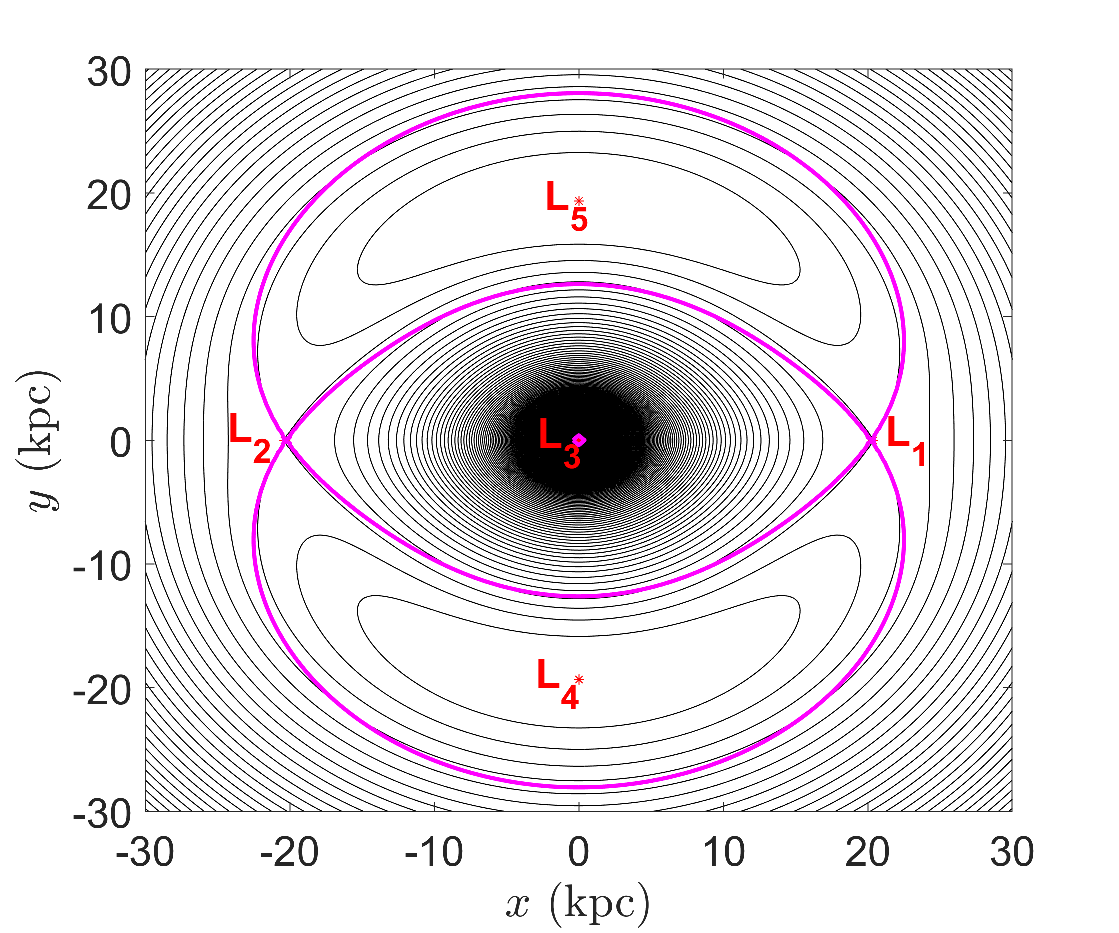}}
\subfigure[Model $2$ (oblate dark halo)]{\label{fig:5.1b}\includegraphics[width=0.49\columnwidth]{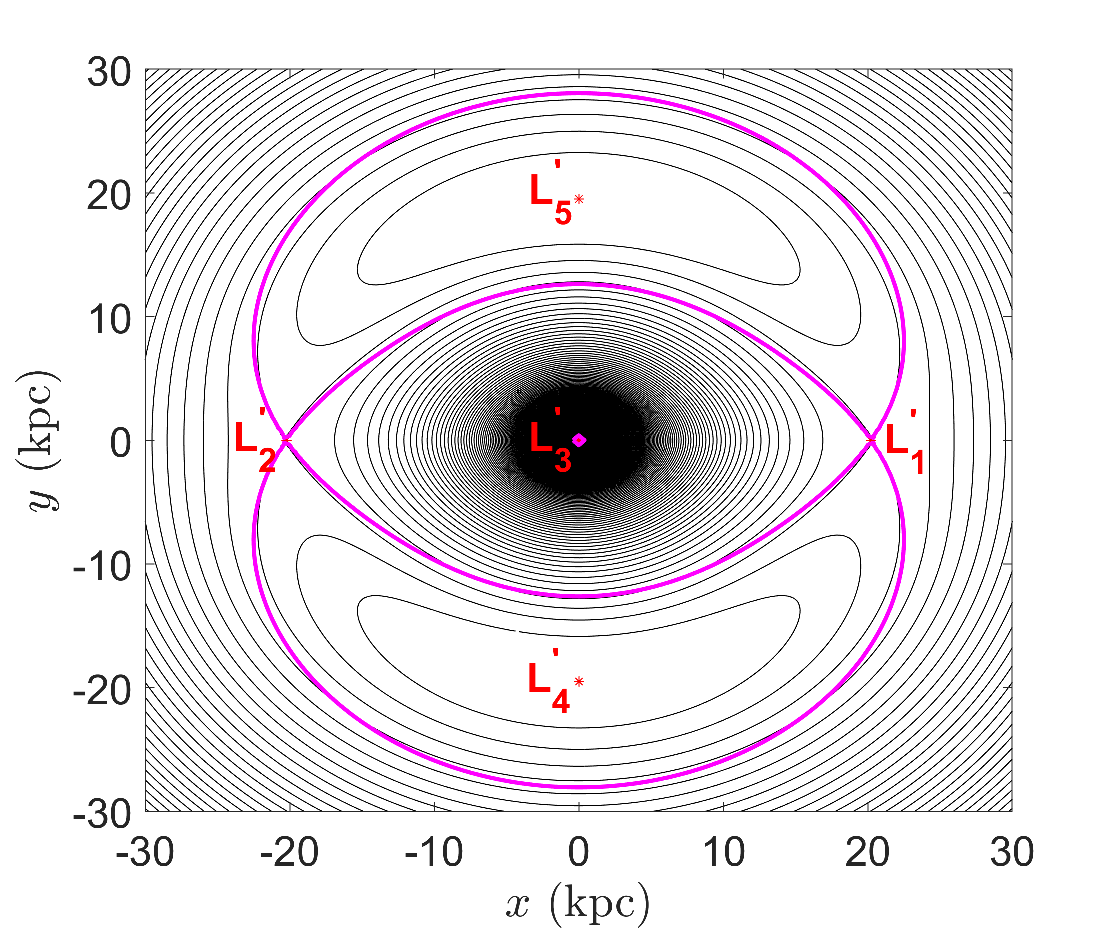}}
\caption{The isoline contours of $\Phi_\text{eff}(x,y,z)$ in the $x - y$ plane for $z = 0$, where the locations of the five Lagrangian points are marked in red, while the contours in magenta are correspond to the energy values of the index-1 saddle point $L_1$ (and $L_1^{'}$).}
\label{fig:5.1}
\end{figure}

The described gravitational model has five Lagrangian points (solutions of the system of Eqs. (\ref{eq:5.5})) in each case, namely $L_1$, $L_2$, $L_3$, $L_4$, $L_5$ for model $1$ and $L_1^{'}$, $L_2^{'}$, $L_3^{'}$, $L_4^{'}$, $L_5^{'}$ for model $2$. The locations of the Lagrangian points and their types for both cases are given in Fig. \ref{fig:5.1} and Table \ref{tab:5.2}.

\begin{table}
\centering
\begin{tabular}{|c|c|c|}
\hline  
Lagrangian Point            & Lagrangian Point                & Type\\  
(Model $1$)                 & (Model $2$)                     &\\                
\hline
\hline
$L_1 \equiv (20.3804,0,0)$  & $L_1^{'} \equiv (20.2337,0,0)$  & Index-1 Saddle\\
$L_2 \equiv (-20.3804,0,0)$ & $L_2^{'} \equiv (-20.2337,0,0)$ & Index-1 Saddle\\
$L_3 \equiv (0,0,0)$        & $L_3^{'} \equiv (0,0,0)$        & centre\\
$L_4 \equiv (0,-19.3505,0)$ & $L_4^{'} \equiv (0,-19.5010,0)$ & Index-2 Saddle\\
$L_5 \equiv (0,19.3505,0)$  & $L_5^{'} \equiv (0,19.5010,0)$  & Index-2 Saddle\\
\hline
\end{tabular}
\caption{Lagrangian point locations and their types.}
\label{tab:5.2}
\end{table} 

\noindent Due to the structural symmetry, it is apparent that the energy values, i.e., the values of the Jacobi integral of motion at $L_1$ (or $L_1^{'}$) and $L_2$ (or $L_2^{'}$), i.e., $E_{L_1}$ (or $E_{L_1^{'}}$) and $E_{L_2}$ (or $E_{L_2^{'}}$) respectively are identical. Similarly, those values at $L_4$ (or $L_4^{'}$) and $L_5$ (or $L_5^{'}$), i.e., $E_{L_4}$ (or $E_{L_4^{'}}$) and $E_{L_5}$ (or $E_{L_5^{'}}$) respectively are identical (see Table \ref{tab:5.3}).

\begin{table}
\centering
\begin{tabular}{|c|c|}
\hline
Energy Value                     & Energy Value \\
(Model $1$)                      & (Model $2$)\\
\hline
\hline
$E_{L_1} / E_{L_2} = -1251.8555$ & $E_{L_1^{'}} / E_{L_2^{'}} = -101.0357$\\
$E_{L_3} = -\infty$              & $E_{L_3^{'}} = 4.4999 \times 10^8$\\
$E_{L_4} / E_{L_5} = -1163.9815$ & $E_{L_4^{'}} / E_{L_5^{'}} = 19.9934$\\
\hline        
\end{tabular}
\caption{Energy values at the Lagrangian points.}
\label{tab:5.3}
\end{table} 

\noindent The following describes the nature of orbits in various energy domains: 
\begin{enumerate}[label=(\roman*)]
\item $E_{L_3} \leq E < E_{L_1}$ (or $E_{L_3^{'}} \leq E < E_{L_1^{'}}$): In this energy domain, stellar orbits follow bound motions inside the barred region.
\item $E = E_{L_1}$ (or $E = E_{L_1^{'}}$): This is the threshold energy for stellar escape through the bar ends.
\item $E > E_{L_1}$ (or $E > E_{L_1^{'}}$): Stellar escape may be possible in this energy domain through the two symmetrical exit channels near $L_1$ and $L_2$, depending upon the initial starting point.
\end{enumerate}

\subsection{NFW versus oblate dark halo}
\label{sec:5.2.3}
\begin{figure}
\centering
\subfigure[Model $1$: Dark halo density ($\rho_\text{h}$) distribution]{\label{fig:5.2a}\includegraphics[width=0.49\columnwidth]{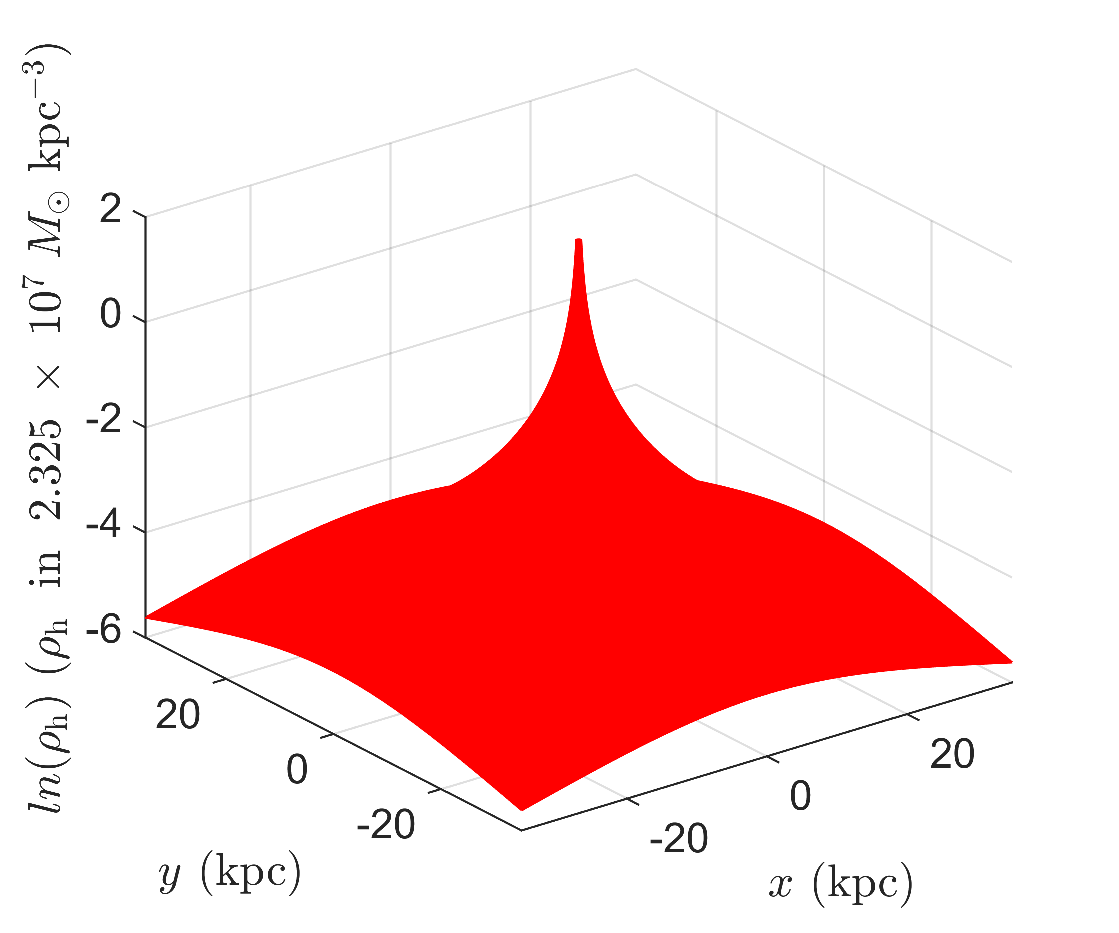}}
\subfigure[Model $2$: Dark halo density ($\rho_\text{h}$) distribution]{\label{fig:5.2b}\includegraphics[width=0.49\columnwidth]{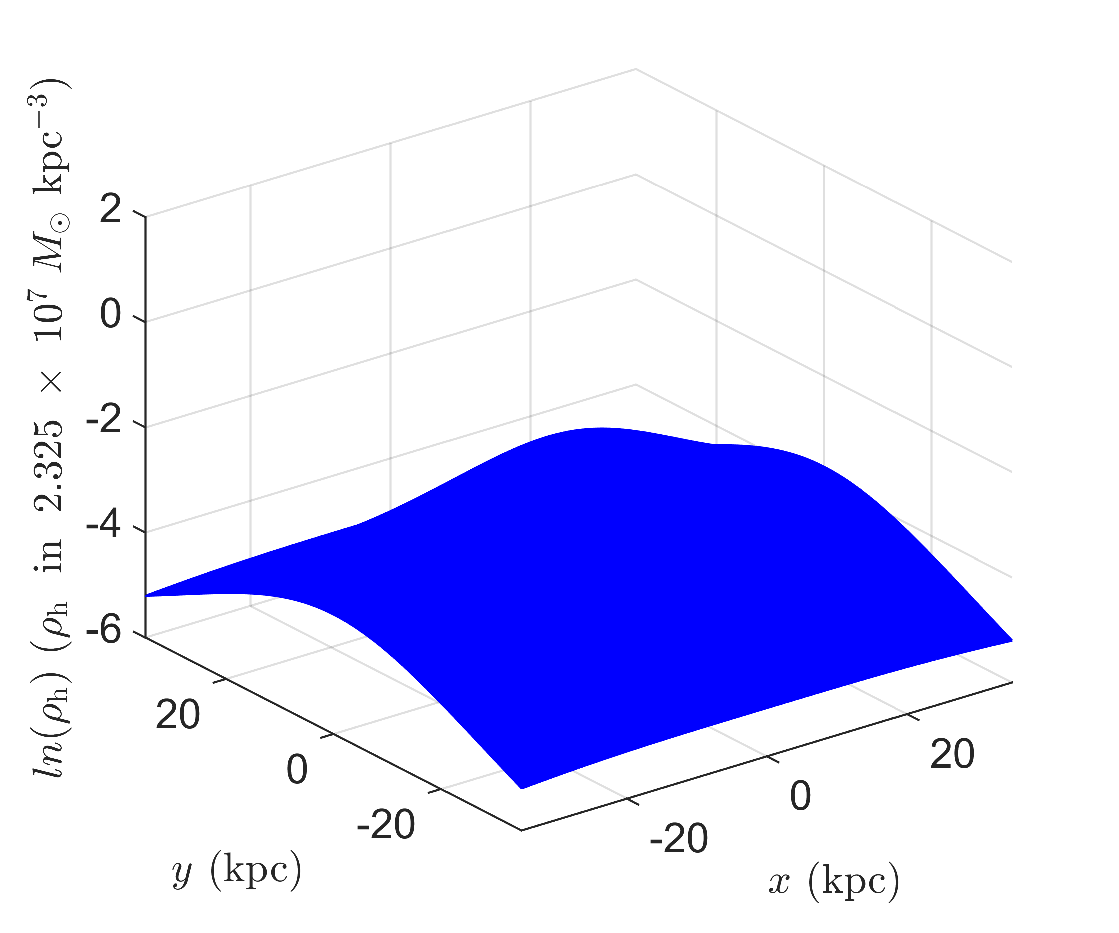}}
\subfigure[Rotation curve]{\label{fig:5.2c}\includegraphics[width=0.49\columnwidth]{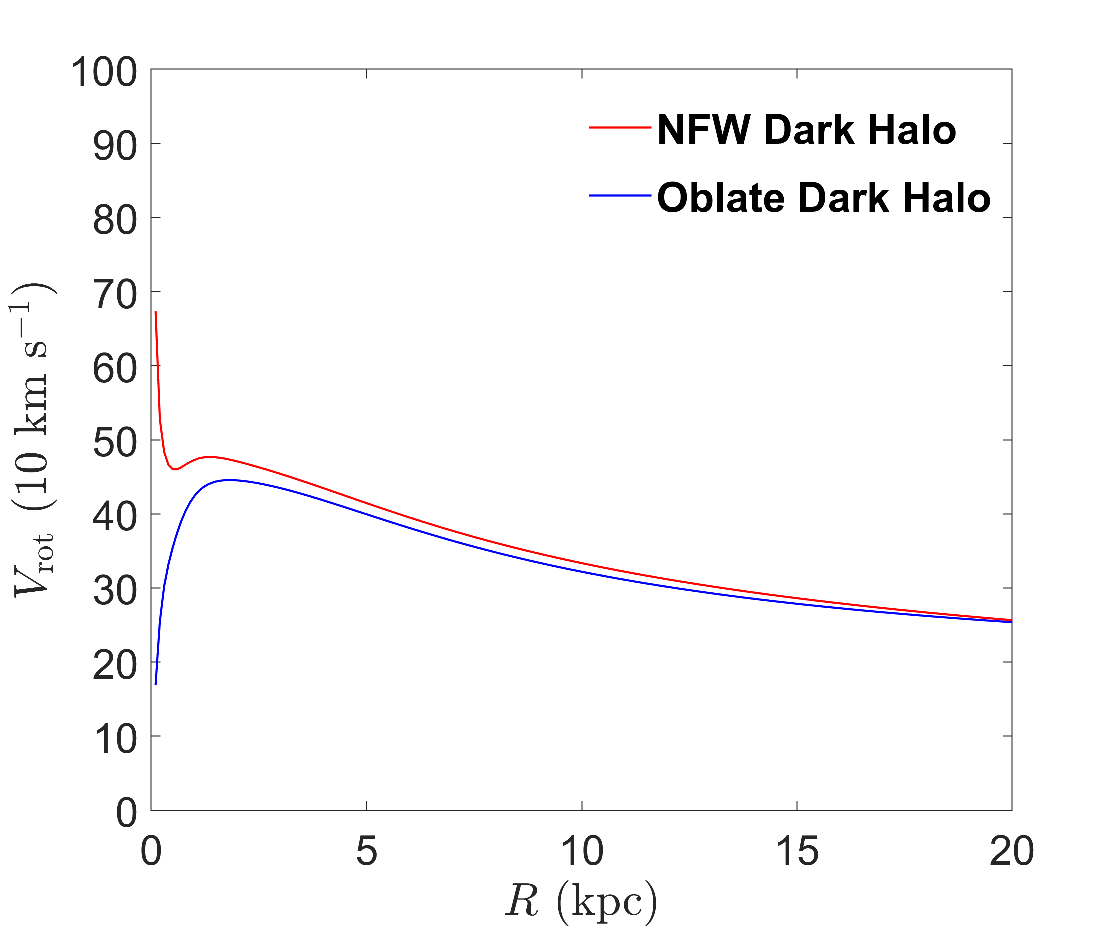}}
\subfigure[Radial force]{\label{fig:5.2d}\includegraphics[width=0.49\columnwidth]{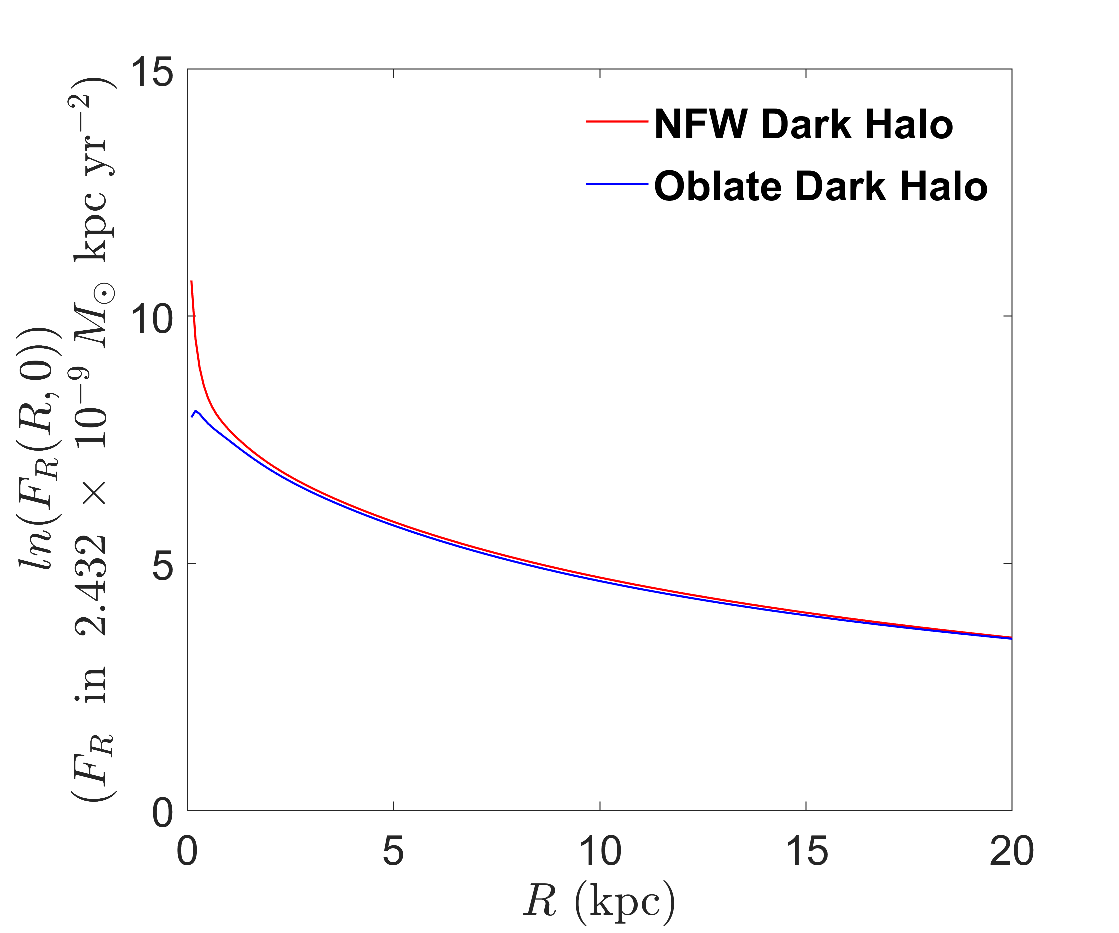}}
\caption{Characteristics: Model $1$ (NFW dark halo) versus model $2$ (oblate dark halo).}
\label{fig:5.2}
\end{figure}

The comparisons between the physical characteristics of model $1$ and $2$ are as follows:
\begin{itemize}[leftmargin=*]
\item Dark halo density: The distribution of dark halo densities ($\rho_\text{h}$) of models $1$ and $2$ are shown in Figs. \ref{fig:5.2a} and \ref{fig:5.2b}, respectively. The dark halo density distributions indicate a cuspy distribution for model $1$ and a flat distribution for model $2$.

\item Rotation curve: Fig. \ref{fig:5.2c} shows the rotation curves ($V_\text{rot} (R,0) = \sqrt{R \frac{\partial}{\partial R} \Phi_\text{t}(R,0)}$ versus $R = \sqrt{x^2 + y^2}$) of both models $1$ and $2$. This figure shows that in the central bulge region ($R < 5$), the rotational velocity is way higher for model $1$ than model $2$. This is due to the accretion effect of the central black hole. The model $1$ has a higher central black hole mass than the model $2$, thus having a greater accretion effect than the model $2$. That is why the excess rotational velocity is observed inside the bulge. However, this trend is nearly similar for both models outside the bulge region ($R > 5$).

\item Radial force: The distribution of the radial force component ($F_R (R,0) = \frac{\partial}{\partial R}\Phi_\text{t} (R,0)$ versus $R = \sqrt{x^2 + y^2}$) of both models $1$ and $2$ is shown in Fig. \ref{fig:5.2d}. The figure shows that the force distributions become steeper as $R \to 0$. For model $1$, the steepness is higher due to a higher central black hole mass than for model $2$. This steepness becomes nearly similar for both models with an increment of $R$.
\end{itemize}

\section{Phase Space Analysis}
\label{sec:5.3}
Let us set $z = 0 = p_z$ to study the stellar orbital and escape dynamics inside the barred region along the plane of the bar. In the energy domain: $E \geq E_{L_1}$ (or $E_{L_1^{'}}$), stellar orbits may escape through the symmetrical exit channels that exist near the bar ends, i.e., $L_1$, $L_2$ (or $L_1^{'}$, $L_2^{'}$) (see Fig. \ref{fig:5.1}). Since the effective potential has structural symmetry about the $y$ axis, the study of tracking the stellar escape dynamics in the phase space is limited to $L_1$ (or $L_1^{'}$). During the analyses, the energy parameter ($E$) is normalised with respect to $E_{L_1}$ (or $E_{L_1^{'}}$) to make it a dimensionless quantity ($C$) \cite{Mondal2021}. This parameter ($C$) has the following form:
\begin{equation}
\label{eq:5.6}
C = \frac{E_{L_1} (\text{or} \; E_{L_1^{'}}) - E}{E_{L_1} (\text{or} \; E_{L_1^{'}})} = \frac{E_{L_2} (\text{or} \; E_{L_2^{'}}) - E}{E_{L_2} (\text{or} \; E_{L_2^{'}})}.
\end{equation} 
$$(\because E_{L_1} = E_{L_2} \; \text{and} \; E_{L_1^{'}} = E_{L_2^{'}})$$ Thus, $C = 0$ becomes the energy threshold corresponding to escape from the barred region. Depending on their starting point in the phase space, orbits may escape if $C > 0$. In the phase space this escaping motion has been studied for the following energy levels: $C = 0.01$ and $C = 0.1$. The orbital maps and Poincaré surface section maps in the $x - y$ and $x - p_x$ planes are visualised for the above energy levels to track the escaping motion in phase space. In these maps, all the chosen initial conditions are restricted within the Lagrange radius, i.e., $x_{0}^2 + y_{0}^2 \leq r_{L_1}^2$ (or $r_{L_1^{'}}^2$), where $r_{L_1}$ (or $r_{L_1^{'}}$) is the radial length of $L_1$ (or $L_1^{'}$) and $(x_0, y_0)$ is an initial condition in the $x - y$ plane. Also, the system of differential Eqs. (\ref{eq:5.4}), i.e., Hamiltonian equations are solved for a given choice of initial conditions to track the trajectory of stellar orbits. The $\tt{ode45}$ package from $\tt{MATLAB}$ programming is used to do it numerically, with a small time step: $\Delta t = 10^{-2}$ time units ($1$ time unit is equivalent to $0.9778 \times 10^8$ yr), to track the time evolution of stellar trajectories up to $10^2$ time units (total orbit integration time) because the age of stellar bars is typically around $10^{10}$ years \cite{Sharma2019}, which is nearly equivalent to $10^2$ time units according to the above-defined scaling relation. The chaotic nature of the stellar orbits is determined with the help of chaos detector MLE \cite{Strogatz2018}, which is defined in Eq. (\ref{eq:2.16}), in which the initial separation between two infinitesimal trajectories is taken as $\delta x(t_0) = 10^{-8}$.

\subsection{Orbital maps}
\label{sec:5.3.1}
The trajectories of stellar orbits are tracked in the $x - y$ plane to visualise the bar-driven escapes along the galactic plane of the bar. Only stellar orbits whose initial condition lies within the energetically allowed region: $\Phi_\text{eff}(x,y,z) < E_{L_1} (\text{or} \; E_{L_1^{'}})$ are taken into consideration. The colour-coded diagrams are plotted for different escape energy values in the $x - y$ plane. In these diagrams, a $256\times256$ grid of initial conditions in the $x - y$ plane within the central barred region is classified according to the nature of stellar orbit starting from a specific initial condition $(x_0,y_0,p_{x_0} = 0,p_{y_0})$, where the $p_{y_0}$ value is determined by Eq. (\ref{eq:5.3}). Here, a stellar orbit starting with an initial condition can be classified into the following categories: (i) escaping from $L_1$ or $L_1^{'}$ (shaded in red), (ii) escaping from $L_2$ or $L_2^{'}$ (shaded in green), and (iii) bounded (shaded in blue).

Also, plotted orbital maps are plotted for a sample initial condition: $(x_0,y_0,p_{x_0},p_{y_0}) \equiv (5,0,15,p_{y_0})$, where the $p_{y_0}$ value is determined by Eq. (\ref{eq:5.3}). This $(5,0,15,p_{y_0})$ is chosen to investigate the escape dynamics in the neighbourhood of $L_1$ (or $L_1^{'}$) \cite{Mondal2021}. There is nothing specific about this initial condition. Any other point in its appropriate neighbourhood will exhibit similar orbital trajectories.

\begin{itemize}[leftmargin=*]	 
\item Model $1$: For the NFW dark halo profile, the classification of orbits is shown in Fig. \ref{fig:5.3} for escape energy values $C = 0.01$ and $0.1$. For $C = 0.01$, it is observed that the barred region is mainly covered by bounded orbits compared to escaping orbits (see Fig. \ref{fig:5.3a}). However, for $C = 0.1$, the number of escaping orbits has been significantly increased (see Fig. \ref{fig:5.3b}). Also, many small escaping islands near the centre for $C = 0.01$ mostly faded as $C$ increased from $0.01$ to $0.1$.
\begin{figure}
\centering
\subfigure[$C = 0.01$]{\label{fig:5.3a}\includegraphics[height=0.4\columnwidth,width=0.49\columnwidth]{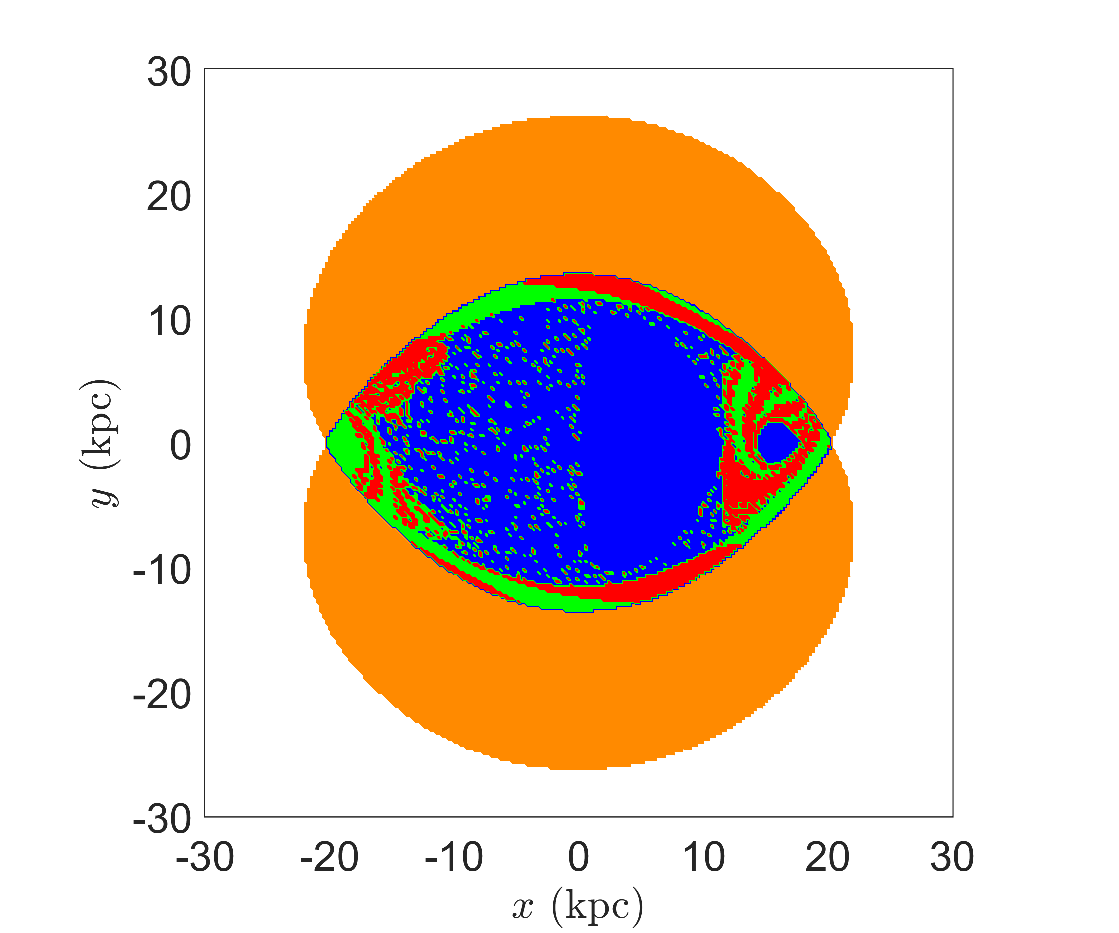}}
\subfigure[$C = 0.1$]{\label{fig:5.3b}\includegraphics[height=0.4\columnwidth,width=0.49\columnwidth]{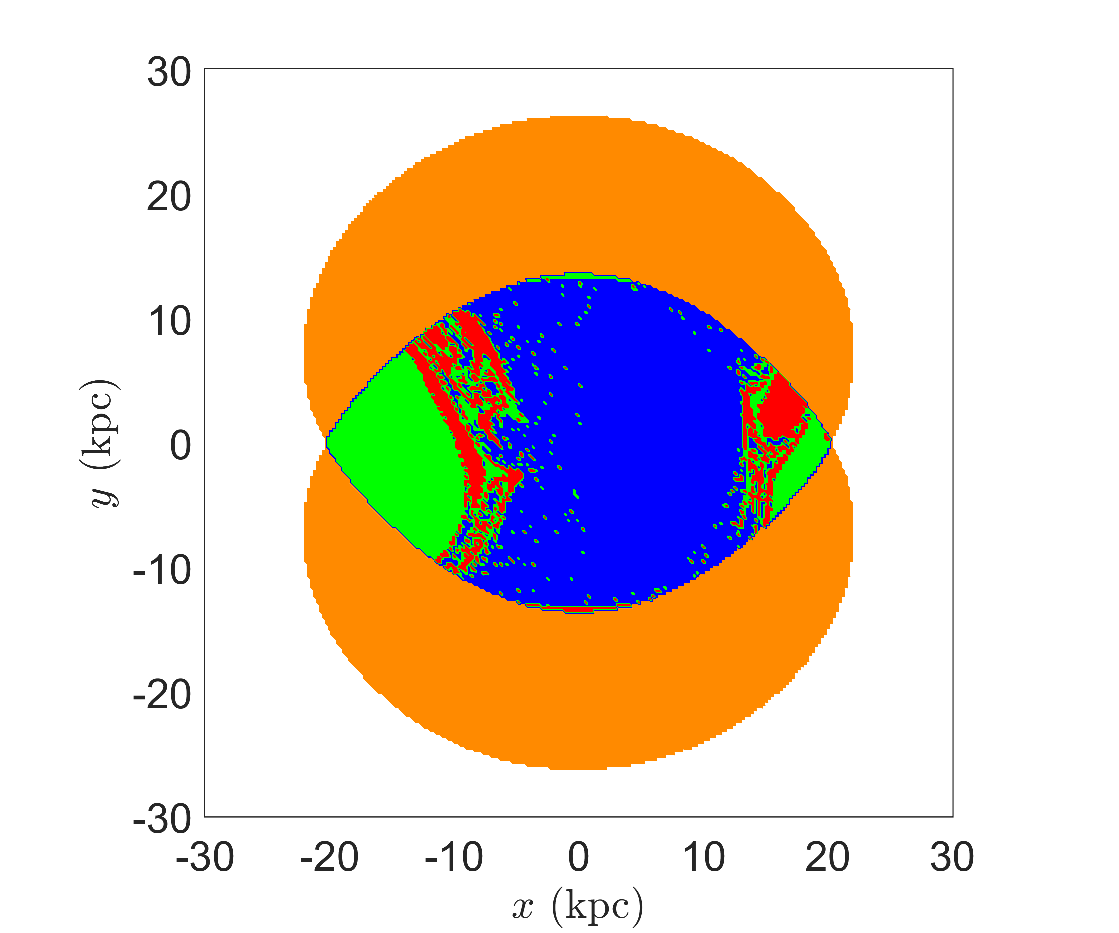}}
\caption{Model $1$: Orbit classification maps according to their final fate: orbits escape through $L_1$ (red), orbits escape through $L_2$ (green), bounded region (blue), energetically forbidden region (orange).}
\label{fig:5.3}
\end{figure}

On the other hand, orbital maps in the $x - y$ plane are plotted in Fig. \ref{fig:5.4} for escape energy values $C = 0.01$ and $0.1$. These figures show that orbits are non-escaping and chaotic for a lower escape energy value, i.e., $C = 0.01$ (see Fig. \ref{fig:5.4a}). In contrast, chaotic escaping motion is observed inside the barred region for a higher escape energy value, i.e., $C = 0.1$ (see Fig. \ref{fig:5.4b}). The MLE values for both the above trajectories are evaluated using Eq. (\ref{eq:2.16}) and listed in Table \ref{tab:5.4}.
\begin{figure}
\centering
\subfigure[$C = 0.01$]{\label{fig:5.4a}\includegraphics[width=0.49\columnwidth]{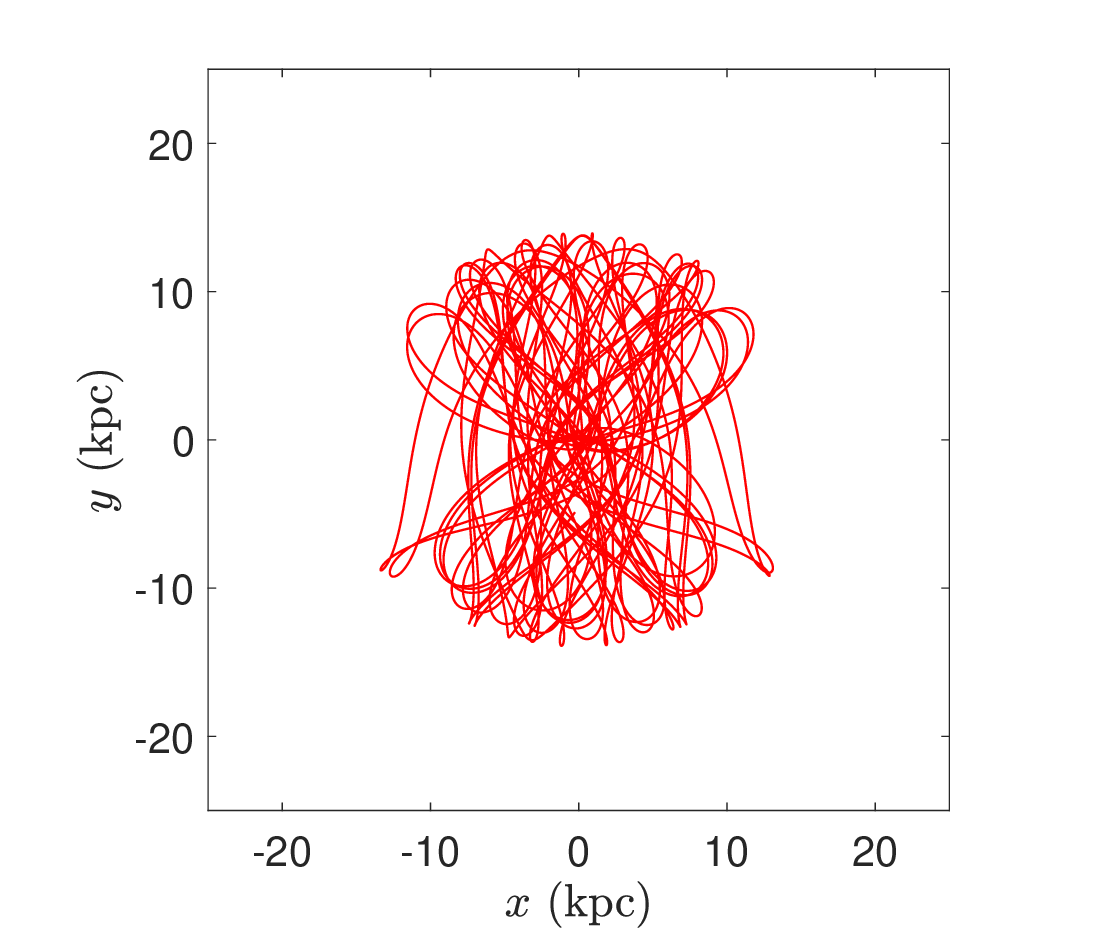}}
\subfigure[$C = 0.1$]{\label{fig:5.4b}\includegraphics[width=0.49\columnwidth]{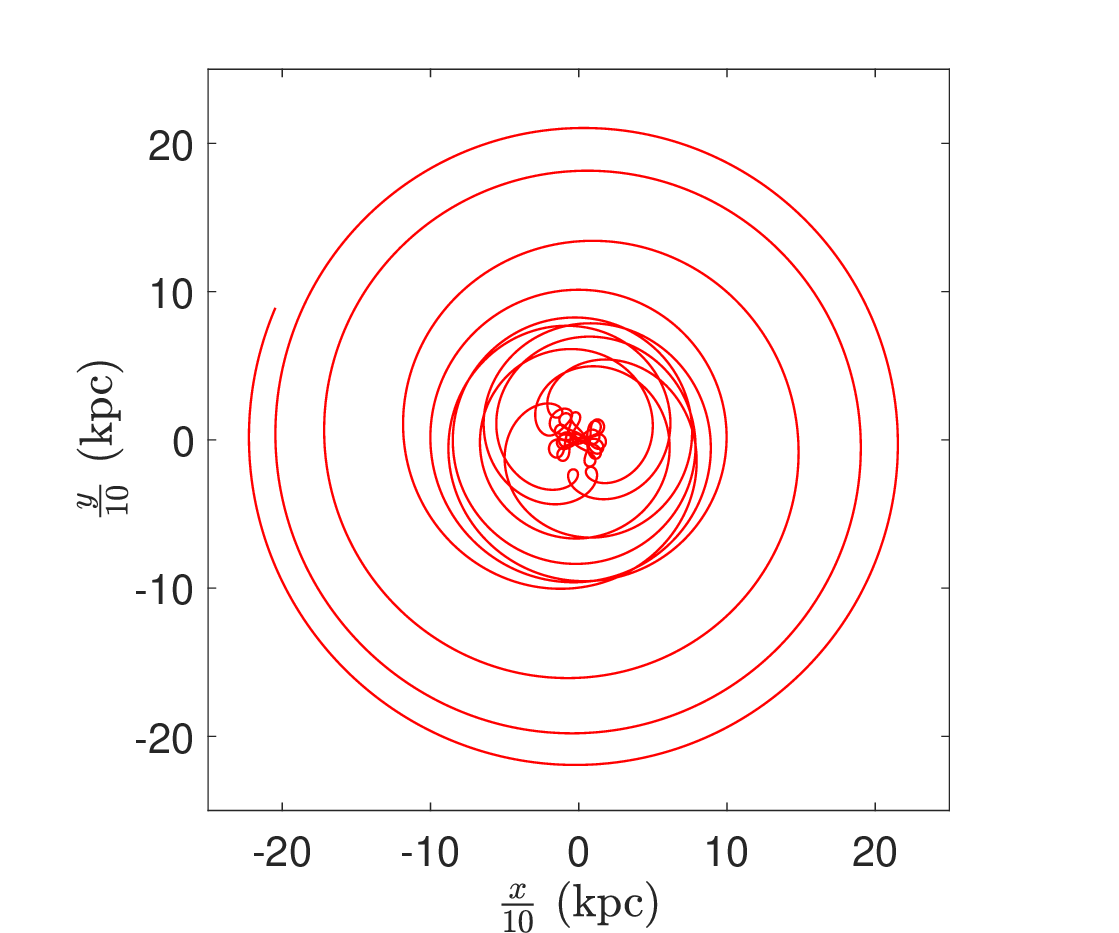}}
\caption{Model $1$: Orbits in the $x - y$ plane for $(x_0,y_0,p_{x_0},p_{y_0})$ $\equiv (5,0,15,p_{y_0})$, where $p_{y_0}$ value is evaluated from Eq. (\ref{eq:5.3}): (a) is non-escaping chaotic orbit, (b) is escaping chaotic orbit.}
\label{fig:5.4}
\end{figure}

\begin{table}
\centering
\begin{tabular}{|c|c|c|c|}
\hline
                                   Initial Condition & $M_\text{bh}$ & $C$    & MLE\\
\hline
\hline
$(x_0,y_0,p_{x_0},,p_{y_0}) \equiv (5,0,15,p_{y_0})$ & $430$         & $0.01$ & $0.2192$\\
                                                     &               & $0.1$  & $0.1693$\\
\hline
\end{tabular}
\caption{Model $1$: MLE value for different values of $C$, where $p_{y_0}$ value is evaluated from Eq. (\ref{eq:5.3}).}
\label{tab:5.4}
\end{table}

\item Model $2$: For the oblate dark halo profile, the classification of orbits is shown in Fig. \ref{fig:5.5} for escape energy values $C = 0.01$ and $0.1$. These figures show that, for $C = 0.01$, bounded orbits, instead of escaping orbits, primarily cover the barred region (see Fig. \ref{fig:5.5a}). However, the number of escaping orbits has substantially risen for $C = 0.1$ (see Fig. \ref{fig:5.5b}). Overall, for each $C$ value, the total number of escaping orbits is lower than model $1$.
\begin{figure}
\centering
\subfigure[$C = 0.01$]{\label{fig:5.5a}\includegraphics[height=0.4\columnwidth,width=0.49\columnwidth]{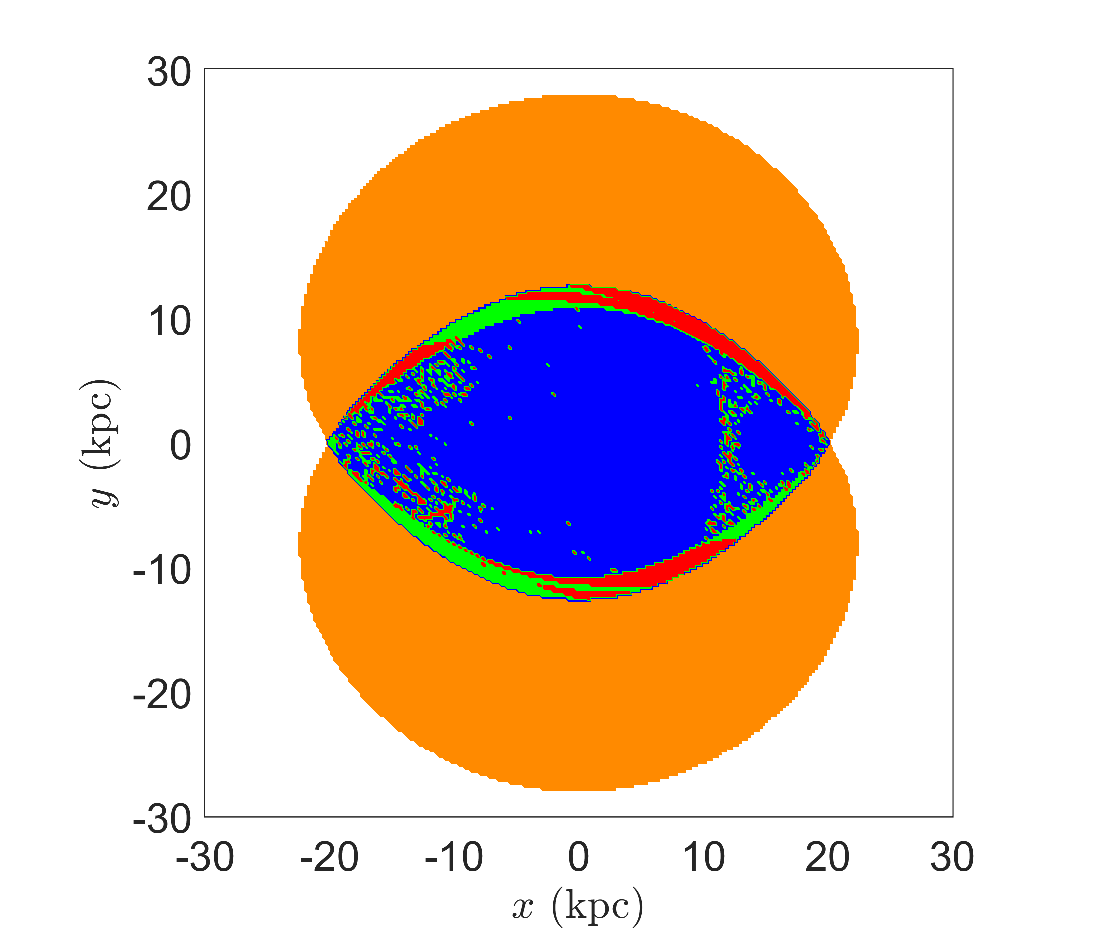}}
\subfigure[$C = 0.1$]{\label{fig:5.5b}\includegraphics[height=0.4\columnwidth,width=0.49\columnwidth]{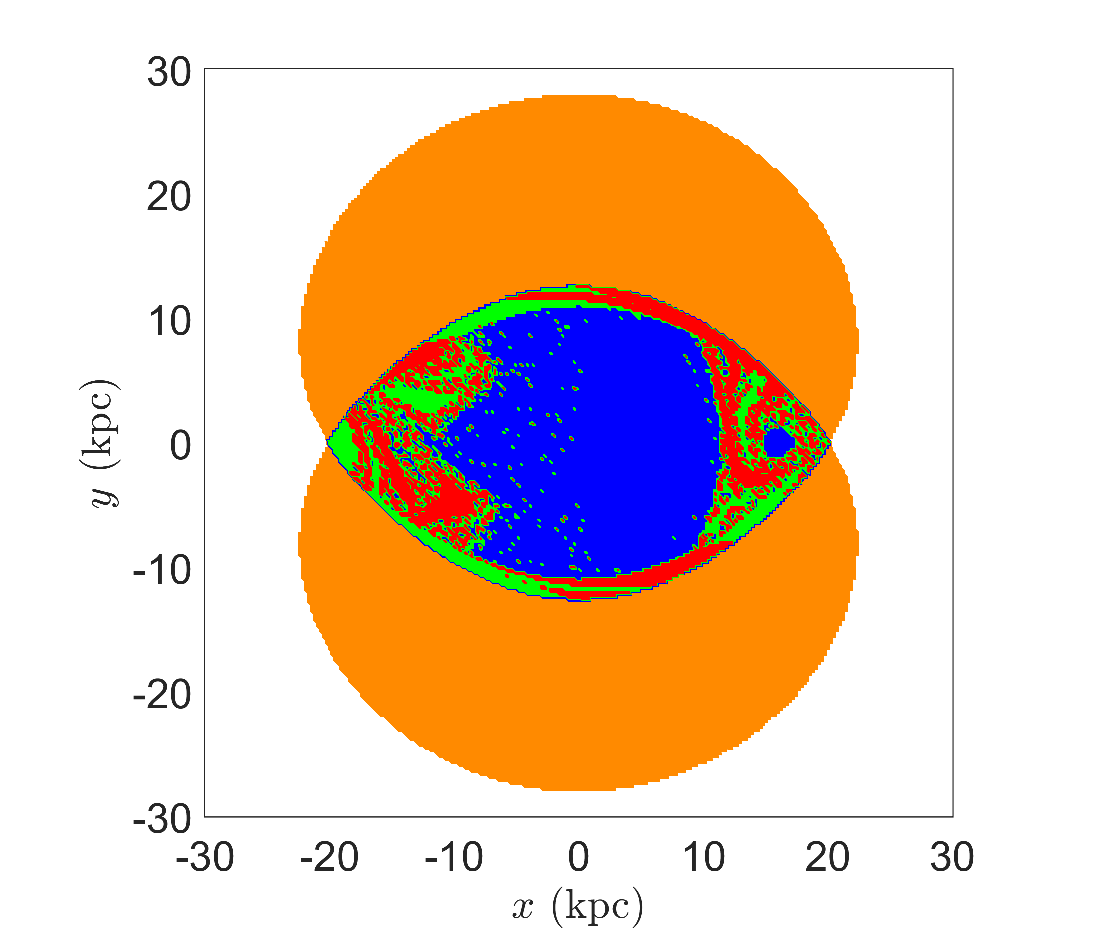}}
\caption{Model $2$: Orbit classification maps according to their final fate: orbits escape through $L_1^{'}$ (red), orbits escape through $L_2^{'}$ (green), bounded region (blue), energetically forbidden region (orange).}
\label{fig:5.5}
\end{figure}

On the other hand, orbital maps in the $x - y$ plane are plotted in Fig. \ref{fig:5.6} for escape energy values $C = 0.01$ and $0.1$. In these figures, it is seen that orbits are non-escaping and chaotic regardless of escape energy values, i.e., orbits remain bounded inside the barred region (see Figs. \ref{fig:5.6a} and \ref{fig:5.6b}). The MLE values for both the above trajectories are evaluated from Eq. (\ref{eq:2.16}) and listed in Table \ref{tab:5.5}.
\begin{figure}
\centering
\subfigure[$C = 0.01$]{\label{fig:5.6a}\includegraphics[width=0.49\columnwidth]{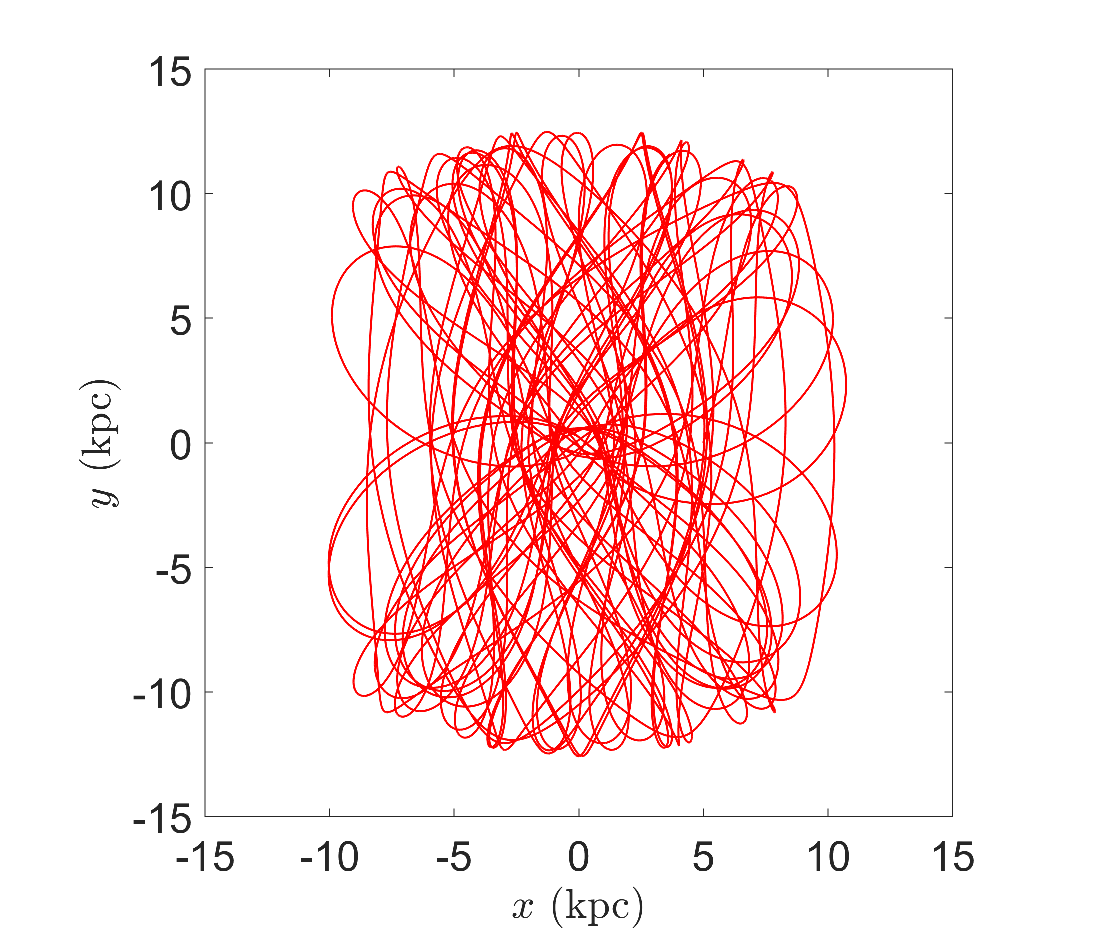}}
\subfigure[$C = 0.1$]{\label{fig:5.6b}\includegraphics[width=0.49\columnwidth]{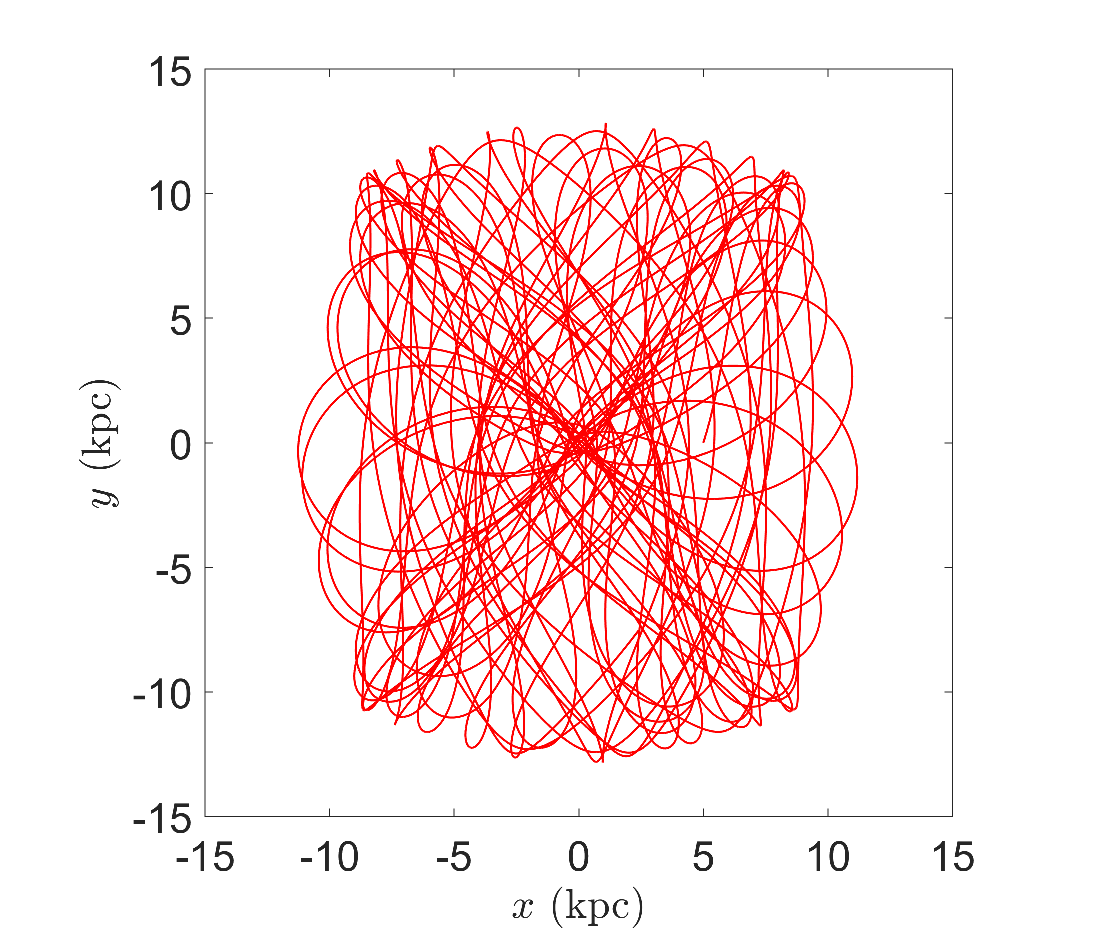}}
\caption{Model $2$: Orbits in the $x - y$ plane for $(x_0,y_0,p_{x_0},p_{y_0})$ $\equiv (5,0,15,p_{y_0})$, where $p_{y_0}$ value is evaluated from Eq. (\ref{eq:5.3}): (a)-(b) are non-escaping chaotic orbits.}
\label{fig:5.6}
\end{figure}

\begin{table}
\centering
\begin{tabular}{|c|c|c|c|}
\hline
                                   Initial Condition & $M_\text{bh}$ & $C$    & MLE\\
\hline
\hline
$(x_0,y_0,p_{x_0},,p_{y_0}) \equiv (5,0,15,p_{y_0})$ & $4.3$         & $0.01$ & $0.2094$\\
                                                     &               & $0.1$  & $0.1944$\\
\hline
\end{tabular}
\caption{Model $2$: MLE value for different values of $C$, where $p_{y_0}$ value is evaluated from Eq. (\ref{eq:5.3}).}
\label{tab:5.5}
\end{table}
\end{itemize}

\subsection{Poincaré maps}
\label{sec:5.3.2} 
\noindent Poincaré surface section maps in the $x - y$ and $x - p_x$ planes are plotted to visualise the bar-driven escapes in phase space. Now, for $z = 0 = p_z$, the phase space becomes four-dimensional. As a result, the surface cross-sections of Poincaré maps will be two-dimensional planes. For all these Poincaré maps, only initial conditions lying within the energetically allowed region: $\Phi_\text{eff}(x,y,z) < E_{L_1} (\text{or} \; E_{L_1^{'}})$ are taken into consideration. For Poincaré maps in the $x - y$ plane (Figs. \ref{fig:5.7} and \ref{fig:5.9}), a $43\times43$ grid of initial conditions has been set up in the $x - y$ plane with step sizes: $\Delta x = 1$ kpc and $\Delta y = 1$ kpc. Only initial conditions within the Lagrange radius ($r_{L_1}$ or $r_{L_1^{'}}$) are considered from this grid. The initial value of $p_x$, i.e., $p_{x_0}$ is considered to be $p_{x_0} = 0$. In addition, the initial value of $p_y$, i.e., $p_{y_0}$ is taken into account as $p_{y_0} > 0$ and further evaluated from Eq. (\ref{eq:5.3}). The chosen surface cross-sections for these Poincaré surface section maps are $p_x = 0$ and $p_y \le 0$ \cite{Ernst2014}. 

Similarly, for Poincaré maps in the $x - p_x$ plane (Figs. \ref{fig:5.8} and \ref{fig:5.10}), a $43\times31$ grid of initial conditions has been set up in the $x - p_x$ plane with step sizes: $\Delta x = 1$ kpc and $\Delta p_x = 10$ km $\text{s}^{-1}$. Here also, only initial conditions within the Lagrange radius ($r_{L_1}$ or $r_{L_1^{'}}$) are considered from this grid. The initial value of $y$, i.e., $y_0$ is considered to be $y_0 = 0$. Moreover, the initial value of $p_y$, i.e., $p_{y_0}$ is considered as $p_{y_0} > 0$ and further evaluated from Eq. (\ref{eq:5.3}). The chosen surface cross-sections for these Poincaré surface section maps are $y = 0$ and $p_y \le 0$ \cite{Ernst2014}.

\begin{itemize}[leftmargin=*]
\item Model $1$: For the NFW dark halo profile, Poincaré surface section maps in the $x - y$ plane are plotted in Fig. \ref{fig:5.7} for escape energy values $C = 0.01$ and $0.1$. In these figures, the following patterns have been identified:
\begin{enumerate}[label=(\roman*)]
\item In both maps, a stability island is identified near $(7, 0)$, corresponding to quasi-periodic stellar motions.

\item The concentration of cross-sectional points inside the corotation region of the bar and disc is more significant for a lower escape energy (i.e., $C = 0.01$) than for a higher escape energy (i.e., $C = 0.1$). On the contrary, the number of cross-sectional points outside that corotation region that signifies the bar-driven escape has increased significantly with an increment of $C$ from $0.01$ to $0.1$.
\end{enumerate}
\begin{figure}
\centering
\subfigure[$C = 0.01$]{\label{fig:5.7a}\includegraphics[height=0.4\columnwidth,width=0.49\columnwidth]{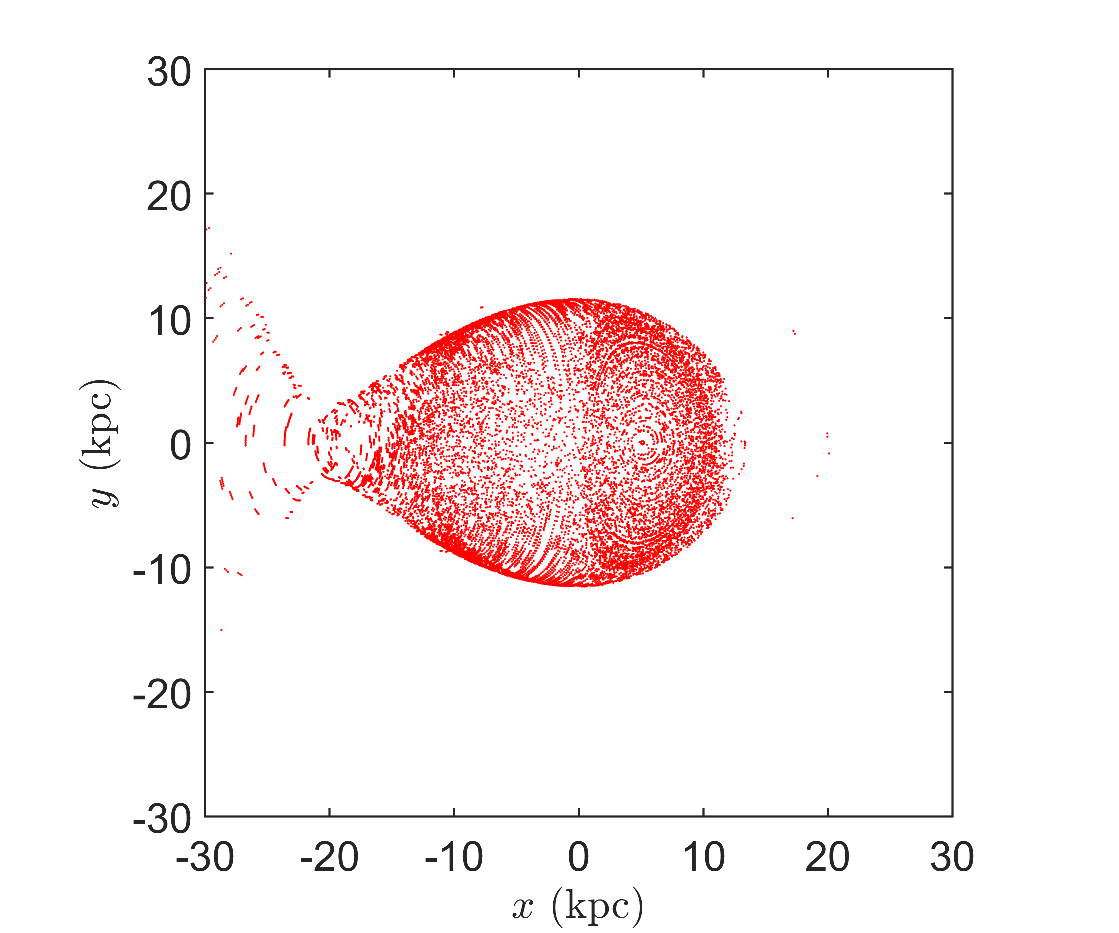}}
\subfigure[$C = 0.1$]{\label{fig:5.7b}\includegraphics[height=0.4\columnwidth,width=0.49\columnwidth]{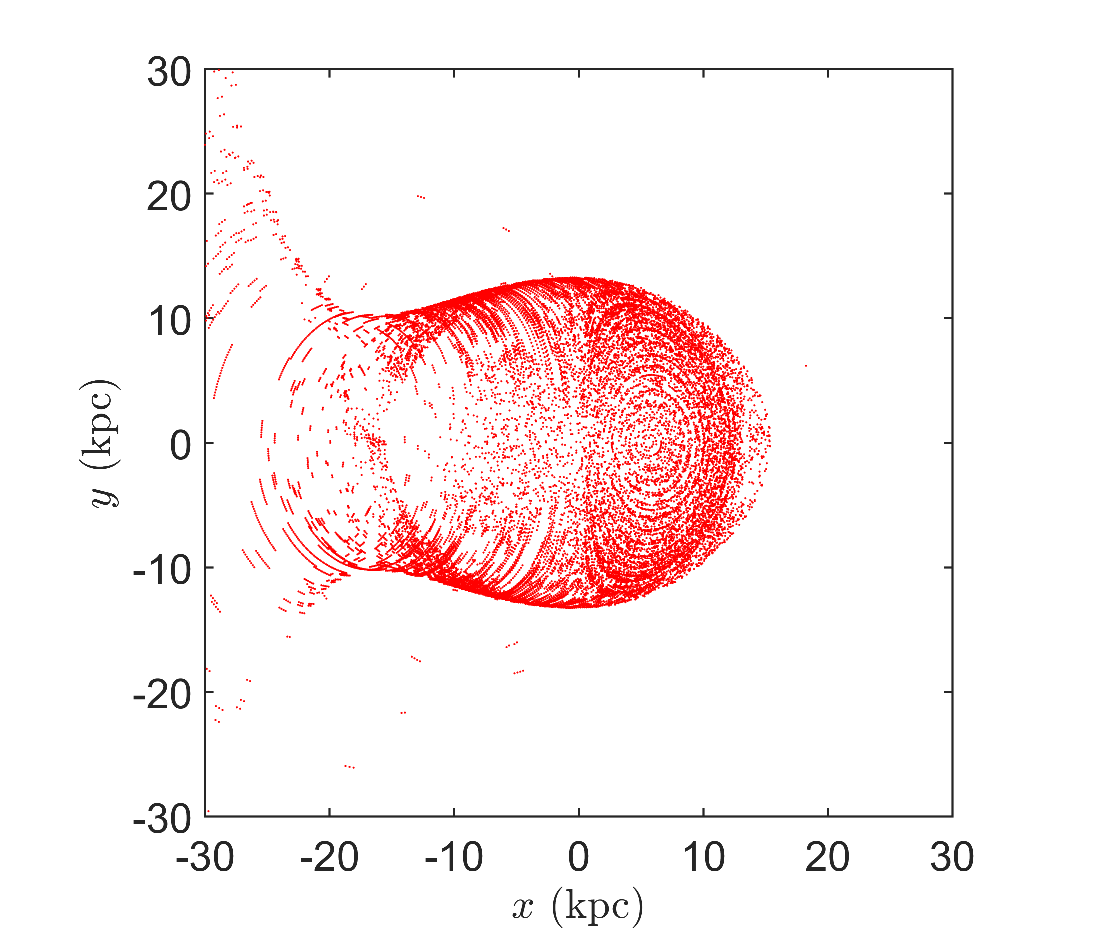}}
\caption{Model $1$: Poincaré maps in the $x - y$ plane with surface sections: $p_x = 0$ and $p_y \leq 0$.}
\label{fig:5.7}
\end{figure}

\noindent Again, for the NFW dark halo profile, Poincaré surface section maps in the $x - p_x$ plane are plotted in Fig. \ref{fig:5.8} for escape energy values $C = 0.01$ and $0.1$. In these figures, the following patterns have been identified:
\begin{enumerate}[label=(\roman*)]
\item A stability island near $(7, 0)$ is identified in both maps, corresponding to quasi-periodic stellar motions as seen earlier in Poincaré surface section maps in the $x - y$ plane.

\item Here also, the concentration of cross-sectional points inside the corotation region of the bar and disc is more significant for a lower escape energy (i.e., $C = 0.01$) than the case with a higher escape energy (i.e., $C = 0.1$). Again, the number of cross-sectional points outside that corotation region has increased with an increment of $C$ from $0.01$ to $0.1$. These trends are identical to those seen in Poincaré surface section maps in the $x - y$ plane.
\end{enumerate}
\begin{figure}
\centering
\subfigure[$C = 0.01$]{\label{fig:5.8a}\includegraphics[height=0.4\columnwidth,width=0.49\columnwidth]{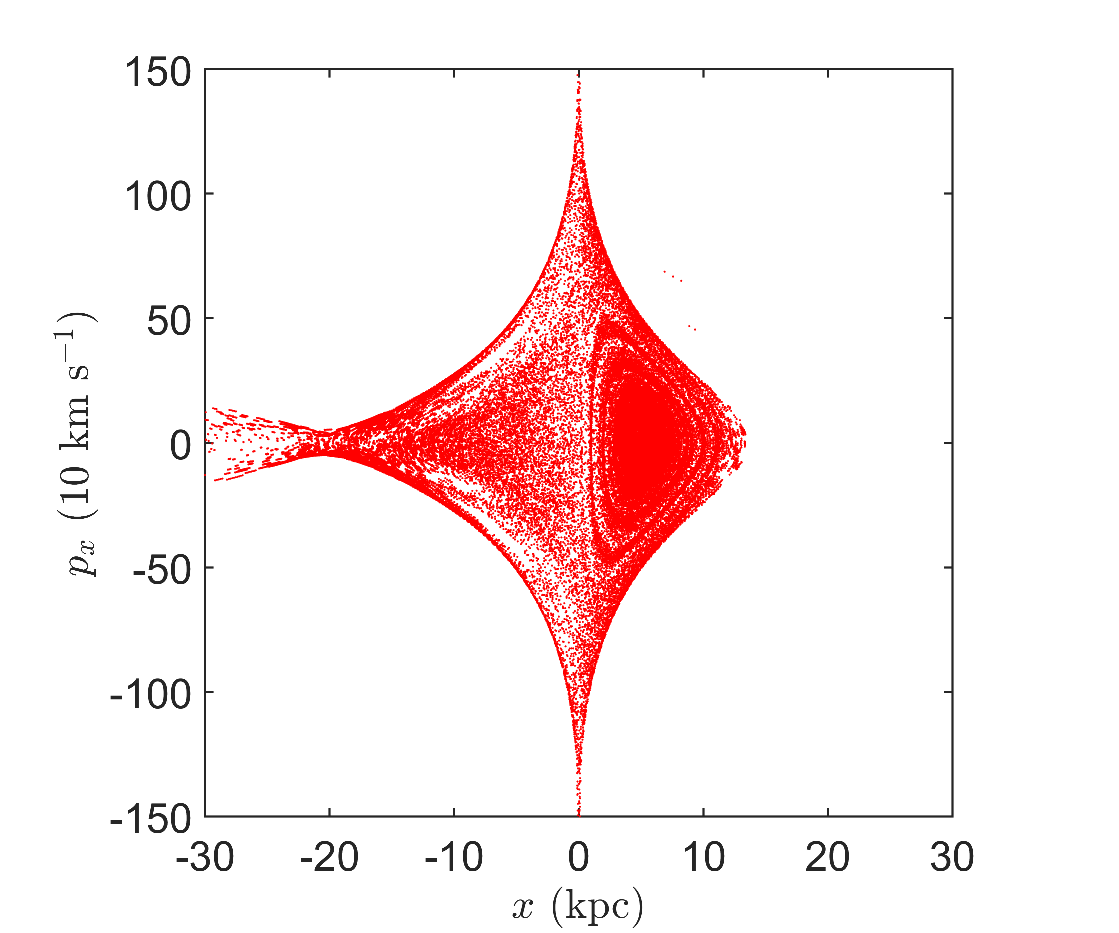}}
\subfigure[$C = 0.1$]{\label{fig:5.8b}\includegraphics[height=0.4\columnwidth,width=0.49\columnwidth]{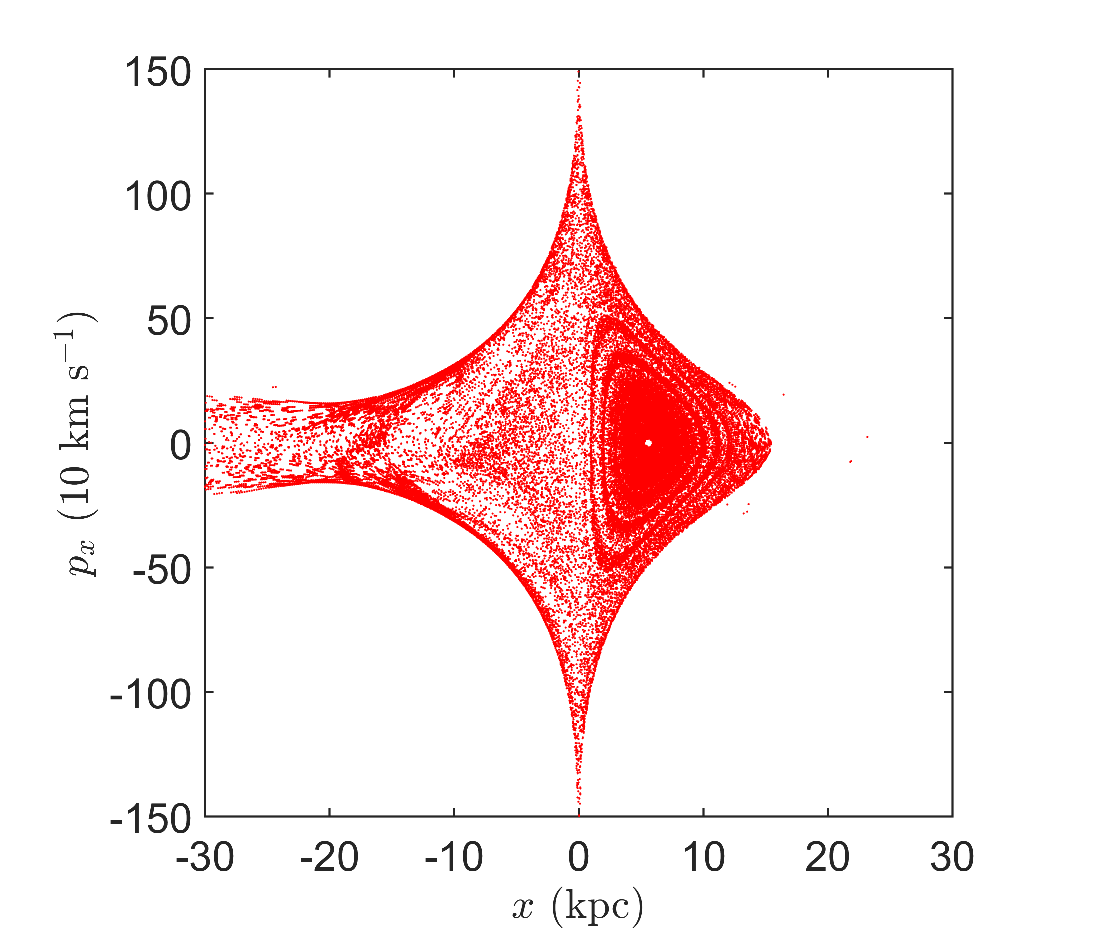}}
\caption{Model $1$: Poincaré maps in the $x - p_x$ plane with surface sections: $y = 0$ and $p_y \leq 0$.}
\label{fig:5.8}
\end{figure}

\item Model $2$: For the oblate dark halo profile, Poincaré surface section maps in the $x - y$ plane are plotted in Fig. \ref{fig:5.9} for escape energy values $C = 0.01$ and $0.1$. The observations regarding these figures are as follows:
\begin{enumerate}[label=(\roman*)]
\item A stability island near $(7, 0)$ is apparent on both maps and connected to quasi-periodic stellar motions.

\item The concentration of cross-sectional points inside the corotation region of the bar and disc is nearly similar for both escape energy values. However, the number of cross-sectional points outside that corotation region has increased with a $C$ increment from $0.01$ to $0.1$. These numbers are far lower than model $1$.
\end{enumerate}
\begin{figure}
\centering
\subfigure[$C = 0.01$]{\label{fig:5.9a}\includegraphics[height=0.4\columnwidth,width=0.49\columnwidth]{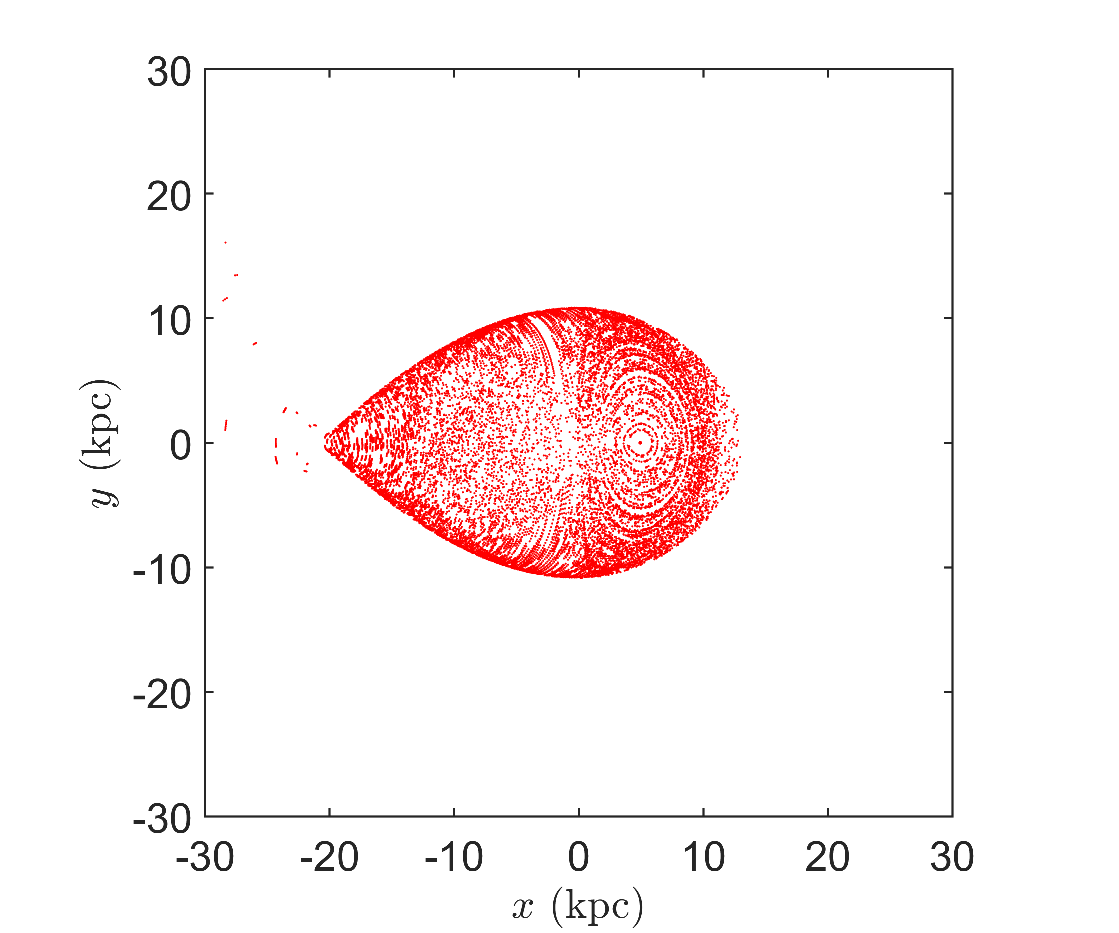}}
\subfigure[$C = 0.1$]{\label{fig:5.9b}\includegraphics[height=0.4\columnwidth,width=0.49\columnwidth]{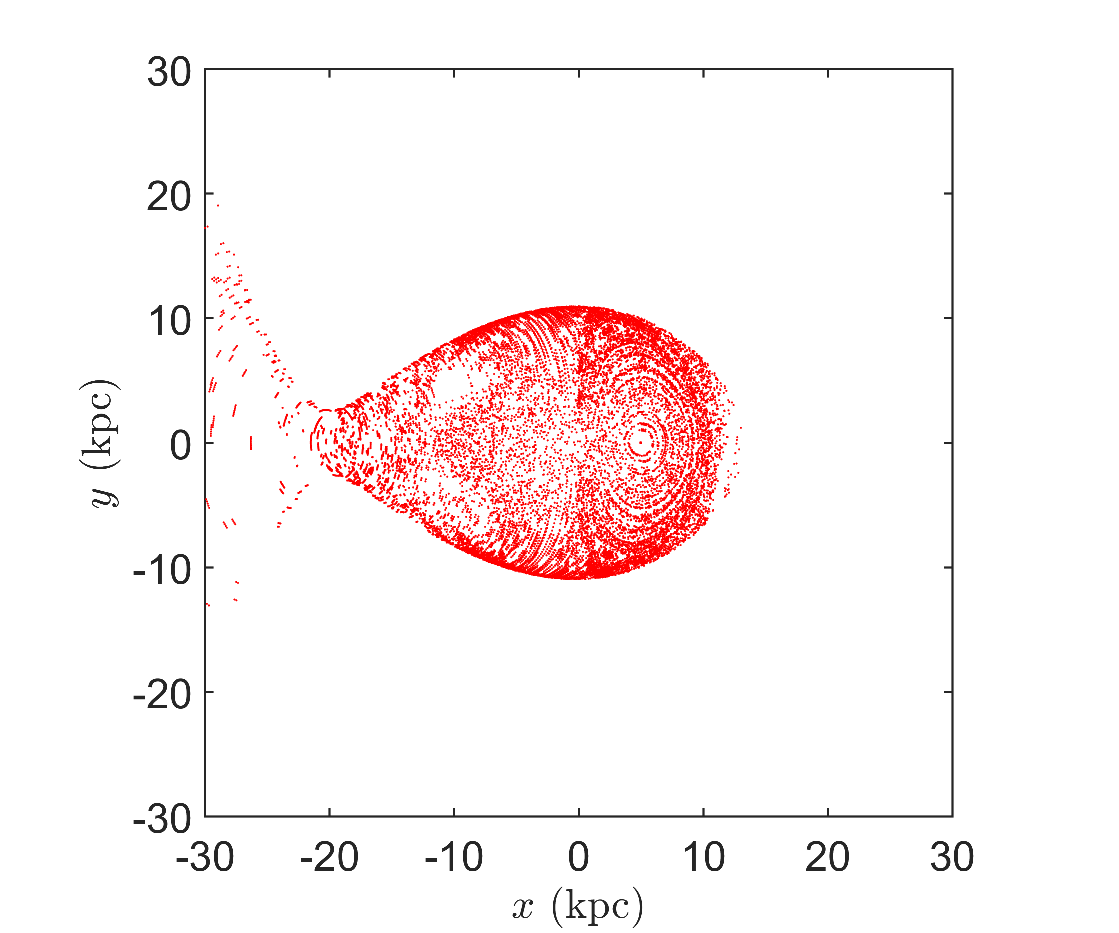}}
\caption{Model $2$: Poincaré maps in the $x - y$ plane with surface sections: $p_x = 0$ and $p_y \leq 0$.}
\label{fig:5.9}
\end{figure}

Again, for the oblate dark halo profile, Poincaré surface section maps in the $x - p_x$ plane are plotted in Fig. \ref{fig:5.10} for escape energy values $C = 0.01$ and $0.1$. Regarding these figures, the observations are as follows:
\begin{enumerate}[label=(\roman*)]
\item In the vicinity of $(7, 0)$ on each map is a stability island linked to quasi-periodic stellar motions. This was also previously visible in Poincaré surface section maps in the $x - y$ plane.

\item Here also, for both escape energy levels, the concentration of cross-sectional points inside the corotation region of the bar and disc is almost identical. Even though the number of cross-sectional points outside that corotation region has increased by an increment of $C$ from $0.01$ to $0.1$, compared to model $1$, these values are considerably lower.
\end{enumerate}
\begin{figure}
\centering
\subfigure[$C = 0.01$]{\label{fig:5.10a}\includegraphics[height=0.4\columnwidth,width=0.49\columnwidth]{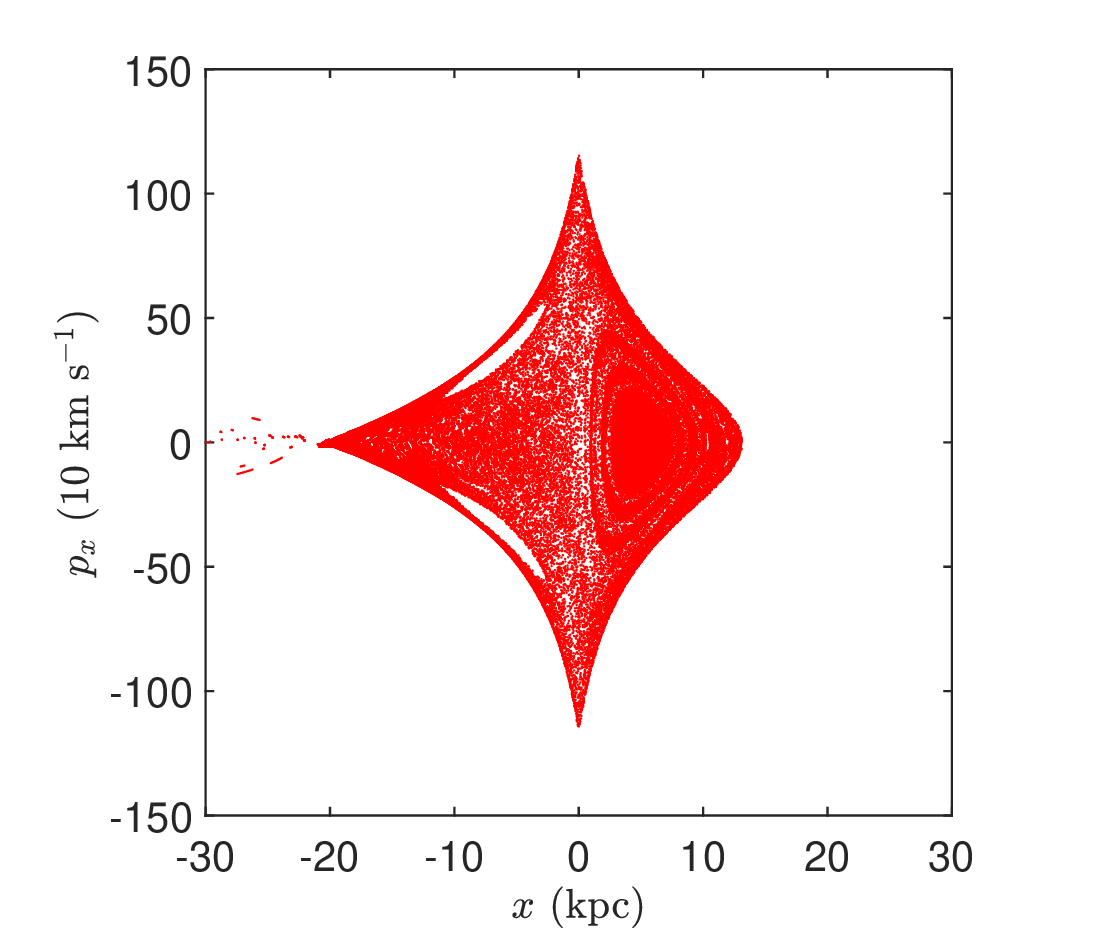}}
\subfigure[$C = 0.1$]{\label{fig:5.10b}\includegraphics[height=0.4\columnwidth,width=0.49\columnwidth]{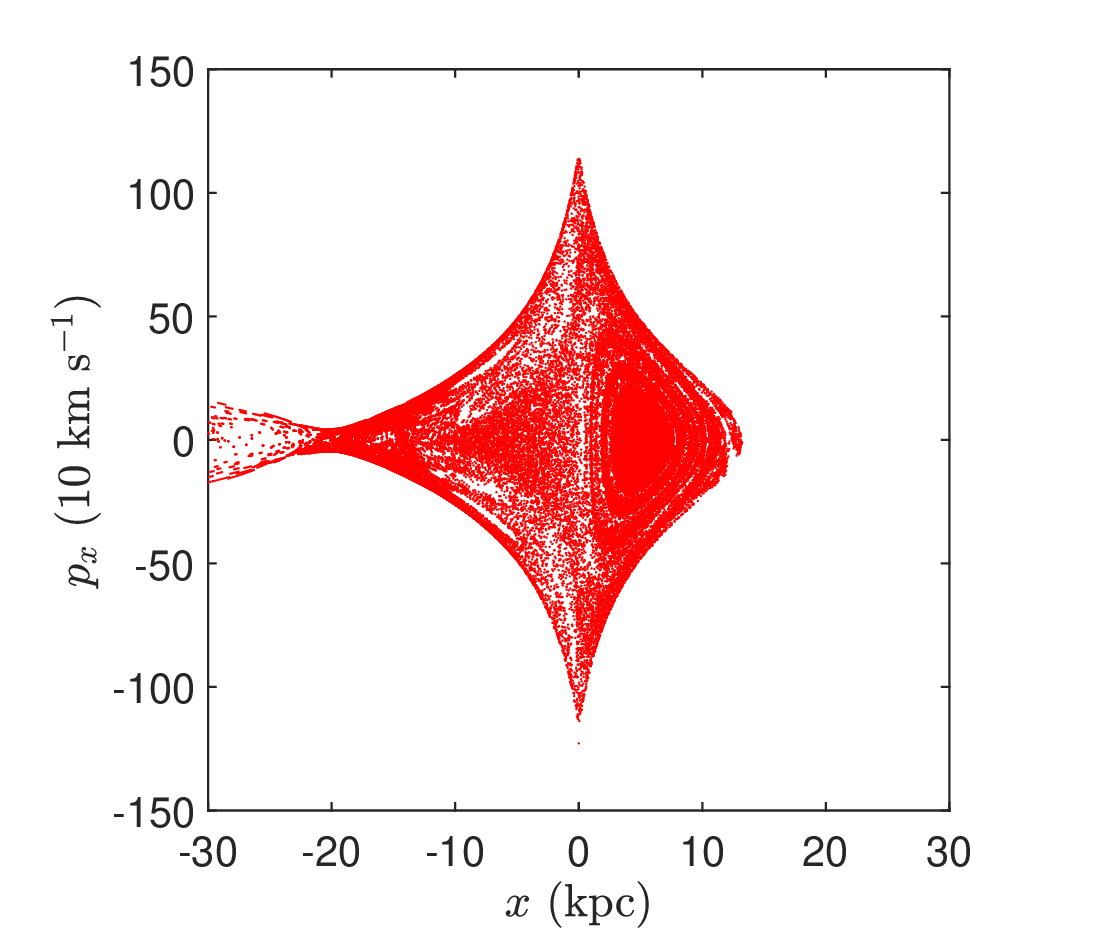}}
\caption{Model $2$: Poincaré maps in the $x - p_x$ plane with surface sections: $y = 0$ and $p_y \leq 0$.}
\label{fig:5.10}
\end{figure}
\end{itemize}

\subsection{Evolution of orbital chaos with the dark halo parameters}
\label{sec:5.3.3}
In this part of this chapter, the evolution of orbital chaos (in terms of MLE) with dark halo parameters like mass, length, and circular velocity is demonstrated. These findings are exciting from the standpoint of parametric studies near $L_1$ (or $L_1^{'}$). This helps to determine the sensitivity of the bar-driven escaping patterns to the dark halo parameters. For MLE calculation for different parametric values, an orbit with the same initial condition: $(x_0,y_0,p_{x_0},p_{y_0}) \equiv (5,0,15,p_{y_0})$ is considered, where the $p_{y_0}$ value is evaluated from Eq. (\ref{eq:5.3}).

\begin{itemize}[leftmargin=*]
\item Model $1$: The evolution of the MLE value at $(5,0,15,p_{y_0})$ with the NFW dark halo concentration parameter ($c_\text{p}$) is plotted in Fig. \ref{fig:5.11a}. Again, the evolution of that same MLE value with the virial mass of the NFW dark halo ($M_\text{vir}$) is plotted in Fig. \ref{fig:5.11b}. In both figures, the overall trend signifies that the MLE values are mostly lower for a higher escape energy ($C = 0.1$) than a lower escape energy ($C = 0.01$).
\begin{figure}
\centering
\subfigure[$c_\text{p}$ versus MLE]{\label{fig:5.11a}\includegraphics[height=0.4\columnwidth,width=0.49\columnwidth]{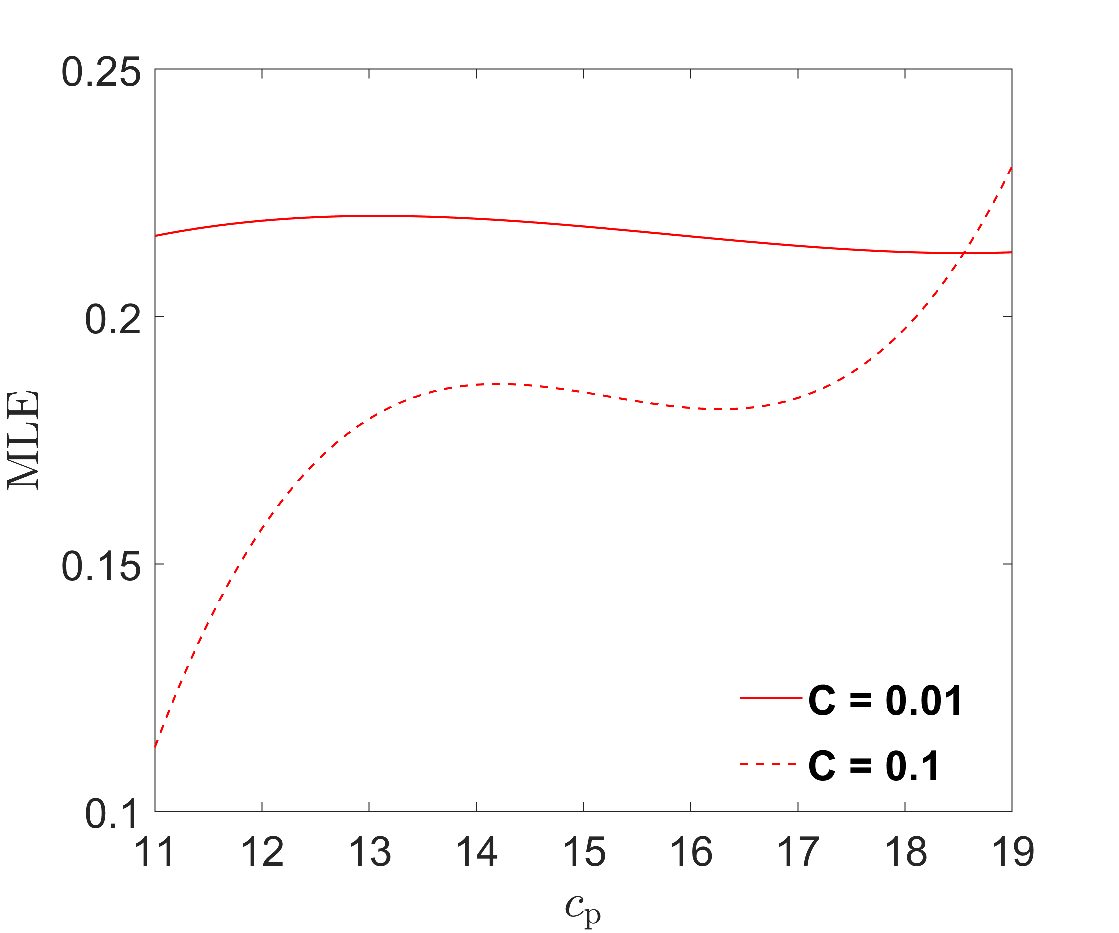}}
\subfigure[$M_\text{vir}$ versus MLE]{\label{fig:5.11b}\includegraphics[height=0.4\columnwidth,width=0.49\columnwidth]{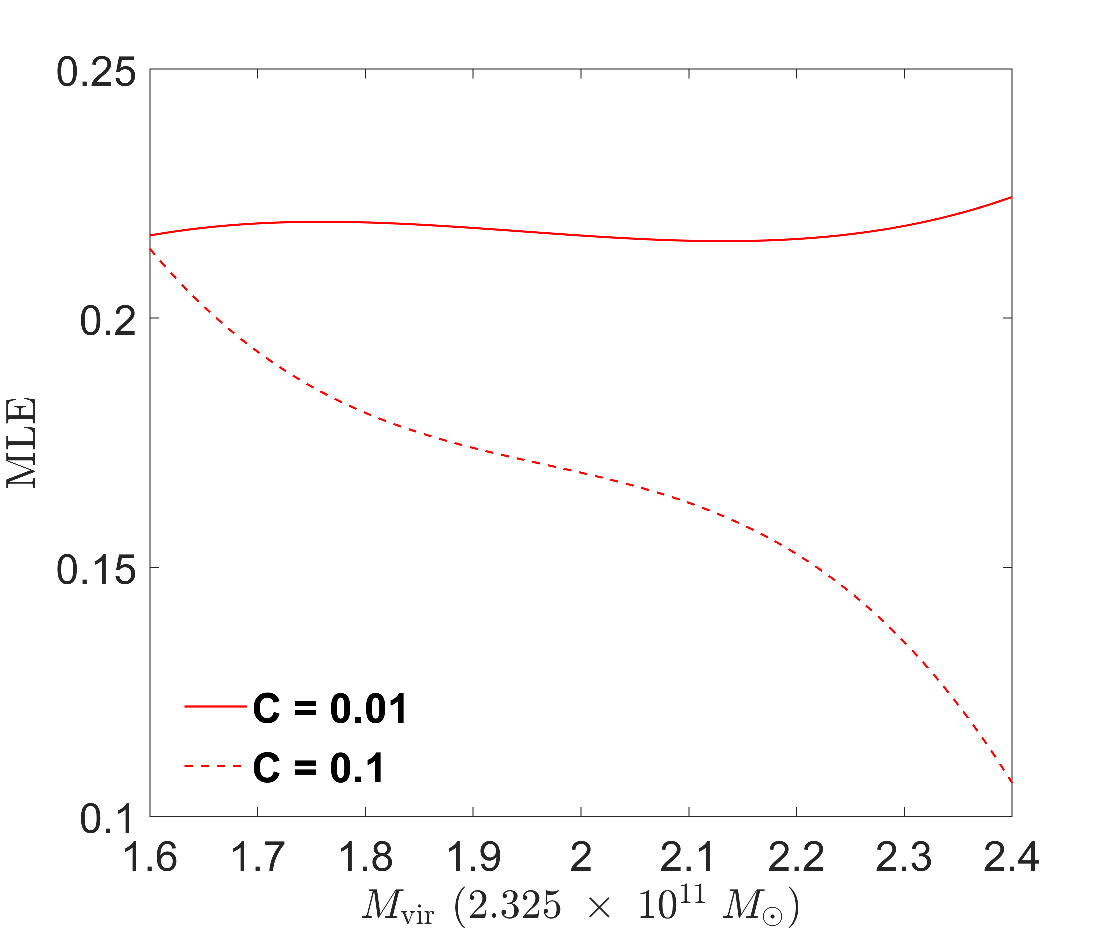}}
\caption{Variation of MLE with the NFW dark halo parameters.}
\label{fig:5.11}
\end{figure}

\item Model $2$: The evolution of the MLE value at $(5,0,15,p_{y_0})$ with the oblate dark halo flattening parameter ($\beta$) is plotted in Fig. \ref{fig:5.12a}. Again, in Fig. \ref{fig:5.12b}, the evolution of that same MLE value with the circular velocity of the oblate dark halo ($v_0$) is plotted in Fig. \ref{fig:5.12b}. In both figures, the overall trend suggests that the MLE values do not vary much with an increment in escape energy value from $C = 0.01$ to $C = 0.01$.
\begin{figure}
\centering
\subfigure[$\beta$ versus MLE]{\label{fig:5.12a}\includegraphics[height=0.4\columnwidth,width=0.49\columnwidth]{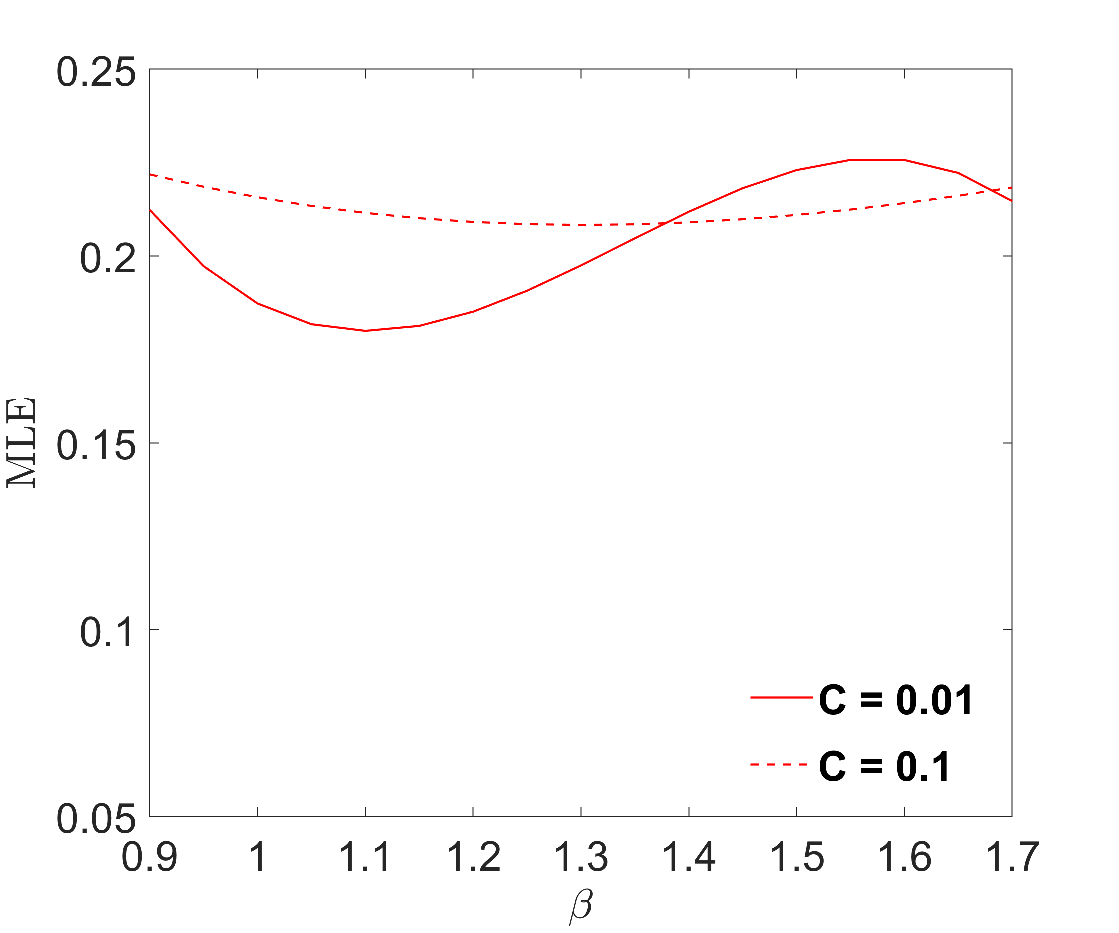}}
\subfigure[$v_\text{0}$ versus MLE]{\label{fig:5.12b}\includegraphics[height=0.4\columnwidth,width=0.49\columnwidth]{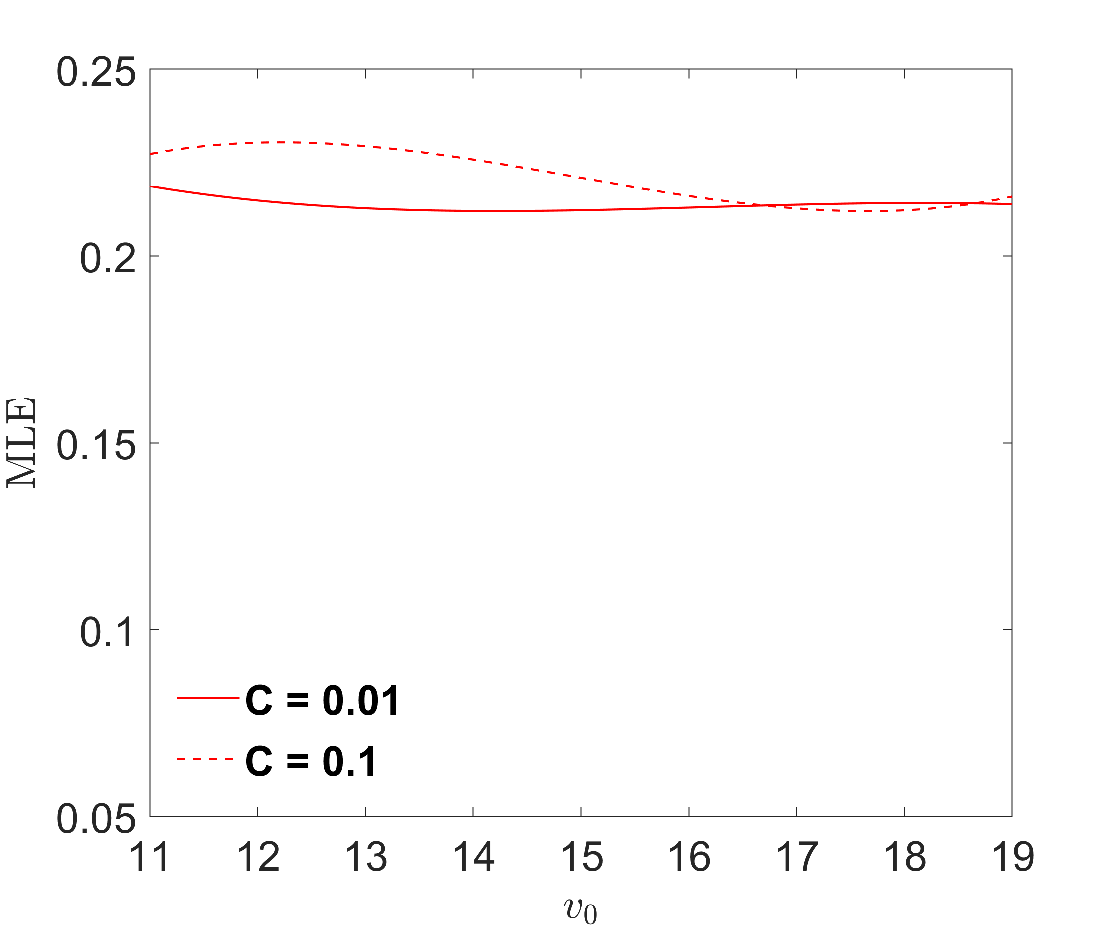}}
\caption{Variation of MLE with the oblate dark halo parameters.}
\label{fig:5.12}
\end{figure}
\end{itemize}

\section{Dark Haloes and Bar-driven Structures}
\label{sec:5.4}
Chaos (or instabilities) within the barred region can spread over time and escape from the bar ends to form structures such as spiral arms or inner disc rings, depending on the bar strength (strong or weak) \cite{Contopoulos2012, Jung2016, Mestre2020, Mondal2021}. Until now, a comprehensive analysis has been provided to describe the influence of dark haloes on the fate of this bar-driven escaping stellar motions in disc galaxies with strong bars.

In this chapter, a five-component gravitational model composed of a central black hole, bulge, bar, disc, and dark halo has been separately analysed for the following two dark halo profiles, namely (i) NFW dark halo (model $1$), and (ii) oblate dark halo (model $2$). These two dark halo profiles are chosen to model massive and low-mass disc galaxies, respectively. The dark halo of model $1$ resembles a cuspy distribution near the centre (see Fig. \ref{fig:5.2a}), while the dark halo of model $2$ resembles a flattened distribution near the centre (see Fig. \ref{fig:5.2b}). The rotation curves of both models are given in Fig. \ref{fig:5.2c}, which shows that, within the bulge region ($R < 5$), stellar rotational velocities are way higher for model $1$ than model $2$. Also, within $R < 5$, the radial force component is much steeper for model $1$ than model $2$ (see Fig. \ref{fig:5.2d}). Due to this steepness, radial forces promote bar-driven escapes more \cite{Mondal2021} in model $1$ than in model $2$.

Further, a classification scheme for orbits in the $x - y$ plane is presented in Figs. \ref{fig:5.3} and \ref{fig:5.5} for models $1$ and $2$, respectively. These figures indicated that the tendency for orbital escape was higher for model $1$ than model $2$. Also, the evolution of a stellar orbit is tracked beginning with the initial condition $(x_0,y_0,p_{x_0},p_{y_0}) \equiv (5,0,15,p_{y_0})$, where the $p_{y_0}$ value is evaluated from Eq. (\ref{eq:5.3}). This initial condition is chosen near $L_1$ (or $L_1^{'}$). For model $1$, this orbit follows both escaping and non-escaping motions (see Fig. \ref{fig:5.4}). The orbit is chaotic and non-escaping, with a high MLE for a lower escape energy value ($C = 0.01$). In comparison, for a higher escape energy value ($C = 0.1$), the orbit is chaotic and escaping with comparatively low MLE (see Table \ref{tab:5.4}). Moreover, corresponding Poincaré surface section maps in the $x - y$ and $x - p_x$ phase planes for different escape energy values are plotted in Figs. \ref{fig:5.7} and \ref{fig:5.8}, respectively. In this model, bar-driven stellar escape is only preferable for high escape energy values. Again, in model $2$, the orbit follows only non-escaping motions (see Fig. \ref{fig:5.6}). Here, the nature of the orbit is chaotic and non-escaping with a high MLE value, regardless of escape energy values (see Table \ref{tab:5.5}). Also, corresponding Poincaré surface section maps in the $x - y$ and $x - p_x$ phase planes for different escape energy values are plotted in Figs. \ref{fig:5.9} and \ref{fig:5.10}, respectively. In model $2$, bar-driven stellar escape is not preferable, irrespective of escape energies. Hence, the nature of the dark halo has an immense influence on bar-driven structure formation via the escape of chaotic orbits. In this analytic setup, such escaping chaotic motions are only feasible in the energy domain $E > E_{L_1}$ (or $E > E_{L_1^{'}}$).

The central regions of giant spirals are violence-prone (i.e., chaotic motion dominated) due to the baryonic feedback (supernovae explosions, shocks, etc.) \cite{Basu1989, Efstathiou2000, Terrazas2017, Mondal2019, Donahue2022}. The mass of the central black hole strongly influences such baryonic feedback. Moreover, the central black hole and the kinetic energy of random motions within the bulge are closely associated with the strength of spiral arms. Galaxies with high mass black holes have spiral arms with small pitch angles and higher kinetic energy transport, which result in tightly wrapped grand design spiral patterns \cite{Seigar2008, Berrier2013, Al2014}. Besides that, a strong bar profile can also escalate the formation of grand design spiral arms if the galaxy hosts a supermassive black hole (SMBH) at the centre \cite{Mondal2021}. Again, dark haloes significantly impact the formation of bars \cite{Debattista2000}. Due to this, dark haloes also have ample influence on forming spiral arms due to bar-driven escape.

\noindent From all these results and analyses, the observations are as follows:
\begin{itemize}[leftmargin=*]
\item Model 1:
\begin{enumerate}[label=(\roman*)]
\item From the colour-coded orbit classification maps (Fig. \ref{fig:5.3}) and orbital trajectories (Fig. \ref{fig:5.4}) in the $x - y$ plane, it is noticeable that NFW dark haloes encourage bar-driven stellar escapes only for high escape energy values. Furthermore, Poincaré surface section maps in the $x - y$ and $x - p_x$ phase planes (Figs. \ref{fig:5.7} and \ref{fig:5.8}, respectively) indicate that there are a large number of orbital escapes from the bar ends. The rate of this escape increases with an increment in escape energy value. Also, for both escape energy values ($C = 0.01$ and $0.1$), the overall escape rate is much greater than model $2$. In this scenario, the escaping orbits leave the disc only for higher escape energies; otherwise, they remain confined within the bar and disc's corotation region.

\item The underlying relation between the NFW dark halo concentration parameter ($c_\text{p}$) and the orbital chaos (in terms of MLE) is shown in Fig. \ref{fig:5.11a}. This figure shows that MLE remains almost fixed for $C = 0.01$ with the increase in $c_\text{p}$. Again, for $C = 0.1$, that trend follows an upward oscillating pattern starting from a very low MLE. Baryonic feedback produced in the central environment is related to these escape energy values ($C > 0$). Here, lower escape energy ($C = 0.01$) signifies an insufficient amount of baryonic feedback, for which chaotic stellar orbits remain trapped within the corotation region of the bar and disc, and this trend remains the same for any $c_\text{p}$ value. In this scenario, escaping stellar motion is not feasible despite having high MLE values. This non-escaping trend is also evident in Fig. \ref{fig:5.4a} (high MLE corresponds to $c_\text{p} = 15$). On the other hand, higher escape energy ($C = 0.1$) signifies a well-sufficient amount of baryonic feedback. In that case, stellar orbits mostly remain quasi-periodic for lower $c_\text{p}$ values, resulting in low MLE. However, as the $c_\text{p}$ value increases, the nature of stellar orbits shifts from quasi-periodic to chaotic. That is why MLE keeps increasing with $c_\text{p}$. Also, beyond a certain $c_\text{p}$ value ($> 18$), this trend leads to much higher MLE values than the case for $C = 0.01$. Escaping stellar motion is feasible even with comparatively low MLE values (weak chaotic motions). This escaping trend is also apparent in Fig. \ref{fig:5.4b} (low MLE corresponds to $c_\text{p} = 15$). Moreover, beyond a specific limit ($c_\text{p} > 18$), more vigorous chaotic motions are also able to escape, which may evolve into much more prominent escape-driven structures. In summary, for NFW dark haloes in massive galaxies, higher dark halo concentrations will help to strengthen the escape-driven patterns, but only for higher escape energy values.

\item Fig. \ref{fig:5.11b} addresses the underlying relationships between the virial mass of the NFW dark halo ($M_\text{vir}$) and the amount of orbital chaos. This figure shows that for $C = 0.01$, MLE values mostly remain unchanged against the increase in $M_\text{vir}$. However, that trend oscillates downward for $C = 0.1$, i.e., low orbital chaos for a higher escape energy. Since lower escape energy ($C = 0.01$) indicates insufficient baryonic feedback, chaotic stellar orbits will not lead to escape from the bar and disc's corotation region. However, higher escape energy ($C = 0.1$) is associated with sufficient baryonic feedback, which helps to generate impulsive gas outflow that leads to the expansion of dark halo \cite{Dutton2016}. With this expansion, as the dark halo mass increases, the nature of stellar orbits shifts from chaotic to quasi-periodic. That is why MLE keeps decreasing. In this scenario, though stellar escape is still possible (see Fig. \ref{fig:5.4b}, where MLE is comparatively low for $M_\text{vir} = 2 \times 10^4$ mass units), chances will diminish as the dark halo mass increases. Now, if there is some additional energy generation source other than these feedbacks, then only orbital chaos will increase, and chances of escape will become prominent again. In summary, for NFW dark haloes in massive galaxies, bar-driven escaping motion will be more promising for higher dark halo masses if there are some additional energy generation sources other than the baryonic feedback generated due to the accretion of the central SMBH.

\item Cosmological \textit{N}-body simulations predicted that dark haloes should have a structure akin to the Navarro-Frenk-White (NFW) profile \cite{Navarro1996}, which has a cuspy distribution near the centre. Sufficient baryonic feedback in the form of gas outflows imparts energy to the orbits of collision-less dark matter particles that lead to the elimination of this central dark matter cusp \cite{Mashchenko2006, Pontzen2012}. This chapter uses the NFW profile, and a high-mass SMBH is used to simulate model $1$'s (massive galaxies) dark halo. Stellar orbits with sufficient escape energies can escape from the bar and disc's corotation region in this setup. This is evident from its phase space analysis for a higher escape energy value. At higher escape energy, the generated outflows are sufficient to eliminate the central dark matter cusp and generate bar-driven escaping patterns in the form of spiral arms. Whereas at lower escape energy, the generated outflows are primarily utilised to eliminate the central dark matter cusp, and for generating further bar-driven escaping patterns, they become insufficient. In this massive galaxy setup, baryonic feedback (in the form of supernovae explosions, shocks, etc.) generated due to the central high-mass SMBH's accretion effect is the central region's only energy generation source. Galaxies should have additional energy generation sources other than the accretion effect of the high-mass SMBH to support higher escape energies. In other words, an excess energy generation source other than the SMBH accretion-driven baryonic feedback is required to eliminate the central dark matter cusp and form bar-driven escaping patterns. Thus, NFW dark haloes in massive galaxies support the formation of grand-design stable spiral arms (as seen in massive disc galaxies) due to a bar-driven escape mechanism only present when their central region is highly energetic. In this scenario, the possible conclusion is that active galaxies might be one of the potential contenders for modelling dark haloes, where NFW profiles may be a better fit. The crucial question is: if so, what is that additional central energy generation source? Regarding this issue, secondary stellar bars may be one of the possible sources of power generation in active galaxies' central engines \cite{Shlosman1989, Shlosman1990}. Observations reveal that only the inner regions of giant spiral galaxies like the Milky Way and M31 are somewhat compatible with the NFW dark haloes. However, this is open to debate \cite{Klypin2002}. Nevertheless, this is not true for all giant spiral galaxies \cite{Flores1994}. The next chapter addresses this issue in the context of bar-driven escaping patterns in double-barred galaxies.
\end{enumerate}

\item Model 2:
\begin{enumerate}[label=(\roman*)]
\item From the colour-coded orbit classification maps (Fig. \ref{fig:5.5}) and orbital trajectories (Fig. \ref{fig:5.6}) in the $x - y$ plane, it is evident that oblate dark haloes do not encourage bar-driven stellar escapes regardless of escape energy values. Furthermore, Poincaré surface section maps in the $x - y$ and $x - p_x$ phase planes (Figs. \ref{fig:5.9} and \ref{fig:5.10}, respectively) indicate a much lower number of orbital escapes than model $1$. In this scenario, the escaping orbits do not leave the disc, irrespective of escape energy values, and remain confined within the corotation region of the bar and disc.

\item The underlying relation between the oblate dark halo flattening parameter ($\beta$) and the orbital chaos (in terms of MLE) is shown in Fig. \ref{fig:5.12a}. This shows that for $C = 0.01$, MLE follows a stable oscillating pattern with the increment in $\beta$, whereas MLE remains almost the same for $C = 0.1$. For both escape energies, overall MLE values are high, and a slight fluctuation is observed for $C = 0.01$, whereas more or less a fixed value persists for $C = 0.1$. Here, the nature of the dark halo is core-dominated and has a flattened distribution near the centre. In this outline, trapped chaotic motions are mainly observed; only at higher escape energies may a small number of stellar orbits be able to cross the corotation region of the bar and disc (see Fig. \ref{fig:5.9} for reference). In summary, the flattening of oblate dark haloes has a minor impact on generating bar-driven structures in low-mass disc galaxies. Also, that impact is only evident for higher escape energies.

\item Fig. \ref{fig:5.12b} addresses the underlying relationship between the circular velocity of the oblate dark halo ($v_0$) and the amount of orbital chaos. From this, it is evident that for $C = 0.01$, MLE is initially high with the increment in $v_0$ and remains nearly the same up to a limit ($v_0 > 16$), then it starts to decline slowly. Whereas for $C = 0.1$, MLE does not vary against the increment in $v_0$. This means trapped chaotic motions are observed at lower escape energy, but only up to $v_0 \sim 16$, and afterwards, orbits shift slowly towards quasi-periodicity. However, with higher escape energy, trapped chaotic motions were consistently observed. So, like the oblate halo flattening parameter, the circular velocity parameter also has a minor impact on generating bar-driven structures in low-mass disc galaxies. Furthermore, that effect is only apparent for higher escape energies.

\item This work uses an oblate profile and a comparative low-mass black hole (than model $1$) to simulate model $2$'s (low-mass disc galaxies) dark halo. Its phase-space analysis shows that irrespective of escape energy values, stellar orbits cannot escape from the bar and disc's corotation region. Here, the dark halo has no central cusp like model $1$; instead, it has a flattened core-dominated distribution near the centre. In this setup, baryonic feedback generated due to the accretion of the central black hole is relatively less than model $1$. Also, such feedback will impart less energy in the form of gas outflows to the orbits of collision-less dark matter particles. That is why central dark matter distribution remains almost core-dominated. Moreover, the formation of bar-driven escaping patterns in the form of spiral arms is less feasible. This is the scenario of S0 and ultra-compact dwarf galaxies, where dark haloes are primarily core-dominated, and less prominent or poor spiral arms are evident in some cases \cite{Seth2014, Davis2018, Roberts2021}.
\end{enumerate}
\end{itemize}    

\indent In this chapter, light is shaded on one of the critical issues of the famous ‘core-cusp problem', which is why NFW (cuspy) dark halo profiles are mostly incompatible in massive disc galaxies. A similar case study for low-mass disc galaxies is also addressed where dark haloes resemble an oblate profile (core-dominated). This study is essential to uncovering the influence of dark haloes on the formation and evolution of spiral arms in both massive and low-mass disc galaxies. The entire study is done from the viewpoint of a bar-driven escape mechanism. Now, from all the results and discussions, the conclusions are summarised as follows:
\begin{enumerate}[label=(\roman*)]
\item Both in massive and low-mass disc galaxies, the bar-driven structure formation process is triggered by the baryonic feedback generated due to the accretion of the central black hole. Besides the stellar bar strength, the shape of the dark halo is another critical parameter that determines the fate of this bar-driven structure formation process.

\item Regarding the ‘core-cusp problem’, the issue of the disappearance of NFW (or cuspy) dark haloes is addressed for massive disc galaxies as predicted by N-body simulations. Their central region has an inadequate energy generation mechanism to support the bar-driven structures (spiral arms) within a NFW dark halo framework. That is why the dark halo structure of massive disc galaxies should not be the NFW profile unless their central regions are highly energetic. In respect of this, active galaxies might be one of the potential contenders where NFW profiles may be a better fit for modelling dark haloes.

\item In low-mass disc galaxies, oblate (or core-dominated) dark haloes well agree with forming less-prominent or poor spiral arms due to the bar-driven escape mechanism. This is evident in S0, ultra-compact dwarf galaxies, etc.
\end{enumerate}

\section*{What's next?}
This chapter discusses how dark haloes and the central black hole play an essential role in generating central baryonic feedbacks that lead to bar-driven structure formations. Moreover, if barred galaxies usually have a central dark matter cusp, then baryonic feedbacks generated in the centre are insufficient to support bar-driven structure formation. However, they can do so for double-barred discs. This is the central theme of the upcoming chapter, where the possibilities of bar-driven structure formation in a double-barred disc have been studied.

%% file: Chapter_6.tex
\chapter{Bar-driven Escapes in Double-barred Discs}
\label{chap:6}

\section{Introduction}
\label{sec:6.1}
Double-barred galaxies are not dynamical freaks in the galactic morphological schema. Nearly 30\% - 40\% of all disc galaxies are known to host secondary bars (also known as nuclear bars) within their primary bars in a nested manner \cite{Erwin2002, Laine2002}. Due to the emergence of modern high-resolution imaging facilities, many double-barred galaxies have been observed in the recent past in both the optical \cite{de1975, Wozniak1995, De2019, De2020} and near-infrared bands \cite{Shaw1993, Mulchaey1997, Greusard2000}. There are several comprehensive catalogues of double-barred galaxies in the literature \cite{Moiseev2001, Erwin2004}, but there are far fewer than for their single-barred counterparts. NGC 1291, NGC 1326, and NGC 1543 are some examples of double-barred galaxies. Also, many galaxies were earlier thought to have double bar-like structures at their centres but were later discovered to be single-barred galaxies \cite{Erwin1999}. Depending on the mass and length scales, a significant fraction of secondary bars are dynamically long-lived \cite{El2003}.

\indent The existence of secondary bars brings lots of insight into the dynamical mechanisms of the galactic central engines, as it opens up many possible theories of energy generation therein. Theoretical models of double-barred galaxies suggest that a secondary bar may increase the gas inflow towards the galactic centre. As a result, the central region becomes extremely energetic (i.e., dominated by chaotic stellar motion). In this way, secondary bars may be one of the possible sources of energy generation in the central engines of the Active Galactic Nucleus (AGN). However, recent observational studies suggest that double-barred discs play only a minor role in AGN fuelling \cite{Martini2001, Erwin2002, Laine2002}. There are two major theories behind the origin of secondary bars. One is the `outside-in formation mechanism' \cite{Shlosman1989, Heller2001}, in which the secondary bar is formed later from materials of the primary bar within the central kpc region. In this instance, the secondary bar experiences a slower pattern speed than the related primary bar. Another is the `inside-out formation mechanism' that derives from \cite{Rautiainen2002} obtained \textit{N}-body simulation results, in which secondary bars form first, contrary to the outside-in theory. The secondary bar rotates faster than the related primary bar in this event.  

\indent The physical properties of the primary and secondary bars are quite different since they evolve together as dynamically decoupled entities for a substantial amount of time inside the galactic disc \cite{Friedli1993, Corsini2003}. This difference is also evident from observing the bimodal distribution of the bar lengths in double-barred galaxies \cite{Laine2002}. The difference is more evident in the case of the periodic orbits associated with both primary and secondary bar formations. For usual single-barred galaxies, the central bar dynamics is mainly governed by a specific class of periodic orbits called $x_1$, $x_2$, etc. The $x_1$ orbits (also known as the backbone of the bar) are elongated parallel to the bar axis within the corotation region of the bar and disc. The $x_2$ orbits are elongated perpendicular to the bar axis and only exist for the higher bar pattern speed. It supports ring-like structures around the bar \cite{Contopoulos1989}. In the case of double-barred galaxies, the concepts of $x_1$ and $x_2$ orbits are replaced by $x_1$ and $x_2$ loops, respectively. Here, $x_1$ loops follow the secondary bars' rotation and are aligned parallel to the primary bars' semi-major axis, whereas $x_2$ loops follow the primary bars' rotation \cite{Maciejewski1997, Maciejewski2000}. Regarding bar resonance, secondary bars are enclosed within the inner Lindblad resonance (ILR) of the related primary bar. From the viewpoint of the structural stability of the bar, this orbital dynamics analysis is more important.

\indent Bar-driven structure formation is an essential aspect of the galaxy's evolutionary process and is related to the phenomenon of escape in open Hamiltonian systems \cite{Aguirre2001}. The bar-driven escapes in single-barred galaxies are extensively discussed in many literatures \cite{Jung2016, Zotos2012, Zotos2020a, Mondal2021}. These studies analysed the role of normally hyperbolic invariant manifolds (NHIMs) located in the vicinity of escape saddles responsible for stellar escapes via escape basins. Furthermore, the strength of the bar determines the fate of escaping orbits in the form of spiral arms or rings. The presence of a secondary bar makes the phase space geometry more complicated than its single-barred counterpart. Thus, bar-driven escapes in double-barred systems have more prospects regarding the galaxy's evolutionary dynamics. However, only a few articles have studied the stellar dynamics in the phase space of double-barred galaxies to date \cite{Zotos2020b, Zotos2020c, Kondratyev2022}.  

\indent The motivation of this chapter is to analyse the possibilities of bar-driven escapes inside a double-barred disc. A six-body gravitational model with a central black hole, bulge, primary and secondary bars, disc, and dark matter halo has been studied to do that. The bar strength in this model resembles that of strong and weak bars, respectively, for primary and secondary bars. The phase space has been analysed using orbital and Poincaré surface section maps. Stellar orbits may escape from the central barred region depending on their regular or chaotic character. The dynamical indicator Smaller Alignment Index (SALI) is used to classify orbits depending on their characteristics \cite{Skokos2001}. The orbital classification scheme is as follows: (i) SALI $< 10^{-8} \rightarrow$ orbit is chaotic; (ii) $10^{-8} \leq$ SALI $\leq 10^{-4} \rightarrow$ orbit is sticky, and (iii) SALI $> 10^{-4} \rightarrow$ orbit is regular. Sticky orbits are a particular class of regular orbits that reveal their true chaotic nature after a long time. The presence of sticky orbits complicates the geometrical complexity of orbital dynamics in phase space. The following are some of the benefits of SALI over other traditional dynamical indicators like the maximal Lyapunov exponent (MLE), the fast Lyapunov indicator (FLI), etc.: (i) It runs much faster due to its simple computational algorithm, and (ii) can detect the presence of weak chaos in the system.

\indent The bar-driven structure formation process in a double-barred disc has been discussed in detail in the following sections from the perspective of stellar motions that might escape the centre region. The primary focus is to find the underlying relationships between the fate of escaping stars and the amount of baryonic feedback produced therein.

\section{Double-barred Galaxy Model}
\label{sec:6.2}
Let us construct a six-component double-barred galaxy model in Cartesian coordinates $(x,y,z)$, whose components are as follows: central black hole, bulge, primary bar, secondary bar, disc, and dark matter halo. Also, let $\Phi_\text{t}(x,y,z)$ and $\rho_\text{t}(x,y,z)$ be the total gravitational potential and corresponding volume density, respectively. The Poisson equation relates this potential-density pair as,
\begin{equation}
\label{eq:6.1} 
\nabla^2 \Phi_\text{t}(x,y,z) = 4 \pi G \rho_\text{t}(x,y,z),
\end{equation}
where $G$ is the gravitational constant, now, let $\Phi_\text{bh}$, $\Phi_\text{B}$, $\Phi_\text{bp}$, $\Phi_\text{bs}$, $\Phi_\text{d}$ and $\Phi_\text{h}$ be the central black hole, bulge, primary bar, secondary bar, disc, and dark matter halo potentials, respectively. Hence,
\begin{equation*}
\begin{split}
\Phi_\text{t}(x,y,z) = (\Phi_\text{bh} + \Phi_\text{B} + \Phi_\text{bp} + \Phi_\text{bs} + \Phi_\text{d} + \Phi_\text{h})(x,y,z).
\end{split}
\end{equation*}

\noindent Let $\vec{\Omega}_{\text{b}_1} \equiv (0,0,\Omega_{\text{b}_1})$ and $\vec{\Omega}_{\text{b}_2} \equiv (0,0,\Omega_{\text{b}_2})$ be the constant rotational velocities (in a clockwise sense along the $z$ - axis) for the primary and secondary bars, respectively, in the rotating reference frame of the double-bar. In general, this system has time-dependent potential and is non-conservative. This time-dependency can be removed from the system if both bars are rotating with the same rotational velocities, i.e., if $\Omega_{\text{b}_1} = \Omega_{\text{b}_2} = \Omega_\text{b}$ (say) \cite{Zotos2020c}. This assumption is physically realistic only if both bars are aligned perpendicular to each other. For such a double-barred system, the effective potential is,
\begin{equation}
\label{eq:6.2}
\Phi_\text{eff}(x,y,z) = \Phi_\text{t}(x,y,z) - \frac{1}{2} \Omega_{\text{b}}^2 (x^2 + y^2).
\end{equation}

\noindent This double-barred galaxy model is a conservative system, so the total energy ($E$) is equal to the system's Hamiltonian ($H$). Thus, for a test particle (star) of unit mass, $H$ is formulated as,
\begin{eqnarray}
\label{eq:6.3}
\begin{split}
H = \frac{1}{2} (p_x^2 + p_y^2 +p_z^2) + \Phi_\text{t}(x,y,z) - \Omega_\text{b} (x p_y - y p_x) = E,
\end{split}
\end{eqnarray}
where $\dot{r} \equiv (x,y,z)$ and $\vec{p} \equiv (p_x,p_y,p_z)$ are the position and linear momentum vector of the test particle at time $t$, respectively. Now, the Hamilton's equations of motion are as follows:
\begin{eqnarray}
\label{eq:6.4}
\begin{split}
\dot{x} = p_x + \Omega_\text{b} y, \; 
\dot{y} = p_y - \Omega_\text{b} x, \;
\dot{z} = p_z,\\
\dot{p}_x = - \frac{\partial \Phi_\text{t}}{\partial x} + \Omega_\text{b} p_y, \;
\dot{p}_y = - \frac{\partial \Phi_\text{t}}{\partial y} - \Omega_\text{b} p_x, \;
\dot{p}_z = - \frac{\partial \Phi_\text{t}}{\partial z},
\end{split}
\end{eqnarray}
where $`\cdot$' $\equiv \frac{\mathrm{d}}{\mathrm{dt}}$. Now, the Lagrangian (or equilibrium) points of this system are solutions of the following equations:
\begin{eqnarray}
\label{eq:6.5}
\begin{split}
\frac{\partial \Phi_\text{eff}}{\partial x} = 0, \;
\frac{\partial \Phi_\text{eff}}{\partial y} = 0, \;
\frac{\partial \Phi_\text{eff}}{\partial z} = 0.
\end{split}
\end{eqnarray}

\subsection{Gravitational potentials}
\label{sec:6.2.1}
\noindent The potential forms of the central black hole, bulge, primary bar, secondary bar, disc, and dark matter halo are as follows: 
\begin{itemize}[leftmargin=*]
\item Central black hole: A non-relativistic pseudo-Newtonian approach is adopted to model the central black hole to avoid relativistic calculations. For that, Paczy{\'n}sky-Wiita potential \cite{Paczynsky1980, Abramowicz2009} is used, $$\Phi_\text{bh}(x,y,z) = - \frac{G M_\text{bh}}{\sqrt{x^2 + y^2 + z^2} - r_\text{s}},$$ where $M_\text{bh}$ is the black hole mass, $r_\text{s} = \frac{2 G M_\text{bh}}{c^2}$ is the Schwarzschild radius, and $c$ is the speed of light.
	
\item Bulge: This model considers a massive, dense stellar bulge rather than a central supermassive black hole (SMBH) to exclude all relativistic effects. For this spherical bulge, the Plummer potential \cite{Plummer1911} is used, $$\Phi_\text{B}(x,y,z) = - \frac{G M_\text{B}}{\sqrt{x^2 + y^2 + z^2 + c_\text{B}^2}},$$ where $M_\text{B}$ is the bulge mass and $c_\text{B}$ is the scale length. 

\item Primary Bar: A strong bar profile (cuspy type) is considered for the primary bar component. For that an anharmonic mass-model potential \cite{Mondal2021} is chosen, $$\Phi_\text{bp}(x,y,z) = - \frac{G M_\text{bp}}{\sqrt{x^2 + a_\text{p}^2 y^2 + z^2 + c_\text{bp}^2}},$$ where $M_\text{bp}$ is the primary bar mass, $a_\text{p}$ is the flattening parameter, and $c_\text{bp}$ is the scale length.

\item Secondary Bar: A weak bar profile (flat type) is considered for the secondary bar component. For that, the Zotos bar potential \cite{Jung2015} is chosen, $$\Phi_\text{bs}(x,y,z) = \frac{G M_\text{bs}}{2 a_\text{s}} \; \ln \left(\frac{y - a_\text{s} + \sqrt{{(y - a_\text{s})}^2 + x^2 + z^2 + c_\text{bs}^2}}{y + a_\text{s} + \sqrt{{(y + a_\text{s})}^2 + x^2 + z^2 + c_\text{bs}^2}} \right),$$ where $M_\text{bs}$ is the secondary bar mass, $a_\text{s}$ is the length of the semi-major axis, and $c_\text{bs}$ is the scale length.

\item Disc: For the flattened and axisymmetric disc component, the Miyamoto and Nagai potential \cite{Miyamoto1975} is used, $$\Phi_\text{d}(x,y,z) = - \frac{G M_\text{d}}{\sqrt{x^2 + y^2 + (k + \sqrt{h^2 + z^2})^2}},$$ where $M_\text{d}$ is the disc mass and $k$, $h$ are the horizontal and vertical scale lengths respectively.

\item Dark matter halo: For the dark matter halo component, the Navarro-Frenk-White (NFW) potential \cite{Navarro1996} is used, which resembles a cuspy profile (see Fig. \ref{fig:5.2a}), $$\Phi_\text{h}(x,y,z) = - \frac{G M_\text{vir}}{\ln (1 + c_\text{p}) - \frac{c_\text{p}}{1 + c_\text{p}}} \frac{\ln (1 + \frac{r}{c_\text{h}})}{r},$$ where $r^2 = x^2 + y^2 + z^2$, $M_\text{vir}$ is the virial mass of the dark matter halo, $c_\text{h}$ is the scale length and $c_\text{p}$ is the concentration parameter.
\end{itemize}

\subsection{Parameter values}
\label{sec:6.2.2}
\noindent Without losing any generality, let us consider $G$ = 1 and adopt the following system of units \cite{Zotos2020c}: unit of length - $1$ kpc, unit of mass - $2.22508 \times 10^{11} M_\odot$, unit of time - $10^6$ year, unit of velocity - $978.564$ km $\text{s}^{-1}$, unit of angular momentum per unit mass - $978.564$ km $\text{s}^{-1}$ $\text{kpc}^{-1}$, unit of energy per unit mass - $9.576 \times 10^5 \; \text{km}^2 \; \text{s}^{-2}$. The values of all physical parameters \cite{Jung2016, Zotos2020c} are listed in Table \ref{tab:6.1}.
\begin{table}
\centering	
\begin{tabular}{|c|c||c|c|}
\hline
Parameter     & Value    & Parameter         & Value\\
\hline
\hline
$M_\text{bh}$ & 0.0449   & $c_\text{bs}$     & 1\\
$c$           & 306.5717 & $\Omega_\text{b}$ & 0.046\\
$M_\text{B}$  & 0.0418   & $M_\text{d}$      & 0.7314\\
$c_\text{B}$  & 0.25     & $k$               & 3\\
$M_\text{bp}$ & 0.3657   & $h$               & 0.175\\
$a_\text{p}$  & 2        & $M_\text{vir}$    & 2.0898\\
$c_\text{bp}$ & 1        & $c_\text{p}$      & 15\\
$M_\text{bs}$ & 0.26     & $c_\text{h}$      & 20\\
$a_\text{s}$  & 8        &                   & \\
\hline
\end{tabular}
\caption{Physical parameter values.}
\label{tab:6.1}
\end{table}

\begin{figure}
\centering
\includegraphics[width=0.5\textwidth]{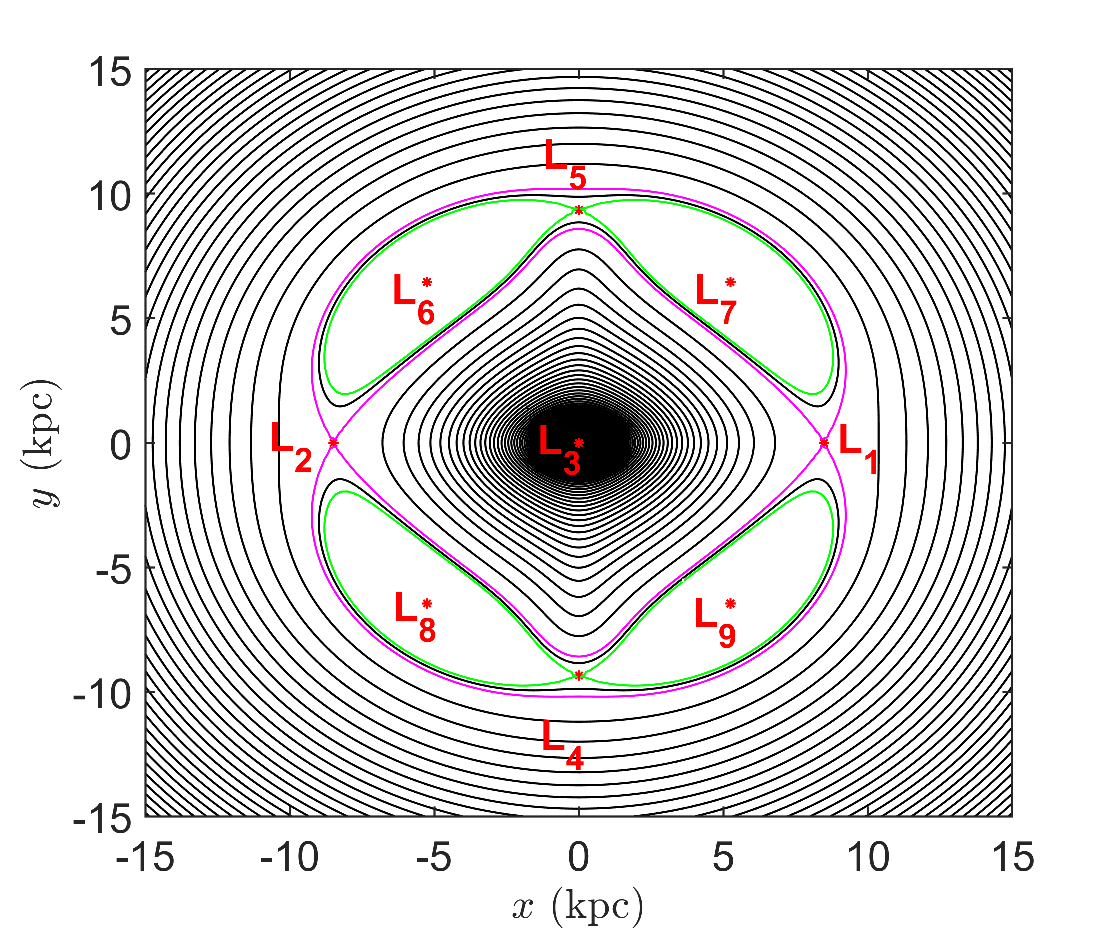}
\caption{The isoline contours of $\Phi_\text{eff}(x,y,z)$ in the $x - y$ plane for $z = 0$, where the locations of the nine Lagrangian points are marked in red. The magenta and green contours correspond to the energy values of the saddle points $L_1$ and $L_4$ respectively.}
\label{fig:6.1}
\end{figure}

\noindent This double-barred system has nine Lagrangian points (solutions of the system of Eqs. (\ref{eq:6.5})), namely $L_{i = 1, 2,\;..., 9}$. Locations and types of the Lagrangian points are given in Fig. \ref{fig:6.1} and Table \ref{tab:6.2}. Also, the energy values of all the Lagrangian points are given in Table \ref{tab:6.3}.

\begin{table}
\centering
\begin{tabular}{|c|c|}
\hline
Lagrangian Point                 & Type\\
\hline
\hline
$L_1 \equiv (8.4867,0,0)$        & Index-1 Saddle\\
$L_2 \equiv (-8.4867,0,0)$       & Index-1 Saddle\\
$L_3 \equiv (0,0,0)$             & Centre\\
$L_4 \equiv (0,-9.3452,0)$       & Index-1 Saddle\\
$L_5 \equiv (0,9.3452,0)$        & Index-1 Saddle\\
$L_6 \equiv (-5.2538,6.4509,0)$  & Index-2 Saddle\\
$L_7 \equiv (5.2538,6.4509,0)$   & Index-2 Saddle\\
$L_8 \equiv (-5.2538,-6.4509,0)$ & Index-2 Saddle\\
$L_9 \equiv (5.2538,-6.4509,0)$  & Index-2 Saddle\\
\hline
\end{tabular}
\caption{Lagrangian point locations and their types.}
\label{tab:6.2}
\end{table}

\noindent  
\begin{table}
\centering
\begin{tabular}{|c|c|}
\hline
Lagrangian Point        & Energy Value\\
\hline
\hline
$L_1 = L_2$             & $E_{L_1} / E_{L_2} = -0.2843$\\
$L_3$                   & $E_{L_3} = -\infty$\\
$L_4 = L_5$             & $E_{L_4} / E_{L_5} = -0.2815$\\
$L_6 = L_7 = L_8 = L_9$ & $E_{L_6} / E_{L_7} / E_{L_8} / E_{L_9} = -0.2720$\\  
\hline
\end{tabular}
\caption{Energy values at the Lagrangian points.}
\label{tab:6.3}
\end{table}

\noindent The nature of orbits across different energy ranges is as follows:
\begin{enumerate}[label=(\roman*)]
\item $E_{L_3} \leq E < E_{L_1}$: In this energy range, stellar orbits are enclosed within the barred region.
\item $E = E_{L_1}$: This is the threshold energy for orbital escapes through the primary bar ends.    
\item $E_{L_1} < E < E_{L_4}$: Stellar orbits may escape through the symmetrical exit channels near $L_1$ and $L_2$ in this energy range.
\item $E = E_{L_4}$: This is the threshold energy for orbital escapes through the secondary bar ends.
\item $E > E_{L_4}$: Stellar orbits may escape through the symmetrical exit channels near $L_4$ and $L_5$ in this energy range.
\end{enumerate}

\subsection{Primary versus double bar}
\label{sec:6.2.3}
\begin{itemize}[leftmargin=*]
\item Rotation curve: The rotation curve ($V_{\text{rot}} = \sqrt{R \frac{\partial \Phi_\text{t}}{\partial R}}$ versus $R =\sqrt{x^2 + y^2}$ for $z = 0$) is given in Fig. \ref{fig:6.2}. From this, it is evident that the secondary bar enhanced nuclear activities inside the barred region, which resulted in an enhancement in rotational velocities compared to its primary bar-only counterpart.
\begin{figure}
\centering
\includegraphics[width=0.5\textwidth]{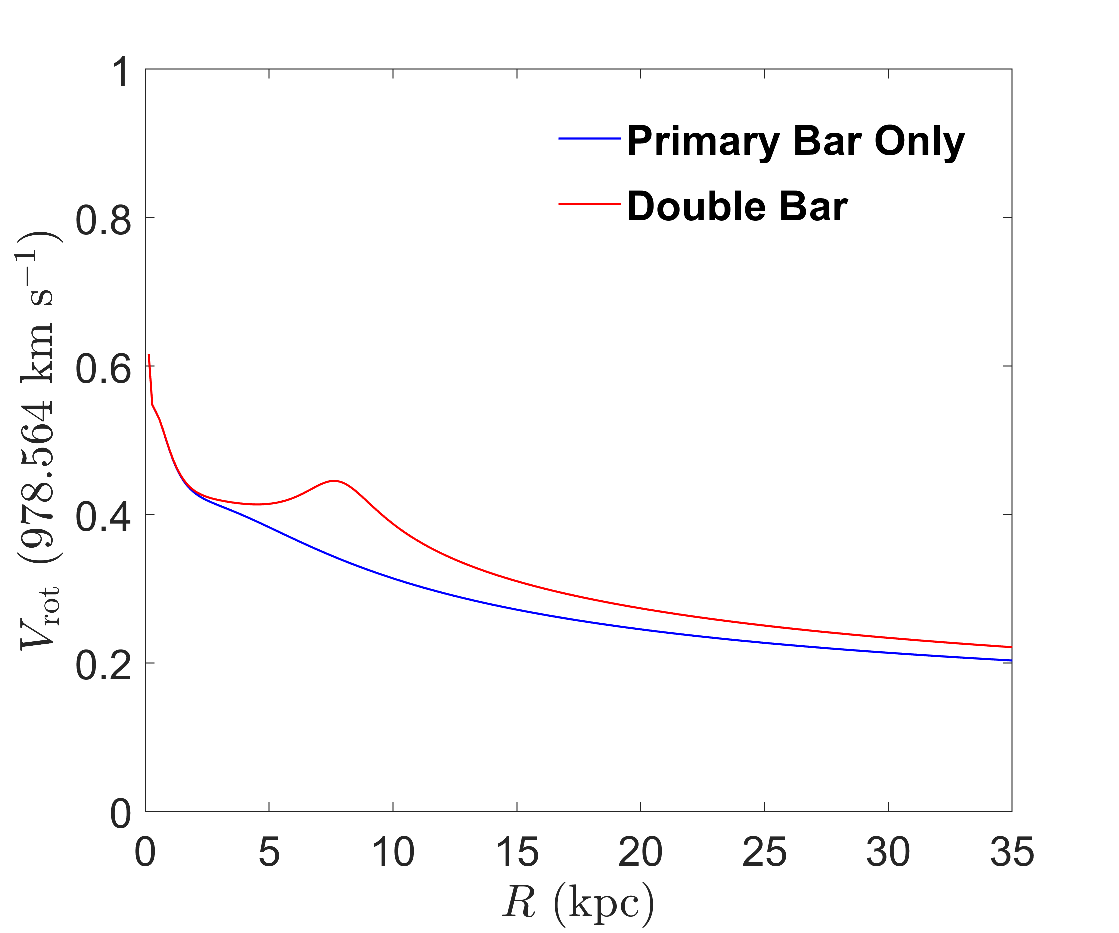}
\caption{Rotation curve.}
\label{fig:6.2}
\end{figure}

\item Force distribution: The distribution of the radial ($F_R = \frac{\partial \Phi_\text{t}}{\partial R}$) and tangential force ($F_\theta = \frac{1}{R} \frac{\partial \Phi_\text{t}}{\partial \theta}$) components of the galactic model with $R \; (= \sqrt{x^2 + y^2})$ for $z = 0$ are given in Figs. \ref{fig:6.3a} and \ref{fig:6.3b}, respectively. For $F_R(R,0)$ versus $R$ plot (Fig. \ref{fig:6.3a}), the presence or absence of the secondary bar does not make any significant difference except in the range $5 < R < 10$, where $F_R(R,0)$ is just slightly higher in the case of the double-bar as compared to its primary bar only analogue. This indicates that in the presence of secondary bars, the radial force component plays a negligible role in enhancing nuclear activities, and this is only observed within the annulus formed by the primary and secondary bars. On the other hand, the presence of a secondary bar makes a significant change in $F_\theta(R,0)$ versus $R$ plot (Fig. \ref{fig:6.3b}) specifically in the range $0 < R < 9$. This plot shows that near the galactic centre ($0 < R < 2$), the tangential force component is somewhat dominant if there is only a primary bar rather than double bars. However, beyond this region ($2 < R < 9$), the tangential force component is way more significant for double bars than the primary bar-only scenario. Now, the tangential force is responsible for driving out the escaped orbits to evolve them into subsequent patterns (similar to ring or spiral arm formations due to orbital escape from the barred region \cite{Mondal2021}). Hence, this indicates that in the gravitational model with cuspy dark matter haloes, some kind of pattern formation mechanism is feasible in the presence of the secondary bars, which is not always feasible if there are no such secondary bar components (see Chapter \ref{chap:5}).
\begin{figure}
\centering
\subfigure[Radial force ($F_R$) versus $R = \sqrt{x^2 + y^2}$ \newline for $z = 0$]{\label{fig:6.3a}\includegraphics[width=0.49\textwidth]{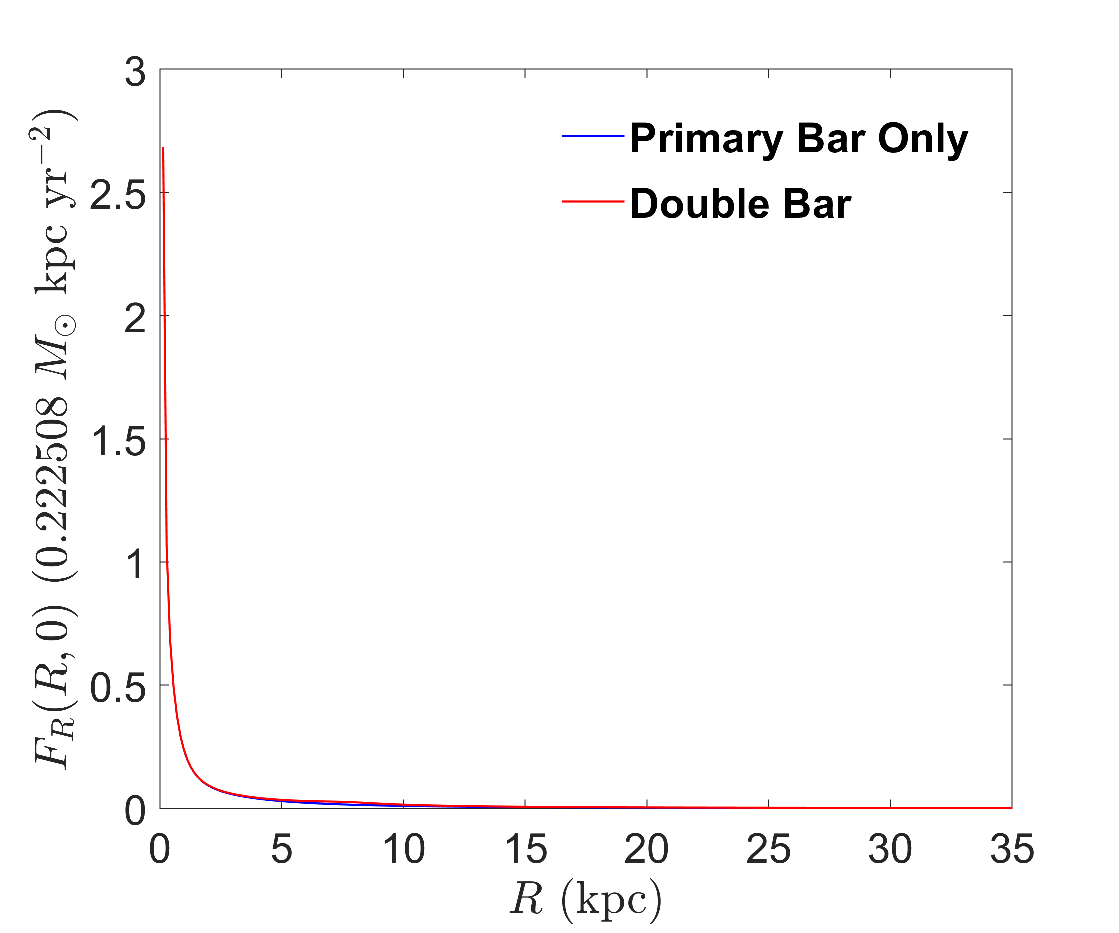}}
\subfigure[Tangential force ($F_\theta$) versus $R = \sqrt{x^2 + y^2}$ for $z = 0$]{\label{fig:6.3b}\includegraphics[width=0.49\textwidth]{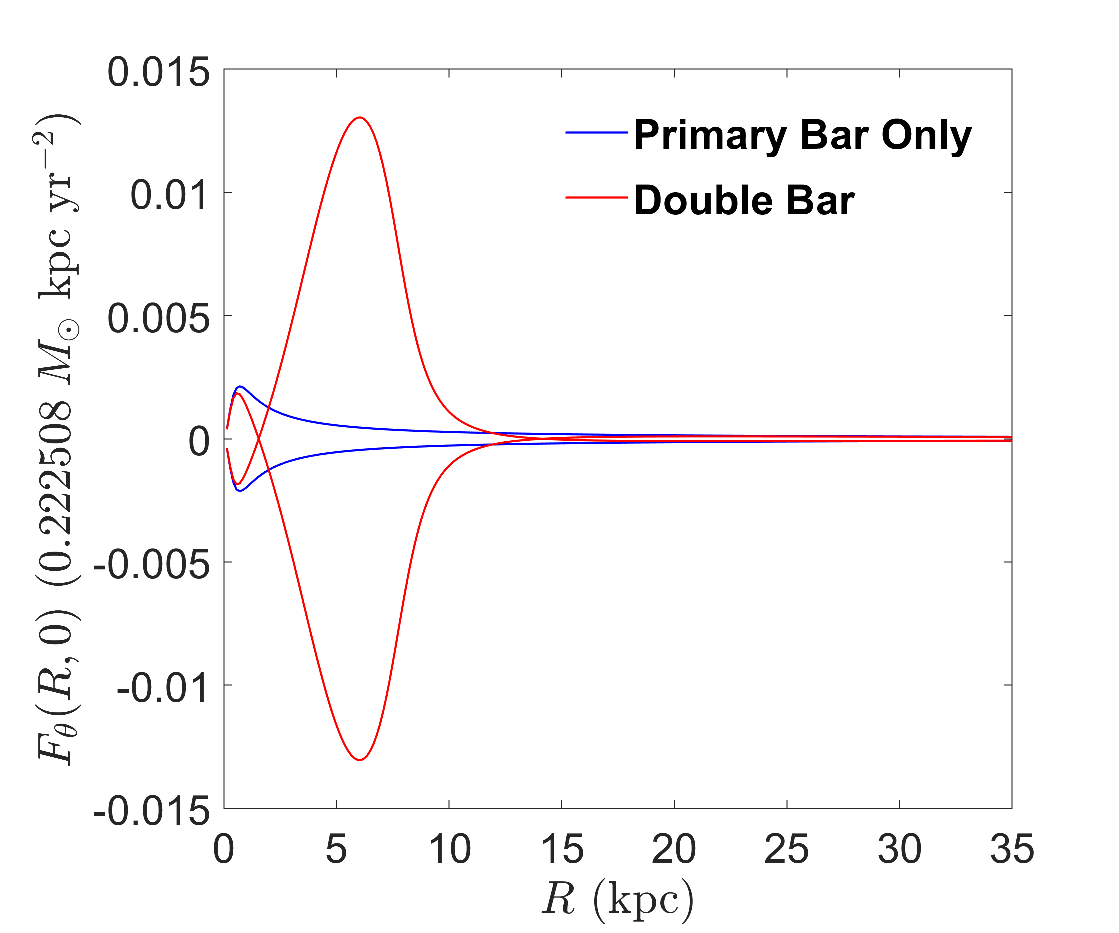}}
\caption{Radial variation of the force components.}
\label{fig:6.3}
\end{figure}
\end{itemize}

\section{Phase Space Analysis}
\label{sec:6.4}
Let us set $z = 0 = p_z$ to study the stellar orbital and escape dynamics inside the barred region along the plane of the bar. Since the effective potential term of this model is symmetric about the $y$ axis and $E_{L_1} = E_{L_2}$, $E_{L_4} = E_{L_5}$. So, studying the orbital motions near only one end of the primary and secondary bars is sufficient. For this reason, the study is only limited to the Lagrangian points $L_1$ and $L_4$. For the energy range $E \geq E_{L_1}$, stellar escape is possible through the symmetrical exit channel near the primary bar's two ends. Similarly, for the energy range $E \geq E_{L_4}$, stellar escape is possible through the symmetrical exit channel near the two ends of the secondary bar. In this model $E_{L_1} < E_{L_4}$, thus $E \geq E_{L_1}$ is sufficient to analyse the entire bar-driven escape mechanism. During the analyses, the energy parameter ($E$) is normalised with respect to $E_{L_1}$ to make it a dimensionless quantity ($C$) \cite{Mondal2021},
\begin{equation}
\label{eq:6.6}
C = \frac{E_{L_1} - E}{E_{L_1}} \; (\because E_{L_1} = E_{L_2}).
\end{equation} 
Thus, depending on the initial condition, stellar orbits may escape through the primary bar ends for $C > 0$. For visualising orbital maps and Poincaré surface section maps in the $x - y$ and $x - p_x$ planes following energy levels, $C = 0.01$ and $C = 0.1$ are used. In these maps, all initial conditions are considered for the region: $x_{0}^2 + y_{0}^2 \leq r_{L_1}^2$, where $(x_0, y_0)$ denotes an initial condition in the $x - y$ plane and $r_{L_1}$ is the radial length of $L_1$. During these analyses, the $\tt{ode45}$ package of $\tt{MATLAB}$ programming is used to solve the system of differential Eqs. (\ref{eq:6.4}), i.e., Hamiltonian equations, for a given choice of initial condition. In the process, a short time step: $\Delta t = 10^{-2}$ time units ($1$ time unit is equivalent to $10^6$ year) is used and the trajectories of stellar orbits are tracked up to $10^4$ time units. This total orbital integration time is equivalent to the age of stellar bars, typically around $10^{10}$ years \cite{Sharma2019}. Moreover, the chaotic nature of the stellar orbits is quantified with the help of the chaos detector SALI \cite{Skokos2001}, which is given in Eq. (\ref{eq:2.20}).

\subsection{Orbital maps}
\label{sec:6.4.1}
\begin{figure}
\centering
\subfigure[$C = 0.01$]{\label{fig:6.4a}\includegraphics[height=0.4\columnwidth,width=0.49\columnwidth]{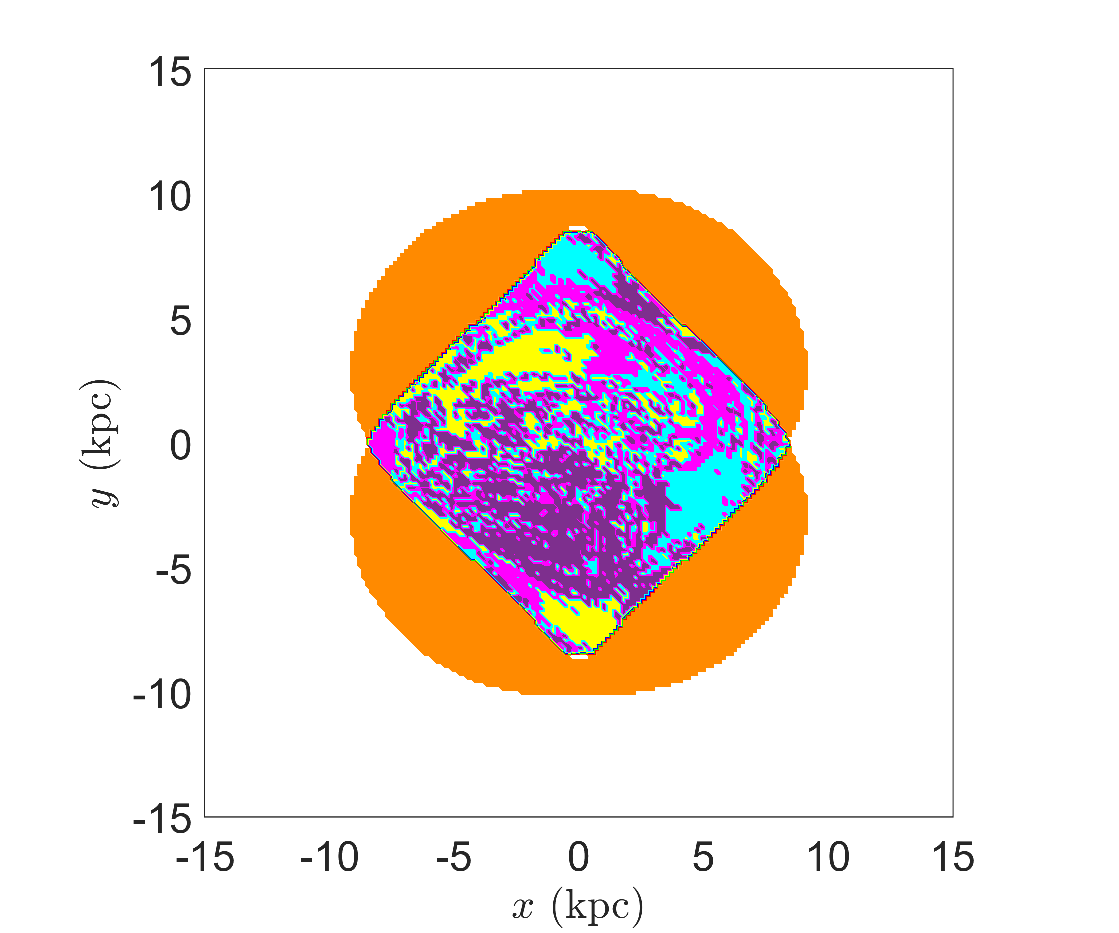}}
\subfigure[$C = 0.1$]{\label{fig:6.4b}\includegraphics[height=0.4\columnwidth,width=0.49\columnwidth]{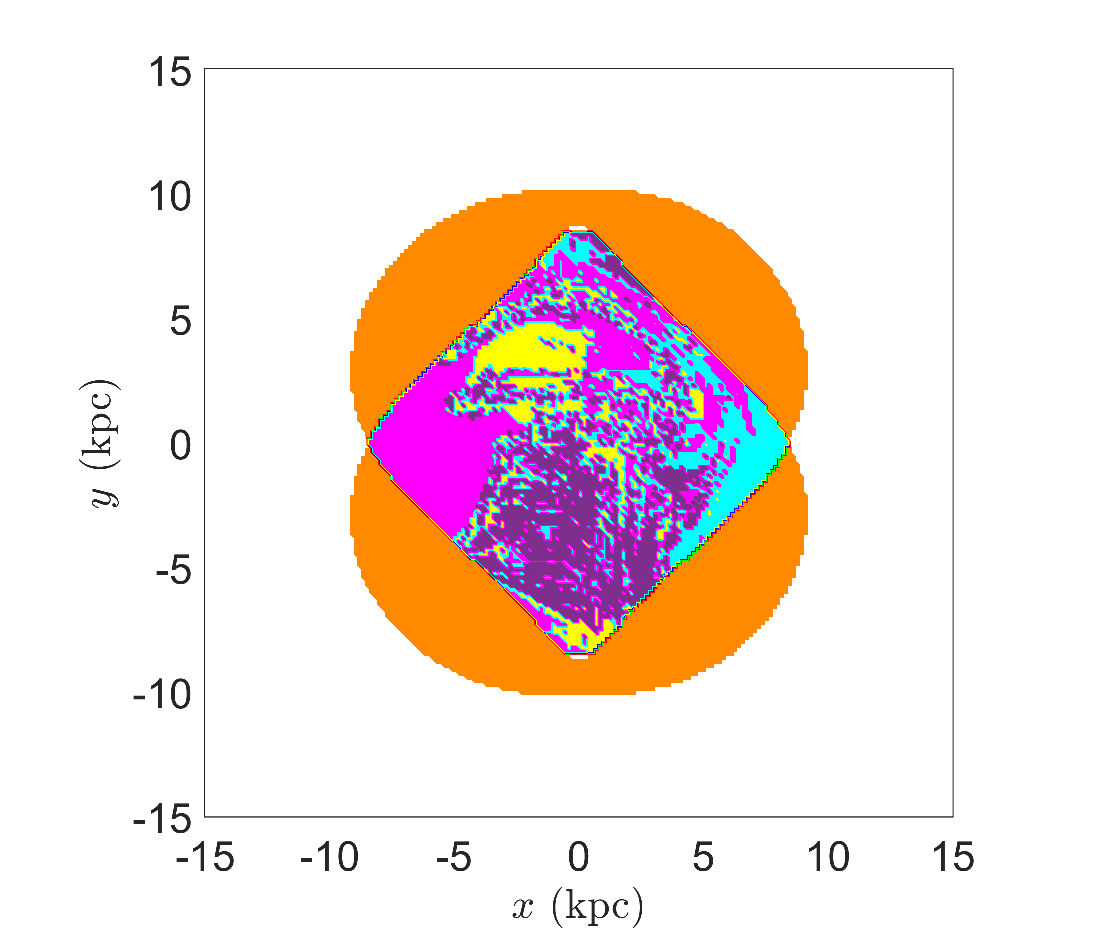}}
\caption{Orbit classification maps according to their final fate: regular non-escaping orbits (blue), trapped sticky orbits (green), trapped chaotic orbits (red), orbits escaped through $L_1$ (violet), orbits escaped through $L_2$ (magenta), orbits escaped through $L_4$ (yellow), orbits escaped through $L_5$ (cyan), energetically forbidden region (orange).}
\label{fig:6.4}
\end{figure}

Colour-coded diagrams in the $x - y$ plane are plotted in Fig. \ref{fig:6.4} to characterise the stellar orbits starting from the central barred region for escape energy values: $C = 0.01$ and $0.1$. In these maps, a $256\times256$ grid of initial conditions in the $x - y$ plane has been constructed, where any initial condition has the following form: $(x_0,y_0,p_{x_0} = 0,p_{y_0})$, where $p_{y_0}$ value is determined by Eq. (\ref{eq:6.3}). Now, a stellar orbit starting from any of these initial conditions on that grid can be classified into any of the following categories: (i) regular non-escaping orbits: identified when SALI $> 10^{-4}$ $\rightarrow$ marked in blue, (ii) trapped sticky orbits: identified when $10^{-8} \leq$ SALI $\leq 10^{-4}$ $\rightarrow$ marked in green, (iii) trapped chaotic orbits: identified when SALI $< 10^{-8}$ $\rightarrow$ marked in red, (iv) orbits which escape through $L_1$ $\rightarrow$ marked in violet, (v) orbits which escape through $L_2$ $\rightarrow$ marked in magenta, (vi) orbits which escape through $L_4$ $\rightarrow$ marked in yellow, and (vii) orbits which escape through $L_5$ $\rightarrow$ marked in cyan. The classification categories (i) - (iii) correspond to bounded motions, and (iv) - (vii) correspond to escaping motions. Figs. \ref{fig:6.4a} and \ref{fig:6.4b} show that the central barred region is almost entirely of escaping motions. Very minor signatures of bounded motions exist near the inner potential boundary. Also, as the escape energy value increased from $C = 0.01$ to $0.1$, the region of escaping motion through the primary bar ends gradually increased. 

\begin{figure}
\centering
\subfigure[$C = 0.01$]{\label{fig:6.5a}\includegraphics[width=0.49\columnwidth]{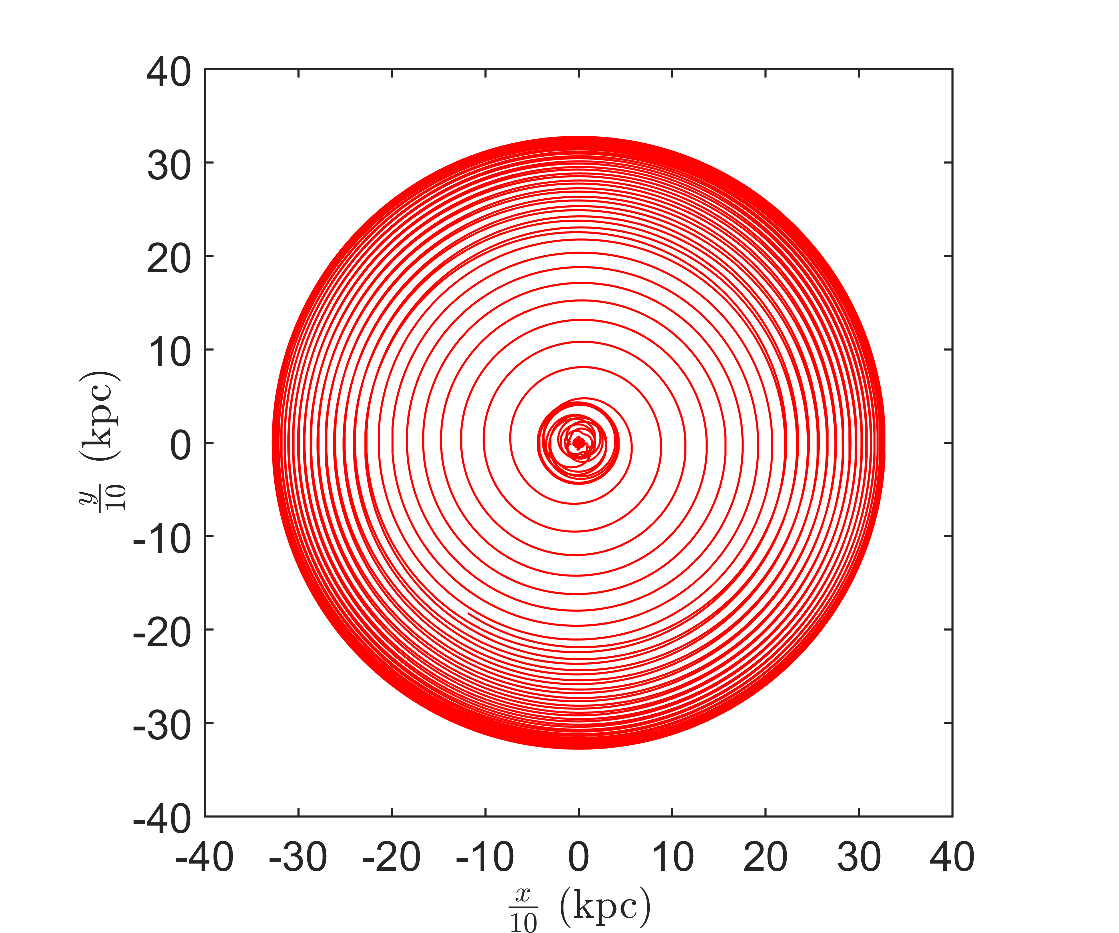}}
\subfigure[$C = 0.1$]{\label{fig:6.5b}\includegraphics[width=0.49\columnwidth]{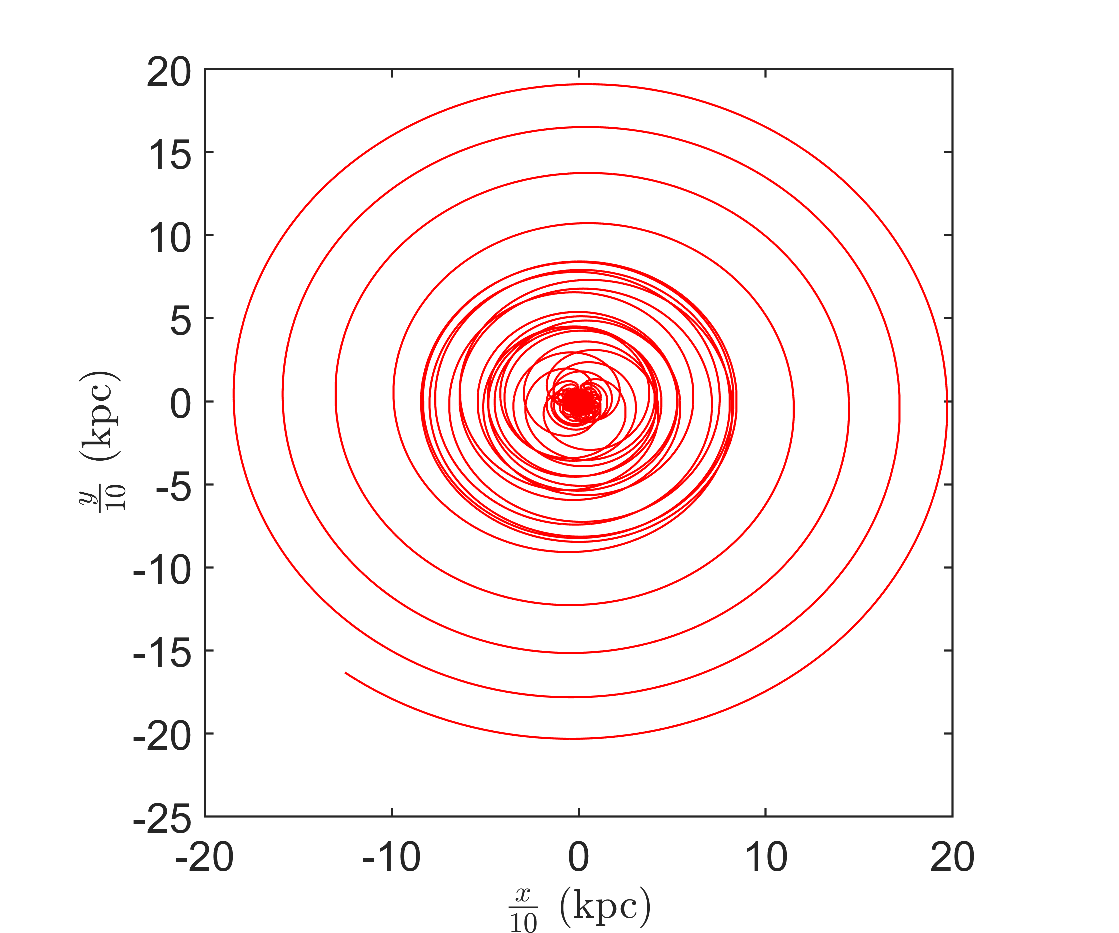}}
\caption{Orbits in the $x - y$ plane for $(x_0,y_0,p_{x_0},p_{y_0})$ $\equiv (5,0,0.1533,p_{y_0})$, where $p_{y_0}$ value is evaluated from Eq. (\ref{eq:6.3}); (a)-(b) escaping chaotic orbit.}
\label{fig:6.5}
\end{figure}

\noindent Furthermore, orbital maps are plotted in Fig. \ref{fig:6.5} for escape energy values $C = 0.01$ and $0.1$ with a test initial condition $(x_0,y_0,p_{x_0},p_{y_0}) \equiv (5,0,0.1533,p_{y_0})$, where the $p_{y_0}$ value is determined by Eq. (\ref{eq:6.3}). This $(5,0,0.1533,p_{y_0})$ is picked out to study the escape dynamics in the neighbourhood of $L_1$ \cite{Mondal2021}. This initial condition is not unique in any way. Any other point in the vicinity will show similar orbital trajectories. This initial condition is the same as the initial condition used in Chapters \ref{chap:4} and \ref{chap:5} when converted into appropriate scaling units. These figures show that stellar orbits are escaping and chaotic in nature for both escape energy values..

\subsection{Poincaré maps}
\label{sec:6.4.2}
\begin{figure}
\centering
\subfigure[$C = 0.01$]{\label{fig:6.6a}\includegraphics[height=0.4\columnwidth,width=0.49\columnwidth]{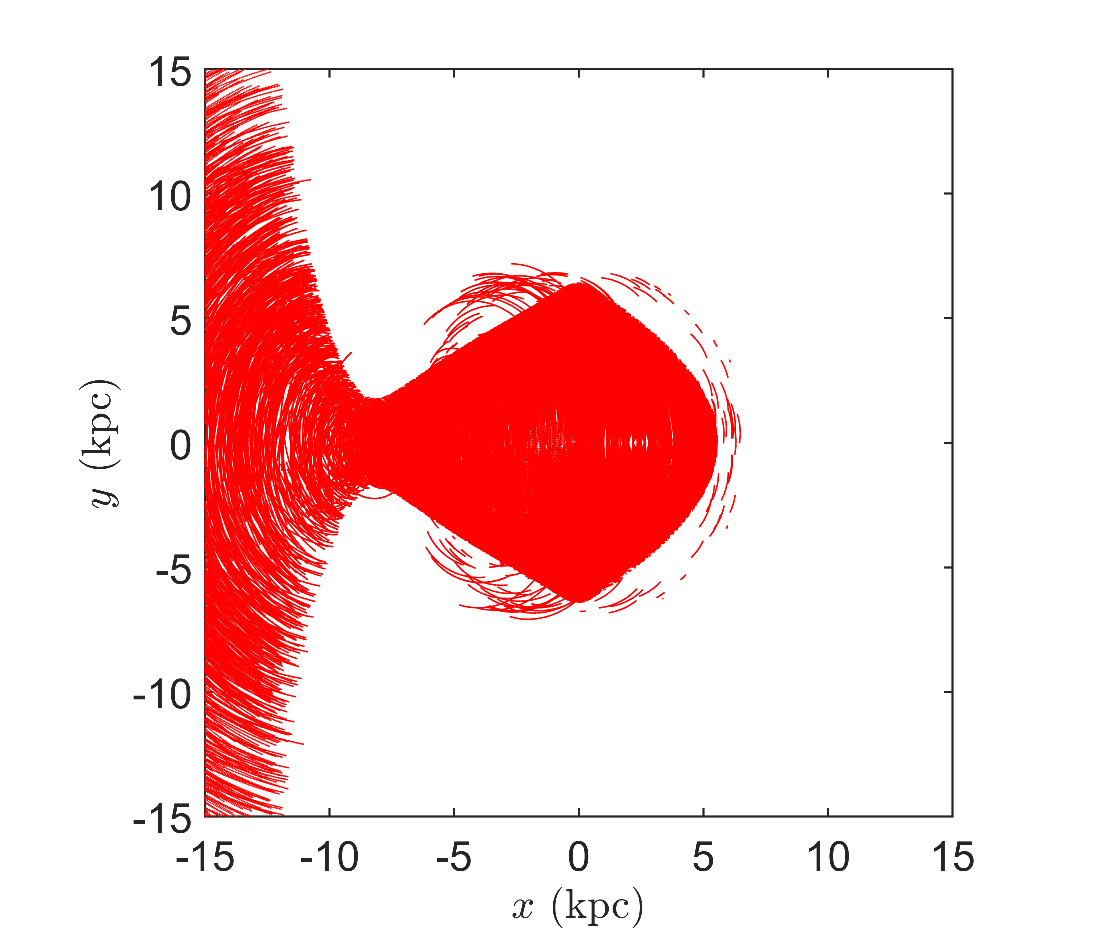}}
\subfigure[$C = 0.1$]{\label{fig:6.6b}\includegraphics[height=0.4\columnwidth,width=0.49\columnwidth]{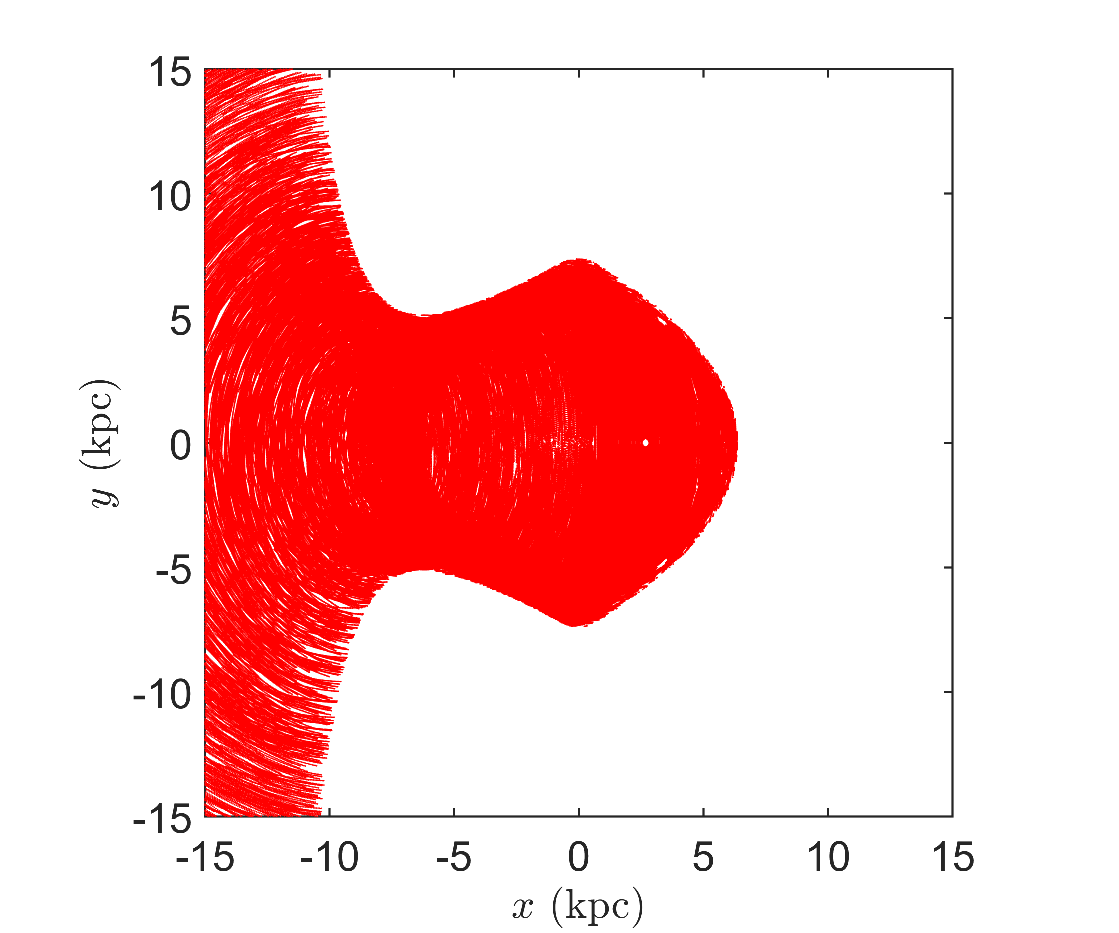}}
\caption{Poincaré maps in the $x - y$ plane with surface sections: $p_x = 0$ and $p_y \leq 0$.}
\label{fig:6.6}
\end{figure}

\noindent Poincaré surface section maps in the $x - y$ and $x - p_x$ planes help visualise the bar-driven escaping motion in phase space. Due to $z = 0 = p_z$, the phase space is reduced from four-dimensional to six-dimensional. As a result, surface cross-sections for Poincaré maps will be two-dimensional planes. In all the plotted Poincaré maps, only initial conditions lying within the energetically allowed region: $\Phi_\text{eff}(x,y,z) < E_{L_1}$ are used. For Poincaré maps in the $x - y$ plane (Fig. \ref{fig:6.6}), a $39\times39$ grid of initial conditions has been set up with step sizes: $\Delta x = 0.5$ kpc and $\Delta y = 0.5$ kpc. This grid considers initial conditions only within the Lagrange radius ($r_{L_1}$). The initial value of $p_x$, i.e., $p_{x_0}$ is considered to be $p_{x_0} = 0$. In addition, the initial value of $p_y$, i.e., $p_{y_0}$ is taken into account as $p_{y_0} > 0$ and further evaluated from Eq. (\ref{eq:6.3}). Thus, the chosen surface cross sections for these Poincaré maps are $p_x = 0$ and $p_y \le 0$ \cite{Ernst2014}.

\begin{figure}
\centering
\subfigure[$C = 0.01$]{\label{fig:6.7a}\includegraphics[height=0.4\columnwidth,width=0.49\columnwidth]{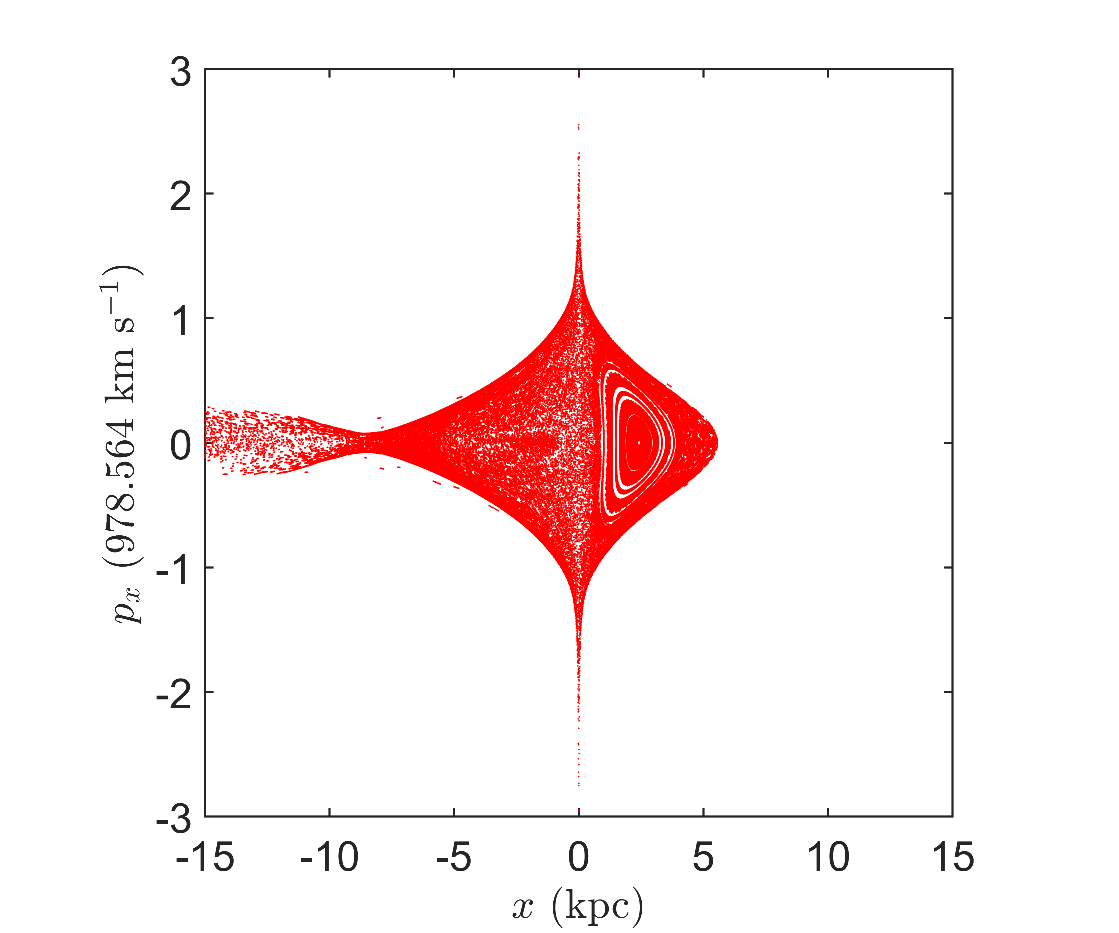}}
\subfigure[$C = 0.1$]{\label{fig:6.7b}\includegraphics[height=0.4\columnwidth,width=0.49\columnwidth]{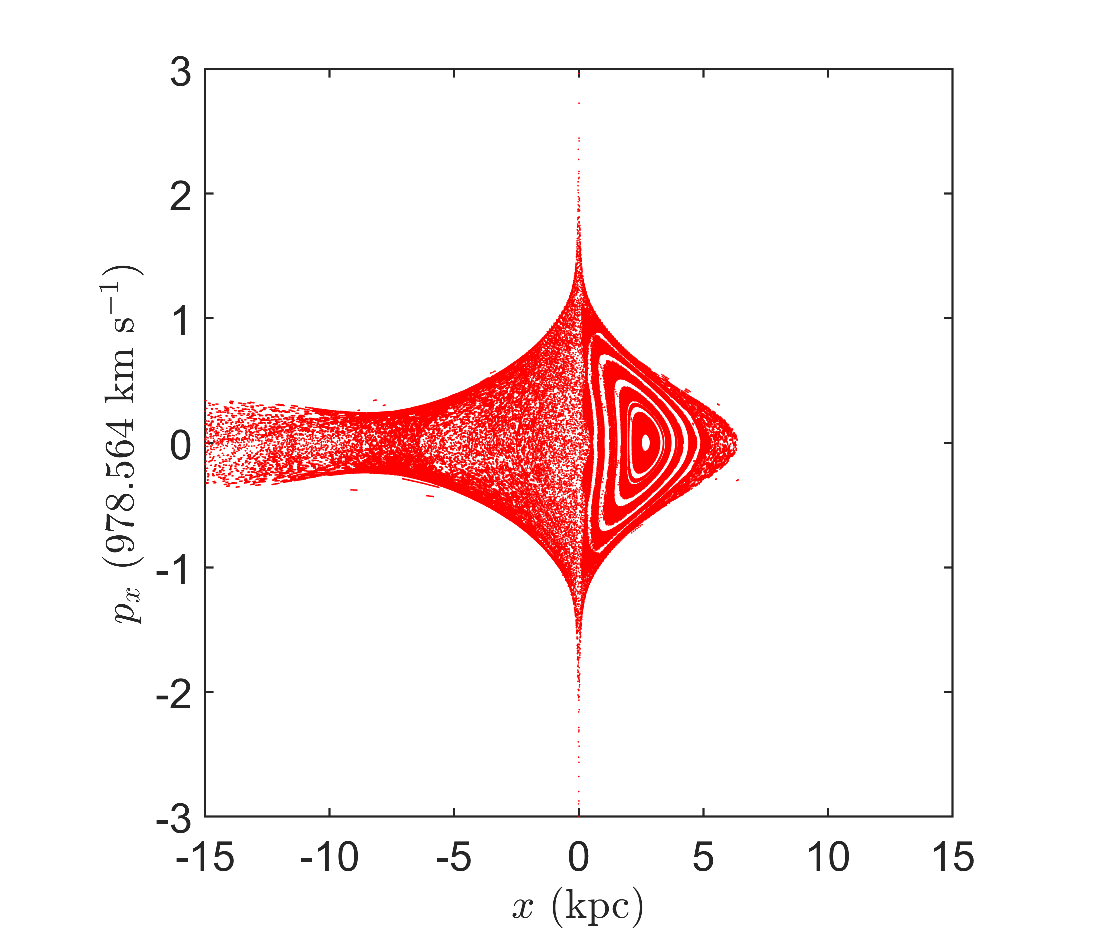}}
\caption{Poincaré maps in the $x - p_x$ plane with surface sections: $y = 0$ and $p_y \leq 0$.}
\label{fig:6.7}
\end{figure}

Similarly, for Poincaré maps in the $x - p_x$ plane (Fig. \ref{fig:6.7}), a $39\times31$ grid of initial conditions has been set up with step sizes: $\Delta x = 0.5$ kpc and $\Delta p_x = 9.78564$ km $\text{s}^{-1}$. Only initial conditions inside the Lagrange radius ($r_{L_1}$) are considered from this grid. The initial value of $y$, i.e., $y_0$ is considered to be $y_0 = 0$. Moreover, the initial value of $p_y$, i.e., $p_{y_0}$ is considered as $p_{y_0} > 0$ and further evaluated from Eq. (\ref{eq:6.3}). Thus, the chosen surface cross sections for these Poincaré maps are $y = 0$ and $p_y \le 0$ \cite{Ernst2014}.

Fig. \ref{fig:6.6} shows Poincaré surface section maps in the $x - y$ plane for the escape energy values. The following trends have been observed in these figures:
\begin{enumerate}[label=(\roman*)]
\item For both escape energy values, the central barred region is entirely covered by the chaotic stellar motions, as no specific patterns (stability islands) are observed.

\item The concentration of cross-sectional points inside the corotation region of the bar and disc is quite large for a lower escape energy (i.e., $C = 0.01$) and increased further with an increment of $C$ from $0.01$ to $0.1$. This escaping trend is much higher when compared to similar figures (see Figs. \ref{fig:4.5}, \ref{fig:4.6}, \ref{fig:5.7} and \ref{fig:5.9}) shown in Chapters \ref{chap:4} and \ref{chap:5}.
\end{enumerate}

\noindent Again, Poincaré surface section maps in the $x - p_x$ plane are shown in Fig. \ref{fig:6.7} for escape energy values $C = 0.01$ and $0.1$. In these figures, the following trends have been identified:
\begin{enumerate}[label=(\roman*)]
\item Here, the central barred region is also mainly covered by chaotic stellar motions. However, some patterns are found in stability islands, which correspond to quasi-periodic stellar motions.

\item The number of cross-sectional points outside the corotation region mentioned above has increased with an increment of $C$ from $0.01$ to $0.1$. These trends are identical to those seen in similar figures (see Figs. \ref{fig:4.5}, \ref{fig:4.6}, \ref{fig:5.8} and \ref{fig:5.10}) of Chapters \ref{chap:4} and \ref{chap:5}. 
\end{enumerate}

\section{AGN and Double-barred Disc}
\label{sec:6}
The central theme of this chapter is to study the orbital and escaping motion inside the central barred region of double-barred galaxies with NFW dark matter haloes. A phase space-based study analysed this on a six-component gravitational model composed of a central black hole, bulge, primary bar, secondary bar, disc, and dark halo. The conclusions from all of these studies are as follows:
\begin{enumerate}[label=(\roman*)]
\item It is apparent from colour-coded orbit classification maps (Fig. \ref{fig:6.4}) and orbital trajectories (Fig. \ref{fig:6.5}) in the $x - y$ plane that double-barred discs inside a NFW dark matter halo always encourage bar-driven stellar escaping motion. Moreover, the amount of escape has been increased with an increment of $C$ from $0.01$ to $0.1$. Poincaré surface section maps in the $x - y$ and $x - p_x$ phase planes (Figs. \ref{fig:6.6} and \ref{fig:6.7}, respectively) also indicate similar phenomena. Overall, the amount of escape is high for both escape energy values compared with similar cases shown in Chapters \ref{chap:4} and \ref{chap:5}.

\item In Chapter \ref{chap:5}, it is already discussed that forming spiral arms resulting from bar-driven escape mechanisms in single-barred galaxies with cuspy dark matter haloes is only possible when their central regions are highly energetic. Active galactic nuclei (AGN) is one such possibility where NFW profiles could fit dark matter halo models better, and bar-driven structure formation is evident in the form of the spiral arms. However, another important question comes into the picture: What are the possible energy sources of AGNs other than the accretion effect of the central black hole? Some scientific literature proposes that secondary stellar bars may be one of the possible sources of power generation in active galaxies' central engines \cite{Shlosman1989, Shlosman1990}. The role of secondary bars in the context of active galactic nuclei has yet to be fully understood and is the subject of ongoing research and investigation \cite{Du2015, Wozniak2015}. The work presented in this chapter supports the studies of \citeauthor{Shlosman1989}, \citeyear{Shlosman1989} \cite{Shlosman1989} and \citeauthor{Shlosman1990}, \citeyear{Shlosman1990} \cite{Shlosman1990}, i.e., it may be possible for secondary nuclear bars to exert gravitational torques that enhance the transport of gas and dust towards the central supermassive black hole, potentially fueling the AGN's activity. As a result, the central region becomes highly energetic, and subsequently, the formation of spiral arms in the form of bar-driven escaping motion becomes feasible.

Observational studies by \citeauthor{Laine2002}, \citeyear{Laine2002} \cite{Laine2002} used a small sample of 112 Seyfert and non-Seyfert galaxies, which seems at first glance to indicate a correlation between double bars and AGN: the fraction of galaxies in their Seyfert sample with two bars is 21\%, versus only 13\% for the non-Seyfert galaxies. Since the sample size is relatively small, the next plan is to build an image identification-based artificial neural network (ANN) model with a much larger sample of active galaxy image data shortly. This is important to validate the correlation between double bars and AGN effectively.
\end{enumerate}

From all the above discussions, the final conclusions are summarised as follows:
\begin{enumerate}[label=(\roman*)]
\item If the dark halo profile of double-barred galaxies is akin to a NFW profile, then the central baryonic feedbacks are sufficient to generate bar-driven patterns in the form of the spiral arms. Active galaxies are one possible example of such systems.

\item Double-barred discs may act as central engines of active galaxies.
\end{enumerate}

%% file: Chapter_7.tex
\chapter{Overall Conclusion and Future Scopes}
\label{chap:7}

\section{Overall Conclusion} 
\label{chap:7.1}
\indent The main focus of this thesis work is to study the effect of chaos or instabilities produced in the central region of barred galaxies. These instabilities result from baryonic feedback and are propagated in the ambient medium like density waves. This central chaos has many prospects for the formation and evolution of galaxies. Firstly, they are responsible for the gravitational collapse of dense molecular clouds that lead to the formation of field stars or star clusters. Secondly, the orbits of these stars escaped from the central disc region through the bar ends and formed galactic structures like spiral arms, inner disc rings, etc. All the research works of this thesis are centred around these above two cases. Let us begin a brief discussion of the research problems and then gradually move towards the final outcomes of the research work.

\indent My first research problem was published with the title \textit{Star formation under explosion mechanism in a magnetized medium} \cite{Mondal2019}. In this work, we investigate the role of the magnetic field in the evolution of dense molecular clouds that formed due to shock waves generated via baryonic feedback within the galactic disc. We have developed a magnetohydrodynamics (MHD) model of star formation due to explosive phenomena (due to baryonic feedback) in the central region of our Milky Way. During the propagation, these explosion-driven shock waves cool and compress the ambient medium in thin, dense shells. Subsequently, these dense shells fragmented into molecular clouds due to gravitational instability. The gravitational collapse has been prevented beyond a threshold mass (known as Jeans mass), and the stars form. We primarily considered the central region of our Milky Way under the influence of a magnetic field for modelling such explosive phenomena and deriving the Jeans mass for gravitational collapsing molecular clouds. It is found that under an inverse variation of temperature with density, a wide range of fragments can be formed. The mass range is enhanced in the presence of a constant and a density-dependent magnetic field. Under suitable physical conditions, a burst of star formation is possible. Overall, a weak magnetic field ($B_0 \sim 1 \mu$G) favours the formation of field stars, which are only feasible to form near the galactic centre ($r \sim 20$ pc). On the other hand, a strong magnetic field ($B_0 \sim 10 \mu$G) favours the formation of massive stars or star clusters and the effect is enhanced at a more considerable distance ($r \sim 1$ kpc). It is also found that rotation of such fragmented clouds of the order of a few km s$^{-1}$ kpc$^{-1}$ might lead to a stable structure.

\indent Now, the density waves generated from instabilities in the central region play a pivotal role in forming stellar bars. This robust galactic structure has an influential effect on the galaxy's evolutionary process. Thus, my subsequent research problem is based on stellar bars and has been published in the following articles with the titles \textit{Role of galactic bars in the formation of spiral arms: a study through orbital and escape dynamics—I} \cite{Mondal2021} and \textit{Fate of escaping orbits in barred galaxies} \cite{Mondal2020}. These works primarily look into the underlying relation between the fate of bar-driven escaping patterns and bar strength. Now, such escapes might lead to structures like spiral arms forming. To analyse this, we have developed a three-dimensional gravitational model for barred galaxies and studied the orbital and escape dynamics of the stars inside the central barred region. This model has four components: bulge, bar, disc and dark matter halo. Moreover, the model has been analysed for two different types of bar profiles: strong and weak. Now, the phenomena of stellar escaping motion through the bar end resembles the phenomena of chaotic scattering observed in open Hamiltonian systems. For this reason, this study has been carried out for an open Hamiltonian system, and thorough numerical investigations have been done to explore stars' regular and chaotic motions in the phase space. The escape mechanism is only near saddle points corresponding to bard ends. Orbital maps in the $x - y$ plane and Poincaré surface section maps in the $x - y$ and $x - p_x$ ($p_x$ is the momentum along $x$ - direction) planes are drawn to capture this escaping motion through the bar ends. Classifications of an orbit as regular or chaotic are done by evaluating its maximal Lyapunov exponent (MLE) value. Also, a detailed analysis was done on the variation of the MLE value in the vicinity of escape saddles with the bar's mass, length, and nature. From the study, it has been found that under suitable physical conditions, the central chaos plays a pivotal role behind the formation of grand design (e.g., giant spiral galaxies) or less prominent spiral patterns (e.g., S0 galaxies) for stronger bars and ring structures (e.g., ring galaxies) for weaker bars.

\indent Again, the nature of dark haloes has an ample influence on the galactic dynamics. That is why these dark haloes too affect the bar-driven structure formation mechanism. This has been studied in detail in my next research problem titled as \textit{Effect of dark matter haloes on the orbital and escape dynamics of barred galaxies}. This work examines the effect of dark matter haloes on the fate of bar-driven escaping patterns in barred galaxies. In this study, a three-dimensional gravitational model with a strong bar component has been considered and examined separately for the following dark halo profiles: NFW and oblate. In both cases, a bar-driven escape mechanism, which corresponds to the bar ends, was identified near the saddle points of the phase space. This bar-driven escaping motion has been analysed via orbital and Poincaré surface section maps. Moreover, with a choice of initial condition in the vicinity of escape saddles, the variation of its MLE values concerning dark halo parameters, such as mass, size, circular velocity and nature, is studied. This is done to estimate the underlying relationships between orbital chaos and the properties of the dark halo. It has been found that if ordinary barred galaxies have a NFW dark halo, central baryonic feedbacks are insufficient to generate bar-driven escaping patterns in the form of spiral arms. They only do so for galaxies with highly energetic centres, like active galaxies. On the other hand, barred galaxies with oblate dark haloes provide a preferable common bar-driven escaping structure formation framework over NFW haloes in both giant spiral and dwarf galaxies. In this case, the presence of a high-mass black hole leads the system to form grand-design spiral arms, and the presence of a low-mass black hole leads the same system to form less-prominent spiral arms.

\indent Till now, we have seen that the NFW dark haloes of ordinary barred galaxies cannot generate spiral patterns due to insufficient central baryonic feedback. Instead, they can only do so for galaxies with extremely energetic centres. This may also be the reason for the disappearance of NFW dark haloes in the ordinary barred spirals, which \textit{N}-body simulations predicted would exist. The works of Shlosman (1989, 1990) \cite{Shlosman1989, Shlosman1990} speculate that the presence of double-barred discs may act as one of the possible sources of power generation in galaxies with extremely energetic centres like active galaxies. This issue has been addressed in my next and final work of this thesis titled as \textit{Orbital and escape dynamics in double-barred galaxies}.

So, from all these research works, the significant outcomes are summarised as follows:
\begin{enumerate}[label=(\roman*)]
\item Slowly rotating molecular clouds with a higher magnetic field lead to the formation of star clusters. 
 
\item Stellar bars with higher strength propel the chaos from the centre, leading to the formation of grand-design spiral patterns.

\item The oblate dark haloes provide a better unified evolutionary framework over the NFW dark haloes regarding bar-driven escaping structure formation in giant spiral and dwarf galaxies.

\item Galaxies with the NFW dark haloes support the formation of spiral arms via bar-driven escaping motion only if they have highly energetic centres like active galaxies. The double-barred disc may be one such possibility.
\end{enumerate}

\section{Future Scopes}
\label{chap:7.2}
\indent So far, my research interest is mainly focused on investigating the role of chaos in the formation and evolution of disc galaxies. Our main motto is to develop an overall framework to describe the formation and growth mechanisms of all the structures in disc galaxies from the viewpoint of chaotic stellar motion or simply chaos. Regarding this, mathematical modelling is done from the perspective of chaotic scattering in open Hamiltonian systems, and conclusions are drawn only from the system's phase space analysis. These all are dynamical astronomy based approaches. We will soon study these topics more extensively using machine learning based computational techniques like deep learning. 

\indent To date, machine learning techniques have been applied to various fields of astrophysics, including the study of galactic dynamics \cite{Ivezic2014, Kremer2017, Ntampaka2015, Cantat2023}. Machine learning based algorithms can identify patterns and correlations in large observational datasets and extract meaningful information from them. Also, machine learning based algorithms can be used to model the gravitational potential of a galaxy and predict its stellar motions. This can provide insight into the distribution of dark matter within a galaxy and the history of its formation. 

\indent Deep learning is a subset of machine learning that performs modelling with the help of \textit{artificial neural networks (ANNs)}. Incorporating deep learning-based models in our upcoming research works, the findings of this thesis work are crucial, as these results will be incorporated into the models to train the ANNs. This study also requires the aid of sophisticated astronomical catalogues like Gaia mission data \cite{Naik2022}. ANN is a deep learning algorithm inspired by the structure and function of biological neural neurons in the mammalian brain. They can be used to model complex systems, such as the galaxy formation process. Galaxies formation exhibits many physical processes, such as gas dynamics, star formation, feedback from supernovae and active galactic nuclei, etc. ANNs can be trained to simulate these processes and predict the properties of galaxies based on their initial conditions and evolution. For example, ANNs can be used to model the formation and evolution of galaxy clusters. By incorporating physical processes such as gas cooling, star formation, and feedback from supernovae and black holes, ANNs can predict the properties of galaxy clusters, such as their mass, size, and distribution of dark matter \cite{Ho2019}. Overall, ANN is a powerful tool that can revolutionize the understanding of galactic dynamics and help us unravel the mysteries of the galaxy formation and evolution process in the near future.

Here are some research problems that I intend to analyse in the near future with the help of machine learning techniques.
\begin{enumerate}[label=(\roman*)]
\item Investigations of the correlation between the properties of supermassive black holes and the central chaos, considering the effect of the dark matter halo on the galaxy's disc.

\item Discuss the possibilities of bar-driven escapes in irregular galaxies: Large Magellanic Cloud (LMC) and Small Magellanic Cloud (SMC). 

\item Discuss the possibilities of bar-driven escapes in dark matter less galaxies: NGC1052-DF2 and NGC1052-DF4.
\end{enumerate}